\documentclass[11pt,a4paper,british]{article}
\usepackage{lmodern}

\usepackage{jcappub}
\usepackage[utf8]{inputenc}
\usepackage[T1]{fontenc}
\usepackage[table]{xcolor}
\usepackage{microtype}
\usepackage{babel}
\usepackage{varioref}
\usepackage{float}
\usepackage{datetime2}
\usepackage{graphicx}
\usepackage{adjustbox}
\usepackage{cancel}
\usepackage[normalem]{ulem}
\usepackage{amssymb}
\usepackage{mathtools}
\usepackage{mathrsfs}
\usepackage{subcaption}
\usepackage{multirow}
\DeclareCaptionLabelFormat{bold}{\textbf{(#2)}}
\usepackage{makecell}
\usepackage{booktabs}
\usepackage{tabularx} 
\usepackage{ragged2e} 
\usepackage{caption} 
\usepackage{placeins}
\usepackage{orcidlink}
\newsavebox{\measurebox}

\makeatletter

\def\hlinewd#1{%
\noalign{\ifnum0=`}\fi\hrule \@height #1 %
\futurelet\reserved@a\@xhline}

\numberwithin{equation}{section}

\pdfoutput=1
\renewcommand\[{\begin{equation}}
\renewcommand\]{\end{equation}} 
\renewcommand*\arraystretch{1.5}

\DeclareMathOperator{\Tr}{tr}

\DeclareMathOperator{\diag}{diag}
\newcommand{\Zt}{\tilde{Z}}
\newcommand{\cs}{c_\text{s}}
\newcommand{\Ztwo}[1]{\mathcal{Z}_2^{#1}}

\newcommand{\Ztwobar}[1]{\overline{\mathcal{Z}}_2^{#1}}
\newcommand{\Ztwoubar}[1]{\underline{\mathcal{Z}}_2^{#1}}

\newcommand{\Stwo}[1]{\mathcal{S}_{2 #1}}
\newcommand{\Stwobar}[1]{\overline{\mathcal{S}}_{2 #1}}
\newcommand{\Stwoubar}[1]{\underline{\mathcal{S}}_{2 #1}}
\newcommand{\Srr}{\Stwo{ij} \dot{r}^i \dot{r}^j}
\newcommand{\nZ}{\overline{\nabla}}
\newcommand{\GZ}{\overline{\Gamma}}
\newcommand{\DZ}{\overline{\Delta}}
\newcommand{\DS}{\underline{\Delta}}
\newcommand{\T}{\mathbb{T}}
\newcommand{\TZ}{\overline{\T}}
\newcommand{\TS}{\underline{\T}}
\newcommand{\perpZ}{\overline{\perp}}
\newcommand{\perpS}{\underline{\perp}}

\newcommand{\KZ}{\overline{K}}
\newcommand{\RS}{\underline{R}}

\newcommand{\detu}{{\det\nolimits_u}}

\newcommand{\Mpl}{M_\text{P}}

\title{Causality and Stability from Acoustic Geometry}
\author[a]{Ignacy Sawicki\orcidlink{0000-0003-2476-9730},}
\author[a,b]{Georg Trenkler\orcidlink{0009-0009-5206-6865}}
\author[a]{and~Alexander~Vikman\orcidlink{0000-0003-3957-2068}}
\affiliation[a]{CEICO, FZU --- Institute of Physics of the Czech Academy of Sciences,\\
Na Slovance 1999/2, 182 00 Prague 8, Czech Republic}
\affiliation[b]{Institute of Theoretical Physics, Faculty of Mathematics and Physics, Charles University,\\ V Holešovičkách 2, 180 00 Prague 8, Czech Republic}

\emailAdd{sawicki@fzu.cz}
\emailAdd{trenkler@fzu.cz}
\emailAdd{vikman@fzu.cz}

\abstract{In scalar-tensor theories with derivative interactions, backgrounds spontaneously break local Lorentz
invariance. We study the motion of perturbations of the scalar, ``phonons'', on these anisotropic time-dependent backgrounds in curved spacetimes. The phonons propagate on null geodesics of an effective acoustic spacetime, which has its own metric and a connection featuring non-metricity with respect to the metric defined by gravity. These acoustic geodesics correspond to motion with four-acceleration in the usual spacetime. We stress the differences and duality between the phonons' canonical four-momenta and four-velocities, and point out analogies with photons in a medium. 

For an arbitrary moving observer, we covariantly define the phonon's energy, relative phase velocity, effective refraction index and mass tensor. We point out that true instabilities (ghosts, gradient) are  observer independent, being identified by the acoustic metric's signature and determinant. However, apparent instabilities, such as complex
phonon energies, can stem from an ill-posed Cauchy problem in certain observer frames.
Negative phonon energies appear for supersonic observers, not indicating true instabilities,
but leading to Cherenkov radiation. We extend this local picture to a global foliation, deriving
the condition for a spatial slice to be a Cauchy surface for a well-posed initial value problem.

The action for perturbations yields an acoustically conserved asymmetric energy-momentum tensor (EMT),
not conserved in the usual spacetime. Yet, with a timelike acoustic Killing vector, this EMT
forms a current conserved in both the acoustic and usual spacetimes, with the acoustic
Hamiltonian functional as its conserved charge. This Hamiltonian is bounded if the foliation's comoving
observer is subsonic. Otherwise, for a Killing vector timelike in both metrics, an alternative conserved charge that bounds motion exists.}

\begin{document}
\maketitle
\flushbottom

\section{Introduction}

The initial conditions of the universe and the nature of the dark sector remain open problems in cosmology. Searches for a solution have resulted in the discovery of a rich set of scalar-tensor theories, such as k-essence~\cite{Armendariz-Picon:1999hyi, ArmendarizPicon:2000ah, ArmendarizPicon:2000dh,Garriga:1999vw}, kinetic gravity braiding~\cite{Deffayet:2010qz,Kobayashi:2010cm} or galileons~\cite{Nicolis:2008in,Deffayet:2009wt} and generalized galileons~\cite{Deffayet:2009mn,Deffayet:2010qz} which feature first and even second-order derivative interactions. Eventually it was realised~\cite{Kobayashi:2011nu} that all these theories belong to the previously discovered  class of Horndeski theories~\cite{Horndeski:1974wa}, which itself was then extended not in the least to include even higher-order derivatives and sufficient degeneracy to not propagate extra degrees of freedom~\cite{Zumalacarregui:2013pma, Gleyzes:2014dya,Langlois:2015cwa, Motohashi:2014opa}. For reviews see refs.~\cite{Langlois:2018dxi,Kobayashi:2019hrl}, but our results are also relevant for other classes of theories, see e.g.~\cite{Clifton:2011jh,Joyce:2014kja,Bull:2015stt,Heisenberg:2018vsk}.

These models are often (almost) shift-symmetric and have solutions in which the derivative terms are large. Such backgrounds spontaneously violate Lorentz invariance, and, what is of particular interest to us here, cause small fluctuations of the scalar field to propagate differently than e.g.~light. When viewed in such a manner, k-essence can be understood as a relativistic perfect superfluid with a non-luminal sound speed for perturbations. This has been exploited e.g.~in cosmology to modify the predictions of standard inflation~\cite{Silverstein:2003hf} or to model the effect of clustering dark energy~\cite{Creminelli:2009mu, Sawicki:2013wja}.  The key object which determines such properties is the acoustic --- or effective --- metric for perturbations, which in cosmology is usually obtained by constructing an effective action for perturbations on the homogeneous cosmological background. As a result of the homogeneity and isotropy of the background universe, this acoustic metric can contain up to two time-dependent parameters, the signs of which describe whether the perturbations are ghosts or not and whether there are gradient instabilities. Either of these is usually considered to be a catastrophic pathology which renders the background unstable on very short timescales and they are used to eliminate such solutions and theories from further consideration~\cite{Joyce:2014kja,Woodard:2015zca}. However, gradient instabilities can be demoted almost to the level of tachyonic instabilities if UV physics can change the dispersion relation on scales parametrically lower than the  cutoff or the strong coupling scale, see e.g.~\cite{ArkaniHamed:2003uy,Babichev:2018twg}. On the other hand, ghost instabilities totally depend on interactions and can be rather benign; for cosmological applications of ghosty Effective Field Theories (EFT) see~\cite{Caldwell:1999ew,Cline:2023cwm,Cline:2024zhs}, while ghosty systems with a finite number of degrees of freedom can even be manifestly stable~\cite{Deffayet:2021nnt,Deffayet:2023wdg,ErrastiDiez:2024hfq}.

The main question we would like to address here is how one should assess the consistency, in particular stability, of general anisotropic backgrounds. Such questions often arise in cosmologically motivated setups such as those involving compact objects in these theories, where screening suppresses the scalar field's interactions --- k-mouflage~\cite{Babichev:2009ee} or Vainshtein~\cite{Vainshtein:1972sx, Kaloper:2011qc}, various exact solutions  e.g.~\cite{Babichev:2012re,Kobayashi:2014eva,Babichev:2017lmw}, EFT setups for black holes~\cite{Franciolini:2018uyq, Hui:2021cpm, Khoury:2022zor, Mukohyama:2022enj}, gravitational-wave emission from binaries~\cite{deRham:2012fw, Dar:2018dra} or gravitational-wave backgrounds in the presence of dark energy~\cite{Creminelli:2019kjy}, but also in apparently unrelated physics --- e.g.~in analogue gravity setups, in which superfluid flows are used as analogues to study curved spacetime and phenomena such as Hawking radiation are modelled by the physics of phonons in this medium~\cite{Barcelo:2005fc, Coviello:2024vht}.

The consistency of some choices of coefficients of operators in these theories and therefore the range of permitted background configurations at low energies has been questioned by appealing to the analyticity of the S-matrix in the UV~\cite{Adams:2006sv, Nicolis:2009qm,Bellazzini:2017fep,Tolley:2020gtv}. For some theories, for reasons that are still not entirely clear, it is possible to obtain similar bounds in the low-energy theory itself by requiring that the phase speed be at most luminal~\cite{Camanho:2014apa}, or a more sophisticated version where time advances with respect to the light cone resolvable within the EFT are forbidden~\cite{CarrilloGonzalez:2022fwg, Serra:2023nrn, Serra:2024tmz} --- usually called ``causality'' bounds. In addition, even on the level of the classical background, in the presence of superluminality there exists a possibility that a time machine --- a background with closed locally future directed signal trajectories --- could be constructed, see e.g.~\cite{Adams:2006sv,Evslin:2011rj}. It is not clear whether such backgrounds can be constructed within the regime of validity of the EFT. In any case, even without gravity, backreaction from quantum corrections appears to prevent such problematic backgrounds from being formed~\cite{Babichev:2007dw, Burrage:2011cr,Kaplan:2024qtf}. Notwithstanding this, the presence of superluminality on the (semi)classical level does not necessarily imply a violation of causality~\cite{Liberati:2001sd,Bruneton:2006gf,Bruneton:2007si,Kang:2007vs,Babichev:2007dw,Geroch:2010da}.

As is usually the case, stability is determined by the response of the backgrounds to small perturbations. The perspective we would like to promote here is that we can abstract the precise model and background since, as we will show,  the physics of interest is contained in the  acoustic metric associated to the particular model and background. In essence, the background configuration is a medium, and the acoustic metric is the covariant encoding of the properties of general media relevant to the propagation of relativistic sound waves. In particular, in the limit of high frequencies, the evolution of small fluctuations occurs along characteristic surfaces of the acoustic spacetime --- the acoustic equivalent of light cones. Using the analogy with standard results in general relativity, we study the properties of the acoustic cone in detail to determine the physical meaning of its properties and geometry and the relation to stable evolution for the fluctuations.  Our discussion is similar to that of~\cite{Babichev:2018uiw, Esposito-Farese:2019vlh} where the relative geometry of the light-cone and acoustic cone was used to determine conditions under which evolution can be stable. We put emphasis on the full 3+1d analysis, revealing that certain aspects remain hidden or at least ambiguous in lower dimensions.

In section~\ref{sec:metric} we demonstrate that the acoustic metric needs to be Lorentzian if the fluctuations are to be a proper degree of freedom described by a hyperbolic system of partial differential equations (PDEs), as may have been expected~\cite{Raetzel:2010je,Khavkine:2012jf}. What is usually called the gradient instability actually signifies a loss of this hyperbolicity and a constraint (elliptical) nature of the equations of motion. Then, we propose that the signature of the acoustic metric  determines whether the fluctuations are ghosts or healthy degrees of freedom. These are coordinate-invariant statements upon which all the observers will agree and which reduce to the usual notion for backgrounds with high symmetry.

High-frequency scalar fluctuations propagate in the acoustic spacetime on its null geo\-de\-sics and only depend on the usual spacetime implicitly. In section~\ref{sec:AcouCones}, we show for the first time that the connection of the acoustic spacetime has nonmetricity with respect to the usual metric of a type that guarantees that vector currents conserved in the acoustic spacetime correspond to ones conserved in the spacetime. We also define an acoustically conserved energy-momentum tensor for fluctuations, which can be used to produce currents conserved in the spacetime whenever the acoustic metric has symmetries.

Hyperbolicity implies that evolution is causal and is associated with the acoustic cone, which is generally different to the light cone. There are in fact two such acoustic cones encoding the same information: one describes rays and the phase velocity of the phonons. The other --- momenta and the dispersion relation. The acoustic metric transforms as a tensor and therefore the acoustic cones and the observables they determine are not invariant. This allows us to discuss two effects which can be confused with physical instabilities, but which are rather only related to coordinate choices.

In section~\ref{sec:CauchySurface}, we discuss when the Cauchy, i.e.\ initial value problem (IVP), is well-posed --- solved for general initial data with a smooth dependence thereof. We show that the IVP is ill-posed if the rays point toward coordinate ``past'', or equivalently when frequencies of some modes are \emph{complex}. Frequently this is misinterpreted as a breakdown of hyperbolicity or at least evidence of ghosts.   In section~\ref{sec:Cerenkov}, we discuss the physics when the observer is moving supersonically: we demonstrate that the appearance of sound horizons and therefore a Mach cone is directly related to the existence of modes with negative (but real) energy \emph{in this frame}.  Both of these coordinate problems can become physical in the presence of a second degree of freedom which interacts with phonons, such as gravity.

In section~\ref{sec:Hamiltonia} we connect to the Hamiltonian functional for perturbations in general curved acoustic spacetimes. We demonstrate that if the chosen time direction is an acoustic Killing vector, this Hamiltonian is a conserved charge not only in the acoustic spacetime, but also in the usual one. For a Lorentzian acoustic metric, this Hamiltonian is bounded from below, provided that i) the IVP is well-posed on the chosen spatial foliation; ii) the chosen time direction is subsonic. One can then use the Hamiltonian to bound the motion of phonons even when they interact with other species moving in the spacetime metric. We then extend this discussion in section~\ref{sec:otherQ}. There we show  that a Killing vector field timelike with respect to the acoustic metrics of all species is sufficient to have a bounded conserved charge for each of them. This then bounds motion even in the supersonic case. This picture matches the geometrical one provided by the cones --- if the ray cones of all degrees of freedom overlap, and none of the fields are ghosts, the motion is bounded and stable.

We complete the paper with section~\ref{sec:Geometry}, where we classify all possible acoustic metrics according to their eigenvectors. We explicitly construct the ray and momentum cones for all classes of Lorentzian acoustic metrics and the associated dispersion relations. Then, in section~\ref{sec:examples}, we illustrate our findings with worked out examples from simple and popular scalar-tensor theories. We close with a discussion and summary of our main results in section~\ref{sec:conclude}.

\section{Acoustic metric: construction, geodesics and hyperbolicity}

\subsection{The eikonal ansatz and the acoustic metric \label{sec:metric}}

We begin by discussing the propagation of short wave-length modes of a scalar field $\phi$ in a general medium provided by a background configuration of the same field  and gravity $g_{\mu\nu}$ and possibly other matter fields $\Psi_I$. This leads to notions of an acoustic metric along with its associated characteristic surfaces, cones of influence and dispersion relations. Here we follow the general consideration from~\cite{LandavshitzII,Courant,Barcelo:2005fc,Moncrief:1980, Unruh:1980cg, Perlick_BOOK,Raetzel:2010je}.

Our general setup comprises some spacetime metric $g_{\mu\nu}$, the dynamics of which is controlled by the theory of gravity, which can be Einstein's general relativity (GR) or some modified gravity. Following the well-established tradition, we call the null-cone (also sometimes called the isotropic cone) of this spacetime metric $g_{\mu\nu}$  the \emph{light cone}, and call the speed of propagation along this cone the \emph{speed of light} which we normalise to unity. However, the reader should keep in mind that while in the case of vanishing backgrounds $\phi$, $\Psi_I$ including the electromagnetic field and spacetime curvature, light must propagate on this light cone, on a general background this \emph{might no longer} be the case. We assume that observers which only interact with gravity move on geodesics of $g_{\mu\nu}$ and that at least some degrees of freedom do propagate on the light cone.\footnote{In a situations when $g_{\mu\nu}$ is demoted from its usual physical meaning, for instance, when no degrees of freedom propagate along its light cone, or when no Lorentz invariant vacuum is  available, or when observers are coupled not to $g_{\mu\nu}$ but something else, one could use as a fiducial metric the effective metric of some other degree of freedom.}
We do not specify the theory of $\phi$ yet, only assuming that the equations of motion for all fields involved are second-order in derivatives. For instance, this theory could be k-essence~\cite{ArmendarizPicon:1999rj, ArmendarizPicon:2000ah, ArmendarizPicon:2000dh}, more general kinetic gravity braiding~\cite{Deffayet:2010qz} or $\phi$ can be non-minimally coupled through derivatives to other fields such as the electromagnetic tensor, see e.g.~\cite{Itzykson:1980rh} and for more recent works e.g.~\cite{Mironov:2024idn,Babichev:2024kfo}. The background $\bar\phi(x^{\mu})$ or backgrounds of other fields $\bar\Psi_I(x^{\mu})$ will in general be not Lorentz invariant and therefore small fluctuations $\pi=\delta\phi$ around it can propagate at speeds different to the speed of light even in the massless (i.e.~gapless) case. One can understand this as a propagation in an effective acoustic spacetime which has essentially all the features of the standard one from the point of view of geometry and geodesics. As we will demonstrate, we are dealing with a theory with multiple metrics. We presume that both the background solution $\bar\phi$ and the perturbed one $\phi = \bar\phi + \pi$ satisfy the equations of motion as do the other fields involved. We can include gravity and $\phi$ and all other fields as elements of $\Psi_I = \bar\Psi_I + \pi_I$, so that $\pi_I=(\pi,\psi_I)$.

Assuming that the perturbations are small and vary on scales much shorter than the background, we can employ the standard eikonal approximation,
\begin{equation}
\pi_I=\Re \, \mathcal{A}_I(x)\exp(i\mathcal{S}(x)/\epsilon)\,,
\label{eq:eikonal}
\end{equation}
where $\Re$ means that we take the real part only. In this ansatz the auxiliary parameter $\epsilon$ is then taken sufficiently close to the limit  $\epsilon\rightarrow0$, to allow us to assume that $\mathcal{A}_I(x)$ varies slowly compared to the phase. In the formal limit $\epsilon\rightarrow0$, the surfaces $\mathcal{S}=\text{const}$ are the characteristic surfaces (or wavefronts) for this linearized system. At the leading order, $\mathcal{O}(\epsilon^{-2})$, the condition that one can find $\mathcal{A}_I(x)$ from the linearized system of second order PDE reads
\begin{equation}
\det\left(\mathcal{P}^{IJ\mu\nu}\partial_{\mu}\mathcal{S}\partial_{\nu}\mathcal{S}\right)=0\,,
\label{eq:general_charact}
\end{equation}
where $\mathcal{P}^{IJ\mu\nu}\partial_{\mu}\partial_{\nu}$ is the so-called principal symbol of the second-order differential operator of the linearized system of equations of motion. In many physically interesting cases, either for particular backgrounds, or even for \emph{all} backgrounds as it is in kinetic gravity braiding (as was demonstrated in~\cite{Deffayet:2010qz}), equation~\eqref{eq:general_charact}  factorises into a product of terms like
\begin{equation}
Z^{\mu\nu}\partial_{\mu}\mathcal{S}\partial_{\nu}\mathcal{S}=0\,,\label{eq:BigG-char}
\end{equation}
where $Z^{\mu\nu}$ is a tensor formed from functions of the background configurations of the spacetime metric, the scalar, all other fields and their derivatives. In the rest of the paper we consider this factorisable situation assuming it is applicable for the fluctuations of the scalar field under consideration, $\pi$. For a more mathematically inclined discussion of the non-factorisable case see~\cite{Raetzel:2010je}. As we will discuss here, $Z^{\mu\nu}$ really acts as an (inverse or contravariant) metric for the fluctuations $\pi$.

Our main goal is to concentrate on this physically relevant, but still relatively simple, case of one factorised scalar degree of freedom to achieve a maximally transparent and physically intuitive discussion. Other factorised degrees of freedom can be added by induction. Also note that any tensor conformally related to $Z^{\mu\nu}$ is equivalent from the point of view of eq.~\eqref{eq:BigG-char}. We will discuss the choice of proper normalisation later, but it has no influence on most of the discussion in this paper.

We can associate a momentum covector to the characteristic surface,\footnote{Strictly speaking, the momentum should be $\partial_\mu \mathcal{S}/\epsilon$ but the auxiliary parameter $\epsilon$ is only used to keep track of orders of expansion and can be set to unity after that.} \begin{equation}
    P_\mu = \partial_\mu \mathcal{S}\,.\label{eq:Pdef}
\end{equation}
The momentum $P_\mu$ is then a null covector for the inverse metric, and the surface \begin{equation}
    Z^{\mu\nu}P_\mu P_\nu = 0 \,,\label{eq:ZPP}
\end{equation}
is a null surface of constant phase $\mathcal{S}$.

We can recover the direction of travel of constant-phase surfaces, and therefore the phase four-velocity, by requiring that on some curve parameterised by $\lambda$ (the ray)
\begin{equation}
    0 = d\mathcal{S} = \partial_\mu \mathcal{S} \,\frac{dx^\mu}{d\lambda}\, d\lambda \,. \label{eq:constS-light}
\end{equation}
Thus $P_\mu \, dx^\mu/d\lambda=0$ and the constant phase surface with momentum $P_\mu$ moves in the direction $dx^\mu/d\lambda$ orthogonal to $P_\mu$. We demand that for any $P_\mu$ there be a unique ray, requiring a linear relationship $dx^\mu/d\lambda=M^{\mu\nu}P_\nu$ with some non-degenerate $M^{\mu\nu}$. By eq.~\eqref{eq:BigG-char} one obtains\footnote{If this proportionality condition were not satisfied, the momentum $P_{\mu}$ would satisfy two independent quadratic equations which would overconstrain the system.} that $M^{\mu\nu} \propto Z^{\mu\nu}$ and the conformal factor can be set correctly by the judicious choice of the $\lambda$ as an affine parameter. We can thus define the ray vector
\begin{equation}
  \frac{dx^\mu}{d\lambda} =  N^\mu \equiv  Z^{\mu\nu} P_\nu\,,\quad \text{so that} \quad N^\mu P_\mu = 0 \,.\label{eq:NullVecsT}
\end{equation}
In particular, a \emph{phonon}, as we colloquially call the quasiparticle which is a quantum of the perturbation $\pi$, has four-momentum  $P^{\mu}$ and a four-velocity proportional to $N^{\mu}$. Note that we assume that $Z^{\mu\nu}$ is dimensionless so that $P^{\mu}$ and $N_{\mu}$ have the same dimension of energy.
When this standard construction is carried out for electromagnetism in vacuum, $g^{\mu\nu}$ appears instead of $Z^{\mu\nu}$ and for a light wave with momentum $p_\mu$, the ray vector is $p^\mu=g^{\mu\nu}p_\nu$. The orthogonality of the ray and the momentum for light is just the statement that the momentum is null, $p^\mu p_\mu=0$ and the ray is just the Poynting vector of the electromagnetic wave.

Instead here, the vectors $N^\mu$ and $P^\mu$ are not coincident and we have the statement of orthogonality for $N^\mu P_\mu =0$. Thus, if one of the two vectors is timelike with respect to $g_{\mu\nu}$, the other is spacelike, cf.~\cite[Chapter XI]{Synge:1960ueh}. In fact the orthogonality conditions can be interpreted as a kind of on-shell condition, defining the direction of travel $N^\mu$ of a momentum mode $P_\mu$.
Propagation is subluminal provided $N^{\mu}$ is $g$-timelike\footnote{We work in the $(-+++)$ signature for the spacetime metric $g_{\mu\nu}$ and use the Planck units $\hbar=c=G_{N}=1$ throughout the paper.} \begin{equation}
g_{\mu\nu}N^{\mu}N^{\nu}<0\,,
\end{equation}
while  superluminal propagation is described by $g$-spacelike wave four-velocities
\begin{equation}
\label{eq:superluminal}
g_{\mu\nu}N^{\mu}N^{\nu}>0\,.
\end{equation}
We stress that owing to eq.~\eqref{eq:constS-light} the four-momentum $P_{\mu}$ is necessarily \emph{$g$-spacelike} for the usual \emph{subluminal} propagation.

When the second metric is introduced, there are now two structures  mapping vectors to covectors and one needs to be careful with notation. In this paper, we will always raise and lower indices using the spacetime metric $g_{\mu\nu}$ and its inverse, $g^{\mu\nu}$, as per usual.  It can easily be seen that $Z_{\mu\nu} = g_{\mu\alpha} g_{\nu\beta} Z^{\alpha\beta}$ is \emph{not} the inverse of $Z^{\mu\nu}$. Rather, provided that $Z^{\mu\nu}$ is not degenerate, a new tensor $S_{\mu\nu}$ exists with
\begin{equation}
    Z^{\mu\rho}S_{\rho\nu} = \delta^\mu_\nu \,, \label{eq:Sdef}
\end{equation}
and the pair $S_{\mu\nu}/Z^{\mu\nu}$ give an alternative to $g_{\mu\nu}/g^{\mu\nu}$ to assign dual one forms (covectors) to vectors and vice versa.
With the definition~\eqref{eq:Sdef}, the equation for $P_\mu$~\eqref{eq:ZPP} can be rewritten as an equation for the rays $N^\mu$,
\begin{equation}\label{eq:ray-cone}
    S_{\mu\nu} N^\mu N^\nu = 0 \,.
\end{equation}
The rays are null vectors of $S_{\mu\nu}$, while the momenta are null covectors of $Z^{\mu\nu}$. As we will see, the respective null surfaces form cones which are distinct from the point of view of the spacetime --- to distinguish them, we will call them the \emph{ray cone} or  \emph{N-cone}~\eqref{eq:ray-cone} and the \emph{momentum cone} or \emph{P-cone}~\eqref{eq:ZPP} respectively. In the end, both of the acoustic cones encode the same information, which we will demonstrate.

We also need to be careful about specifying the meaning of timelike, spacelike and null. We will use the prefix $g$-, $Z$- and $S$- (e.g.~$Z$-timelike) to specify with respect to which metric the (co)-vectors are timelike/spacelike.
Introducing $S_{\mu\nu}$ gives a simple formula
\begin{equation}
    P_{\mu}=S_{\mu\nu}\,N^{\nu} \,, \label{eq:P_from_N}
\end{equation}
inverting the relation~\eqref{eq:NullVecsT}. Later we are going to illustrate our results plotting $N^{\mu}$ and
\begin{equation}
    P^{\mu}=S^{\mu}_{\nu}\,N^{\nu} \,,  \label{eq:Pup_from_N}
\end{equation}
so that the linear operator $S^{\mu}_{\nu}$ can be thought of as playing the role of an \emph{effective mass tensor} (up to normalisation factor related to the norm of $N^\mu$) relating  the four-velocity with canonical four-momentum even for gapless waves. We discuss this tensor in section~\ref{sec:Acou+Obs}, showing that $\det S^\mu_\nu>0$ is required for the existence of acoustic cones. Later, we also discuss another effective mass concept commonly used in condensed matter physics~\eqref{eq:reciprocal_M}. This object measures the inertial properties of the phonon moving along the geodesics --- i.e.~how difficult is it for an external force to accelerate the phonon. It is important to clarify that even though these effective masses are useful for different physical situations the phonons do not have a rest mass --- for subluminal motion~\eqref{eq:NullVecsT} guarantees that the energy as measured in the rest frame of the phonon identically vanishes.

\subsection{Acoustic geodesics and nonmetricity} \label{sec:AcouCones}
Let us now make the claim that $S_{\mu\nu}/Z^{\mu\nu}$ are really an (inverse) metric more concrete by illustrating that we can replicate the whole geometrical machinery of general relativity.

We can define a covariant derivative compatible\footnote{Compatibility with $Z^{\mu\nu}$ implies compatibility with $S_{\mu\nu}$.} with $Z^{\mu\nu}$
and give it a torsion-free connection; $\nZ_\alpha Z^{\mu\nu}=0$ (e.g.~\cite{Babichev:2007dw,deRham:2014wfa}).
Applying this acoustic covariant derivative to~\eqref{eq:ZPP}, we obtain two equations
as an analogue of the geodesic equation,
\begin{equation}
N^{\mu}\nZ_\mu N^{\lambda}=0 \qquad\text{and}\qquad N^{\mu}\nZ_{\mu}P_{\nu}=0\,, \label{eq:geodeisc}
\end{equation}
where we have used the fact that $P_{\mu}$ is a derivative of a scalar
and multiplied by $Z^{\lambda\nu}$ to obtain the first equation
from the second. We should interpret the first equation~\eqref{eq:geodeisc}
as meaning that the ray vectors are parallel transported along themselves and therefore, when integrated, give the $Z$-null geodesics of the acoustic metric $Z^{\mu\nu}$. The second equation implies that momentum covectors are parallel transported in $Z^{\mu\nu}$ along their associated rays.
Note that there are no such equations for the momentum \emph{vector} $P^{\mu}=g^{\mu\nu}P_{\nu}$ or for parallel transport along $P^{\mu}$ or for the ray covector $N_\mu$.

Using the standard procedure for metric-compatible connections, we can find an explicit expression for
the acoustic Christoffel symbols in the derivative $\nZ_{\mu}$ (such a connection was constructed for k-\emph{essence} in ref.~\cite{Gangopadhyay:2012dz})
\begin{equation}
\GZ_{\mu\nu}^{\alpha}=\frac{1}{2}Z^{\alpha\beta}\left(\partial_{\mu}S_{\beta\nu}+\partial_{\nu}S_{\mu\beta}-\partial_{\beta}S_{\mu\nu}\right)\,.
\label{eq:Chrstoffel}
\end{equation}
The difference between the acoustic and the usual Christoffel symbols is given by the disformation tensor
\begin{equation}
\label{eq:disformation}
L_{\phantom{\alpha}\mu\nu}^{\alpha}=\GZ_{\mu\nu}^{\alpha}-\Gamma^{\alpha}_{\mu\nu}=\frac{1}{2}Z^{\alpha\beta}\left(\nabla_{\mu}S_{\beta\nu}+\nabla_{\nu}S_{\beta\mu}-\nabla_{\beta}S_{\mu\nu}\right)\,,
\end{equation}
where $\nabla_{\mu}$ is the usual covariant derivative compatible with the gravitational spacetime metric: $\nabla_{\alpha}g_{\mu\nu}=0$.

Using this formula for the disformation tensor $L_{\phantom{\alpha}\mu\nu}^{\alpha}$ it is straightforward to demonstrate that the acoustic Weyl transformation
\begin{equation}
\label{eq:Weyl_trans}
S_{\mu\nu}\rightarrow \Omega^2(x) \,S_{\mu\nu}\,,\qquad
Z^{\mu\nu}\rightarrow \Omega^{-2}(x) \,Z^{\mu\nu}\,,\qquad N^\mu\rightarrow \Omega^{-2}(x)\, N^\mu\,,
\end{equation}
leaves the form of both geodesic equations~\eqref{eq:geodeisc} invariant. Thus, the four-momentum $P_{\mu}$ has zero conformal weight --- remains invariant --- while the four-velocity $N^{\mu}$ transforms as the contravariant metric --- has conformal weight two. Acoustic Weyl invariance is important, as the eikonal formalism fixes $Z^{\alpha\beta}$ up to a conformal factor only.

\sloppy{Acoustic geodesics generically do not map to the usual spacetime geodesics, as $N^{\mu}\nabla_{\mu}N^{\beta}=-N^{\nu}L_{\phantom{\alpha}\nu\alpha}^{\beta}N^{\alpha}$. Thus $N^{\mu}$ is not transported parallel to itself in the usual spacetime sense, so that also the spacetime norm of $N^\mu$ is not conserved under geodesic transport, $N^{\mu}\partial_{\mu}\left(N^{\alpha}N_{\alpha}\right)=-2N^{\alpha}N^{\mu}L_{\phantom{\alpha}\mu\alpha}^{\lambda}N_{\lambda}$, and similarly for the spacetime norm of $P_\mu$. Therefore normalising $N^\mu$ does not have the utility that normalising four-velocities has in the usual case. For $g$-timelike acoustic geodesics one can nonetheless introduce the unit vector (which is also Weyl invariant)}
\begin{equation}
\label{eq:N_normal}
    \mathscr{N}^{\mu}=N^{\mu}/\sqrt{-N^{\alpha}N_{\alpha}}\,,
\end{equation}
in terms of which the acceleration for the acoustic geodesic can be calculated as
\begin{equation}
a^{\mu}\equiv\mathscr{N}^{\alpha}\nabla_{\alpha}\mathscr{N}^{\mu}=-\left(\delta_{\nu}^{\mu}+\mathscr{N}^{\mu}\mathscr{N}_{\nu}\right)L_{\phantom{\alpha}\alpha\beta}^{\nu}\,\mathscr{N}^{\alpha}\mathscr{N}^{\beta}\,.
\label{eq:acceleration}
\end{equation}
In particular, phonons (or photons in a medium) propagate with the acceleration above. It is important to stress that momentum transport or the relativistic Newton's law takes a different form,
\begin{equation}
\label{eq:Newton_Law}
\mathscr{N}^{\lambda}\nabla_{\lambda}P_{\mu}=L_{\phantom{\alpha}\mu\nu}^{\lambda}P_{\lambda}\mathscr{N}^{\nu}\,.
\end{equation}
This difference is due to the non-conservation of the effective mass in the relation connecting the normalized four-velocity of the wave $\mathscr{N}^{\mu}$ and its four-momentum
\begin{equation}
    P^{\mu}=\sqrt{-N^{\alpha}N_{\alpha}}\,S^{\mu}_{\nu}\,\mathscr{N}^{\nu}\,.
\end{equation}

The derivative $\nZ_{\mu}$ is not compatible with the spacetime
metric and gives the nonmetricity tensor $Q_{\alpha\mu\nu}$ according to
\begin{equation}
\label{eq:nonmetricity}
\nZ_{\alpha}g_{\mu\nu}=Q_{\alpha\mu\nu} = - L_{\mu\,\alpha\nu} - L_{\nu\,\alpha\mu}\,.
\end{equation}
Following~\cite{McCrea:1992wa,Hehl:1994ue}, we expand the acoustic nonmetricity tensor into
\begin{equation}
\label{eq:nonmetric_decompose}
Q_{\alpha\mu\nu}=g_{\mu\nu}W_{\alpha}+\cancel{Q}_{\alpha\mu\nu}\,,
\end{equation}
where $\cancel{Q}_{\alpha\mu\nu}$ is trace-free in indices $\mu,\nu$ and
$W_{\alpha}$ denotes the Weyl vector
\begin{equation}
W_{\alpha}=\frac{1}{4}g^{\mu\nu}\nZ_{\alpha}g_{\mu\nu}\,.
\label{eq:Weyl_vector}
\end{equation}
The acoustic nonmetricity in~\eqref{eq:nonmetricity} has
\begin{equation}
W_{\alpha}=-\frac{1}{4}Z^{\mu\nu}\nabla_{\alpha}S_{\mu\nu}=-\frac{1}{4}\partial_{\alpha}\ln\left|\det S_{\mu}^{\nu}\right|\,,
\label{eq:WeylNonMetricity}
\end{equation}where $S^\alpha_\beta$ is defined in eq.~\eqref{eq:Pup_from_N} and discussed around eq.~\eqref{eq:ZSud_def}.\footnote{It is important to stress that $\det S_{\nu}^{\mu}\equiv\varepsilon^{\alpha\beta\gamma\sigma}\varepsilon_{\alpha'\beta'\gamma'\sigma'}S_{\alpha}^{\alpha'}S_{\beta}^{\beta'}S_{\gamma}^{\gamma'}S_{\sigma}^{\sigma'}/4!$ is a scalar quantity, contrary to $\det S_{\mu\nu}\equiv-g\,\varepsilon^{\alpha\beta\gamma\sigma}\varepsilon^{\alpha'\beta'\gamma'\sigma'}S_{\alpha\alpha'}S_{\beta\beta'}S_{\gamma\gamma'}S_{\sigma\sigma'}/4!$, where $\varepsilon_{\alpha\beta\gamma\sigma}$ denotes the totally antisymmetric Levi-Civita \emph{tensor}, and as usual $g=\det g_{\mu\nu} $ see e.g.~\cite[pg.~250]{LandavshitzII}.}  An implication of eq.~\eqref{eq:WeylNonMetricity} is the simple relation between the acoustic and spacetime divergences of a vector:
\begin{equation}
    \nZ_\mu V^\mu = \frac{1}{\sqrt{\det S^\alpha_\beta}}\nabla_\mu \left(\sqrt{\det S^\alpha_\beta}\,V^\mu \right)\,. \label{eq:AcouDiv}
\end{equation}

We can continue this geodetic picture by deriving the geodesic deviation
equation --- again, the equation only exists for the closely separated
geodesics with tangent ray vectors $N^{\mu}$ and separation vector
$\eta^{\mu}$, and not for $P_{\mu}$. The derivation proceeds
as usual, giving
\begin{equation}
N^{\mu}\nZ_{\mu}\left(N^{\nu}\nZ_{\nu}\eta^{\alpha}\right)=\mathcal{R}[Z]_{\phantom{\alpha}\mu\nu\beta}^{\alpha}N^{\mu}N^{\nu}\eta^{\beta},\label{eq:geo-dev}
\end{equation}
with $\mathcal{R}[Z]_{\phantom{\alpha}\mu\nu\beta}^{\alpha}$ the
Riemann curvature tensor of the $S_{\mu\nu}$/$Z^{\mu\nu}$
metric formed from Christoffel symbols defined by eq.~\eqref{eq:Chrstoffel}.

\enlargethispage{-\baselineskip}

Furthermore, even though the Lie derivative $\mathsterling_{\xi}$ is insensitive to the connection, the acoustic Killing equation for vector field $\xi^{\mu}$ generating the symmetry, $\mathsterling_{\xi}S_{\alpha\beta}=0$, reads\footnote{The difference from the standard Killing equation $\nabla_{\mu}\xi_{\nu}+\nabla_{\nu}\xi_{\mu}=0$ is caused by our convention to raise and lower indices with the spacetime metric $g_{\mu\nu}$, not compatible with the acoustic covariant derivative $\nZ_\mu$.
}
\begin{equation}
     S_{\nu\alpha}\nZ_\mu \xi^\alpha  +  S_{\mu\alpha}\nZ_\nu \xi^\alpha = 0\,.
     \label{eq:acosutic_Killing}
 \end{equation}
 Acoustic Killing vector fields (KV) satisfying~\eqref{eq:acosutic_Killing} allow us to construct quantities conserved along acoustic geodesics. In particular, using geodesic equation~\eqref{eq:geodeisc} for momentum transport one can check that
\begin{equation}
N^{\mu}\partial_{\mu}\left(P_{\alpha}\xi^{\alpha}\right)=0\,.
\label{eq:P_0_conserv}
\end{equation}
Hence, $P_{\alpha}\xi^{\alpha}$ is conserved along geodesics.

Interestingly, in the discussion of propagation of photons in a medium with refraction index $n>1$ there is the more than  century-old Abraham-Minkowski controversy, for reviews see e.g.~\cite{Leonhardt,Pfeifer:2007zz,Enigma}. The controversy is in the ambiguity in the definition of the photon momentum assuming its energy is $E$. Namely, Minkowski proposed momentum $p_M\equiv nE$~\cite{Minkowski} while Abraham proposed $p_A\equiv E/n$~\cite{Abraham}. In this way the $P^{\mu}_M\equiv (E,nE)$ and spacelike, while $P^{\mu}_A\equiv (E,E/n)$ is collinear with photon four-velocity and timelike. In fact $g_{\mu\nu}P^{\mu}_A P^{\mu}_M=0$. Of course, in this discussion it was \emph{assumed} that $E$ is conserved even when the photon enters the medium. As we have seen above, in our case $P_{\mu}$ is: i) spacelike for subluminal propagation ii) in the presence of time-translation symmetry along the KV $\xi^{\alpha}=\delta^{\alpha}_0$ it is $P_0=P_{\alpha}\xi^{\alpha}$ which is conserved iii) finally, by construction, $P_{\mu}$ is a canonical momentum $P_{\mu}=\partial_\mu \mathcal{S}$. From these properties we conclude that the four-momentum $P_{\mu}$ should be identified with the Minkowski momentum, while the ray vector $N^{\mu}$ corresponds to the proper choice of Abraham momentum, differing from Abraham's definition only by a space-time dependent normalisation factor. Crucially $N_{\mu}\xi^{\mu}$ is \emph{not} conserved along acoustic geodesics, thus it is the Minkowski four-momentum $P_{\mu}$ which is responsible for conserved quantities.

Given the rederivation of the all the standard GR machinery for $Z^{\mu\nu}$, we are really dealing with a theory with two metrics: (i) $g_{\mu\nu}/g^{\mu\nu}$, and (ii) $S_{\mu\nu}/Z^{\mu\nu}$ (two inequivalent tensors with respect to $g_{\mu\nu}$, but really just a metric and its inverse).

\subsection{Acoustic metric signature: hyperbolicity and ghosts}\label{sec:signature}
If  $Z^{\mu\nu}$ is to be a metric, it must be non-degenerate
and therefore the consideration for $S_{\mu\nu}$ is equivalent. To
describe a causal structure a metric must have Lorentzian signature. The same is required to allow for a well-posed formulation of the Cauchy problem, i.e.~for  the initial value problem (IVP) for $\phi$. This
is necessary so that the differential operator describing the propagation
of perturbations be hyperbolic. This is then equivalent to the existence of cones of influence. We will recover this standard result for the spacetime metric also for the acoustic spacetime, setting up the discussion without making reference to the spacetime metric.

Let us choose an arbitrary vector $W^\mu$. We only require that it not be null with respect to $S_{\mu\nu}$ and we do not normalise it. We associate a covector $u_\mu$ to it,
\begin{equation}
   u_\mu \equiv S_{\mu\nu} W^\mu\,,\qquad S_{\mu\nu}W^\mu W^\nu = -D \neq 0 \,, \label{eq:W-def}
\end{equation}
so that
\begin{equation}
   D=-Z^{\alpha\beta}u_{\alpha} u_{\beta} \,. \label{eq:D-def}
\end{equation}
We can now define a projector
\begin{equation}
\perpZ^\mu_\nu=\delta_{\nu}^{\mu}+\frac{W^{\mu}u_{\nu}}{D}\,,
\label{eq:perp_bar}
\end{equation}
onto a subspace orthogonal to $W^\mu$ and the associated induced inverse metric on this subspace, $\DZ^{\mu\nu}=Z^{\alpha\beta}\perpZ^\mu_\alpha \perpZ^\nu_\beta$. To be specific, this hypersurface is orthogonal in the $Z$-metric
\begin{equation}
    \DZ^{\mu\nu} \equiv Z^{\mu\nu} + \frac{W^\mu W^\nu}{D}\,,\qquad \perpZ^\mu_\nu \equiv  \DZ^{\mu\lambda} S_{\lambda\nu} \,. \label{eq:DZ_def}
\end{equation}
where the expressions here allow for the arbitrary normalisation of $W^\mu$. We shall call this projection orthogonal in $Z$ the $Z$-frame.

The momentum covector can be decomposed,
\begin{align}
    P_\mu &= \frac{\omega_Z}{D}u_\mu + \KZ_\mu\,,\quad\text{with}\quad \KZ_\mu\equiv \perpZ^\nu_\mu P_\nu\,, \label{P-Zdecomp}
\end{align}
and we can carry this through to the characteristic equation~\eqref{eq:BigG-char},
\begin{align}
    &Z^{\mu\nu}P_\mu P_\nu = -\frac{1}{D}\left(\omega_Z^2 - D \DZ^{\mu\nu} \KZ_\mu \KZ_\nu\right) =0 \,.\label{eq:P-cone-Zdecomp}
\end{align}
If there exists any such vector $W^\mu$ that the tensor $D\DZ^{\mu\nu}$ is positive definite then the characteristic surface described by~\eqref{eq:P-cone-Zdecomp} is a cone. This is only possible if the signature of the acoustic metric is Lorentzian --- $(3,1)$ or $(1,3)$ --- and then $W^\mu$ is $S$-timelike. Equivalently, $u_\mu$ is $Z$-timelike, a covector lying inside the cone.\footnote{Note that we are not guaranteed that $D>0$ even when $U^\mu$ is $S$-timelike --- this depends on which of the two Lorentzian signatures $S/Z$ have.}

Since $Z^{\mu\nu}$ and $S_{\mu\nu}$ are inverses, they have the same signature. We are nonetheless still left with two possible hyperbolic signatures. We \emph{define} the \emph{ghost} as having the acoustic metric $Z^{\mu\nu}$ of the \emph{opposite signature} to the one of the fiducial spacelike metric $g^{\mu\nu}$. We assume that, at least some standard healthy degrees of freedom propagate in the usual spacetime metric, for lack of a better term we call such a standard degree of freedom --- a \emph{non-ghost}. Which one is which is just a convention, but for this paper:
\begin{itemize}\label{def:ghosts}
    \item Signature (3,1) (mostly plus) represents a healthy non-ghost degree of freedom,
    \item Signature (1,3) (mostly minus) is the invariant definition of a ghost.
\end{itemize}
For both of these cases, and only for these cases, the determinant of the metric $Z$ (and $S$) is negative and the null surface~\eqref{eq:P-cone-Zdecomp} is a cone and causal evolution is possible.

The null surfaces of metrics with other signatures are not cones. Usually this sort of pathological situation is referred to as a gradient instability. It means the differential operator is no longer hyperbolic and the system cannot be solved as an initial value problem. Attempting to do so leads to exponentially growing modes which rapidly dominate the solution.

The cone eq.~\eqref{eq:P-cone-Zdecomp} is quadratic in $\omega_Z$ --- there are two roots which build the two \emph{nappes} of the cone --- the future and past. In the $Z$-frame one root is positive and one negative. If two acoustic metrics differ only by the overall sign (implying a switch of signatures between (3,1) and (1,3)), the cones are the same. The difference is that since $N^\mu =Z^{\mu\nu}P_\nu$, the upper nappe of the ray cone is mapped by the acoustic metric to the lower nappe of the P-cone for a ghost, as opposed to the upper-to-upper mapping for a non-ghost. This is a Lorentz-invariant geometrical statement valid in any frame and we propose should be considered the defining difference between ghosts and non-ghosts. Since they have acoustic cones,  ghosts are proper dynamical degrees of freedom with normal causal evolution --- it is only that their four-momenta are taken from the nappe opposite to that of the non-ghosts.

In solving the Cauchy problem and in any consideration of causality, one has to select the as future one of the nappes of the ray cone.\footnote{Note that any two distinct $S$-null vectors $N_{1}^{\mu}$ and $N_{2}^{\mu}$ belonging to the same nappe of their cone have a negative product in acoustic geometry, i.e.~$S_{\mu\nu}N_{1}^{\mu}N_{2}^{\nu}<0$, for non-ghosts. Conversely, $S_{\mu\nu}N_{1}^{\mu}N_{2}^{\nu}>0$ implies that the vectors lie in opposite nappes. By the maps~\eqref{eq:NullVecsT} and~\eqref{eq:P_from_N}, these inequalities apply also to the covectors $P_{1,2\mu}$ and their metric $Z^{\mu\nu}$ in the same manner. For ghosts, these inequalities are reversed.\label{ref:CommonNappe}} For a single isolated degree of freedom, any such choice is fine. Whatever we call the future also defines, through the acoustic metric, the choice of relevant future nappe of the P-cone and therefore the sign of energies of the modes.

However, in the presence of a second metric, e.g.~the spacetime metric $g_{\mu\nu}$ and other matter fields propagating in it, the choice of future must be consistent between all the degrees of freedom. Thus one is forced to designate as the future nappe of the acoustic N-cone that cone half which has overlap with what is designated  as the future nappe of the light cone. Then the mapping between the P-cone and ray-cone nappes through the acoustic metric, determines also the relative energies of the modes of the different degrees of freedom and makes ghosts have physical implications.
A situation in which the ray cone overlaps both nappes is acausal (see section~\ref{sec:acausal}). On the other hand, if the N-cone does not overlap with the light cone at all, one cannot uniquely select the future nappe and there are two non-equivalent time orientations (see sections~\ref{sec:otherQ} and~\ref{sec:Cerenkov}).

The covector $u_\mu$ defined in eq.~\eqref{eq:W-def} describes a surface $\Sigma_u$, $u_\mu = \partial_\mu \Sigma_u$ with induced metric $\DZ^{\mu\nu}$. For Lorentzian $Z^{\mu\nu}$, if $\Sigma_u$ is $Z$-spacelike, i.e.~outside of the cone, it provides a spatial hypersurface on which initial values can be set up for the Cauchy problem. We will use the shorthand that $u_\mu$ is a \emph{good Cauchy frame} for the scalar when this is the case, i.e.~whenever in eq.~\eqref{eq:P-cone-Zdecomp} the tensor
\begin{equation}
\label{Z-inducedmetric}
\mathcal{Z}_2^{\mu\nu}\equiv D \DZ^{\mu\nu} = Z^{\mu\alpha}u_\alpha \, Z^{\nu\beta}u_\beta - (Z^{\alpha\beta} u_\alpha u_\beta)\, Z^{\mu\nu} \succ 0 \,,
\end{equation}
where we use the symbol $\succ$ to mean positive definite. This condition is quadratic in $Z^{\mu\nu}$ and therefore not sensitive to the overall sign of the metric. Eq.~\eqref{Z-inducedmetric} is purely spatial with respect to $u^{\mu}$: $\mathcal{Z}_{2}^{\mu\nu}u_{\mu}=0$. This tensor was obtained in~\cite{Nicolis:2004qq} as a test for gradient instabilities --- see our discussion on page~\pageref{thing:Z2disc}. We can also write an analogous expression for $S_{\mu\nu}$ which will be useful later,
\begin{equation}\label{eq:Stwomunu}
\Stwo{\mu\nu}\equiv S_{\mu\alpha}u^{\alpha} \, S_{\nu\beta}u^{\beta} - (S_{\alpha\beta} u^\alpha u^\beta)\, S_{\mu\nu}\,.
\end{equation}

We stress that unless we already know the signature, one cannot identify which of the directions is timelike by testing the norm of just one of the vectors. Absent prior knowledge of the ghost status of the background, we have to first determine whether we are in a good Cauchy frame, and therefore whether $\DZ^{\mu\nu}$ is spatial or not, and only then check for ghostness.

\subsection{Action and the acoustic energy-momentum tensor} \label{sec:Action_EMT}
Let us illustrate our construction.
Given a local action $S[\Psi_I]$ describing the dynamics of the fields $\Psi_I$ containing no higher than their second derivatives, we obtain as the equations of motion a system,
\begin{equation}
\mathcal{E}_I(\nabla\nabla\Psi_J, \nabla\Psi_J, \Psi_J) =  0 \,.\label{eq:eom}
\end{equation}
Linearising the above and potentially performing the diagonalisation of the kinetic term as discussed around eq.~\eqref{eq:BigG-char} yields for the fluctuation of the scalar field:
\begin{equation}
     \Zt^{\mu\nu}\nabla_\mu\nabla_\nu \pi + V_\pi^\mu\nabla_\mu \pi + M^2_\pi \pi= V_J^\mu \nabla_\mu\psi_J + \mu_J \psi_J \,,\label{eq:EoMlin}
\end{equation}where the tensors $\Zt^{\mu\nu}$, $V_\pi^\mu$, $M_\pi^2$, $V_J^\mu$ and $\mu_J$ depend on background quantities only and the fields $\psi_J$ represent the small fluctuations of the other degrees of freedom, while $\nabla_\mu$ is the usual covariant derivative compatible with the spacetime metric $g_{\mu\nu}$. Since equation~\eqref{eq:EoMlin} originates from a local action, then $V_\pi^\mu$ can only be of the form\footnote{For timelike $V_\pi^\mu$ this term is similar to friction, which from an action can only appear as fictitious, related to an explicit time-dependence of the kinetic term, i.e.\ the metric, like it is the case in cosmology. Otherwise one cannot obtain friction from the usual local action.}
\begin{equation}
    V_\pi^\mu = \nabla_\nu\Zt^{\mu\nu}\,, \label{eq:eom-vmu}
\end{equation}
which we will assume from here on.

We are free to change the normalisation of $\Zt^{\mu\nu}$ by an overall background-dependent conformal rescaling without changing the leading eikonal approximation~\eqref{eq:ZPP} along with causality and stability. When $\Zt^{\mu\nu}$ is Lorentzian and non-singular, choosing
\begin{equation}
\label{eq:Killing_V}
    Z^{\mu\nu}\equiv (\det\Zt^{\alpha}_{\beta})^{-1/2}\Zt^{\mu\nu}\,, \quad \text{or equivalently} \quad \Zt^{\mu\nu} = (\det Z^{\alpha}_{\beta})^{-1/2}Z^{\mu\nu}\,,
\end{equation}
 we rewrite the equation of motion for perturbations~\eqref{eq:EoMlin} as a sourced (or mixed) Klein-Gordon equation
\begin{equation}
    \overline{\Box} \pi + \overline{M}^2 \pi = \overline{V}_J^\mu \nZ_\mu \psi_J + \overline{\mu}_J \psi_J  \,, \label{eq:EoM-lin-Z}
\end{equation}
where $\overline{\Box}\equiv Z^{\mu\nu} \nZ_\mu \nZ_\nu$
being the d'Alembert operator in the curved \emph{acoustic} spacetime, with acoustic covariant derivative $\nZ_\mu$ compatible with the new acoustic metric $Z^{\mu\nu}$. Here the barred quantities are rescaled by the scalar  $\sqrt{\text{det}Z_{\nu}^{\mu}}$ as $\overline{M}^2=\sqrt{\text{det}Z_{\nu}^{\mu}}\,M_{\pi}^{2}$,  etc.
 In what follows, we will deal mostly with the high-frequency limit of the dynamics. The choice of normalisation $Z^{\mu\nu}$ does not affect the conclusions. The effective background dependent mass $\overline{M}^2$ and mixing terms on the right of eq.~\eqref{eq:EoM-lin-Z} do not contribute in this limit. However, the $\overline{V}_J^\mu$ and $\overline{\mu}_J$ terms would lead to quasiparticle oscillations (similar to neutrino oscillations in the Standard Model for $\overline{\mu}_J$, or Primakoff~\cite{Primakoff:1951iae} or Gertsenshtein~\cite{Gertsenshtein:1962kfm} effects for the kinetic mixing $\overline{V}_J^\mu$) between phonon $\pi$ and other species $\psi_J$. This would result in the non-conservation of flux. At the subleading order in eikonal, beyond geometric optics, both the effective mass $\overline{M}^2$ and the mixing terms $\overline{\mu}_J$ and $\overline{V}^\mu_J$ would contribute.

To simplify our discussion and to concentrate on acoustic geometry, let us neglect these terms. In this simplified case equation of motion~\eqref{eq:EoM-lin-Z} arises from the quadratic action for fluctuations,
\begin{equation}
S_{2}=-\frac{1}{2}\int d^{4}x\sqrt{-S}\,
Z^{\mu\nu} \, \partial_{\mu}\pi \, \partial_{\nu}\pi \,,\label{eq:quadaction}
\end{equation}
where as usual $S\equiv \det (S_{\mu\nu})$ is the metric determinant of the Lorentzian covariant acoustic metric, so that $S=\det S_{\mu\nu}=\det g_{\mu\lambda}\det S^{\lambda}_{\nu}\equiv g\det S^{\lambda}_{\nu}$. This action is still interesting for physical applications. In particular, it is well known that fluctuations in gapless k-\emph{essence}/$P(X)$ are described by this action, see e.g.~\cite{Babichev:2007dw} and older papers for the irrotational superfluid~\cite{Moncrief:1980,Bilic:1999sq}. Moreover, cosmological scalar perturbations of general, not only shift-symmetric, kinetic gravity braiding are also described\footnote{The factorisation and the description above may fail, see e.g.~\cite{Mironov:2023bdq}.} in this way, see e.g.~\cite{Dobre:2017pnt}. Clearly the equation of motion in this case is just a wave equation
\begin{equation}
\overline{\Box}\pi\equiv Z^{\mu\nu} \nZ_\mu \nZ_\nu \pi=0\,.
\label{eq:acoustic_wave}
\end{equation}
Descending to the next order in the eikonal approximation~\eqref{eq:eikonal}, $\mathcal{O}(\epsilon^{-1})$, in the wave equation above we obtain a transport equation for the amplitude $\mathcal{A}$
\begin{equation}
\nZ_{\mu}\left(\left|\mathcal{A}\right|^{2}N^{\mu}\right)=N^{\mu}\partial_{\mu}\left|\mathcal{A}\right|^2+\nZ_{\mu}N^{\mu}\,\left|\mathcal{A}\right|^2=0\,,\label{eq:flux-cons}
\end{equation}
which is just a statement of flux conservation in the acoustic metric --- the change in the intensity
$\left|\mathcal{A}\right|^2$ along the direction of propagation $N^{\mu}$ is determined by the divergence of the bundle of rays in the \emph{acoustic} metric $Z^{\mu\nu}$.
In the eikonal approach, the amplitude is transported with a material derivative containing the group velocity for the wave (see e.g.~\cite[pg.~367]{thorne_modern_2017}). Since in eq.~\eqref{eq:flux-cons} we have the same $N^\mu\partial_\mu$ as obtained in the  definition of the phase four-velocity~\eqref{eq:NullVecsT}, the phase and group velocities are always equal. Let us remark here that while the normalisation of $Z^{\mu\nu}$ does not influence the speed of propagation and causality, since the integral curves of $N^\mu$ are independent of it, it does seemingly affect the conservation of the flux, since $N^\mu$ is inside the derivative in eq.~\eqref{eq:flux-cons}. However, a change of normalisation is a background-dependent conformal transformation of the acoustic metric~\eqref{eq:Killing_V} also requires the redefinition of the acoustic covariant derivative $\nZ_\mu$ and the amplitude.\footnote{In theories where one knows the action like~\eqref{eq:quadaction} the normalization is fixed. Here we want to make a detour to a more general case when the action and therefore the normalisation are unknown.} The amplitude transport equation~\eqref{eq:flux-cons} remains invariant under the acoustic Weyl transformations~\eqref{eq:Weyl_trans} if one assigns conformal weight one to the amplitude: $\mathcal{A}\rightarrow\Omega^{-1}\mathcal{A}$.
Moreover, since the nonmetricity is Weyl-integrable, eq.~\eqref{eq:WeylNonMetricity}, we can even exchange the amplitude for a charge density $\rho$ which is conserved in the \emph{spacetime} itself, while moving along the acoustic geodesics $N^\mu$,
\begin{equation}
    \nabla_\mu (\rho N^\mu) = 0\,,\qquad \text{where}\qquad\rho\equiv \sqrt{\det (S^\alpha_\beta)} \,|\mathcal{A}|^2 \,, \label{eq:eiko-charge}
\end{equation}
so that the conformal weight of $\rho$ is $(-2)$ to compensate for the conformal weight of $N^{\mu}$. For $g$-timelike $N^{\mu}$ one can then write the above flux-conservation equation as $\nabla_{\mu}\left(\varrho\,\mathscr{N}^{\mu}\right)=0$, where we used~\eqref{eq:N_normal} and defined $\varrho\equiv\rho \sqrt{-N^{\alpha}N_{\alpha}} $. The Weyl-invariant quantity $\varrho$ corresponds to the phonons' density in the wave as measured in the wave's proper frame.
In general cases, even if the principal symbol factorises at $\mathcal{O}(\epsilon^{-2})$ and a diagonal basis for the fields $\Psi_I$ can be picked, the flux-conservation equation from $\mathcal{O}(\epsilon^{-1})$ may remain mixed (e.g.~see the recent work in refs.~\cite{Ezquiaga:2020dao, Menadeo:2024uoq}). The flux in~\eqref{eq:flux-cons} would then be not quite conserved without affecting the stability and causality arguments of this paper.

Since eq.~\eqref{eq:eiko-charge} is the high-frequency approximation to~\eqref{eq:acoustic_wave} which is in turn an approximation to the perturbed equation of motion~\eqref{eq:EoM-lin-Z}, $\rho N^\mu$ is just the shift current carried by small fluctuations in the high-frequency limit and approximately conserved whenever the effective mass $\overline{M}^2$ and mixing terms in eq.~\eqref{eq:EoM-lin-Z} can be neglected, though $\rho$ is \emph{not} the shift charge.

For a discussion of the physics of the lowest order in eikonal, beyond geometrical optics, and the complications arising from kinetic and mass mixing see e.g.~\cite{Menadeo:2024uoq}.

On the other hand, by varying this action with respect to the acoustic metric,\footnote{Even though $Z^{\mu\nu}$ is not our dynamical variable.} we can obtain an acoustic energy-momentum tensor (EMT) cf.~\cite{Moncrief:1980}:
\begin{equation}
\TS_{\mu\nu} = -\frac{2}{\sqrt{-S}}\, \frac{\delta S_2}{\delta Z^{\mu\nu}} = \partial_\mu \pi \partial_\nu \pi - \frac{1}{2} S_{\mu\nu} Z^{\alpha\beta}\partial_\alpha \pi \partial_\beta \pi\,, \label{eq:emtZ}
\end{equation}
where we have used $\delta Z = -Z S_{\mu\nu} \delta Z^{\mu\nu}$. One needs to be careful with raising and lowering indices here, so we are using the notation
\begin{equation}\label{eq:TZS}
    \TS_{\mu\nu} \equiv S_{\mu\lambda}\T^\lambda_\nu\,,\quad \text{and}\quad \TZ^{\mu\nu}\equiv Z^{\mu\lambda} \T^\nu_\lambda \,.
\end{equation}
It is the \emph{nonsymmetric} tensor
\begin{equation}
\T^\mu_{\nu}=Z^{\mu\lambda}\partial_\lambda \pi \,\partial_\nu \pi - \frac{1}{2} \delta^{\mu}_{\nu} \,\, Z^{\alpha\beta}\partial_\alpha \pi \partial_\beta \pi\,, \label{eq:T_correct}
\end{equation}
that would be obtained through the Noether procedure (see~\cite{Babichev:2018uiw}) and it is the one which, on equations of motion, is covariantly conserved with respect to $\nZ_{\mu}$,
\begin{equation}
\label{eq:EMT_conserv}
    \nZ_\mu \T^\mu_\nu = \overline\Box \pi\, \partial_\nu \pi = 0 \,.
\end{equation}
Notice that the form of the EMT is as for a canonical scalar field with the complications of the background and non-linear kinetic terms appearing only through the acoustic metric. The dynamics for small fluctuations in arbitrary scalar-field theories is as for a canonical scalar field with the space-time metric replaced with the acoustic metric. The presence of this metric is the key for appearance of asymmetry in $\T^{\mu\nu}$ obtained from a generally covariant action.

We also note that the tensor $\TZ^{\mu\nu}$ of eq.~\eqref{eq:TZS} is also covariantly conserved in the acoustic spacetime, but this is \emph{not} the case for $\TS_{\mu\nu}$. It is worth mentioning that the Abraham-Minkowski controversy also extends to the correct choice of the EMT for electromagnetic waves.

Substituting the eikonal ansatz~\eqref{eq:eikonal} and averaging over phase cycles, leads to the expression
\begin{equation}\label{T=NP}
    \left< \T^\mu_\nu \right> = | \mathcal{A} |^2 N^\mu P_\nu\,,
\end{equation}
i.e.~that in the eikonal limit of the acoustic EMT represents the flux $\mathcal{|A|}^2$ of acoustic momentum  $P_\nu$ moving along the null vectors $N^\mu$. This EMT corresponds to the Minkowski EMT of the electromagnetic wave in media~\cite[eq.~(2.59) and eq.~(6.40)]{Anile_BOOK}.\footnote{We are thankful to Vladimír Karas for pointing out this useful reference.}
The acoustic conservation of $\T^\mu_\nu$ implies the conservation of $\left< \T^\mu_\nu \right>$. However, instead of the acoustic wave equation~\eqref{eq:acoustic_wave} one should assume that both the geodesic equation~\eqref{eq:geodeisc} (in form of momentum transport) and the amplitude transport equation~\eqref{eq:flux-cons} hold, as
\begin{equation}\label{eq:aver_conserved}
\nZ_\mu \left< \T^\mu_\nu \right> =\nZ_{\mu}\left(\left|\mathcal{A}\right|^{2}N^{\mu}\right)\,P_{\nu}+\left|\mathcal{A}\right|^{2}N^{\mu}\overline{\nabla}_{\mu}P_{\nu}\,.
\end{equation}
We would like to stress that even though we obtained~\eqref{T=NP} using the action for perturbations~\eqref{eq:quadaction} which is assumed to be valid in UV and in IR, the physically intuitive meaning of $\left< \T^\mu_\nu \right>$ allows one to consider this as a valid acoustic EMT in the UV also without requiring the existence of any action. Raising the index with the acoustic metric gives $\left<\TZ^{\mu\nu}\right>=|\mathcal{A}|^2N^\mu N^\mu$, which would be the natural conserved tensor for the construction of acoustic angular momentum density, and which is symmetric as is usually required.

The following properties of $\left< \T^\mu_\nu \right>$ are definitely worth mentioning: (i) it is traceless $\left< \T^\mu_\mu \right>=0$, due to~\eqref{eq:NullVecsT}; (ii) contrary to $\TS_{\mu\nu}$ and $\TZ^{\mu\nu}$ from~\eqref{eq:TZS}, the acoustic EMT with indices lowered, $\left< \T_{\mu\nu} \right>= \left< g_{\mu\lambda}\T^\lambda_\nu \right>$, and raised $\left< \T^{\mu\nu} \right>=\left<g^{\mu\lambda} \T^\nu_\lambda \right>$ are asymmetric revealing the spontaneous violation of local Lorentz symmetry due to the presence of the background; (iii) its conformal weight with respect to acoustic Weyl transformations is four. This discussion of the acoustic EMT is also applicable to propagation of high-frequency waves in other systems, in particular, to propagation of electromagnetic waves in plasma and more general media, see e.g.~\cite{Ehlers_I,Ehlers_II,Bicak_Hadrava, deBoer:2017ing}.

It should be stressed that the covariant  conservation of the acoustic EMT using $\nZ_\mu$ generically does not imply the conservation of the acoustic EMT with respect to $\nabla_{\mu}$. Indeed, using~\eqref{eq:disformation}  on equation of motion~\eqref{eq:EMT_conserv}, the spacetime non-conservation of the acoustic EMT reads
\begin{equation}
\nabla_{\mu}\T_{\nu}^{\mu}=\frac{1}{2}Z^{\alpha\beta}\left(\mathbb{T_{\alpha}^{\mu}}\,\nabla_{\nu}S_{\beta\mu}-\mathbb{T_{\nu}^{\mu}}\,\nabla_{\mu}S_{\alpha\beta}\right)\,.
\end{equation}
However, in the presence of a symmetry generated by an acoustic Killing vector field $\xi^\mu$ this non-conserved acoustic EMT still induces conserved currents and charges. The acoustic Killing equation~\eqref{eq:acosutic_Killing}
together with~\eqref{eq:EMT_conserv} imply the acoustic conservation of the corresponding current
 \begin{equation}
     \nZ_\mu \bar{J}^\mu = 0\,\quad \text{where}\quad\bar{J}^\mu \equiv -\mathbb{T}^\mu_\nu \xi^\nu\,.
     \label{eq:consJ_acoustic}
 \end{equation}
 Then by virtue of relation~\eqref{eq:AcouDiv} for divergences of vectors, a rescaled version of it is conserved in the usual spacetime
 \begin{equation}
      \nabla_\mu J^\mu = 0\,,\quad \text{with}\quad J^\mu \equiv \sqrt{\det S^\alpha_\beta}\,\bar{J}^\mu \,, \label{eq:consJ}
 \end{equation}
and is invariant with respect to acoustic Weyl transformations~\eqref{eq:Weyl_trans}: $J^{\mu}$ has conformal weight zero. Note that the other symmetric forms of EMT, $\left<\TZ^{\mu\nu}\right>$ and $\left<\TS_{\mu\nu}\right>$ from~\eqref{eq:TZS}, have conformal weights six and two respectively, so that they are incapable of forming a Weyl-invariant current $J^{\mu}$ from~\eqref{eq:consJ}. This current transports the conserved quantity $P_{\mu}\xi^{\mu}$, see eq.~\eqref{eq:P_0_conserv}, along the acoustic geodesics. The Hamiltonian for perturbations is a particular example of the conserved charge associated with the current $J^{\mu}$. Whether this charge bounds motion depends on whether it itself is bounded. We discuss this further in section~\ref{sec:Hamiltonia}.

\section{Acoustic physics: causality, stability and horizons}\label{sec:twometrics}

We now turn to the core of this paper: the geometry of the acoustic cone from the point of view of some observer which defines their frames and coordinates with respect to the spacetime metric $g_{\mu\nu}$. In the natural $Z$-frame we defined in eq.~\eqref{eq:perp_bar}, the acoustic cone is isotropic and the discussion usually applied to the spacetime metric is valid. We shall see that introducing a second metric uncovers new features.

We will begin by discussing what an observer would see locally, in particular showing that from their point of view --- in the $g$-frame defined by the spacetime metric and the four-velocity of the chosen observer --- the ray cone and the P-cone are no longer the same surface ---  for example, if one is $g$-spacelike, the other is $g$-timelike. We discuss the dispersion relation as perceived by a local observer and the phase velocities of the wave fronts as resulting from the geometry of these cones.

The two cones are nonetheless dual to each other and their geometry encodes the same information. We  demonstrate that the good choice of frame in which information propagates only into the future  is equivalent to requiring that the P-cones cover the frame's spatial hypersurface, i.e.~the energy is real for modes with arbitrary spatial momentum. This choice of a good Cauchy frame allows us to determine in the standard manner if the scalar is a ghost.

In the $g$-frame, there is a separate set of conditions which determine whether information can propagate in all the directions of the frame's spatial hypersurface --- whether or not sound horizons of the scalar are present for an observer. We prove that this is equivalent to having negative-energy modes be available to this observer. The emission of Cherenkov radiation is possible for a source at rest in a frame in which there is a sound horizon.

We also extend the local frame picture to a global one by introducing the standard foliation and showing that if the conditions for a good Cauchy frame are satisfied at every point on a spatial slice for the foliation's normal frame observer, then the spatial slice is a Cauchy surface for the scalar fluctuations.

We then discuss the relation of this foliation to the boundedness of the Hamiltonian for fluctuations, showing that its boundedness is completely determined by two conditions: (i) that the spatial slice is a Cauchy surface and that (ii) the foliation's comoving observers are subsonic with respect to the fluctuations.

\begin{figure}
\begin{subcaptionblock}[B]{0.55\textwidth}
\centering
\includegraphics[width=0.8\textwidth]{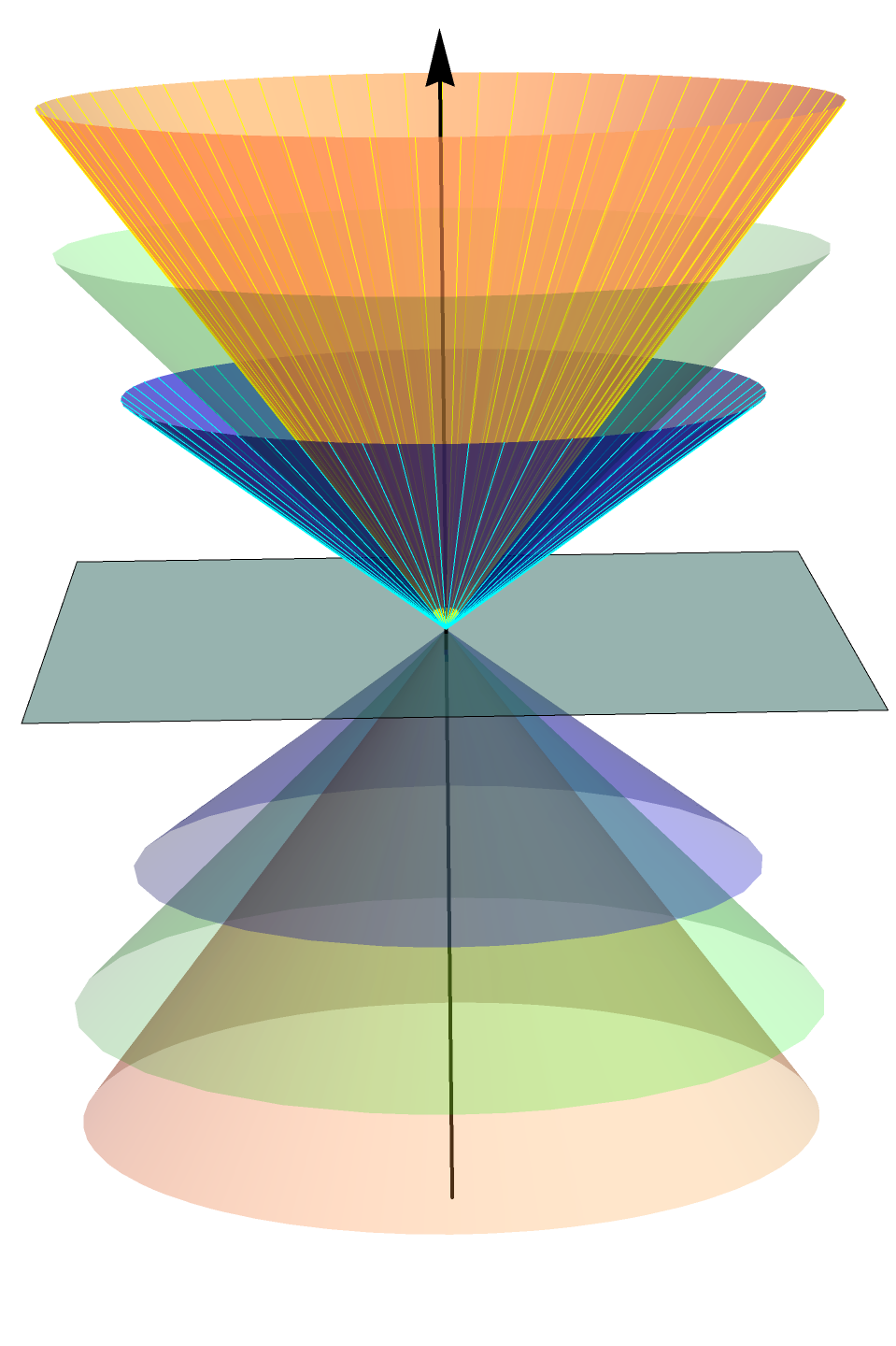}
\caption{\label{fig:isotropic_GoodCauchy_cones}}
\end{subcaptionblock}\quad
\begin{subcaptionblock}[B]{0.35\textwidth}
\centering
\includegraphics[width=0.8\textwidth]{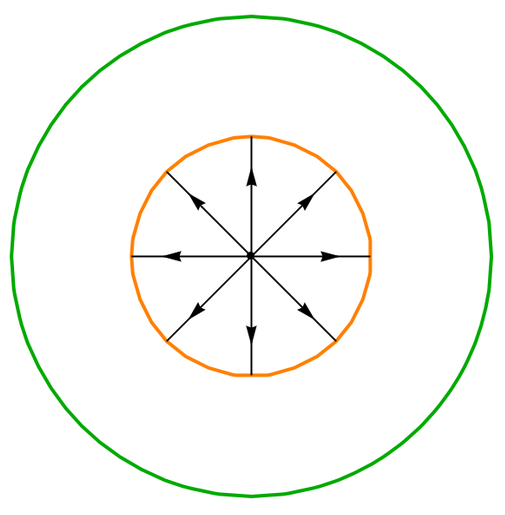}
\caption{\label{fig:phaseVelGoodCauchy} }\includegraphics[width=0.8\textwidth]{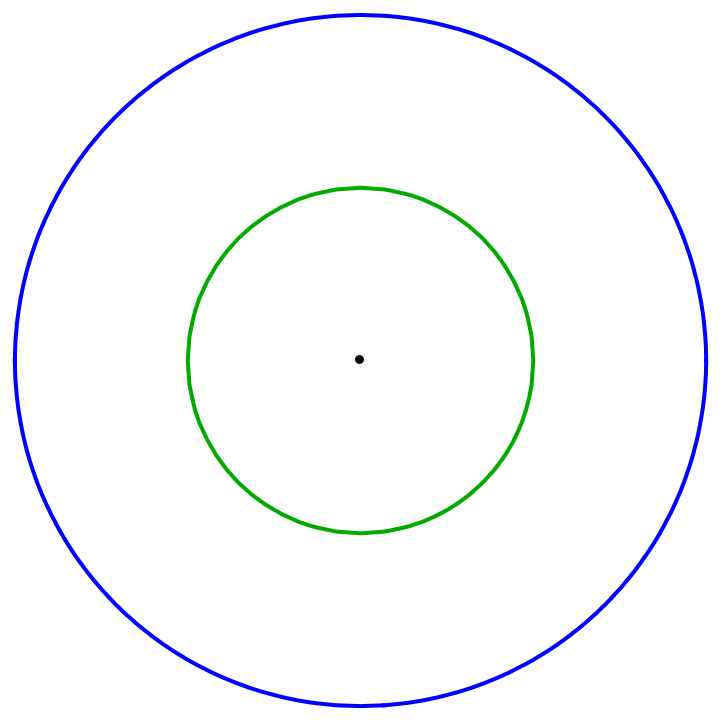}
\medskip
\caption{\label{fig:dispRelGoodCauchy} }
\end{subcaptionblock}
\caption{\looseness=-1 Relative geometry of the acoustic cones with respect to the light cone in the rest frame of an isotropic medium with a \emph{subluminal} sound speed.  (a) The light cone is in green. The acoustic ray cone in orange is inside the light cone and is centred on the observer's worldline $u^\mu$ in this frame. A selection of ray vectors is highlighted on the future nappe of the ray cone. In dark blue we plot the cone formed by the momentum \emph{vectors} $P^\mu=g^{\mu\nu}P_\nu$. The P-cone is $g$-spacelike and also centred on $u^\mu$ in this frame. We have highlighted some momenta in the upper nappe of the P-cone corresponding to the future-facing ray vectors --- the association between these two nappes implies that the scalar is not a ghost. (b) Phase velocity of light rays (green) and the outgoing scalar modes (orange) plotted as the change in the position of a wavefront in the chosen frame. The medium is at rest for the observer, so the phase velocity is isotropic around the origin. (c) Wave-vector surface ($n_\mu$, see eq.~\eqref{eq:n-def}) for the observer at rest: light in green, scalar in blue. The momentum vectors are spacelike for subluminal sound speeds and centred on the observer. All wave vectors come from the upper P-nappe for a non-ghost phonon.\label{fig:isotropic_GoodCauchy}}
\end{figure}

\subsection{The acoustic metric and an observer \label{sec:Acou+Obs}}

Let us introduce an observer with four velocity $u^\mu$, associated to the usual matter sector --- this means that we will normalise $u^\mu$ using the spacetime metric in the usual way $g_{\mu\nu} u^\mu u^\nu =-1$. We then have the usual projector onto this observer's spatial hypersurface
\begin{equation}
    h^\mu_\nu \equiv \delta^\mu_\nu + u^\mu u_\nu\,. \label{eq:h-proj}
\end{equation}
This defines the $g$-frame orthogonal in the usual fashion with respect to the metric $g$ as opposed to $Z$. We can now decompose the momentum and ray vectors analogously to~\eqref{P-Zdecomp},
\begin{align}\label{eq:PN-gdecomp}
    P_\mu &= \omega u_\mu + k_\mu\,,& k_\mu &= h^\nu_\mu P_\nu \,,  \\
    N^\mu&= \mho u^\mu + \dot{r}^\mu\,,& \dot{r}^\mu &= h^\mu_\nu N^\nu \,, \notag
\end{align}
with $\omega\equiv -P_\mu u^\mu$ the frequency of the mode with four momentum $P_\mu$ as would be measured by an observer with $u^\mu$. Thus, the corresponding phonon has the \emph{Minkowski energy} $\omega$ and spatial momentum $k_\mu$. If $u^{\mu}$ is the KV then $\omega$ is conserved along acoustic geodesics, as would be the $k_\mu$ is there is a $g$-spacelike acoustic KV. On the other hand, $\mho\equiv -N^\mu u_\mu $ is the \emph{Abraham energy} of the phonon and would not be conserved along acoustic geodesics.

We can similarly decompose the characteristic equation~\eqref{eq:ZPP}, to obtain the explicit dispersion relation as seen in the frame $u_\mu$ as its roots, $\omega_\pm (k)$.
\begin{equation}
    \omega_\pm (k)= -\frac{1}{Z^{uu}} \left(Z^{u\nu} k_\nu \pm \sqrt{\mathcal{Z}_2^{\mu\nu}k_\mu k_\nu}\right) \,. \label{eq:dispersion-munu}
\end{equation}
We have used the index $u$ to signify contracting with $u_{\mu}$, such that $Z^{u\nu} \equiv Z^{\mu\nu}u_{\mu}$ and as before $Z^{uu} \equiv Z^{\mu\nu}u_\mu u_\nu =-D$ and similarly for $S_{\mu\nu}$.
Note the appearance of $\mathcal{Z}_2^{\mu\nu}$, defined in eq.~\eqref{Z-inducedmetric}, in the square root.
Moreover, we would like to emphasise that $\mathcal{Z}_2^{\mu\nu}$ is \emph{quadratic} in components of $Z^{\mu\nu}$ therefore $\omega_\pm(k)$ is invariant under Weyl transformations $Z^{\mu\nu}\rightarrow\Omega^{-2}(x) Z^{\mu\nu}$. This property is crucial as $\omega_\pm$ is a physical observable, while the acoustic metric in many cases can only be found up to a conformal factor in the leading eikonal approximation. On the other hand, the frequency is a first degree homogeneous function of spatial momenta, as $\omega(\zeta k)=\zeta\omega(k)$. This is also a physically crucial property, as due to the scaling symmetry of the leading eikonal approximation the four-momenta $P_{\mu}$ are also defined only up to rescaling
\begin{equation}
\label{eq:momentum_rescaling}
P_{\mu}\rightarrow \zeta P_{\mu}\,,
\end{equation}
where $\zeta=const>0$. The good Cauchy condition~\eqref{Z-inducedmetric} guarantees that there is a real solution $\omega_\pm$ for eq.~\eqref{eq:dispersion-munu} for any spatial momentum $k_\mu$. We discuss how this is related to the usual notion of a Cauchy surface in section~\ref{sec:CauchySurface}.

Note also that we can relate the energy $\omega$ as seen by an observer to the Abraham energy $\mho$ for this observer,
\begin{equation}\label{eq:bromance}
    \mho = -Z^{uu}\omega - Z^{u\mu}k_\mu,\qquad \omega = -S_{uu}\mho - S_{u\mu}\dot{r}^\mu.
\end{equation}
and the dependence on $k_\mu$/$\dot{r}^\mu$ prevents a simple replacement of $\mho$ for $\omega$ whenever the medium is not at rest in the frame of $u^\mu$. In fact using these in the dispersion relation~\eqref{eq:dispersion-munu} yields,
\begin{equation}\label{eq:mhopm-munu}
    \mho_\pm(k) = -N^\mu u_\mu = -Z^{\mu\nu} P_\nu u_\mu = \pm \sqrt{\Ztwo{\mu\nu}k_\mu k_\nu}\,.
\end{equation}
This expression illustrates the meaning of the two roots $\omega_\pm$. The Abraham energy $\mho_+>0$ always independently of the sign of $Z^{uu}$ --- even for a ghost --- since it is part of the ray, it corresponds to the direction of propagation of signals. When $\Ztwo{\mu\nu}\succ 0$, the future nappe of the ray cone corresponds to $\mho_+$ and it is constructed from a single $\omega_+$ branch of the dispersion relation. For a non-ghost $\omega_+>0$, while $\omega_+<0$ for a ghost. If $\Ztwo{\mu\nu}\nsucc0$, the  modes $k_\mu$ for which $\Ztwo{\mu\nu}k_\mu k_\nu<0$ give complex solutions $\omega$, which signifies that they are not in the P-cone --- they do not propagate. The future nappe of the N-cone is now constructed by both the $\mho_+>0$ and $\mho_-<0$ roots, with the $\mho_-$ part below the surface $\Sigma_u$; correspondingly some roots $\mho_+>0$ now construct the \emph{lower} nappe of the N-cone. Nonetheless, the roots $\omega_{\pm}$ constructing the upper nappe of the P-cone may still be all positive; this is an independent condition --- see section~\ref{sec:Cerenkov}. It is worth noting that, contrary to $\omega$, the quantity $\mho$ is not invariant under conformal transformations of the acoustic metric $Z^{\mu\nu}\rightarrow\Omega^{-2}(x) Z^{\mu\nu}$, but similarly to the frequency it is a homogeneous function of rescalings $\mho(\zeta k)=\zeta\mho(k)$.

There is a dual ``dispersion relation'' for the Abraham energy arising from the ray cone~\eqref{eq:ray-cone},
\begin{equation}\label{eq:dispersion-abr-munu}
    \mho_{[\pm]} (\dot{r})= -\frac{1}{S_{uu}} \left(S_{u\nu} \dot{r}^\nu \pm \sqrt{\Stwo{\mu\nu}\dot{r}^\mu \dot{r}^\nu}\right) \,.
\end{equation}
where $\Stwo{\mu\nu}$ was defined in eq.~\eqref{eq:Stwomunu} and expressed in terms of Abraham momentum $\dot{r}^\mu$ instead of Minkowski momenta $k_\mu$. We will discuss this further in section~\ref{sec:Cerenkov}, noting now that the $[\pm]$ subscript indicates that the roots identification here differs from the $\pm$ choice of eq.~\eqref{eq:mhopm-munu}.

For a wave with ray $N^\mu$ we can then define the speed of the wave \emph{relative} to the observer or the phase three-velocity as\footnote{Cf.~\cite[eq.~(6.11), page 114]{Perlick_BOOK}.}
\begin{equation}
\label{AcPhaseVel}
    v_\text{p}^\mu \equiv -\, \frac{h^\mu_\nu N^\nu }{N^\alpha u_\alpha} = \frac{\dot{r}^\mu}{\mho} \,.
\end{equation}
It is amusing to note that the relative velocity is the ratio of the Abraham momentum to Abraham energy reproducing the simple relativistic formula $\mathbf{v}=\mathbf{p}/E$.  The ray equation $S_{\mu\nu}N^{\mu}N^{\nu}=0$ eq.~\eqref{eq:ray-cone} can now be expressed as a constraint for phase velocity,
\begin{equation}\label{eq:V_ph_constraint}
S_{\mu\nu}v_{\text{p}}^{\mu}v_{\text{p}}^{\nu}+2S_{u\mu}v_{\text{p}}^{\mu}+S_{uu}=0\,.
\end{equation}
Calculating the square of this phase velocity yields
\begin{equation}
\label{V_ph_square}
    v_\text{p}^2= 1+\frac{N^{\mu}N_{\mu}}{\left(N^{\alpha}u_{\alpha}\right)^{2}}=1+\frac{g_{\alpha\beta}\,Z^{\alpha\mu}Z^{\beta\nu}\,P_{\mu}P_{\nu}}{\left(Z^{\mu\nu}P_{\nu}u_{\mu}\right)^{2}} \,,
\end{equation}
and confirms that a $g$-timelike $N^{\mu}$ implies $v_\text{p}^2<1$, a $g$-spacelike $N^{\mu}$ implies $v_\text{p}^2>1$, while $g$-null rays propagate with $v_\text{p}^2=1$ relative to all observers. It is worth noting that this simple formula~\eqref{V_ph_square} looks rather differently from the classical expression~\cite[eq.~(19), page 374]{Synge:1960ueh}. The relative four velocity $v_\text{p}^\mu(u)$ is always purely spatial $v_\text{p}^\mu u_{\mu}=0$ and is well defined even when the speed of the waves is superluminal. It is useful to note that one can rewrite the definition of the relative four velocity~\eqref{AcPhaseVel} in the form of a boost
 \begin{equation}
 \label{eq:superboost}
    N^{\mu}=\left(u^{\mu}+v_{\text{p}}^{\mu}\right)\sqrt{\frac{-N^{\alpha}N_{\alpha}}{1-v_\text{p}^{2}}}\,,\qquad\text{so that}\qquad
    \mho=\sqrt{\frac{-N^{\alpha}N_{\alpha}}{1-v_\text{p}^{2}}}\,,
 \end{equation}
 which is also valid for a superluminal $N^{\mu}$ and which reduces to the standard boost, if it is subluminal
 \begin{equation}
 \label{eq:boost}
    \mathscr{N}^{\mu}=\frac{u^{\mu}+v_{\text{p}}^{\mu}}{\sqrt{1-v_\text{p}^{2}}}\,,
 \end{equation}
 where we used definition~\eqref{eq:N_normal} for the normalised $\mathscr{N}^\mu$.

In analogy with eq.~\eqref{AcPhaseVel},  we can also define a similar object based on the momentum four-covector,
\begin{equation}
    n_\mu \equiv -\, \frac{h^\nu_\mu P_\nu}{u^\alpha P_\alpha} = \frac{k_\mu}{\omega}\,, \label{eq:n-def}
\end{equation}
for which the on-shell condition $N^\mu P_\mu=0$  is equivalent to
\begin{equation}\label{eq:vpn}
    v_\text{p}^{\mu} n_\mu = 1\,.
\end{equation}
$n_\mu$ gives the wave-vector surface in three dimensions, equivalent to the dispersion relation, and can also be seen as a direction-dependent refractive index. In fact this object is the key quantity in the theory of propagation of electromagnetic waves in anisotropic media, see~\cite[pg.~334]{Landafshitz_V8}.
It is instructive to calculate
\begin{equation}
\label{eq:n_square}
n^{2}=1+\frac{P^{\mu}P_{\mu}}{\left(P_{\alpha}u^{\alpha}\right)^{2}}\,,
\end{equation}
which perfectly well matches the so-called \emph{medium equation} from~\cite[eq.~(21), page 376]{Synge:1960ueh}. Note that only in the case of isotropic propagation in the preferred isotropic frame are $n^{\mu}$ and $v_\text{p}^{\mu}$ are collinear. Hence, only in this particular case would the reciprocal relation~\eqref{eq:vpn} reproduce~\cite[eq.~(19), p.~374]{Synge:1960ueh}, see more around eq.~\eqref{eq:usual_vp}. While these observer-dependent objects are the ones usually discussed in wave optics, and are proper tensors, they nonetheless transform non-trivially under a change of observer. As a result, operations such as addition of three-velocities are not particularly natural.

It is worth mentioning that the phase velocity $v_\text{p}^{\mu}$ and the direction-depended refractive index $n_{\mu}$ are both physical observables, and are both Weyl invariant and independent of the rescaling of the four-momentum $P_{\mu}$~\eqref{eq:momentum_rescaling} and of a similar rescaling of the rays $N^{\mu}$.

Again using the decompositions~\eqref{eq:PN-gdecomp} and the on-shell condition $N^\mu P_\mu=0$ we have that an observer $u^\mu$ would see the rays and momenta as
\begin{equation}
    \dot{r}^\mu k_\mu = \omega\mho \,. \label{eq:vp-dir}
\end{equation}
In general it is possible for the phase velocity (Abraham momentum) to have an antiparallel \emph{component} to the mode's momentum, $v_\text{p}^\mu k_\mu <0$. When this happens, the Minkowski energy of the mode is negative in the frame of the observer, $\omega<0$. This implies a negative refractive index, a phenomenon exploited in metamaterials and as we show later, also related to Cherenkov radiation (see section~\ref{sec:Cerenkov}).  Alternatively the mode's ray points toward the past and the Abraham energy is negative, $\mho<0$ (see the section on bad Cauchy frames~\ref{sec:CauchySurface}). Note that changing the signature of $Z^{\mu\nu}$ to a ghost one does not change any of these properties --- for the same ray vector, both the energy $\omega$ and the spatial momentum $k_\mu$ are reversed and therefore the phase velocity is the same as for the healthy mode. We illustrate a simple configuration of the P- and ray cones in figure~\ref{fig:isotropic_GoodCauchy}.
Using~\eqref{T=NP} we can express the acoustic energy density measured by an observer $u^{\mu}$ as cf.~\cite[eq.~(6.41)]{Anile_BOOK}
\begin{equation}\label{eq:epsilon_u}
    \varepsilon (u)=\left< \T^\mu_\nu \right> u^{\nu}u_{\mu}= | \mathcal{A} |^2 \omega \mho \,,
\end{equation}
symmetrically involving both the Minkowski and the Abraham energies, so that~\eqref{T=NP} can be written as
\begin{equation}\label{T=NP_expressed}
    \left< \T^\mu_\nu \right> = \frac{\varepsilon}{\omega \mho}\, N^\mu\, P_\nu\,,
\end{equation}
resembling the standard formula for the EMT for the electromagnetic wave~\cite[(48.15)]{LandavshitzII}.

With the definition of phase velocity~\eqref{AcPhaseVel}, the operator $N^\mu \nZ_\mu$ in the equation for transport of momentum~\eqref{eq:geodeisc} can be reinterpreted as a material derivative for the phase in the frame of the observer $u^\mu$, with the phase velocity playing the role of the flow velocity,
\begin{equation}
    N^\mu \nZ_\mu = \mho \left(u^\mu \nZ_\mu + v_\text{p}^\mu \nZ_\mu \right) \,.
\end{equation}
When $Z^{\mu\nu}$ is sufficiently constant, the covariant derivatives reduce to partial derivatives and we recover the standard expression for a material derivative.

With the projector~\eqref{eq:h-proj} and definition~\eqref{Z-inducedmetric}, we can also rewrite the tensor $\DZ^{\mu\nu}$ as
\begin{equation}
    \DZ^{\mu\nu} = -\frac{\Ztwo{\mu\nu}}{Z^{uu}} = \left ( Z^{\alpha\beta} -\frac{Z^{u\alpha}Z^{u\beta}}{Z^{uu}}\right)h^\mu_\alpha h^\nu_\beta \,. \label{eq:Schur}
\end{equation}
We can see that, in the $g$-frame, $\DZ^{\mu\nu}$ and $\Ztwo{\mu\nu}$ are spatial and that $\DZ^{\mu\nu}$ is in fact the Schur complement of the $u^\mu u^\nu$ block of the metric $Z^{\mu\nu}$. This observation allows us to use some standard results for determinants and inverses, in particular,
\begin{equation}
    \det Z^{\mu\nu} = Z^{uu}\detu \DZ^{\mu\nu} =  -(Z^{uu})^{-2}\detu \Ztwo{\mu\nu} \,. \label{eq:Schur-det}
\end{equation}
where $\detu$ signifies that the determinant is taken in the three-dimensional subspace orthogonal to $u^\mu$:
\begin{equation}
\label{eq:3D_Det}
\text{det}_{u}\mathcal{M}^{\mu\nu}\equiv\frac{h}{3!}\,u_{\mu}\varepsilon_{\,\,\alpha\beta\gamma}^{\mu}\,u_{\nu}\varepsilon_{\,\,\alpha'\beta'\gamma'}^{\nu}\mathcal{M}^{\alpha\alpha'}\mathcal{M}^{\beta\beta'}\mathcal{M}^{\gamma\gamma'}\,\,,
\end{equation}
where $h$ is the determinant of the projector $h^{\mu\nu}$ from~\eqref{eq:h-proj} calculated in coordinates where $u_{\mu}\propto \delta^0_{\mu}$. This formula is applicable for any tensor orthogonal to $u_{\mu}$, so that $\mathcal{M}^{\alpha\beta}u_{\beta}=u_{\alpha}\mathcal{M}^{\alpha\beta}=0$.
\paragraph{The general acoustic metric.}\label{sec:GeneralAcousticMetric}
Given a $g$-timelike $u^\mu$, any general symmetric (2,0) tensor including the acoustic metric $Z^{\mu\nu}$ can be decomposed as
\begin{equation}
    Z^{\mu\nu} = - D u^\mu u^\nu + C h^{\mu\nu}  - u^\mu q^\nu - u^\nu q^\mu  + \sigma^{\mu\nu} \,, \label{eq:GenZ}
\end{equation}
with, a spatial vector $u_\mu q^\mu =0$ with norm $q^2\equiv q_\mu q^\mu \geq0$ and a symmetric, spatial and traceless tensor $\sigma^{\mu\nu}$, $u^\mu \sigma_{\mu\nu} = \sigma^\mu_\mu = 0$. While this construction may appear unnecessarily general, all of these terms are present in the acoustic metric of the kinetic gravity braiding scalar-tensor theory even in the natural unitary-gauge coordinates (see section~\ref{sec:KGB}) and therefore can be concretely realised. This is a well-defined set of models featuring consistent backgrounds  for which it is not possible to boost to the rest frame of the medium, where $q^\mu$ would vanish.\footnote{In the frame with $q^\mu=0$, $S_{u\mu}=0$ also.} We need the fully relativistic approach proposed here to understand such cases at all.

Using decomposition~\eqref{eq:GenZ}, we can rewrite the characteristic equation~\eqref{eq:ZPP} as a direct analogue of the Fresnel equation for the refractive index~\eqref{eq:n-def} as used in crystal optics, see~\cite[pg.~334]{Landafshitz_V8},
\begin{equation}
\label{eq:Fresnel}
\left(Ch^{\mu\nu}+\sigma^{\mu\nu}\right)n_{\mu}n_{\nu}+2n_{\mu}q^{\mu}=D\,,
\end{equation}
which implicitly defines the dispersion relation and which is dual to the phase velocity constraint~\eqref{eq:V_ph_constraint}. It is interesting to note that due to the ``drag'' $q^{\mu}$ this Fresnel equation possesses a linear term in refractive index $n_{\mu}$. This linear term can be removed by the shift ${n}_{\mu}=\bar{n}_{\mu}+c_{\mu}$, provided $\det\left(\mathcal{Z}_{2}^{\mu\nu}-q^{\mu}q^{\nu}\right)\neq0$.

We can define a matrix
\begin{equation}
    Z^\mu_{\nu} \equiv Z^{\mu\alpha} g_{\nu\alpha}\,,\qquad S^\mu_\nu = (Z^{-1})^\mu_\nu \label{eq:ZSud_def}
\end{equation}
which is an operator and has the advantage that its determinant transforms as a scalar.\footnote{We note the apparent similarity of this matrix to the combination $g^{\mu\lambda}f_{\lambda\nu}$ the square root of which appears as the fundamental new object in massive gravity~\cite{deRham:2010kj} and in bimetric theory~\cite{Hassan:2011zd}.} Then since $\det Z^{\mu}_{\nu}=g\,\det (Z^{\mu\nu})$, if $g_{\mu\nu}$ is Lorentzian and itself has cones as characteristic surfaces, then the requirement of the existence of the acoustic cone discussed in section~\ref{sec:signature} is equivalent to\footnote{In principle $g^{\mu\nu}$ could be singular, while $Z^{\mu\nu}$ is not, in which case, the determinant~\eqref{eq:detZ-updown} would diverge and yet this would not signify an issue for $Z^{\mu\nu}$. For the purpose of this work, we are interested in spacetimes without singularities, so we will not complicate the discussion to include such edge cases (see ref.~\cite{Gangopadhyay:2011iu} for such a set up).} \begin{equation}
    \det Z^{\mu}_{\nu} > 0 \,.\label{eq:detZ-updown}
\end{equation}
with the determinant for the general acoustic metric given by
\begin{align}
    \det(Z^{\mu}_{\nu}) &=  D C^3 +C^2 q^2-\frac{1}{2}( q^2+D C )\sigma^{\mu\nu}\sigma_{\mu\nu}+ \label{eq:detZ-gen} \\
    &\quad +\frac{1}{3} D \sigma^{\mu\nu}\sigma_\mu^{\rho}\sigma_{\rho\nu}+ q^\mu q^\nu \left(\sigma_\mu^\rho\sigma_{\rho\nu}-C \sigma_{\mu\nu}\right) \,. \notag
\end{align}
We also have the relation $\det(S^\mu_\nu) \det (Z^\alpha_\beta) = 1$. We will henceforth assume that acoustic cones exist and therefore condition~\eqref{eq:detZ-updown} is satisfied.

In the frame $u_\mu$ we then have
\begin{align}\label{genZ2}
    \Ztwo{\mu\nu}&=D \left( C h^{\mu\nu} + \sigma^{\mu\nu} \right)+ q^\mu q^\nu \,,
\end{align}
and the dispersion relation~\eqref{eq:dispersion-munu} becomes
\begin{equation}
\label{eq:w_q}
    \omega_\pm = \frac{q^\mu k_\mu}{D} \pm \frac{\sqrt{\Ztwo{\mu\nu} k_\mu k_\nu}}{D}\,.
\end{equation}
For the phase velocity we have
\begin{equation}
\label{eq:v_p}
    v_\text{p}^\mu = \frac{C k^\mu +\sigma^{\mu\nu}k_\nu +\omega q^\mu} { \omega D  - q^\alpha k_\alpha} = \frac{q^\mu}{D} \pm \frac{\Ztwo{\mu\nu} k_\nu} {D \sqrt{\Ztwo{\alpha\beta}k_\alpha k_\beta}} \,.
\end{equation}
We can also recover this phase three-velocity from the usual definition of the \emph{group}-velocity, $v_\text{gr}^\mu$, when starting from the dispersion relation~\eqref{eq:dispersion-munu},
\begin{equation}
\label{eq:v_gr}
    v_\text{gr}^\mu\equiv\frac{\partial\omega(k)}{\partial k_\mu} = v_\text{p}^\mu\,.
\end{equation}
As discussed on page~\pageref{eq:flux-cons}, the group velocity for this class of theories is equal to the phase velocity since they are dispersionless and we see this result here.
The presence of the tensor $\sigma^{\mu\nu}$ or/and the ``drag'' $q^\mu$ makes the phase velocity deviate from the direction of $k^\mu$. Only for such media and in such frames where both $q^{\mu}$ and $\sigma^{\mu\nu}$ vanish, one obtains from~\eqref{eq:v_p} and~\eqref{eq:w_q} that
\begin{equation}
\label{eq:usual_vp}
\text{if }\quad\sigma^{\mu\nu} =0\,,\quad q^\mu = 0, \quad \text{then } \quad v_{\text{p}}^{\mu}=\frac{\omega(k)h^{\mu\nu}k_{\nu}}{h^{\alpha\beta}k_{\alpha}k_{\beta}}=\frac{\omega(k)}{k}\,\frac{k^{\mu}}{k}=\sqrt{\frac{C}{D}}\,\frac{k^{\mu}}{k}\,,
\end{equation}
recovering the often used definition of the phase velocity. In particular, even for perfect fluids this restricted expression~\eqref{eq:usual_vp} works only in the rest frame of the medium.

It is worth mentioning that Minkowski and Abraham energies are connected through Doppler-like relations as~\eqref{eq:bromance} and can be rewritten as
\begin{equation}\label{eq:more_bromance}
\mho=\omega \left(D  - q^\alpha n_\alpha\right)\,,\qquad\text{and}\qquad \omega=\mho\left(-S_{uu}-S_{u\mu}v^\mu_\text{p}\right)\,,
\end{equation}
so that for $D=1$ and vanishing drag, $q^{\mu}=0$, the energies are equal. In general, using~\eqref{eq:dispersion-munu} and~\eqref{eq:mhopm-munu}
we can write the first relation in a compact form
\begin{equation}\label{eq:MhoOmega_Z}
    \mho_\pm =\sigma \omega_\pm \mathfrak{n}\,,\qquad \mathfrak{n} \equiv \sqrt{\Ztwo{\mu\nu}n_\mu n_\nu}\,,
\end{equation}
with $\sigma=+1$ for non-ghosts and $\sigma=-1$ for ghosts. We can thus interpret the tensor $\Ztwo{\mu\nu}$ as a metric on the space of refractive indices as observed in the frame $u_\mu$ with the frame-dependent norm $\mathfrak{n}$. Using eq.~\eqref{genZ2} and~\eqref{eq:Fresnel} we obtain
\begin{equation}
\label{eq:square_nfrac}
\mathfrak{n}^{2}=D^{2}-2Dn_{\mu}q^{\mu}+\left(q^{\mu}n_{\mu}\right)^{2}\,.
\end{equation}
Thus, $\mathfrak{n}=|D|$ for the vanishing drag, $q^{\mu}=0$. Without an action, one could imagine rescaling the metric to set $D=1$ in the rest frame to absorb this difference between the energies $\omega$ and $\mho$. Even in such a case, it would reappear immediately upon a boost to another frame. If we have a full action for perturbations, the normalisation of the metric is fixed. Thus, generically one cannot set $\mathfrak{n}$ to 1 and so the conversion factor between Minkowski and Abraham energy is not the refractive index. There also exists a dual relation~\eqref{eq:MhoOmega_S} in terms of phase velocities which we discuss in~\ref{sec:Cerenkov}.

Following the standard solid-state physics approach, see e.g.~\cite[pg.~33]{Callaway}, one can also define the ``reciprocal effective mass tensor'' as
\begin{equation}
\label{eq:reciprocal_M}
\left(M^{-1}\right)^{\mu\nu}\equiv\frac{\partial^{2}\omega_{\pm}}{\partial k_{\mu}\partial k_{\nu}}=\frac{\partial v_\text{gr}^{\mu}}{\partial k_{\nu}}=\pm\frac{\left(\mathcal{Z}_{2}^{\mu\nu}\mathcal{Z}_{2}^{\alpha\beta}-\mathcal{Z}_{2}^{\mu\alpha}\mathcal{Z}_{2}^{\nu\beta}\right)k_{\alpha}k_{\beta}}{D\left(\mathcal{Z}_{2}^{\alpha\beta}k_{\alpha}k_{\beta}\right)^{3/2}}\,.
\end{equation}
Contrary to the ``effective mass'' eq.~\eqref{eq:Pup_from_N}, this tensor is responsible for the inertia of the phonon in case an external force acts to influence its motion. It is worth noting that this tensor is (i) symmetric, (ii) purely spatial  $\left(M^{-1}\right)^{\mu\nu}u_{\nu}=0$ due to~\eqref{Z-inducedmetric}, (iii) transverse $\left(M^{-1}\right)^{\mu\nu}k_{\nu}=0$ to the three-momentum $k_{\mu}$ and (iv) invariant under conformal transformations $Z^{\mu\nu}\rightarrow\Omega Z^{\mu\nu}$. Thus, for forces along the spatial momenta of the phonon or in the limit of very high spatial momenta, the inertia of the phonon diverges. Furthermore, for an arbitrary spatial covector $e_{\mu}$ due to positive-definiteness of eq.~\eqref{Z-inducedmetric} and the Cauchy-Bunyakovsky-Schwarz inequality one obtains
\begin{equation}
\left(\mathcal{Z}_{2}^{\mu\nu}\mathcal{Z}_{2}^{\alpha\beta}-\mathcal{Z}_{2}^{\mu\alpha}\mathcal{Z}_{2}^{\nu\beta}\right)k_{\alpha}k_{\beta}e_{\mu}e_{\nu}=\left(\mathcal{Z}_{2}^{\mu\nu}e_{\mu}e_{\nu}\right)\left(\mathcal{Z}_{2}^{\alpha\beta}k_{\alpha}k_{\beta}\right)-\left(\mathcal{Z}_{2}^{\mu\nu}e_{\mu}k_{\nu}\right)^{2}\geq0\,,
\end{equation}
where equality is only possible for $e_{\mu}\propto k_{\mu}$. Hence the sign of $\left(M^{-1}\right)^{\mu\nu}e_{\mu}e_{\nu}$ is the same as the sign of $\pm D$. As one can foresee, in a good Cauchy frame the forward P-cone nappe is given by $\omega_+$, see eq.~\eqref{eq:N0vsZ2}, thus the sign in~\eqref{eq:reciprocal_M} is ``$+$''. Further, for a non-ghost and a good Cauchy frame $D>0$ so that this ``reciprocal effective mass tensor'' is a non-negative contravariant second rank tensor. On the other hand, for a ghost in a good Cauchy frame $D<0$ so that this ``reciprocal effective mass tensor'' is negative definite.

The Abraham energy and phase velocity is only defined for those modes for which $\Ztwo{\mu\nu} k_\mu k_\nu>0$, i.e.~for those for which the Minkowski energy $\omega$ is real; as we will show, the others do not propagate. We stress that this non-propagation of some modes $k_\mu$  is observer-dependent and is the outcome of having acoustic cones which are not invariant with respect to Lorentz boosts.

The condition of positive definiteness of $\Ztwo{\mu\nu}$~\eqref{Z-inducedmetric} is then equivalent to the statement that all its tensor invariants are positive, namely:
\begin{align}
    \Tr\Ztwo{}&=3 D C  + q^2 \label{eq:Z2invs}>0\,, \\
     (\Tr\Ztwo{})^2-\Ztwo{} {}_{\mu\nu} \Ztwo{\mu\nu}&=D\left(6\alpha C^2+4C q^2 -2q^\mu q^\nu \sigma_{\mu\nu}-\alpha \sigma^{\mu\nu}\sigma_{\mu\nu}\right)>0\nonumber\,, \\
    {\detu} \Ztwo{\mu\nu}&=- D^2 \det (Z^{\mu\nu})>0 \,, \notag
\end{align}
with the last equality resulting from the Schur-complement relationship~\eqref{eq:Schur} and always satisfied for a Lorentzian $Z^{\mu\nu}$. As we will demonstrate in the next section, these conditions together with the hyperbolicity condition~\eqref{eq:detZ-updown} are then a sufficient and necessary condition for $u_\mu$ to be a good Cauchy frame.

For completeness, the inverse of $\Ztwo{\mu\nu}$ is
\begin{align}\label{genZ2inv}
D(\det Z^{\alpha}_{\beta} )\left(\mathcal{Z}_2^{-1}\right)^{\mu\nu}&=\left(C(q^2+D C)-q^\rho \sigma_{\rho\kappa} q^\kappa -\frac{D}{2}\sigma_{\rho\kappa}\sigma^{\rho\kappa}\right)h^{\mu\nu}\\
&\quad -D q^\mu q^\nu +2q^{(\mu}\sigma^{\nu)\rho}q_\rho-(q^2+D C)\sigma^{\mu\nu} + D \sigma^{\mu\rho}\sigma^\nu_\rho\,, \notag
\end{align}
and this expression can be used to calculate $S_{\mu\nu}$ using the standard results involving the Schur complement.

\subsection{Acoustic metric and coordinates}\label{sec:coords}

For a local discussion, it is sufficient and more efficient to remain in the tensor language of the previous section. However, to discuss the global picture, we must extend the frame results to the foliation and it is necessary to connect to coordinates.  We thus take a detour to flag some particularities which must be taken care of if the choice of foliation is \emph{not} synchronous.

Let us foliate the spacetime in hypersurfaces  $\Sigma_t$ of constant time, $t=\text{const}$ using the Arnowitt-Deser-Misner (ADM) decomposition~\cite{Arnowitt:1962hi,Poisson}, expressing the spacetime line element
\begin{equation}
    ds^2 =g_{\mu\nu}dx^{\mu}dx^{\nu}=  - \alpha^2 dt^2 + \gamma_{ij} (dx^i + \beta^i dt)(dx^j + \beta^j dt)\,,
    \label{eq:ADM_metric}
\end{equation}
with $\alpha$ the lapse, $\beta^i$ the shift vector and $\gamma_{ij}$ the spatial metric and $\beta_i\equiv \gamma_{ij}\beta^j$. The standard ADM expressions for the inverse metric are then
\begin{equation}\label{eq:ADM_results}
g^{00}=-\frac{1}{\alpha^{2}}\,,\quad
g^{0i}=\frac{\beta^{i}}{\alpha^{2}}\,,\quad g^{ik}=\gamma^{ik}-\frac{\beta^{i}\beta^{k}}{\alpha^{2}}\,,
\end{equation}
with
\begin{equation}\label{eq:ADM_dets}
    \gamma^{ik}\gamma_{kj}=\delta_j^i\,,\quad \det g = -\alpha^2 \gamma\,,\ \text{and}\  \gamma\equiv \det \gamma_{ij}>0.
\end{equation}

The ADM coordinates naturally define \emph{two particular observers} at each spacetime point: the \emph{comoving} observer (CMO) whose four-velocity $V^{\mu}$ follows the time flow ($V^{\mu}\propto \delta_{0}^{\mu}$) and the \emph{normal frame} observer (NFO) whose four-velocity $U^{\mu}$ follows the normal to the hypersurface of constant time ($U_{\mu}\propto \partial_{\mu}t= \delta_{\mu}^{0}$). In this way for the comoving observer one has
\begin{equation}
V^{\mu}=\frac{1}{\sqrt{-g_{00}}}\,\delta_{0}^{\mu}=\frac{1}{\sqrt{\alpha^{2}-\beta^{2}}}\,\delta_{0}^{\mu}\,,\quad\text{and} \quad V_{\mu}=\frac{g_{\mu0}}{\sqrt{-g_{00}}}=\frac{\left(\beta^{2}-\alpha^{2},\beta_{i}\right)}{\sqrt{\alpha^{2}-\beta^{2}}}\,,
\label{eq:V}
\end{equation}
where $\beta^{2}=\beta^{i}\beta_{i}$, while the NFO is described by
\begin{equation}
U_{\mu}=-\frac{1}{\sqrt{-g^{00}}}\,\delta_{\mu}^{0}=-\alpha\,\delta_{\mu}^{0}\,,\quad\text{and} \quad U^{\mu}=-\frac{g^{\mu0}}{\sqrt{-g^{00}}}=\frac{1}{\alpha}\left(1,-\beta^{i}\right)\,,
\label{eq:U}
\end{equation}
where we used ADM results~\eqref{eq:ADM_results}. For the NFO, the induced metric that is constructed with the projector orthogonal to $U^\mu$,~\eqref{eq:h-proj}, is then coincident with the ADM spatial metric $h_{ij}=\gamma_{ij}$. When $\beta^i\neq0$, these two observers do not coincide and neither of them moves on a geodesic. In the usual spacetime, all observers are $g$-timelike vectors and therefore no implication arises as a result of this subtle difference. We shall see that the difference is key for acoustic spacetimes.

The simple form of $V^\mu$ and $U_\mu$ allows us to relate the components of any tensor to scalars formed with the basis of one of these special observers, but the presence of the lapse and shift means that a component may be observed by neither the NFO nor CMO. In particular, the components of any vector $\mathscr{V}^{\mu}$ are given by $\mathscr{V}^{0}=-\sqrt{-g^{00}}\,\mathscr{V}^{\mu}U_{\mu}$ and $\mathscr{V}_{0}=-\sqrt{-g_{00}}\,\mathscr{V}_{\mu}V^{\mu}$, while for a tensor $\mathscr{T}^{\mu\nu}$ one can write $\mathscr{T}^{00}=-g^{00}\,\mathscr{T}^{\mu\nu}U_{\mu}U_{\nu}$, $\mathscr{T}_{00}=-g_{00}\,\mathscr{T}_{\mu\nu}V^{\mu}V^{\nu}$ and finally $\mathscr{T}_{0}^{0}=\sqrt{g^{00}g_{00}}\,\mathscr{T}_{\mu}^{\nu}U_{\nu}V^{\mu}$. Clearly one can continue in the same fashion for an arbitrary type of tensor quantity. Only for the synchronous frame with $\alpha=1$, $\beta_i=0$ are the components measured\footnote{In particular, in expressions like~\eqref{eq:3D_Det} one should induce the 3D totally antisymmetric Levi-Civita tensor, $\epsilon^{\alpha\beta\gamma}$, from the 4D one $\varepsilon^{\mu\alpha\beta\gamma}=-\left[\mu\alpha\beta\gamma\right]/\sqrt{-g}$, where $\left[\mu\alpha\beta\gamma\right]$ is the permutation symbol, following the contraction with NFO  $\epsilon^{\alpha\beta\gamma}=U_{\mu}\varepsilon^{\mu\alpha\beta\gamma}=\left[0\alpha\beta\gamma\right]/\sqrt{\gamma}$.} by an observer, as both observers do coincide in this case, but such a frame is not possible to set up everywhere in a general spacetime.

We will now relate the observer-dependent decomposition for $P_\mu$ and $N^\mu$ eqs.~\eqref{eq:PN-gdecomp} to the components. The energy and spatial momentum of a phonon observed by the NFO $U^\mu$ is given by
\begin{align}\label{eq:Pnfo}
    \omega_U &= -\frac{P_0-\beta^i P_i}{\alpha} = \alpha P^0 \\
    k^U_i &= P_i, \qquad k_0^U =\beta^i k_i^U, \qquad  (k^U)^i = \gamma^{ij} k^U_j,\notag
\end{align}
 with the last relation confirming that $\gamma_{ij}$ is the metric for the spatial vectors. The corresponding quantities observed by the CMO $V^\mu$ can be expresses as
\begin{align}\label{eq:Pcmo}
    \omega_V &= -\frac{P_0}{\sqrt{\alpha^2-\beta^2}} = \sqrt{\alpha^2-\beta^2} \left(P^0 - \frac{\beta_i P^i}{\alpha^2-\beta^2}\right),\\
    k_V^i &= P^i,\qquad k_V^0 = \frac{\beta_i k_V^i}{\alpha^2-\beta^2},\qquad (k^V)^i = \left(\gamma^{ij}- \frac{\beta^i \beta^j}{\alpha^2}\right)k^V_i.  \notag
\end{align}
which demonstrates that the metric for the CMO's spatial vectors is not $\gamma_{ij}$, but rather modified by $\beta_i$ terms. The two spatial slices are misaligned and therefore different momenta and energies are measured by the two observers, with
\begin{equation}
    k^U_i= k^V_i + \frac{\beta_i \omega_V}{\sqrt{\alpha^2-\beta^2}},\qquad k_V^i = k_U^i - \frac{\beta^i}{\alpha}\omega_U.
\end{equation}

It is important to stress that only in the presence of an acoustic Killing vector field $\xi^{\mu}$ is there a conservation law for momentum~\eqref{eq:P_0_conserv} along an acoustic geodesic. If the Killing vector is aligned with a spatial direction $\xi^\mu = \delta^\mu_i$ then the momentum \emph{component} $P_i$ is conserved along geodesics. By eq.~\eqref{eq:Pnfo} this corresponds to the spatial momentum as observed by the NFO $k_i^U$, but not to that of the CMO, since the spatial slice of the CMO is misaligned with the slicing. This makes $k_i^U$ the natural label for modes as it does not change during free evolution on a translation-invariant background. Moreover, we will demonstrate in section~\ref{sec:CauchySurface} that when the spatial slice of the NFO is a good Cauchy surface, all the modes $k_i^U$ are propagating modes with the mode's energy $\omega_U$ and $\omega_V$ taking real values.

If we have a stationary spacetime with a timelike KV, aligning the coordinate frame so that $\xi^{\mu}=\delta_0^{\mu}$, it is the component $P_0=P_{\mu}\xi^{\mu}$ which is conserved. This is neither $\omega_U$ nor $\omega_V$. However, the energies $-P_0$ and $\omega_V$ differ only by a positive spacetime-dependent redshift factor $\sqrt{\alpha^2-\beta^2}$ and therefore their signs are always the same. $\omega_U$ contains an additional contribution owing to the motion of the NFO and is $P^0$ up to a lapse factor.

A similar discussion applied to the decomposition of $N^\mu$ yields
\begin{equation} \label{eq:nfoN}
     \mho_U = \alpha N^0, \quad \dot{r}^i_U = N^i + \beta^i N^0 = \gamma^{ij}\dot{r}^U_j, \qquad \dot{r}_i^U = N_i,\qquad \dot{r}^0_U = 0.
\end{equation}
From the point of view of the NFO, the ray points to the future whenever the Abraham energy $\mho_U>0$, which has the same sign as the $N^0$ component. However, the motion of the NFO with respect to these coordinates implies that the phase velocity~\eqref{AcPhaseVel} with respect to the NFO, while still spatial, is not parallel to $N^i$ but rather obtains a contribution from the shift,
\begin{equation}\label{eq:vpi-NFO}
    v_{\text{p},U}^i = \frac{\dot{r}^i_U}{\mho_U} = \frac{N^i}{\alpha N^0} + \frac{\beta^i}{\alpha}, \qquad v_{\text{p},U}^0 = 0.
\end{equation}
On the other hand, for the CMO $V^\mu$, we have
\begin{align}\label{eq:N-cmo}
    \mho_V &=\sqrt{\alpha^2-\beta^2} \left(N^0 - \frac{\beta_i N^i}{\alpha^2-\beta^2}\right) = -\frac{N_0}{\sqrt{\alpha^2-\beta^2}}\\
    \dot{r}^i_V &= N^i ,\qquad \dot{r}^0_V = \frac{\beta_i N^i}{\alpha^2-\beta^2}. \notag
\end{align}
The phase velocity relative to $V^\mu$ is much simpler when expressed in term of $N_0$ as opposed to $N^0$,
\begin{equation}\label{eq:vpi-CMO}
    v_{\text{p},V}^i = \frac{\dot{r}^i_V}{\mho_V} = -\frac{\sqrt{\alpha^2-\beta^2}N^i} {N_0},\qquad v_{\text{p},V}^0 = \beta_i v_\text{p}^i .
\end{equation}
and $v_{\text{p},V}^i$ is parallel to $N^i$, but, in the presence of a shift, the phase velocity four-vector is no longer spatial for this observer.

The temporal component of the acoustic EMT in the eikonal limit~\eqref{T=NP} in terms of the observer quantities is,
\begin{equation}\label{eq:T00}
    \T^0_0 = |\mathcal{A}^2| N^0 P_0 = -\frac{\sqrt{\alpha^2 - \beta^2}}{\alpha}|\mathcal{A}|^2 \mho_U \omega_V \,.
\end{equation}
As we discuss in section~\ref{sec:Hamiltonia}, the Hamiltonian density for fluctuations is proportional  to this component of the acoustic EMT. Our general approach, differentiating between $N^\mu$ and $P_\mu$, uncovers that the Hamiltonian density is generically not actually the energy density as observed by any one observer, but rather it is a combination of two types energies observed by two different observers: the Abraham energy observed by the NFO $\mho_U$ rescaled by the lapse, determining if the propagation of modes is into the future of the NFO, and $P_0$ --- the would-be conserved momentum component conserved along geodesics and proportional to the Minkowski energy as observed by the CMO $\omega_V$. Only for static spacetimes, when $\beta^i$ can be zero everywhere, does this subtle difference vanish between the observers vanish. And only in the rest frame of the medium $Z^{0i}=0$ is the EMT proportional to $\omega^2$.

Let us now discuss the components of the acoustic metric connecting them to the decomposition~\eqref{eq:GenZ} with respect to the NFO $U_\mu$ and the associated frame:
\begin{equation}
    Z^{UU} = \alpha^{2}Z^{00},
\end{equation}
while $\Ztwo{U\mu\nu}$ --- the metric on refractive indices for the NFO $U_\mu$ given by eq.~\eqref{Z-inducedmetric} is purely spatial and the lapse factors out,
\begin{equation}
    \Ztwo{Uij} = \alpha^{2} \Ztwobar{ij}\,,\qquad \Ztwo{U00}=0\,,\qquad \Ztwo{U0i}=0\,,
    \label{eq:Z_2_NFO}
\end{equation}
with the components
\begin{equation}\label{eq:Z2bar}
    \Ztwobar{ij}\equiv Z^{0i}Z^{0j}-Z^{00}Z^{ij}.
\end{equation}
The local question of whether $\Sigma_t$ is a good Cauchy frame for the NFO reduces to $\Ztwobar{ij}\succ 0$. In section~\ref{sec:CauchySurface} we discuss how to extend this local notion to the whole $\Sigma_t$ surface.

Note that while the tensor $\Ztwo{U\mu\nu}$ is spatial in these coordinates, $\mathcal{Z}_{2\mu\nu}^U$ is \emph{not}. Nonetheless, since $k_U^0=0$, we have that
\begin{equation}\label{eq:Zup=Zdown}
    \alpha^2 \Ztwobar{ij} k^U_i k^U_j = \Ztwo{U\mu\nu}k^U_\mu k^U_\nu = \mathcal{Z}^U_{2\mu\nu} k_U^\mu k_U^\nu = \mathcal{Z}^U_{2ij}k_U^i k_U^j.
\end{equation}
Thus the good-Cauchy condition $\Ztwobar{ij}\succ 0$ is equivalent to the positivity condition for the spatial components of the spacetime tensor $\mathcal{Z}^U_{2ij}\succ 0$.

Given these expressions, the Minkowski dispersion relation~\eqref{eq:dispersion-munu} as observed by the NFO becomes
\begin{equation}\label{eq:dispersion-nfo}
    \omega_{U,\pm} = -\frac{1}{\alpha Z^{00}}\left( \left(Z^{0i}+Z^{00}\beta^i\right)k^U_i \pm \sqrt{\Ztwobar{ij}k^U_i k^U_j}\right)\,.
\end{equation}
However, the equivalent expression for $P_0/\omega_V$ is
\begin{equation}\label{eq:dispersion-P0}
    \omega_{V,\pm} =\frac{-P_{0,\pm}}{\sqrt{\alpha^2-\beta^2}} =  -\frac{1}{\sqrt{\alpha^2-\beta^2}Z^{00}}\left(Z^{0i}k^U_i \pm  \sqrt{\Ztwobar{ij}k^U_i k^U_j}\right)\,,
\end{equation}
The reality of $\omega_U$ and $\omega_V$ is controlled by the same positivity requirement on $\Ztwobar{ij}$, since the modes $k_i^U$ either propagate or not, and just their energy is boosted when the observer changes. However, the sign of the energy $\omega_U$ and $\omega_V$ for a particular mode $k_i^U$ can differ in the presence of $\beta^i$,
\begin{equation}
    \frac{\omega_V}{\omega_U}< 0 \qquad\Leftrightarrow\qquad \frac{\beta^i}{\alpha} \frac{k^U_i}{\omega_U} > 1 \qquad\Leftrightarrow\qquad \frac{N^i}{\alpha N^0} \frac {k^U_i}{\omega_U} < 0
\end{equation}
with the last expression arising from eq.~\eqref{eq:vpn}. Once the shift is large enough so that $N^i$ and $k_i^U$ have antiparallel components, the energies $\omega_V$ and $\omega_U$ have opposite signs.

We can also compute the dispersion relation for $\omega_V$ in terms of the   spatial momentum components as defined by the CMO, $k_i^V$. Since $V^\mu\propto \delta^\mu_0$, it helps start by lowering the indices on the acoustic inverse metric $Z^{\mu\nu}$ in the characteristic equation~\eqref{eq:ZPP}, $Z_{\mu\nu}P^\mu P^\nu=0$ and then proceed to transform $k_V^i=P^i$ to $k^V_i$ using eqs.~\eqref{eq:Pcmo},
\begin{equation}\label{eq:dispersion-cmo}
    \omega_{V,\pm} = -\frac{\sqrt{\alpha^2 -\beta^2}}{Z_{00}} \left(Z_0^i k_i^V \pm  \sqrt{\Ztwoubar{ij}k_i^V k_j^V} \right)
\end{equation}
with the metric for refractive indices for the CMO $V^\mu$, $\Ztwo{V\mu\nu}$ eq.~\eqref{Z-inducedmetric},
\begin{equation}\label{eq:Ztwoubar}
    \Ztwo{Vij} = \frac{1}{\alpha^2-\beta^2}\Ztwoubar{ij},\qquad     \Ztwoubar{ij} \equiv Z_0^i Z_0^j - Z_{00} Z^{ij}
\end{equation}
Note that only $\mathcal{Z}^V_{2\mu\nu}$ is spatial but not $\Ztwo{V\mu\nu}$. We can still exploit $k_0^V = 0$ to relate
 \begin{equation}
     \mathcal{Z}^{V}_{2ij}k^i_V k^j_V= {\Ztwo{V}}_{\mu\nu} k_V^\mu k_V^\nu = \Ztwo{V\mu\nu}k^V_\mu k^V_\nu = \Ztwo{Vij}k_i^Vk_j^V.
 \end{equation}
We thus find that the reality of $\omega_V$ is determined by the positivity of $\Ztwoubar{Vij}$, or equivalently $\mathcal{Z}^{V}_{2ij}\succ0$. This is a \emph{different} condition to $\Ztwo{Uij}\succ0$. This should have been expected, since the spatial slice for the CMO is in general different than that of the NFO and therefore whether it is a good Cauchy frame is an independent question.

\enlargethispage{-2\baselineskip}

This begs the question which of the three dispersion relations, $\omega_U(k^U)$ eq.~\eqref{eq:dispersion-nfo}, $\omega_V(k^U)$ eq.~\eqref{eq:dispersion-P0} or $\omega_V(k^V)$ eq.~\eqref{eq:dispersion-cmo} is the closest to the one that would be actually measured in an experiment. We re-iterate that the confusion only arises in the presence of a non-zero shift $\beta^i$, i.e.~for spacetimes which are not static. For a detector not moving in its own coordinates, one would expect that the observed energy would be $\omega_V$ --- this is a local measurement in the frame, but which is closely related to $P_0$, the momentum component conserved along geodesics if the detector's time is a KV. On the hand, the determination of spatial momenta could occur either through some measurement of a wavelength by a set of nearby observers, or through some sort of scattering process. The first is a measurement performed on the spatial slice, while the second is related to the spatial momentum conserved in the presence of a translation symmetry, $P_i$. Both of these are related to the frame and therefore to the spatial vectors of the NFO, $k_i^U$. We would thus conclude that the dispersion relation~\eqref{eq:dispersion-nfo}  would be the most physical and closest related to would-be conserved quantities, despite mixing quantities as measured by two different physical observers.

We will use expressions involving the components of the covariant metric $S_{\mu\nu}$, which can be written as
\begin{equation}
S_{\mu\nu}=\left(\begin{array}{cc}
\left(Z^{00}\right)^{-1}\left(1-Z^{0a}\left({\Ztwobar{}}^{-1}\right)_{ab}Z^{0b}\right) & \,\,\quad\left({\Ztwobar{}}^{-1}\right)_{mk}Z^{0m}\\
\left({\Ztwobar{}}^{-1}\right)_{im}Z^{0m} & -Z^{00}\left({\Ztwobar{}}^{-1}\right)_{ik}
\end{array}\right)\,,\label{eq:S_components}
\end{equation}
where the inverse  $({\Ztwobar{}}^{-1})_{ik}$ is defined as usual, $({\Ztwobar{}}^{-1})_{ij}\Ztwobar{jk}=\delta_{i}^{k}$ with $\Ztwobar{jk}$ given by~\eqref{eq:Z2bar}.

The components of $S_{\mu\nu}$ are most easily extracted using the CMO $V^\mu$ with e.g.~$S_{VV}\equiv S_{\mu\nu}V^\mu V^\nu = (\alpha^2-\beta^2)^{-1} S_{00}$. Scalars formed with the NFO velocity, e.g.~$S_{uu}\equiv S_{\mu\nu} u^\mu u^\nu$ are instead easily expressed in terms of the components of the tensor $S^{\mu\nu}=S_{\alpha\beta}g^{\alpha\mu}g^{\beta\nu}$,
\begin{equation}
    S_{UU} = \alpha^2 S^{00} = \alpha^{-2} \left(S_{00} + 2 S_{0i}\beta^i +  S_{ij} \beta^i \beta^j \right). \label{eq:Sup}
\end{equation}
In fact we have
\begin{align}\label{eq:S00S0i}
S^{0}_j &= -\frac{Z^{00}}{\alpha^2}\left (\frac{Z^{0i}}{Z^{00}}+\beta^i\right) ({\Ztwobar{}}^{-1})_{ij}, \\
S^{00} &= \frac{Z^{00}}{\alpha^4} \left[\left(Z^{00}\right)^{-2} - \left (\frac{Z^{0i}}{Z^{00}}+\beta^i\right) ({\Ztwobar{}}^{-1})_{ij} \left (\frac{Z^{0j}}{Z^{00}}+\beta^j\right) \right]. \notag
\end{align}
The difference between such objects is the presence of the shift $\beta^i$ together with the $Z^{0i}$ terms --- this is just the effect of the shift representing the relative velocity between the two observers. Nonetheless, we will see that depending on the subject of interest, objects formed from components of either $S^{\mu\nu}$ or $S_{\mu\nu}$ will appear. In particular, $\Stwo{\mu\nu}$ an analogue of $\Ztwo{\mu\nu}$ defined in eq.~\eqref{eq:Stwomunu} will be of importance in section~\ref{sec:Cerenkov}.
Contrary to $\Ztwo{\mu\nu}$, when $u_\mu$ is the NFO, $\Stwo{\mu\nu}^U$ is not spatial --- $\mathcal{S}_2^{U\mu\nu}$ however is. Nonetheless, the spatial part can be expressed as
\begin{equation}\label{eq:S2bar}
    \Stwo{ij}^U = \alpha^2\Stwobar{ij},\qquad  \Stwobar{ij}\equiv S^{0}_{i}S^{0}_{j}-S^{00}S_{ij}.
\end{equation}
On the other hand, when constructed with the CMO $V^\mu$, $\mathcal{S}_{2\mu\nu}$ is spatial and expressable as
\begin{equation}\label{eq:S2ubar}
    \Stwo{ij}^V = \frac{1}{\alpha^2-\beta^2}\Stwoubar{ij},\qquad  \Stwoubar{ij} \equiv  S_{0i}S_{0j}-S_{00}S_{ij}.
\end{equation}
similarly to the discussion leading to eq.~\eqref{eq:Ztwoubar}.

\subsection{Acoustic metric and Cauchy surface \label{sec:CauchySurface}}

We now would like to ask when a $g$-frame, orthonormal in the metric $g_{\mu\nu}$ can be used to evolve the scalar fluctuations. We take a $g$-timelike velocity $u^\mu$ with projector~\eqref{eq:h-proj} as the induced metric on $\Sigma_u$. We will immediately go to the frame of the NFO of a foliation, since we will show this allows us to extend the discussion away from just a local one to a global condition on the slice $\Sigma_t$ being a Cauchy surface.

Taking the NFO of the slicing $U^\mu$ as the observer, the Abraham dispersion relation equation~\eqref{eq:dispersion-abr-munu} becomes
\begin{equation}
    (\alpha\mho_U)^2 S^{00} + 2(\alpha\mho_U) S^0_i \dot{r}_U^i + S_{ij}\dot{r}_U^i \dot{r}_U^j =0\,, \label{eq:raycone_g}
\end{equation}
with the Abraham energy $\mho_U = -N^\mu U_\mu$. The $0$ indices in $S_{\mu\nu}$ are raised as a result of the conversion from $N^i$ to $\dot{r}_U^i$ through eq.~\eqref{eq:nfoN} and absorbing the shift through relations~\eqref{eq:Sup}.
The ray cone describes the motion of phase and therefore the propagation of wavefronts. To be able to set up the Cauchy problem in some coordinates with $\Sigma_t$ as the hypersurface for arbitrary initial conditions, information must not propagate into the coordinates' past, i.e.~the upper nappe of the ray cone must be completely above $\Sigma_t$, $\mho>0$ (since $\alpha>0$). This is only possible if the ray cone does not intersect $\Sigma_t$ anywhere but the origin, i.e.\
\begin{equation}
    S_{ij} \dot{r}_U^i \dot{r}_U^j = 0 \quad \Rightarrow \quad \dot{r}_U^i = 0 \,. \label{eq:Sijrr}
\end{equation}
This implies that $S_{ij}$ must be either positive or negative definite, since otherwise $S_{ij}\dot{r}_U^i \dot{r}_U^j=0$ is itself a cone of spatial directions on which $\Sigma_t$ is cut.

If at a point the ray cone does intersect $\Sigma_t$ along directions $\dot{r}_*^i$, we are dealing with a bad Cauchy frame at this location  and propagation of information is instantaneous along $\dot{r}^i_*$ or in \emph{this particular frame} even into the \emph{coordinate} past. This means that we are not free to choose any arbitrary set of  initial conditions. However, if in a different frame the situation is normal --- there exists a good Cauchy frame at all --- as a result of general covariance of the underlying theory, the solution obtained there, appropriately transformed, must also be the solution in the bad Cauchy frame. By requiring that this problem does not occur anywhere on the slice, choosing a good Cauchy frame for the NFO everywhere, we ensure that the slice is a Cauchy surface.

Since $\DZ^{\mu\nu}$ is a Schur complement~\eqref{eq:Schur} and spatial, it is the inverse of the spatial part of the metric $S_{\mu\nu}$,
\begin{equation}
    \DZ^{\mu\lambda} \left(S_{\alpha\beta} h^\alpha_\lambda h^\beta_\nu \right)= h^\mu_\nu \,, \label{eq:DZs-cov}
\end{equation}
while hyperbolicity implies that $Z^{uu}\det_u\DZ^{\mu\lambda}=\det Z^{\mu\nu} <0$ for any $u^\mu$. Rewriting this in ADM coordinates for the NFO, as in section~\ref{sec:coords}, this tensor identity in terms of components becomes
\begin{equation}\label{eq:SijZ2inv}
    S_{ik} \Ztwobar{kj}= -Z^{00}\delta^k_i
\end{equation}
with $\Ztwobar{ij}$ defined in eq.~\eqref{eq:Z2bar}, while hyperbolicity in  components is $Z^{00}\det \DZ^{ij} <0$. Since $\det S_{ij} = (\det \DZ^{ij})^{-1}$, the condition for propagation into the future eq.~\eqref{eq:Sijrr}, selects a particular sign for the eigenvalues of the definite matrix $S_{ij}$ and therefore the good-Cauchy condition in these coordinates is
\begin{equation}
    \frac{1}{Z^{00}}S_{ij} \prec 0, \label{eq:GoodCauchySij}
\end{equation}
independent of whether we have a ghost not. If satisfied at every point on the slice, $\Sigma_t$ is a Cauchy surface for \emph{both} ghosts and healthy degrees of freedom.

Let us now recover what the above requirement means from the point of view of the P-cone. In the $Z$-frame we obtained the induced metric $\DZ^{\mu\nu}$ on the hypersurface $\Sigma_u$~\eqref{eq:DZ_def}. Eq.~\eqref{eq:SijZ2inv} implies that the good Cauchy condition~\eqref{eq:GoodCauchySij} for the NFO can be expressed as
\begin{equation}\label{eq:GoodCauchyP}
    \Ztwobar{ij} \equiv Z^{0i}Z^{0j} - Z^{00} Z^{ij} \succ 0\,,
\end{equation}
where we again stress, this is no longer a spacetime tensor condition, but valid when we pick the NFO as the observer, owing to~\eqref{eq:Z_2_NFO}. However this is now a requirement that needs to be satisfied at every point on the spatial slice of the foliation.  The IVP for the scalar is well posed when the NFO is $Z$-timelike everywhere. This allows us to choose arbitrary initial conditions on the spatial slice.

Equivalently, on a Cauchy surface, the roots of the dispersion relation for the \emph{NFO} eq.~\eqref{eq:dispersion-nfo} are real at every point for every spatial momentum vector $k_i^U$. Note that having the same sign for both roots $\omega_{U,\pm}$ is a symptom of neither a ghost nor a Cauchy-frame problem --- see section~\ref{sec:Cerenkov}.

In terms of cone geometry relative to an arbitrary observer, the reality of $\omega$ for all $k_\mu$ can be phrased as  the P-cone covering $\Sigma_u$ completely, or equivalently the $u_\mu$ being inside the P-cone, as was discussed in~\cite{Babichev:2018uiw}. However, since the scalar could be a ghost, the overall sign of $Z^{\mu\nu}$ is a priori unknown and whether $u_\mu$ is $Z$-timelike cannot be determined by testing for the sign of $Z^{\mu\nu}u_\mu u_\nu$.We can extend the local frame condition to a condition on the whole slice by choosing the foliation's NFO as the frame's observer.

Thus, provided $\det Z^\mu_\nu>0$ and therefore the acoustic metric is Lorentzian, we have the following statements for the foliation:
\begin{itemize}
\label{Conditions_Well_IVP}
\item If $\Ztwobar{}$ is positive definite, $\Sigma_t$ is a Cauchy surface, and $Z^{00}<0$ means that the scalar healthy and $Z^{00}>0$ implies it is a ghost --- as per usual;
\item If $\Ztwobar{}$ is not positive definite everywhere, $\Sigma_t$ is not a Cauchy surface. In the region where the test fails, $Z^{00}<0$ means that the scalar is a ghost and $Z^{00}>0$ implies it is healthy --- the opposite to the usual case. This is so since the chosen $U_\mu$ is $Z$-spacelike.
\end{itemize}
Thus prior to answering whether the acoustic metric implies that the scalar is a ghost, one must first check the status of the Cauchy surface in the chosen coordinates. In any case, this bad slicing cannot be used to evolve the system forward and an alternative must be found.

\label{thing:Z2disc} The matrix $\Ztwobar{ij}$ was already discussed in~\cite{Nicolis:2004qq}, where it was referred to as a Lorentz-invariant condition for avoiding gradient instabilities. The gradient instability appears when the system is not hyperbolic and there is no cone
at all. Here, rather, condition~\eqref{eq:GoodCauchyP} is a statement about the (Minkowski) energies as seen by the NFO of a foliation and is a condition defining the spatial slice as a Cauchy surface. It is therefore not a Lorentz invariant quantity but a statement about the particular foliation chosen. Irrespective of whether $\Ztwobar{ij}$ satisfies positivity conditions, there are other observers and other frames that could be used to construct a different --- better --- foliation.

When superluminality is present, at least a part of the P-cone is $g$-timelike. Then there is no guarantee that even if condition~\eqref{eq:GoodCauchyP}
is satisfied in one frame, it will be so in another one related through
a Lorentz boost. Provided that we are not in the acausal situation we discuss on page~\pageref{sec:acausal}, we are guaranteed that there will
be at least one frame where condition~\eqref{eq:GoodCauchyP} is
true.

\begin{figure}
\begin{subcaptionblock}[T]{0.65\textwidth}
\centering
\includegraphics[width=0.8\textwidth]{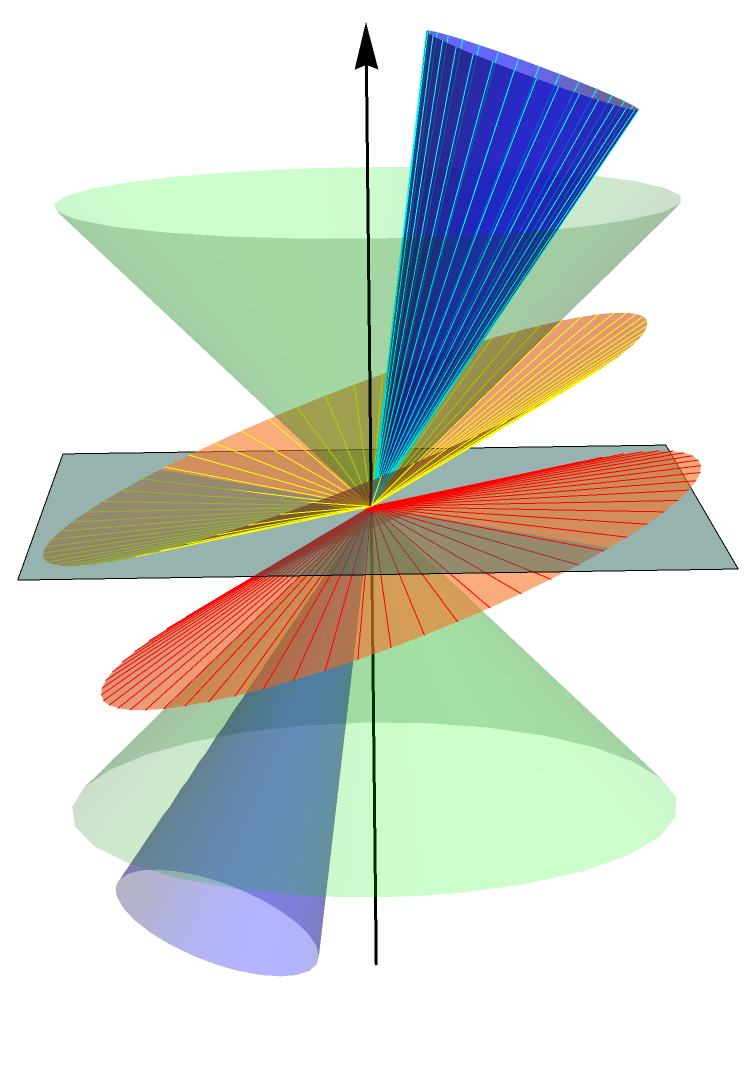}
\caption{\label{fig:conesBadCauchy}}
\end{subcaptionblock}\hfill
\begin{subcaptionblock}[T]{0.35\textwidth}
\centering
\includegraphics[width=0.85\textwidth]{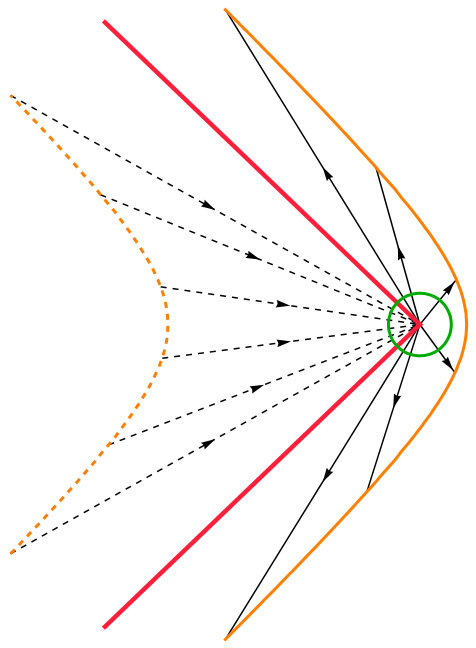}
\caption{\label{fig:phaseVelBadCauchy}}\includegraphics[width=0.85\textwidth]{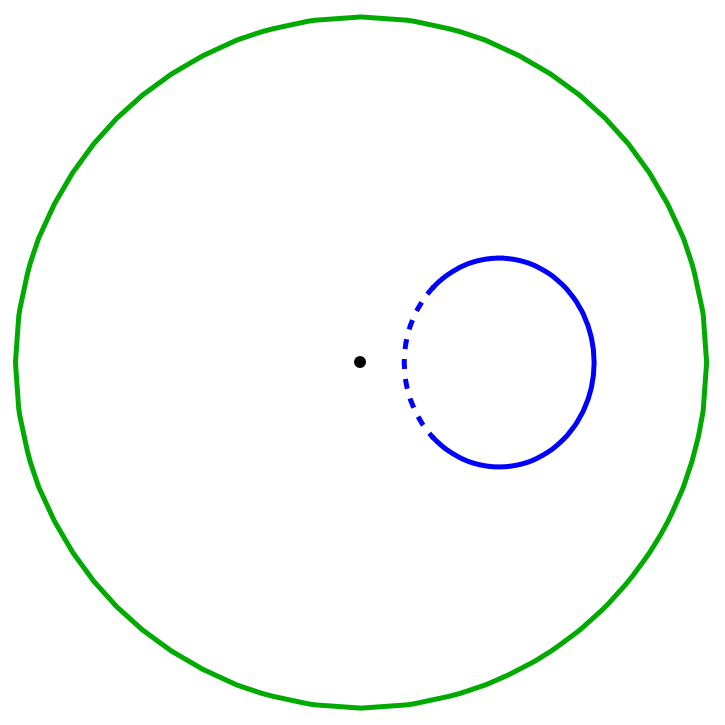}
\caption{\label{fig:dispRelBadCauchy}}
\end{subcaptionblock}
\caption{\looseness=-1 Appearance of cones in a bad Cauchy frame for a medium with isotropic \emph{superluminal} sound speed in its own rest frame. Colour coding of surfaces as in figure~\ref{fig:isotropic_GoodCauchy}. (a) The ray cone (orange) is $g$-spacelike. Selected rays in the future nappe are marked in yellow; those in the past nappe are marked in red. In this bad Cauchy frame, the ray cone cuts the spatial hypersurface (gray plane) and the future nappe propagates information into the \emph{coordinate} past; conversely, the past nappe crosses into the \emph{coordinate} future. For proper Lorentz covariance of solutions, in this frame the modes corresponding to the complete future nappe (yellow) should be selected for the retarded Green's function. For non-ghosts, the acoustic metric maps the upper N-nappe to the complete upper P-nappe (blue with highlighted modes), even in this frame. The observer's world line is outside the P-cone and the P-cone does not cover every spatial momentum. Thus the initial conditions cannot be set arbitrarily on the spatial hypersurface and therefore it is not a Cauchy surface. (b) Motion of phase in this frame (phase velocity). Outgoing modes from the upper N-nappe propagating into the coordinate past, appear as an incoming wavefront (dashed), absorbed during the production of the pulse. In the remaining directions,  an outgoing wavefront is produced (solid orange). The two branches are separated by the (red) spatial cone of directions with instantaneous propagation speed. (c) The wave-vector surface is $g$-timelike and does not contain the frame's energy/time direction. This means that some of the momentum directions are not covered by the P-cone, and therefore some spatial momentum modes \emph{of this frame} do not propagate at all.  The complete upper P-nappe is constructed from both the roots of the dispersion relation~\eqref{eq:dispersion-munu}: $\omega_+$ (solid) maps onto the coordinate future rays ($\mho>0$, solid wavefront in figure~\ref{fig:phaseVelBadCauchy} and $\omega_-$ (dashed) maps onto the (dashed) apparently incoming rays ($\mho<0$).  \label{fig:BadCauchy}}
\end{figure}

\paragraph{Bad Cauchy frame.}
Let us give a brief overview of what changes when we are in a bad Cauchy frame for the scalar. For the NFO, eq.~\eqref{eq:mhopm-munu} becomes
\begin{equation}\label{eq:N0vsZ2}
\mho_{U,\pm} = \alpha N^0_\pm =\pm \alpha\sqrt{\Ztwobar{ij}k_ik_j}\,.
\end{equation}
We can see now that when $\Sigma_t$ is a good Cauchy surface $\Ztwobar{ij}\succ0$, the rays with $\mho_U>0$ all come from the upper nappe of the ray cone. Intersections of the ray cone with constant time surfaces (i.e.~the wavefronts) are ellipsoids which may or may not contain the origin (see figure~\ref{fig:phaseVelGoodCauchy} and section~\ref{sec:Cerenkov}). In a bad Cauchy frame, with $\Ztwobar{ij}\nsucc 0$, a part of the  upper N-nappe points toward the coordinate past: as a result, the wavefronts are no longer closed --- see figure~\ref{fig:BadCauchy} for an illustration.  Moreover, the lower N-nappe also has a part pointing to positive $\mho$, moving to the coordinate future. Since $\mho_U=0$ corresponds to $\Ztwobar{ij}k^U_ik^U_j=0$, each of the nappes of the ray cones is constructed by two branches separated by $\mho_U=0$, and the momenta for which $\Ztwobar{ij}k^U_ik^U_j<0$ and are not in the ray cone at all.

To solve for the evolution one must pick the correct retarded  Green's function. In a good Cauchy frame, this is just given by the \emph{upper} nappe of the ray cone, which is future-facing, $\mho_U>0$. The Green's function should transform continuously under Lorentz boosts even when they are large enough to make the frame a bad Cauchy frame, i.e.~where the upper N-nappe faces partially into the coordinate past, $\mho_U<0$. Thus we should still continue to construct the retarded  Green's function from the complete upper N-nappe to maintain the correct covariance of the solutions. The logic of setting up the Green's function in this way was demonstrated in ref.~\cite{Babichev:2007dw}. In a bad Cauchy frame, one might be tempted to construct the retarded  Green's function from the coordinate-future parts (positive Abraham energy part) of both the N-nappes (see for example ref.~\cite{Dubovsky:2005xd}), but appealing to the geometry of the cones shows that this would lead to an inequivalent solution and in fact is a source of apparent instabilities if one tries to do it. The complication is that one needs to be careful to include the correct branches of the dispersion relation and not to attempt to include the modes for which $\Ztwobar{ij} k^U_i k^U_j <0$ --- they do not propagate at all, or equivalently, are not in the ray cone.

Despite the simple geometrical picture above, what is seen by an observer in a bad Cauchy frame is related to the root structure and therefore not trivial --- for the modes with momenta
\begin{equation}
    \Ztwobar{ij} k^U_{*i} k^U_{*j} = 0\,, \label{eq:fast-cone}
\end{equation}
as a result of eq.~\eqref{eq:vp-dir}, the phase speed diverges on a cone of spatial directions orthogonal to the momentum cone~\eqref{eq:fast-cone}, $\dot{r}_U^i k^U_{*i} =0$. Moreover, inside this spatial cone, the phase speed is directed in the opposite sense to that given by the $\dot{r}_U^i$ (since $\mho_U<0$).  We illustrate this in figure~\ref{fig:BadCauchy}.

The same tensor $\Ztwo{\mu\nu}$ controls the energy difference between the two roots~\eqref{eq:dispersion-munu} of the dispersion relation. The dispersion relation for the energy observed by the NFO~\eqref{eq:dispersion-nfo} gives
\begin{equation}
\omega_{U,+}-\omega_{U,-}=-\frac{2}{\alpha Z^{00}}\sqrt{\Ztwobar{ij}k^U_ik^U_j}\,.\label{eq:omegadifference}
\end{equation}
In a good Cauchy frame, each of the roots corresponds to a different P-nappe. Acoustic metrics for non-ghosts map future-facing rays onto the what is usually called the forward-moving upper nappe of the P-cone formed solely by $\omega_+$. For ghosts --- $\omega_{U+}$ still maps to the future-facing N-nappe but constructs the lower P-nappe --- $\omega_{U+}<0$ and therefore it is sometimes said that ghosts move backward in time. This is not the correct interpretation, since the direction of motion is related to the ray and Abraham energy and not the momentum cone. $\mho$ has the same sign for both ghosts and non-ghosts. Also note that the Minkowski energy $\omega_{U\pm}$ can have the opposite sign to the expected for some modes even in a good Cauchy frame --- we describe this effect related to supersonic motion and negative frequencies in section~\ref{sec:Cerenkov}.

The frame is a bad Cauchy frame whenever the $U_\mu$ direction is not $Z$-timelike.  Then, $Z^{00}$ has the opposite sign to the usual one and intersections of constant energy ($\omega_U$) surfaces with the P-cone (i.e.~the dispersion relation) do not include the momenta for which $\Ztwobar{ij} k^U_{i}k^U_{j}<0$ (see figure~\ref{fig:dispRelBadCauchy}). The modes which \emph{in this frame} propagate  instantaneously  have momenta $k^U_{*i}$, eq.~\eqref{eq:fast-cone} and $\omega_{U+}=\omega_{U-}$, forming the outer edge of the projection of the P-cone onto $\Sigma_t$. The momenta with $\Ztwobar{ij}k^U_ik^U_j<0$ are then not in the P-cone at all and $\Sigma_t$ is not fully covered. In this bad Cauchy frame, both the roots $\omega_{U,\pm}$ form parts of both the P-nappes --- the future P-nappe contains both what would \emph{naively} be called forward- and backward moving modes. This results in two branches for phase speeds for the future modes, one outgoing (formed by the usual $\omega_{U.+}$ branch) and one apparently incoming (from  $\omega_{U,-}$) as shown in figure~\ref{fig:BadCauchy}. We have to include the whole future P-nappe (i.e~the lower P-nappe for ghosts). This again is the natural geometrical construction --- the Fourier transform of the Green's function is constructed from a single P-nappe and even when we boost to a bad Cauchy frame, this is still the case. The roots arrange themselves in such a manner that the future N-nappe is constructed exactly by the same single P-nappe in any frame.

The presence of frames in which the Cauchy problem is ill-posed provides a loophole in the argument from~\cite{Sawicki:2012pz} on instability of generic systems violating the null-energy-condition (NEC). There, it was shown that even if an NEC-violating theory is linearly stable and free of ghost-like perturbations, it still admits configurations with negative energies unbounded from below. Indeed, this result implicitly used the assumption that all observers are equivalent. However, when NEC is violated in a theory with superluminality, the maximal possible boost in which the observer's frame is still a good Cauchy frame corresponds to motion along the P-cone, $Z^{00}=0$. This would give the most negative energy density as measured by an observer for whom causality is meaningfully defined. This energy density is finite, provided the P-cone is inside of the light cone in the direction of the NEC violation. Clearly an observer/source cannot freely create data which make the Cauchy problem ill-posed. Thus, superluminality can save us from non-perturbative instabilities caused by the unbounded negative energies of the whole system. However, the reason is not the same as in~\cite{Dubovsky:2005xd}. This issue requires a detailed case-by-case study.

\paragraph{Momentum-space volume.}
The same $\Ztwobar{ij}$ also appears when we integrate out the $P_0$ direction of the Lorentz-invariant momentum-space volume forcing it to be on shell,
which appears in any phase-space integration (e.g.~\cite{weinberg_quantum_2013}).
In our notation the standard expression for the integral over four-momentum of a quantity $\mathcal{O}(P)$ is given as
\begin{equation}
I=\int\frac{d^{4}P}{(2\pi)^{4}}H(N^0)\delta\left(N^\mu P_\mu\right)\mathcal{O}(P),\label{LI-Pvolume}
\end{equation}
where the delta function ensures the momenta are on-shell in the acoustic
metric and $H(N^ 0)$ is the Heaviside function picking out the future part of the of the ray cone, where the rays should be thought of as functions of the momenta, $N^\mu=Z^{\mu\nu}P_\nu$.

On a foliation which is a Cauchy surface,  $N^0>0$ for all the rays of the future nappe, with $N^{0}$ given by eq.~\eqref{eq:N0vsZ2}. This then implies $\Ztwobar{ij}\succ0$, and this integration can be performed in the standard manner, giving
\begin{equation}
I=\int\frac{d^{3}k^U}{(2\pi)^{3}}\frac{1}{\mho_{U+}}\mathcal{O}(\omega(k^U),k^U_i)=\int\frac{d^{3}k^U}{(2\pi)^{3}}\frac{1}{\alpha\sqrt{\Ztwobar{ij}k^U_ik^U_j}}\mathcal{O}(\omega_U(k^U),k^U_i)\,,\label{eq:P--measure}
\end{equation}
and the remaining spatial momentum integration can proceed without
any further restrictions. $\Ztwobar{ij}$ appears here quite naturally as the spatial metric induced on $\Sigma_t$ by $Z^{\mu\nu}$. Note that this integral is perfectly well-behaved for ghosts.

However, if the chosen foliation is not a  Cauchy surface, the P-cone does not cover all the spatial momenta at every point and no on-shell modes exist for some $k^U_i$. The $P_0$ integration leads to a restricted domain for the spatial momenta, $\Ztwobar{ij}k ^U_ik^U_j>0$. Moreover, mirroring the previous discussion, care would need to be taken to only pick the momenta corresponding to the upper ray nappe. Taking all this into account, setting up any computation in a bad Cauchy frame would be at least extremely non-standard if not impossible.

\paragraph{Frames and gauges.}

The previous discussion demonstrated that not all frames are good for evolving a system causally. Problems can appear in the presence of superluminality and anisotropy.

One of the corollaries of this is that the unitary gauge might fail.
In cosmology, the unitary gauge is a frequently deployed simplification when describing
e.g.~physics during inflation or for dark energy, in which the slicing
is chosen so that scalar perturbations $\pi$ are zero. This
is equivalent to constructing the foliation with the normal frame defined by the gradient of the scalar,
\begin{equation}
u_{\mu}=-\frac{\partial_{\mu}\phi}{\sqrt{2X}}\,,\qquad 2X\equiv - \partial^\alpha \phi \partial_\alpha \phi \label{eq:comovframe}
\end{equation}
with the requirement that the scalar field gradient be $g$-timelike.
This is now readily extended to formulate effective field theories
for the scalar, by enumerating all operators compatible with the remaining rotational symmetry on the spatial slice
in this frame, for example for inflation~\cite{Cheung:2007st} or dark energy~\cite{Gubitosi:2012hu}.

The unitary gauge is perfectly safe on isotropic backgrounds. However, when the background configuration is sufficiently inhomogeneous (e.g.\ large spatial derivatives $\partial_i X$), the tensor  $\Ztwo{\mu\nu}$ in the frame~\eqref{eq:comovframe} can stop being positive definite --- for large enough $q^\mu$ or $\sigma^{\mu\nu}$ one of the invariants~\eqref{eq:Z2invs} can become negative. In such a case, the EFT description defined in the unitary gauge breaks down. However, it is not true that the underlying covariant
theory has broken down --- this problem is the result of the breakdown
of the unitary gauge itself. Provided that on this anisotropic background we still have $\det Z^\mu_\nu>0$ and therefore the system remains hyperbolic, there exists a good Cauchy frame in which we could evolve the scalar with such a background successfully. We demonstrate on an explicit example in section~\ref{sec:fail_unitary}, that it is possible to pick a theory in which an  anisotropic background is perfectly causal (hyperbolic) and a non-ghost and yet the unitary gauge is a bad Cauchy surface.

\paragraph{Multiple degrees of freedom.}

The setup presented above allows for an easy generalisation to multiple degrees of freedom. In principle, every field has its own acoustic metric. When the backgrounds are irrelevant, it is the usual spacetime metric. If the principal symbol~\eqref{eq:general_charact} is factorisable --- it takes the form of another tensor such as $Z^{\mu\nu}$. Locally, the question of whether it is possible to sensibly evolve the fields together boils down to whether there exists a choice of frame
in which the upper nappes of all the ray cones are in the future
and the lower nappes of all the ray cones are in the past everywhere. Since the relative geometry of cones is Lorentz-invariant, the existence of such a good choice of frame is observer independent.

Equivalently, we need to find a common covector $u_\mu$ which is inside the P-cones of all the degrees of freedom. Condition~\eqref{Z-inducedmetric} needs to be satisfied for each of the inverse acoustic metrics simultaneously. Since at least gravity is always present and presumably not a ghost, we have already satisfied this condition for all the degrees of freedom for which the spacetime metric is the acoustic metric by choosing $u_\mu$ to be $g$-timelike, $u_\mu u^\mu = -1$. If there is no superluminality for any of the fields, then any $g$-timelike $u_\mu$ (or $g$-spacelike $\Sigma_u$) gives a good Cauchy frame.

This can then be extended to a global question by asking if there exists a slicing such that the velocity of the NFO is $Z$-timelike everywhere for all degrees of freedom. Notice however, that when event horizons are present, the foliation provided by the ADM decomposition might not a good one inside the horizon. For example in Schwarzschild static coordinates, $g^{00}>0$ inside the horizon and the NFO is $g$-spacelike there.

In general, even if there is superluminality and a $u_\mu$ common to the interior of all the P-cones (or some $\Sigma_u$ exterior to all the ray cones) can be chosen, there is locally at least one set of coordinates in which the evolution can be calculated in the standard manner. In other frames, some of the fields may appear to evolve acausally, but this is just a question of trying to set up the Cauchy problem on a surface which is not a good Cauchy surface and not all possible initial conditions are allowed. The true solution is related to the one from the good Cauchy frame by a boost.

\paragraph{Relation to well-posedness.}
The question of well posedness of quasi-linear partial differential equations is usually approached in the first-order formalism. We will demonstrate here that the conditions for weak hyperbolicity for the scalar field are the same as those for choosing a good Cauchy frame for a hyperbolic operator.

The usual approach (we follow~\cite{Sarbach:2012pr}) is to start with the linearised second-order equation of motion~\eqref{eq:EoMlin} in some chosen set of coordinates,
\begin{equation}
    \partial^2_t \pi + 2B^i \partial_t\partial_i \pi -A^{ij}\partial_i \partial_j\pi =0\,,  \label{eq:EoMforFO}
\end{equation}
with $B^i\equiv Z^{0i}/Z^{00}$ and $A^{ij}\equiv-Z^{ij}/Z^{00}$ where we have assumed that constant-time surfaces are not characteristic ($Z^{00}\neq 0$), we can perform the factorisation as in eq.~\eqref{eq:general_charact}, and we have already dropped the lower-derivative terms not important for the high-frequency limit relevant for causality and well-posedness and we are neglecting gravity so as not to deal with its inherent constraint structure.

The standard procedure then calls for defining $w\equiv \partial_t \pi$, taking a Fourier transform in the spatial directions and rewriting eq.~\eqref{eq:EoMforFO} as a first-order system for the state vector $\vec{u}=(|k|\pi,w)$, with $|k|$ the magnitude of the spatial momentum:
\begin{equation}
    \partial_t \vec{u} = P(ik_i) \vec{u}\quad\text{with}\quad
    P(ik_i) = |k|\begin{pmatrix}\begin{array}{cc}
            0 & 1 \\
            -A^{ij}\hat{k}_i \hat{k}_j & -2iB^i \hat{k}_j
            \end{array}\end{pmatrix}
\end{equation}
with $\hat{k}_i \equiv k_i/|k|$. Given the preceding discussion, we note that the Fourier transform for the fluctuation field $\pi$ are only well defined if the ray cones originating from any point on the spatial hypersurface do not intersect it anywhere but their origins.

The system~\eqref{eq:EoMforFO} is then weakly hyperbolic whenever the eigenvalues of the principal symbol $P(ik)$ are imaginary, i.e.
\begin{equation}\label{eq:wellposed}
    \lambda_\pm = -i\frac{Z^{0i}\hat{k}_j}{Z^{00}}\pm \frac{i}{Z^{00}}\sqrt{\Ztwobar{ij}\hat{k}_i \hat{k}_j}
\end{equation}
By comparing this expression with the dispersion relation~\eqref{eq:dispersion-P0}, we can immediately see that the eigenvalues of the principal symbol should be identified with the energies of the modes,
\begin{equation}
    \lambda_\pm = i \frac{\omega_\pm}{|k|}\,.
\end{equation}
Weak hyperbolicity therefore is exactly the same requirement as the P-cone's covering the spatial hypersurface, i.e.~that we are on a good Cauchy surface of a hyperbolic PDE. Usually to establish weak hyperbolicity, one assumes that the chosen spatial coordinates are spacelike with respect to any possible ray cones and then the weak hyperbolicity establishes that the system was hyperbolic in the first place. With the possibility of superluminality, we do not a priori know that a $g$-spacelike surface that we pick for the IVP is also $S$-spacelike. In our setup, the coordinate-invariant condition on the determinant of the acoustic metric~\eqref{eq:detZ-updown} establishes the hyperbolicity of the linearised equation~\eqref{eq:eom} and the existence of cones while the usual weak hyperbolicity condition then confirms that the chosen coordinates are good to evolve the system.

A much more difficult question is whether the full linearised system containing the scalar and gravity is well posed. If it is possible to factorise the acoustic metrics~\eqref{eq:general_charact} for some chosen background, the principal symbol for the combined state vector of would be block diagonal and therefore the conclusions for our linearised scalar equation are independent of those of gravity. Our requirement that the P-cones of all the fields have a common timelike eigenvector is equivalent to the necessary condition that the system for all the fields is weakly hyperbolic. Then as a result of the block-diagonal form, strong hyperbolicity can be ascertained for each field separately.

The well-posedness of the full non-linear system is even more difficult to assess and beyond the modest aims of this paper. Already the  equation of motion for kinetic gravity braiding is not of the form covered by the Leray theorem (e.g.~see~\cite[pg.~252]{wald_general_2009}). Nonetheless, some headway has been made confirming this desirable property for Horndeski theories, e.g.~\cite{Kovacs:2020ywu}.

\label{sec:acausal}\paragraph{Truly acausal setups.} Let us now turn to sound-cone configurations which are truly acausal  --- no choice of coordinates exists which would be a good Cauchy frame, or in which the complete differential operator is weakly hyperbolic. In particular, this occurs whenever the P-cones do not overlap, not having even one vector that would be timelike for both the metrics, e.g.\ $g^{\mu\nu}$
and $Z^{\mu\nu}$. This means that there is no spatial hypersurface
which would be covered by both the cones and in \emph{all frames} the
energies $\omega_{\pm}$ of some modes of at least one of the fields are complex.

\begin{figure}
\begin{subfigure}[t]{0.49\textwidth}
\centering
\includegraphics[height=0.8\textwidth]{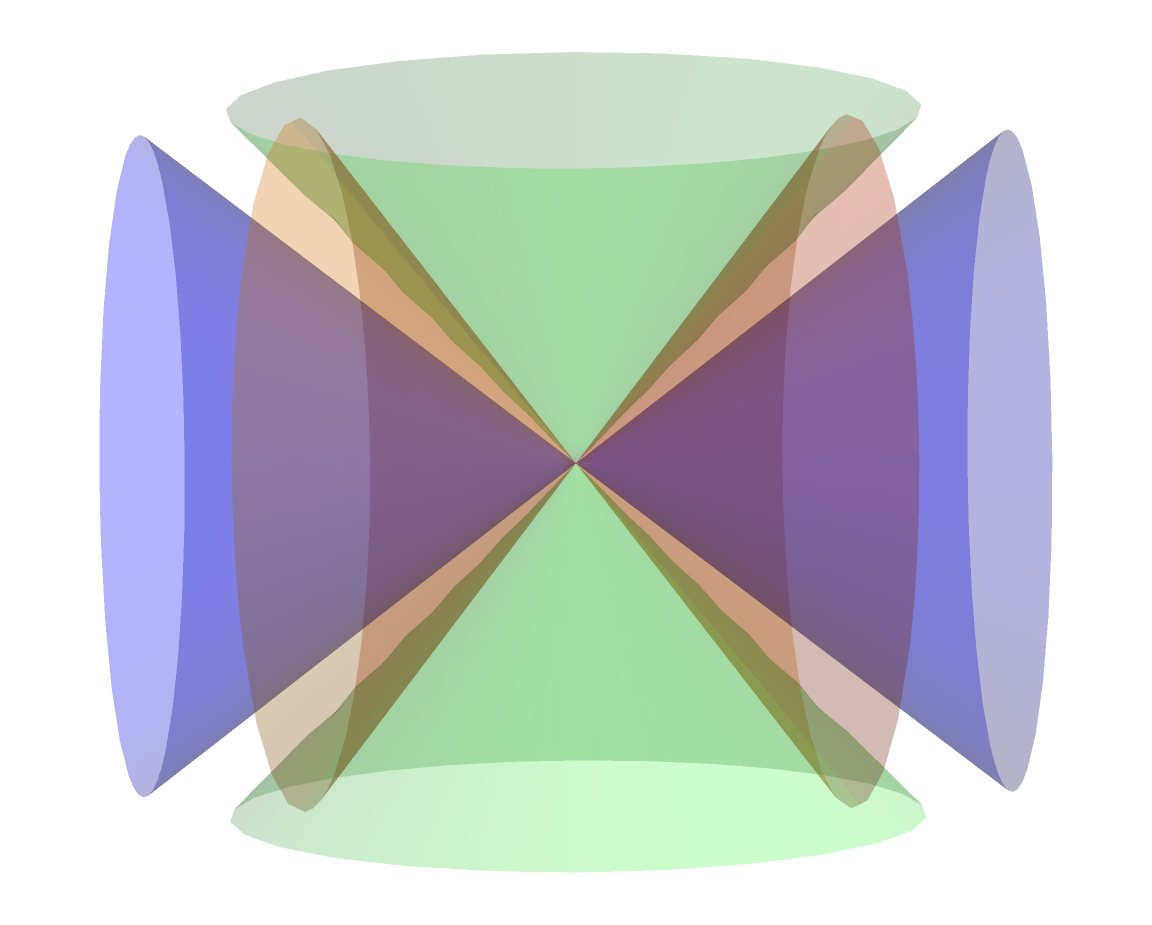}
\caption{\label{fig:ConeAcausal}}
\end{subfigure}\hfill
\begin{subfigure}[t]{0.49\textwidth}
\centering
\includegraphics[height=0.8\textwidth]{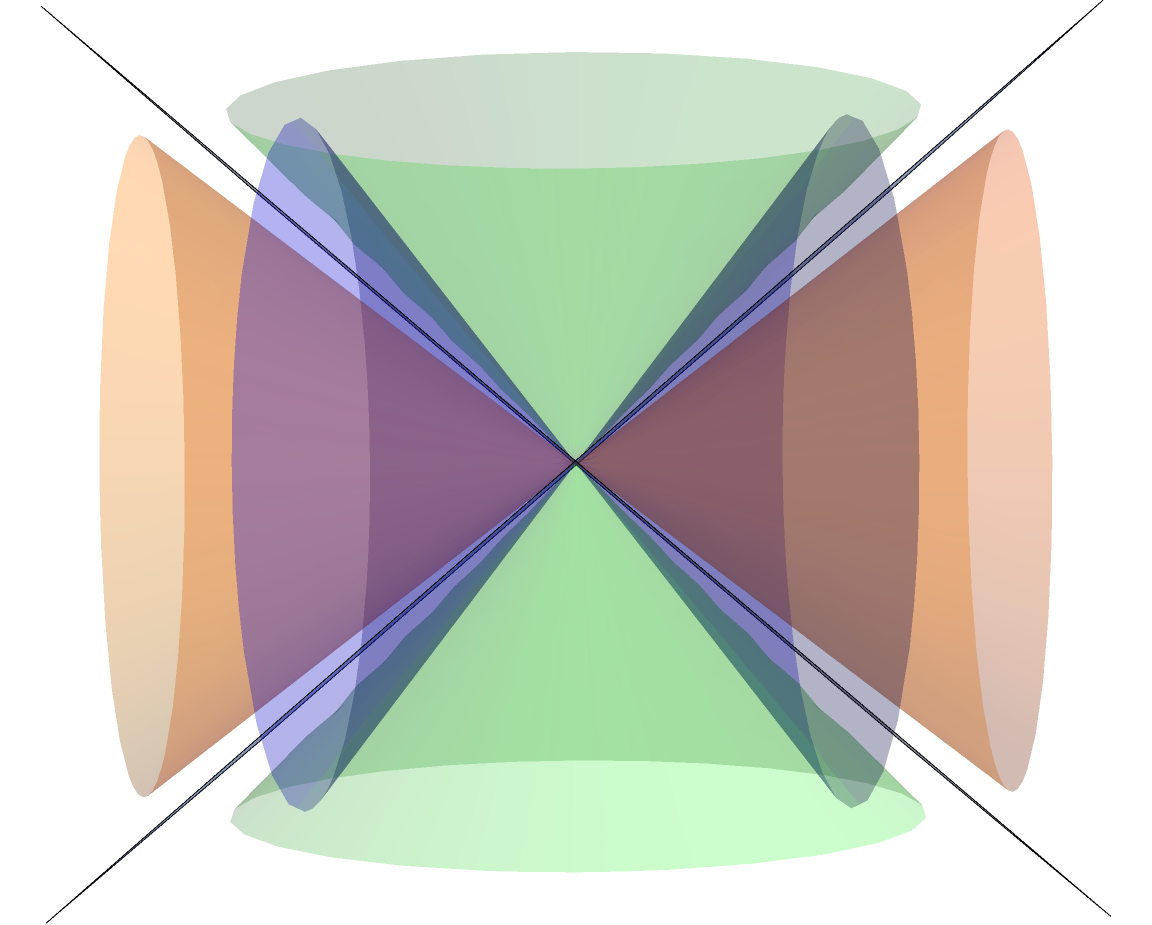}
\caption{\label{fig:ConeNoIntersection}}
\end{subfigure}
\caption{Cone configurations in which one of the acoustic cones has no common vectors with the light cone. Colours as in figure~\ref{fig:isotropic_GoodCauchy}. (a) \emph{Truly acausal configuration}: the future acoustic ray cone (orange) overlaps with both the future and past lightcone nappes (green). There is no spatial hypersurface outside of both the ray cones. Thus in \emph{any} possible frame information propagates both into the future and past and the initial value problem cannot be set up. Equivalently, the acoustic P-cone (blue) does not overlap with the lightcone and there is no common timelike-direction for both of them. Evolving this system is impossible. See page~\pageref{sec:acausal}. (b) \emph{Necessarily transonic configuration}: the acoustic ray cone is completely $g$-spacelike.  Equivalently, the acoustic P-cone intersects both the upper and lower nappes of the lightcone --- in any possible frame, there are always negative energy modes available for both the fields. There are two disjoint classes of spatial hypersurfaces which identity different nappes of the acoustic ray cone as the future. Since no vector common to interiors of both ray cones exists, there is no bounded charge (see section~\ref{sec:otherQ}).  A Cherenkov-like emission process is kinematically allowed from any source and spontaneously. See page~\pageref{sec:transonic}. \label{fig:ConesIb}}
\end{figure}

This pathological setup is equivalent to the situation when the ray cone of one degree of freedom intersects both the future and past of the second (see figure~\ref{fig:ConeAcausal}). No hypersurface exterior to both the cones can be  found.

To elucidate the acausality, let us imagine an experiment where a grid of detectors is set up to coordinize the spacetime of some observer $u^\mu$. The origin of both the space and time coordinate is set to the event of producing a scalar pulse at the location of the observer. A detector upon the passing of the scalar-wave pulse through it responds by sending a \emph{light} signal back to the observer which encodes the triggered detector's coordinates. The observer can then use this information to reconstruct the path taken by the pulse.

In a good Cauchy frame, the reconstructed path has positive time and space coordinates, in the usual manner. In a bad Cauchy frame, for a pulse sent in a direction $\dot{r}^i$ which cuts the spatial hypersurface, $S_{ij} \dot{r}^i \dot{r}^j <0$, the reconstructed time coordinate will be negative. This gives the apparently incoming phase velocity discussed in figure~\ref{fig:phaseVelBadCauchy}. However, the signal from any detector is always received by the observer \emph{after} the pulse is produced and the problem is only related to the reconstructed coordinates and not to causal ordering. In the necessarily acausal setup, the light signal from the detectors arrives \emph{before} the pulse is produced and therefore there is no well-defined causal ordering of events. This is the pathological setup.

When more degrees of freedom are present, it is possible to construct situations in which there exist common Cauchy surfaces for pairs of the fields, but not one for all the degrees of freedom together. This situation is also pathological.

Even if the local frame is a good Cauchy frame, it might prove impossible to extend the Cauchy surface to the whole spatial slice. This depends on the  presence of closed (acoustic) null curves. When eq.~\eqref{eq:GoodCauchyP} is satisfied everywhere on a spatial slice for all degrees of freedom, we are guaranteed that we are not on a background with a closed (acoustic) null curve --- a time machine. This is of course not a guarantee that the background would not nonetheless evolve toward creating such a pathology. However, if a closed null curve were to appear during evolution, there would no longer exist a slice which is a Cauchy surface in some region.

An interesting direction for further study would be to understand whether it is even possible within some effective description to evolve into an acausal one from good initial conditions. The effective theory of fluctuations appears to become strongly coupled whenever $Z^{\mu\nu}u_\mu u_\nu \rightarrow 0$, since canonically normalising the fluctuations causes the interaction terms to diverge. As long as one can change the foliation to remove this singularity, this is just a frame/coordinate problem. However, if the background evolves to the vicinity of true acausality, no such frame changes exist anymore. Since one should be able to reduce the evolution of the background over a small-enough time step to that of the fluctuations, such an acausal configuration should never be reached within the region of validity of the theory. This is in spirit similar to the setup in ref.~\cite{Kaplan:2024qtf}, where it is argued that any space-dependent background of a single field which contains a closed null geodesic curve in the acoustic metric would lead to new irremovable divergences and therefore its formation would be prevented by divergent quantum corrections.

\paragraph{Summary.}
We have demonstrated that even if the acoustic cone exists, it is possible to choose a frame in which the Cauchy problem cannot be solved. To be able to evolve the system in the usual manner, we need to make sure that the $g$-spatial hypersurface $\Sigma_u$ is also $S$-spacelike. This ensures that signals propagate into the frame's future (Abraham energy is positive). We showed that this is equivalent to picking a $Z$-timelike covector $u_\mu$ to define our frame. The sign of $Z^{uu}$ depends on \emph{both} whether the surface is a good Cauchy frame and on whether the degree of freedom is a ghost. We have shown that $u_\mu$ is $Z$-timelike iff the tensor $\Ztwo{\mu\nu}$ is positive definite. In a good Cauchy frame,  $Z^{uu}>0$ implies we have a ghost, while in a bad Cauchy frame, this is exactly a non-ghost. When multiple degrees of freedom are present, these conditions must be satisfied for all of them simultaneously. If they cannot (there is no timelike covector which common to all inverse metrics) then it is impossible to set up initial conditions and evolve. This is a truly acausal situation which is pathological. We have also extended this local picture to a global one showing that the spatial slice of a foliation is a Cauchy surface only if the good Cauchy frame condition is satisfied at every point on the slice for the frame's normal frame observer.

\subsection{Positivity of Hamiltonian\label{sec:Hamiltonia}}

We have thus far discussed the requirement for the existence of acoustic cones
and their geometric configuration consistent with a unique causality
and the possibility of formulating the IVP. However, the usual discussion about instability
focuses on  the fact that
the Hamiltonian for perturbations is not bounded from below. These
two properties are closely related (but not identical) which we will demonstrate here. As usual we will foliate the spacetime with $g$-spacelike equal time $t$ hypersurfaces $\Sigma$ equipped with coordinates $\mathbf{x}$.
The Lagrange functional corresponding to the quadratic action for perturbations~\eqref{eq:quadaction}
\begin{equation}
L[\pi]=-\frac{1}{2}\int_{\Sigma} d^{3}\mathbf{x}\,\sqrt{-S}\,Z^{\mu\nu}\partial_{\mu}\pi\,\partial_{\nu}\pi\,,
\end{equation}
defines the canonical momentum through the variational derivative with respect to $\dot \pi\equiv \partial_t \pi$
\begin{equation}
\label{eq:Pi}
\Pi= \frac{\delta L}{\delta\dot{\pi}}=-\sqrt{-S} \left(Z^{00}\dot{\pi}+Z^{0i}\partial_{i}\pi\right)\,.
\end{equation}
Then the Hamiltonian functional given by\footnote{Here we assume that $\dot\pi$ is expressed through $\pi$ and $\Pi$ using~\eqref{eq:Pi}.}
\begin{equation}
H[\pi,\Pi]=\int_{\Sigma} d^{3}\mathbf{x}\,\Pi\,\dot{\pi}-L\,,
\end{equation}
takes the form
\begin{equation}
H=\frac{1}{2}\int_{\Sigma} d^{3}\mathbf{x}\sqrt{-S}\left(\frac{Z}{Z^{00}}\left(\Pi+\sqrt{-S}\,Z^{0i}\partial_{i}\pi\right)^{2}+Z^{ij}\partial_{i}\pi\,\partial_{j}\pi\right)\,.\label{eq:Hamiltonian}
\end{equation}
It is straightforward to check that
\begin{equation}
\label{eq:H_T}
H=-\int_{\Sigma} d^{3}\mathbf{x}\,\sqrt{-S}\,\T^0_0\,,
\end{equation}
where the acoustic EMT, $\T^{\mu}_\nu$, is given by~\eqref{eq:TZS}.
In the usual manner, only when the chosen time slicing corresponds to an acoustic Killing vector $\xi^\mu\partial_\mu = \partial_t$, is this Hamiltonian a conserved charge, as we have discussed around eq.~\eqref{eq:consJ}. We note that any non-conservation --- being related to the scales with which the acoustic metric changes --- might nonetheless be irrelevant to the problem at hand for modes of sufficiently high frequency. It is worth repeating that $\T^0_0=\sqrt{g^{00}g_{00}}\,\T^{\mu}_{\nu}\,U_{\mu}\,V^{\nu}$, so that the Hamiltonian density does not correspond to the energy density measured by either the CMO or the NFO. In particular, in the high-frequency limit, the EMT reduces to eq.~\eqref{T=NP} so that the Hamiltonian density is proportional to the product of frequency $\omega_V$, as measured by CMO, and $\mho_U$, as measured by the NFO, see~\eqref{eq:T00}. Note that the difference between the NFO and the CMO is the reason for the requirement for the existence of \emph{two} timelike vector fields in the stability analysis of ref.~\cite{Babichev:2018uiw}. In our language this corresponds to requiring the existence of a foliation with a Z-timelike NFO and an S-timelike CMO.

Rescaling the current $\bar{J}^\mu$ conserved in the acoustic metric, gives a current conserved in the spacetime in the usual sense, see eq.~\eqref{eq:consJ}. The conserved charge, however,  is \emph{invariant} with respect to this rescaling
\begin{equation}
H=\int_{\Sigma} d^{3}\mathbf{x}\sqrt{-S}\bar J^{0}=\int_{\Sigma} d^{3}\mathbf{x}\sqrt{-g} J^{0}\,.
\label{eq:invariant_Charge}
\end{equation}
For the conservation of the Hamiltonian we have
\begin{equation}
\label{eq:energy_conservation}
\frac{dH}{dt}=-\oint_{\partial\Sigma} d^{2}\sigma_{i}\,\sqrt{-S}\,\bar J^{i}=-\oint_{\partial\Sigma} d^{2}\sigma_{i}\,\sqrt{-g}\, J^{i}\,,
\end{equation}
where the integral is taken over the 2d boundary of the hypersurfaces $\Sigma$
\begin{equation}
\label{eq:current}
\bar J^{i}=-\T^i_{0}=-Z^{i\alpha}\dot{\pi}\,\partial_{\alpha}\pi\,.
\end{equation}
Without the timelike acoustic Killing vector, the above Hamiltonian is not conserved.

To be bounded from below, the Hamiltonian~\eqref{eq:Hamiltonian} requires that $Z^{00}<0$ and that $Z^{ij}\succ0$ (positive definite). For a more detailed analysis, we use the Schur complement relations~\eqref{eq:Schur},~\eqref{eq:Schur-det} and definition~\eqref{eq:Z2bar} of $\Ztwobar{ij}$ to re-express the Hamiltonian~\eqref{eq:Hamiltonian} as
\begin{equation}
\label{Hamil-Z2}
H=\frac{\sigma}{2}\int_{\Sigma}\frac{d^{3}\mathbf{x}}{\sqrt{\det\Ztwobar{}}}\left(\frac{\det\Ztwobar{}}{\left(Z^{00}\right)^{2}}\,\Pi^{2}+\Ztwobar{ij}\,\partial_{i}\pi\,\partial_{j}\pi\right)-\int_{\Sigma} d^{3}\mathbf{x}\left(\frac{Z^{0i}}{Z^{00}}\right)\Pi\,\partial_{i}\pi\,,
\end{equation}
where
\begin{equation}
\sigma=-\frac{\left|Z^{00}\right|}{Z^{00}}\,,\qquad \text{and}\qquad \det\Ztwobar{}=\det\Ztwobar{ij}\,,
\end{equation}
and we assumed that $Z^{00}$ does not change the sign along the hypersurface $\Sigma$.
As implied by eq.~\eqref{eq:Schur-det}, for a hyperbolic system $\det\Ztwobar{}>0$ always. Only if the Cauchy problem is well-posed on $\Sigma$ is Hamiltonian mechanics meaningful. By the discussion of sections~\ref{sec:coords} and~\ref{sec:CauchySurface}, this requires that the matrix $\Ztwobar{ij}\succ0$. Thus the first integral in~\eqref{Hamil-Z2} is definite: it is positive for non-ghosts ($\sigma=+1$), or negative for ghosts ($\sigma=-1$) and therefore bounded from one side. In both these cases, the dispersion relation~\eqref{eq:dispersion-nfo} implies that there are no linear instabilities, i.e.~frequencies $\omega(k)$ are real for all spatial momenta $k_i$. The second term does not depend on $\sigma$.

If for a hyperbolic system with superluminality we have chosen a foliation where the IVP is ill posed, $\Ztwobar{ij}$ is not positive definite, and has signature $(-,-,+)$. The dispersion relation~\eqref{eq:dispersion-munu} will demonstrate linear instabilities (complex energies) for some wave vectors. The bullet-points discussion on page \pageref{Conditions_Well_IVP} \emph{in this} pathological situation implies:
\begin{itemize}
 \item for non-ghosts $\sigma=-1$ and negative energies are associated with the kinetic term and the gradient energy along \emph{one} principal spatial direction. The dispersion relation~\eqref{eq:dispersion-munu} meanwhile implies that the other two principal spatial directions are associated with a linear instability
 \item for ghosts $\sigma=+1$ the negative energies are associated with gradients in \emph{two} principal spatial directions, as in the dispersion relation, while the kinetic energy is positive.
\end{itemize}
These would-be linear instabilities are artifacts of the incorrect choice of foliation and cannot be exploited by a local source, see~\cite{Babichev:2007dw}.  Nonetheless, the preferred symmetric frame (e.g.~a spherically symmetric and static foliation) may be a frame where the Cauchy problem is ill-posed as discussed in~\cite{Babichev:2017lmw,Babichev:2018uiw}. Then performing the analysis in 2d, for a ghost one might miss the negative energies completely. Indeed, one could be satisfied that the radial and kinetic terms are positive and therefore miss both, the linear instability in two other directions and the ghost nature of the field $\pi$. This can be crucial for investigating stability of spherically symmetric objects, see e.g.~\cite{Babichev:2017lmw, Babichev:2018uiw, Takahashi:2019oxz,Khoury:2022zor}.

The Hamilton equations of motion corresponding to eq.~\eqref{Hamil-Z2} are
\begin{align}
\label{eq:Hamilton_EoM}
  \dot \Pi&=-\frac{\delta H}{\delta\pi}=\partial_{i}\left(\,\sigma\,\frac{\Ztwobar{ij}\,\partial_{j}{\pi}}{\sqrt{\det\Ztwobar{}}}-\frac{Z^{0i}}{Z^{00}}{\Pi}\right)\,,\notag\\
  \dot \pi&=\frac{\delta H}{\delta\Pi}=\frac{\sqrt{\det\Ztwobar{}}}{\left(Z^{00}\right)^{2}}\,\sigma\,{\Pi}-\frac{Z^{0i}}{Z^{00}}\,\partial_{i}{\pi}\,.
\end{align}
with $\sigma$ selected as above. Substituting $\Pi$ from the second equation into the first one obtains the usual wave equation~\eqref{eq:acoustic_wave}.

Further, one can show that the Hamiltonian density
\begin{equation}
\label{eq:Ham_density}
\mathcal{H}=\frac{1}{2}\frac{\det\Ztwobar{}}{\left(Z^{00}\right)^{2}}\,\Pi^{2}+\frac{1}{2}\Ztwobar{ij}\,\partial_{i}\pi\,\partial_{j}\pi-\sqrt{\det\Ztwobar{}}\left(\frac{Z^{0i}}{Z^{00}}\right)\Pi\,\partial_{i}\pi\,,
\end{equation}
as a function of $\Pi$ and $\partial_i\pi$ can violate convexity even on a correct Cauchy surface and in the ghost-free case for which this expression is written above. If convexity is violated, the Hamiltonian density fails to be  bounded from below.
The second derivatives are
\begin{equation}
   \frac{\partial^2\mathcal{H}}{\partial\Pi^{2}}=\frac{\det\Ztwobar{}}{\left(Z^{00}\right)^{2}}\,, \qquad \frac{\partial^2\mathcal{H}}{\partial\Pi\,\partial\partial_{i}\pi}=-\sqrt{\det\Ztwobar{}}\left(\frac{Z^{0i}}{Z^{00}}\right)\,,\qquad \frac{\partial^2\mathcal{H}}{\partial\partial_{i}\pi\,\partial\partial_{j}\pi}=\Ztwobar{ij}\,.
\end{equation}
Taking into account the Sylvester criterion and positive definiteness of $\Ztwobar{ij}$ one finds that $\mathcal{H}$ is convex, if the determinant of the matrix of second derivatives $\hat{\mathcal{H}}''$ is positive. Using the Schur formula one obtains that
\begin{equation}
\label{eq:small_z}
\det\hat{\mathcal{H}''}=\left(\frac{\det\Ztwobar{}}{Z^{00}}\right)^{2}\left(1-z^{2}\right)\,,\qquad\text{where}\qquad z^2=(\Ztwobar{-1})_{ij}Z^{0i}Z^{0j}\,,
\end{equation}
with the inverse  $(\Ztwobar{-1})_{ij}$ defined as usual\label{text:boundedHam}, $(\Ztwobar{-1})_{ik}\Ztwobar{kj}=\delta_{i}^{j}$, see eq.~\eqref{genZ2inv}. Hence for $z^2>1$, i.e.~for sufficiently large $Z^{0i}$, the Hamiltonian density is not convex and is not bounded from below. As we demonstrate in section~\ref{sec:Cerenkov}, $z^2>1$ corresponds to the comoving observer $V^\mu$ moving supersonically, i.e.~the four-velocity $V^\mu$ being S-spacelike. There we also discuss that a particle in this supersonic rest frame can spontaneously emit Cherenkov radiation, see figure~\ref{fig:SoundHorizon}. Thus, the appearance of Cherenkov radiation is in one to one correspondence with the unboundedness of the acoustic Hamiltonian in a good Cauchy frame.

The above unboundedness is entirely owing to the second term of eq.~\eqref{Hamil-Z2},
\begin{equation}
\label{eq:mixingPipi}
\Phi=\int_{\Sigma} d^{3}\mathbf{x}\left(\frac{Z^{0i}}{Z^{00}}\right)\Pi\,\partial_{i}\pi\,,
\end{equation}
mixing momentum with the field. It is instructive to calculate the time evolution of this term giving the negative energies. For simplicity one can assume stationary acoustic geometry. Taking time derivatives of $\Pi$ and $\pi$ and using Hamilton equations of motion~\eqref{eq:Hamilton_EoM} one obtains
\begin{align}
\label{eq:dotPHI}
\frac{d{\Phi}}{dt}&=\int_{\Sigma}d^{3}\mathbf{x}\,\frac{Z^{0i}}{Z^{00}}\left[\partial_{k}\left(\frac{\Ztwobar{kj}\,\partial_{j}\pi}{\sqrt{\det\Ztwobar{}}}\right)\,\partial_{i}\pi+\Pi\partial_{i}\left(\frac{\sqrt{\det\Ztwobar{}}}{\left(Z^{00}\right)^{2}}\,\Pi\right)\right]
\\
&\quad -\oint_{\partial\Sigma}d^{2}\sigma_{i}\,\frac{Z^{0i}}{Z^{00}}\Pi\frac{Z^{0k}}{Z^{00}}\partial_{k}\pi\,,\notag
\end{align}
where on the way we have utilized Gauss's theorem. Furthermore, it is useful to consider sufficiently small spatial volumes where the acoustic geometry is almost constant in comparison with the high-frequency perturbations. In this approximation of almost constant acoustic geometry this expression takes the form of the surface integral:
\begin{align}
\label{eq:dotPHI_surface}
\frac{d\Phi}{dt}\simeq \oint_{\partial\Sigma}d^{2}\sigma_{i}\,&\left[\frac{Z^{0i}}{Z^{00}}\left(\frac{1}{2}\frac{\sqrt{\det\Ztwobar{}}}{\left(Z^{00}\right)^{2}}\,\Pi^{2}-\Pi\frac{Z^{0k}}{Z^{00}}\partial_{k}\pi\right)\right. \\
&~+\left.\frac{Z^{0k}}{Z^{00}}\frac{\Ztwobar{ij}\,\partial_{j}\pi\partial_{k}\pi}{\sqrt{\det\Ztwobar{}}}-\frac{1}{2}\frac{Z^{0i}}{Z^{00}}\frac{\Ztwobar{kj}\,\partial_{j}\pi\partial_{k}\pi}{\sqrt{\det\Ztwobar{}}}\right]\,.\notag
\end{align}
Thus, we have confirmed that time evolution of $\Phi$ is given entirely by the data on the boundary $\partial\Sigma$ under our approximation, as was observed in~\cite{Nicolis:2004qq}. Moreover, energy conservation~\eqref{eq:energy_conservation} implies then that, in this high-frequency approximation, the evolution of the first $\sigma$-dependent term in~\eqref{Hamil-Z2} for $H$ is also given by a surface integral, i.e.~entirely by boundary data. Naively one could think that we can specify boundary data which would fix or even forbid the growth of $\Phi$ providing in this way a lower bound for the Hamiltonian. However, for the only case in which unbounded negative energies are possible, i.e.~in the supersonic case, a part of the boundary $\partial\Sigma$ becomes S-spacelike and corresponds to the \emph{future} of the evolution. It is not physical to impose boundary data in the future in the IVP and therefore the boundary character of the time derivative of $\Phi$ does not save the system from unbounded negative energies. One can illustrate the peculiarities of the supersonic regime by considering static solutions.

\paragraph{Static waves.}\label{sec:H_min}Let us find static configurations of perturbations $(\bar{\pi},\bar{\Pi})$ for $\sigma=+1$.
Expressing $\bar{\Pi}$ from the second Hamilton equation~\eqref{eq:Hamilton_EoM} and plugging into the first we obtain
\begin{equation}
\label{eq:Would_be_Laplace}
\partial_{i}\left(\left(\Ztwobar{ij}-Z^{0i}Z^{0j}\right)\frac{\partial_{j}\bar{\pi}}{\sqrt{\det\Ztwobar{}}}\right)=0\,.
\end{equation}
This second order PDE can either be elliptic\footnote{Note that only the static equation of motion can be elliptic, while we assume that the equation of motion is a usual hyperbolic acoustic wave equation~\eqref{eq:acoustic_wave}.} allowing only for the trivial solutions completely determined by the boundary data\footnote{In simple topology if this equation is elliptic everywhere vanishing boundary conditions imply $\bar{\pi}=0$ and correspondingly $\bar{\Pi}=0$.}  or be hyperbolic --- in which case  nontrivial solutions are possible. These solutions would be waves ``propagating'' not in the four-dimensional spacetime, but just inside the three-dimensional spatial foliation. Crucially the type of this equation can change from region to region in the Cauchy hypersurface $\Sigma$. It may happen that an elliptic region has holes where equation~\eqref{eq:Would_be_Laplace} is hyperbolic.

Let us introduce the spatial covariant derivative $\vec{\nabla}_{i}$ compatible with $\Ztwobar{ij}$ so that
\begin{equation}
\label{eq:Covariant_nabla_Z2}
\vec{\nabla}_{k}\Ztwobar{ij}=0\,.
\end{equation}
Recall that a foliation corresponding to the well-posed Cauchy problem implies that $\Ztwobar{ij}$ is positive definite and corresponds to a proper euclidean contravariant metric in 3d space $\Sigma$. Now we can raise and lower Latin indices using this metric or its inverse. Furthermore, it is convenient introduce a unit 3d spatial vector
\begin{equation}
Z^{i}=\frac{Z^{0i}}{z}\,,\qquad \text{where as in~\eqref{eq:small_z} we use} \qquad z^2=({\Ztwobar{}}^{-1})_{ij}Z^{0i}Z^{0j}\,, \label{eq:z2def}
\end{equation}
along with the associated orthogonal projector
\begin{equation}
\mathcal{P}^{ik}=\Ztwobar{ik}-Z^{i}Z^{k}\,,
\end{equation}
and decomposition of the covariant derivative
\begin{equation}
\label{eq:Decompose}
\vec{\nabla}^{k}=Z^{k}Z^{i}\vec{\nabla_{i}}+\mathcal{P}^{ki}\vec{\nabla_{i}}\equiv Z^{k}\vec{\nabla}_{Z}+\vec{\nabla}^{k\bot}\,.
\end{equation}
Using this 3d covariant notation one can write~\eqref{eq:Would_be_Laplace} as
\begin{equation}
\label{eq:Cov_EoM_pP}
\vec{\nabla}_{i}\left(\left(\Ztwobar{ij}-z^{2}Z^{i}Z^{j}\right)\partial_{j}\bar{\pi}\right)=0\,,
\end{equation}
which using~\eqref{eq:Decompose} expands to
\begin{equation}
\label{eq:Can_be_Wave}
\left(z^{2}-1\right)\vec{\nabla}_{Z}^{2}\bar{\pi}-\vec{\nabla}_{i}^{\bot}\vec{\nabla}^{i\bot}\bar{\pi}+\vec{\nabla}_{i}\left((z^{2}-1)Z^{i}\right)\,\vec{\nabla}_{Z}\bar{\pi}+\left(\vec{\nabla}_{i}^{\bot}\bar{\pi}\right)\vec{\nabla}_{Z}Z^{i}=0\,.
\end{equation}
This equation is elliptic for $z^2<1$ and hyperbolic --- for $z^2>1$. Thus, for $z^2>1$ this is a wave equation with ``time'' in direction along $Z^i$ and ``speed of propagation'' $1/\sqrt{z^2-1}$. We show in section~\ref{sec:Cerenkov} that $z^2$ is related to the supersonic motion of the observer, see eq.~\eqref{eq:S00_z}. The same object appears in the partition function for phonons obtained in~\cite{Kourkoulou:2022doz}. It is worth mentioning that a similar emergence of Lorentz signature from disformally transformed euclidean metrics has been considered in~\cite{Mukohyama:2013ew,Kehayias:2014uta}.

Now we insert the solution $(\bar{\pi},\bar{\Pi})$ into the Hamiltonian~\eqref{Hamil-Z2} to obtain
\begin{equation}
\label{eq:Ham_value}
\bar{H}=\frac{1}{2}\int_{\Sigma}\frac{d^{3}\mathbf{x}}{\sqrt{\det\Ztwobar{}}}\,\left(\Ztwobar{ij}-z^{2}Z^{i}Z^{j}\right)\,\partial_{i}\bar{\pi}\partial_{j}\bar{\pi}\,,
\end{equation}
which on the equation of motion~\eqref{eq:Cov_EoM_pP} becomes just a total derivative,
\begin{equation}
\label{eq:Ham_boundary}
\bar{H}=\frac{1}{2}\int_{\Sigma}\frac{d^{3}\mathbf{x}}{\sqrt{\det\Ztwobar{}}}\,\vec{\nabla}_{i}\left(\left(\Ztwobar{ij}-z^{2}Z^{i}Z^{j}\right)\,\bar{\pi}\partial_{j}\bar{\pi}\right)=\frac{1}{2}\oint_{\partial\Sigma} d^{2}\sigma_{i}\,\frac{\left(\Ztwobar{ij}-z^{2}Z^{i}Z^{j}\right)}{\sqrt{\det\Ztwobar{}}}\,\bar{\pi}\partial_{j}\bar{\pi}\,.
\end{equation}
The Hamiltonian for these static solutions can be non-vanishing, its value is completely determined by data on the 2d boundary $\partial\Sigma$ of the 3d Cauchy hypersurface $\Sigma$. Note that this is true regardless of the nature, hyperbolic or elliptic, of equation~\eqref{eq:Can_be_Wave}.

The type of equation~\eqref{eq:Can_be_Wave} dictates the type of boundary data needed. In particular, in the hyperbolic case, the problem of finding the static configuration or --- ``frozen wave'' --- can be ill-posed when trying to provide boundary data in the ``future'' along $Z^i$. One can speculate that, the most interesting situation occurs when in different regions of $\Sigma$ equation~\eqref{eq:Can_be_Wave} has different types. Suppose regions --- ``holes'' --- where equation~\eqref{eq:Can_be_Wave} is hyperbolic are immersed in a larger region where it is elliptic. Solving an elliptic equation in this larger region starting from the external boundary $\partial \Sigma$ still requires data on internal boundaries separating hyperbolic and elliptic regimes. Crucially the boundary data are usually not provided on such internal boundaries. On the other hand, nontrivial solutions inside hyperbolic regions can play a role of charges which provide the internal boundary data. Thus, it seems that even trivial data on $\partial \Sigma$ may not preclude the existence of nontrivial static solutions in $\Sigma$, provided there are hyperbolic regions inside. Of course, these static solutions, if they exist, do extremise the Hamiltonian functional due to vanishing of both functional derivatives in~\eqref{eq:Hamilton_EoM}.
Further, it is useful to note that one can also extremise the local Hamiltonian density~\eqref{eq:Hamiltonian},
as a function of $\Pi$ and $\partial_i\pi$. In that case conditions for extremum are the second equation from~\eqref{eq:Hamilton_EoM} and the first equation from there without the partial derivative. Thus, the local extremum (saddle point) is reached on zero eigenvectors of $Z^{ik}$ as
\begin{equation}
\left(\Ztwobar{ij}-z^{2}Z^{i}Z^{j}\right)\partial_{j}\tilde{\pi}\propto Z^{ij}\partial_{j}\tilde{\pi}=0\,.
\end{equation}
For positive definite $Z^{ik}$ there are only trivial solutions. However, even nontrivial configurations $\tilde{\pi}$ existing only for $z>1$ have vanishing Hamiltonian~\eqref{eq:Ham_value}. Clearly $\tilde{\pi}$ also satisfy~\eqref{eq:Cov_EoM_pP}. However, these algebraic solutions $\tilde{\pi}$ build a subclass among $\bar{\pi}$. The key difference between these configurations is that $\tilde{\pi}$ may not satisfy boundary conditions, while $\bar{\pi}$ is capable of that.

\subsection{Is motion bounded by charges other than the Hamiltonian? \label{sec:otherQ}}
As we saw in section~\ref{sec:Hamiltonia}, there exist slicings in which the acoustic Hamiltonian is not bounded from below and therefore does not confine the evolution of the fluctuations. Could a different charge exist which nonetheless does so? In this section we generalise the discussion of the Hamiltonian, showing that if there exists an $S$-timelike acoustic Killing vector, the Noether charge associated to it is bounded and therefore it confines the motion of perturbations.
Rather than adapting the slicing to the Killing vector as in section~\ref{sec:Hamiltonia}, let us pick an arbitrary frame $u_\mu$ and the Killing vector $\xi^\mu$, and form the current $\bar{J}^\mu_{\xi} \equiv -\mathbb{T}^\mu_\nu \xi^\nu$,
following discussion around equation~\eqref{eq:consJ_acoustic}. This covariant conservation equation yields
\begin{equation}
\label{eq:covariant_charge}
\frac{dQ_\xi}{dt}\equiv\frac{d}{dt}\int_{\Sigma}d^{3}\mathbf{x}\sqrt{-S}\,\bar{J}_{\xi}^{0}=-\oint_{\partial\Sigma}d^{2}\sigma_{i}\,\sqrt{-S}\,\bar{J}_{\xi}^{i}\,.
\end{equation}
Using~\eqref{eq:consJ} this relation can be rewritten as a conservation of \emph{the same} charge $Q_\xi$ in the usual spacetime
\begin{equation}
\label{eq:covariant_charge_usual}
\frac{dQ_\xi}{dt}\equiv\frac{d}{dt}\int_{\Sigma}d^{3}\mathbf{x}\sqrt{-g}\,J_{\xi}^{0}=-\oint_{\partial\Sigma}d^{2}\sigma_{i}\,\sqrt{-g}\,J_{\xi}^{i}\,.
\end{equation}
Thus, the conserved Noether charge $Q_\xi$ associated with the symmetry of the acoustic metric is given by
\begin{equation}\label{eq:Qxi}
    Q_\xi \equiv - \int_{\Sigma}\!d^3\mathbf{x}\, \sqrt{-S} \,\,\T^0_\nu \xi^\nu = -\frac{1}{2}\int_{\Sigma}\!d^3\mathbf{x}\, \sqrt{-S}\, \frac{\xi^0 }{Z^{00}}\, Q^{\mu\nu}\partial_\mu\pi \partial_\nu\pi \,,
\end{equation}
and the components of the quadratic form $Q^{\mu\nu}$ can be written as
\begin{align}
    Q^{00} &= \left(Z^{00}\right)^2  \,,& Q^{0i} &= Q^{i0} = -Z^{00} v^i \equiv \left(Z^{00}\right)^2 \frac{\xi^i}{\xi^0}\,, \label{eq:Xdef}  \\
    Q^{ij} &  \equiv \Ztwobar{ij} + {v^i v^j} - w^i w^j\,, &\text{with }\quad w^i  &\equiv Z^{0i} + {v^i}\,. \notag
\end{align}
Note that, contrary to the Hamiltonian, this charge is an \emph{explicitly time-dependent} integral of motion i.e.~up to the boundary term in~\eqref{eq:covariant_charge}
\begin{equation}
\frac{dQ_{\xi}}{dt}=\partial_{t}Q_{\xi}+\int_{\Sigma}d^{3}\mathbf{x}\left(\frac{\delta Q_{\xi}}{\delta\pi}\dot{\pi}+\frac{\delta Q_{\xi}}{\delta\Pi}\dot{\Pi}\right)=0\,.
\end{equation}
We will require that our slicing is such that  everywhere the frame $u_\mu$ is a good Cauchy frame, $\Ztwobar{ij}\succ0$ (see section~\ref{sec:coords}). Otherwise, we are not necessarily free to pick arbitrary gradients of $\pi$. This implies that we have $Z^{00}<0$ for a non-ghost and $Z^{00}>0$ for a ghost. Let us choose $\xi^0>0$ without loss of generality.

Then $Q_\xi$ is bounded only if $Q^{\mu\nu}$ is definite everywhere. Since it contains the positive definite $\Ztwobar{ij}$, we actually require that $Q^{\mu\nu} \succ 0$. To find conditions for positive definiteness it is convenient to represent
\begin{equation}
Q^{\mu\nu}=\left(\begin{array}{cc}
\left(Z^{00}\right)^2 & -Z^{00} v^{i} \\
- Z^{00} v^{j} &  \Ztwobar{ij}+{v^{i}v^{j}} \\
\end{array}\right)-\left(\begin{array}{cc}
0 & 0 \\
0 & w^{i}w^{j}
\end{array}\right)\,,
\end{equation}
where the last quadratic form is positive semidefinite, while the first one is positive definite. Indeed, its determinant is given\footnote{Similarly to eq.~\eqref{eq:z2def} here $v^{2}\equiv\left({\Ztwobar{}}^{-1}\right)_{ik}v^{i}v^{k}$. Thus we use $\left({\Ztwobar{}}^{-1}\right)_{ik}$ as a 3d covariant metric so that the same rule applies for other scalar products as well as for raising and lowering indices.} by $\left(Z^{00}\right)^{2}\cdot\det\left(\Ztwobar{ij}+{v^{i}v^{j}}\right)/{(1+v^{2})}$, while $\Ztwobar{ij}+{v^{i}v^{j}}$ is positive definite, so that Sylvester's criterion is satisfied.\footnote{Usually this  criterion is formulated as positivity for all upper right corner determinates. However, it can be shown that for an $n\times n$ matrix the positivity of any nested sequence of $n$ principal minors is equivalent to positive definiteness, see~\cite{Matrix_Analysis}.} The second quadratic form is maximal on vectors proportional to $(\beta, w^{i})$, with arbitrary $\beta$. On these vectors the first quadratic form is minimal for $\beta={w_{i}v^{i}}/{Z^{00}}$. Thus, these vectors provide the minimum of the quadratic form $Q^{\mu\nu}$, with the value $w^{2}(1-w^{2})$. Requiring positivity of this minimum yields
\begin{equation}
    w^{2}=w^i ({\Ztwobar{}}^{-1})_{ij} w^j < 1 \,. \label{eq:wZw}
\end{equation}
Using the relationship~\eqref{eq:S_components} between $S_{\mu\nu}$ and $\Ztwobar{ij}$  and re-expressing $w^i$ in terms of $\xi^i$, one can show that eq.~\eqref{eq:wZw} is equivalent to
\begin{equation}\label{eq:Sxixi}
    \frac{-Z^{00}}{(\xi^0)^2} S_{\mu\nu}\xi^\mu \xi^\nu < 0 \,.
\end{equation}

We have thus proven that whenever there exists an acoustic Killing vector which is $S$-timelike, there also exists a charge $Q_\xi$ which is conserved and bounded, and taking its extremal value in the absence of fluctuations,  $\partial_\mu\pi =0$. For non-ghosts $Q_\xi\geq0$, while for ghosts $Q_\xi\leq 0$.

The meaning of this result is that, in the presence of multiple degrees of freedom, if there exists a Killing vector field timelike with respect to the acoustic metrics $S_{\mu\nu}$ of every field  and all the fields are non-ghosts, then there is a conserved charge, bounded  to be positive for each field separately, $Q_\xi\geq0$. Even in a frame in which the Hamiltonian itself is unbounded and in which negative energy modes can be produced, the boundedness and conservation of $Q_\xi$ will eventually arrest this instability. For example, in the frame of a massive particle travelling supersonically through a medium, negative energy modes exist and can be produced, as we discussed in the previous section~\ref{sec:Hamiltonia}. However, these modes will nonetheless  carry a  positive charge $Q_\xi$ and deplete it from the massive particle. The instability will stop at the latest when the particle's $Q_\xi$ has been fully transferred to the phonons.

While strict conservation requires the presence of a Killing vector, for modes of high enough frequencies and momenta, the acoustic metrics can be considered effectively constant, and therefore there are approximately conserved charges associated with all directions. In this local limit,  if there exists even one vector inside all the ray cones ($S$-timelike) and one covector inside all the P-cones ($Z$-timelike), and all the $S$ metrics have the same signature, this is enough to give a good Cauchy surface and a positive bound $Q_\xi$ for each non-ghost field and therefore guarantee that fast instabilities are arrested eventually. Conversely, if the ray cones do not overlap at all, no such bounded charge related to spacetime symmetries exists and the instability cannot be arrested in this way. This is the case for the necessarily transonic configuration of figure~\ref{fig:ConeNoIntersection}, see also the discussion  on page~\pageref{sec:transonic}. This local discussion reflects the construction described in ref.~\cite{Babichev:2018uiw}.

Note that since ghosts have $Q_\xi\leq 0$, in a mixed ghost-non-ghost setup motion is not bounded. However, in classical field theory the corresponding runaway can be rather slow and benign, see e.g.~\cite{Smilga:2013vba,Damour:2021fva,Gross:2020tph,Deffayet:2025lnj}.

\subsection{Acoustic metric and sound horizons\label{sec:Cerenkov}}
In constructing the $Z$-frame in section~\ref{sec:signature} we chose not to use the spacetime metric. Nonetheless, in the presence of two metrics, there are two independent ways of mapping vectors to covectors and therefore we could have chosen a different construction. We can alternatively start off from a four \emph{vector} $u^\mu=g^{\mu\nu}u_\nu=S^\mu_\nu W^\nu$.
\begin{equation}
    v_\mu \equiv S_{\mu\nu}u^\nu\,,\quad u^\mu v_\mu \equiv S_{uu} \neq 0 \,, \label{eq:v-def}
\end{equation}
where the $u$ subscript is a contraction with $u^\mu$. We can now define a different projector $\perpS^\mu_\nu$, onto the subspace orthogonal to $u^\mu$, and the associated induced metric on this subspace, $\DS_{\mu\nu}$.
\begin{equation}
    \DS_{\mu\nu} \equiv S_{\mu\nu} - \frac{v_\mu v_\nu}{S_{uu}}\,,\quad \perpS^\mu_\nu \equiv  \DS_{\mu\lambda} Z^{\lambda\nu}  \label{eq:DS_def}
\end{equation}
It may be somewhat surprising, but this projector is not the same as that defined in the $Z$-frame in eq.~\eqref{eq:DZ_def}, $\perpZ^\mu_\nu \neq \perpS^\mu_\nu$, since they are projecting orthogonally to distinct vectors $W^\mu$ and $u^\mu$ respectively. Similarly, the two induced metrics are \emph{not} inverses of each other, $\left(\DS_{\mu\nu}\right)^{-1} \neq \DZ^{\mu\nu}$. We will call the frame based on $v_\mu$ with the induced spatial metric~\eqref{eq:DS_def} the $S$-frame.

Note that in eq.~\eqref{eq:v-def} we could have started with an arbitrary vector $u^\mu$ unrelated to $W^\mu$ of eq.~\eqref{eq:W-def}. Proceeding as we have illustrates that  the distinction between the $S$-frame defined here and the $Z$-frame defined in section~\ref{sec:signature} survives even when the two vectors are closely to related to each other.

As in section~\ref{sec:signature}, we can decompose the rays~\eqref{P-Zdecomp} in the $S$-frame
\begin{align}
    N^\mu &= -\frac{\mho_S}{S_{uu}} u^\mu + \dot{\RS}^\mu\,,\quad\text{with}\quad \dot{\RS}^\mu\equiv \perpS^\mu_\nu N^\nu \,, \label{eq:N-Sdecomp}
\end{align}
which provides also the natural decomposition for the ray null surface,
    \begin{equation}
      S_{\mu\nu}N^\mu N^\nu = \frac{1}{S_{uu}}\left(\mho_S^2 + S_{uu} \DS_{\mu\nu}\dot{\RS}^\mu \dot{\RS}^\nu\right) =0 \,. \label{eq:N-cone-Sdecomp}
\end{equation}
$\mho_S$ is real in the $S$-frame whenever the metric $S_{\mu\nu}$ has Lorentzian signature and the induced metric $\DS_{\mu\nu}$ is spatial, in which case there is a ray pointing in every spatial direction $\dot\RS^\mu$ --- there is no sound horizon.

The situation now is analogous to that described in section~\ref{sec:Acou+Obs}, with the metrics $Z^{\mu\nu}$ and $S_{\mu\nu}$ exchanged. We can perform the decomposition in the $g$-frame of an observer $u^\mu$, eq.~\eqref{eq:PN-gdecomp}, to obtain the equation for the Abraham energy $\mho$ valid locally, eq.~\eqref{eq:dispersion-abr-munu}, for convenience written again
\begin{equation}\label{eq:mho_frame}
    \mho_{[\pm]} =-\frac{1}{S_{uu}} \left(S_{u\nu}\dot{r}^\nu \pm  \sqrt{\Stwo{\mu\nu}\dot{r}^\mu \dot{r}^\nu}\right)\,,
\end{equation}
with the tensor $\Stwo{\mu\nu}$ defined in eq.~\eqref{eq:Stwomunu} and the equivalent of eq.~\eqref{eq:mhopm-munu},\footnote{We have a condition equivalent to eq.~\eqref{eq:Schur-det},
    $\det S_{\mu\nu} = -(S_{uu})^{-2}\detu \Stwo{\mu\nu}$ giving $\det_u \Stwo{\mu\nu} > 0$ for a hyperbolic acoustic metric.\label{detS2-footnote}} \begin{equation}\label{omegapm-munu}
    \omega_{[\pm]} = -S_{\mu\nu}N^\mu u^\nu = \pm \sqrt{\Stwo{\mu\nu}\dot{r}^\mu \dot{r}^\nu}\,.
\end{equation}
Then we obtain an ``inverse'' relation to eq.~\eqref{eq:MhoOmega_Z} for the Abraham and Minkowski energies
\begin{equation}\label{eq:MhoOmega_S}
\sigma \omega_{[\pm]}=\mho_{[\pm]}\mathfrak{v_\text{p}}\,,\qquad \mathfrak{v}_\text{p}\equiv \sqrt{\Stwo{\mu\nu}v^\mu_{\text{p}} v^\nu_{\text{p}}}\,.
\end{equation}
This allows us to identify $\Stwo{\mu\nu}$ as the metric on the space of phase velocities for the observer $u^\mu$ with norm $\mathfrak{v}_\text{p}$. We recover the naive scalar relationship
\begin{equation}
    \mathfrak{v}_\text{p} \mathfrak{n} = 1\,,
\end{equation}
but only when the two metrics are positive definite, $\Ztwo{\mu\nu}\succ 0$, $\Stwo{\mu\nu}\succ 0$. Otherwise there is a sign difference for some of the modes.

The meaning of the $[\pm]$ branch subscript in eqs.~\eqref{eq:mho_frame} and~\eqref{omegapm-munu} is distinct from that of the dispersion relation~\eqref{eq:dispersion-munu} and its $\pm$ branch subscript: the set of solutions is clearly the same, but eq.~\eqref{omegapm-munu} implies that the $[+]$ branch selects only positive observable energies $\omega_{[+]}>0$, while the Abraham energy $\mho_{[+]}$ is now not definite.  When $\Stwo{\mu\nu}\succ0$ there are rays in every spatial direction of the frame $\dot{r}^\mu$ and $u^\mu$ is subsonic ($S$-timelike). Then each cone nappe is constructed from a single $[\pm]$ branch. The future nappe of the ray cone is constructed by the $\mho_{[+]}$ roots and mapped by the acoustic metric to the $\omega_{[+]}>0$ branch of the momentum cone for non-ghosts (all outgoing energies are positive for the observer); for ghosts --- the future nappe of the ray cone is constructed by the $\mho_{[-]}>0$ branch which is mapped onto the $\omega_{[-]}<0$ branch of the P-cone.

However, when $\Stwo{\mu\nu} \nsucc 0$, $\mho_{[\pm]}$ is complex for some of the directions $\dot{r}^\mu$ ---  this implies that the ray cone does not cover these directions, and the phonons do not propagate along them. A sound horizon would be observed by this supersonic $u^\mu$ ($u^\mu$ is $S$-spacelike) along the conical surface $\Stwo{\mu\nu}\dot{r}_*^\mu \dot{r}_*^\nu =0$  with phonons propagating solely inside it. The future N-nappe is constructed by both the roots $\mho_{[+]}$ and $\mho_{[-]}$ (both positive if $\Sigma_u$ is a good Cauchy frame). For a non-ghost, the acoustic metric then still maps this future N-nappe to the upper P-nappe, but the modes belonging to the $\mho_{[-]}$ branch correspond to negative energies $\omega_{[-]}<0$, while $\mho_{[+]}$ map onto $\omega_{[+]}>0$.

\begin{figure}
\begin{subcaptionblock}[T]{0.65\textwidth}
\centering
\includegraphics[width=0.8\textwidth]{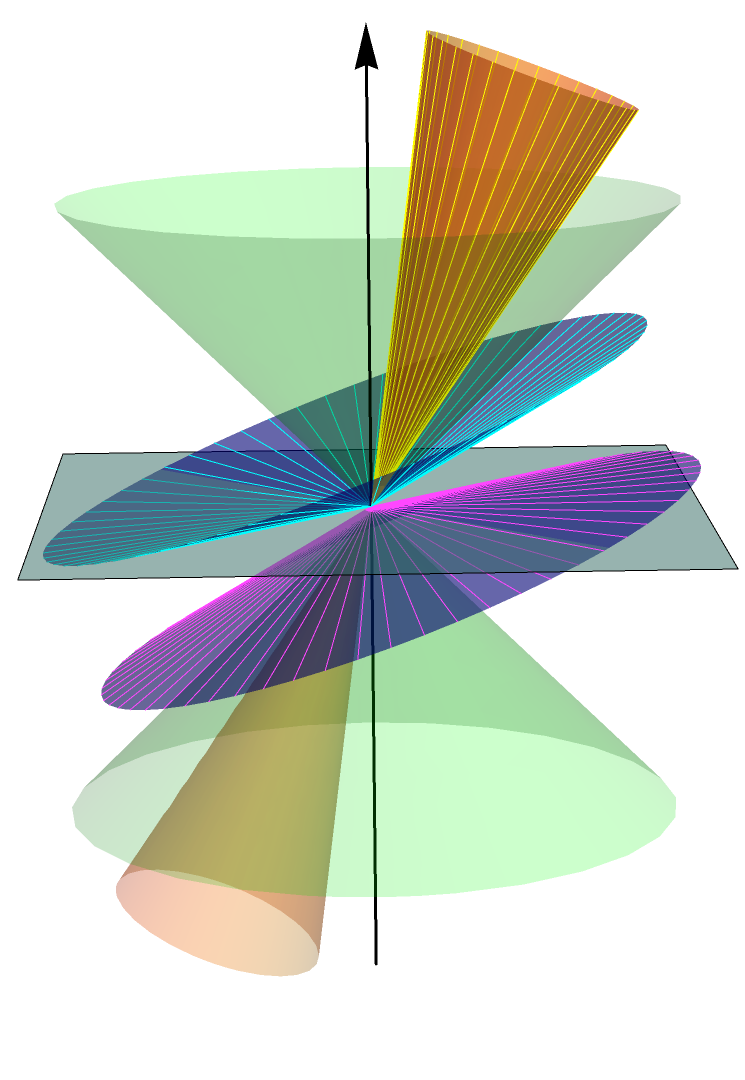}
\caption{\label{fig:conesSoundHorizon}}
\end{subcaptionblock}\hfill
\begin{subcaptionblock}[T]{0.350\textwidth}
\centering
\includegraphics[width=0.9\textwidth]{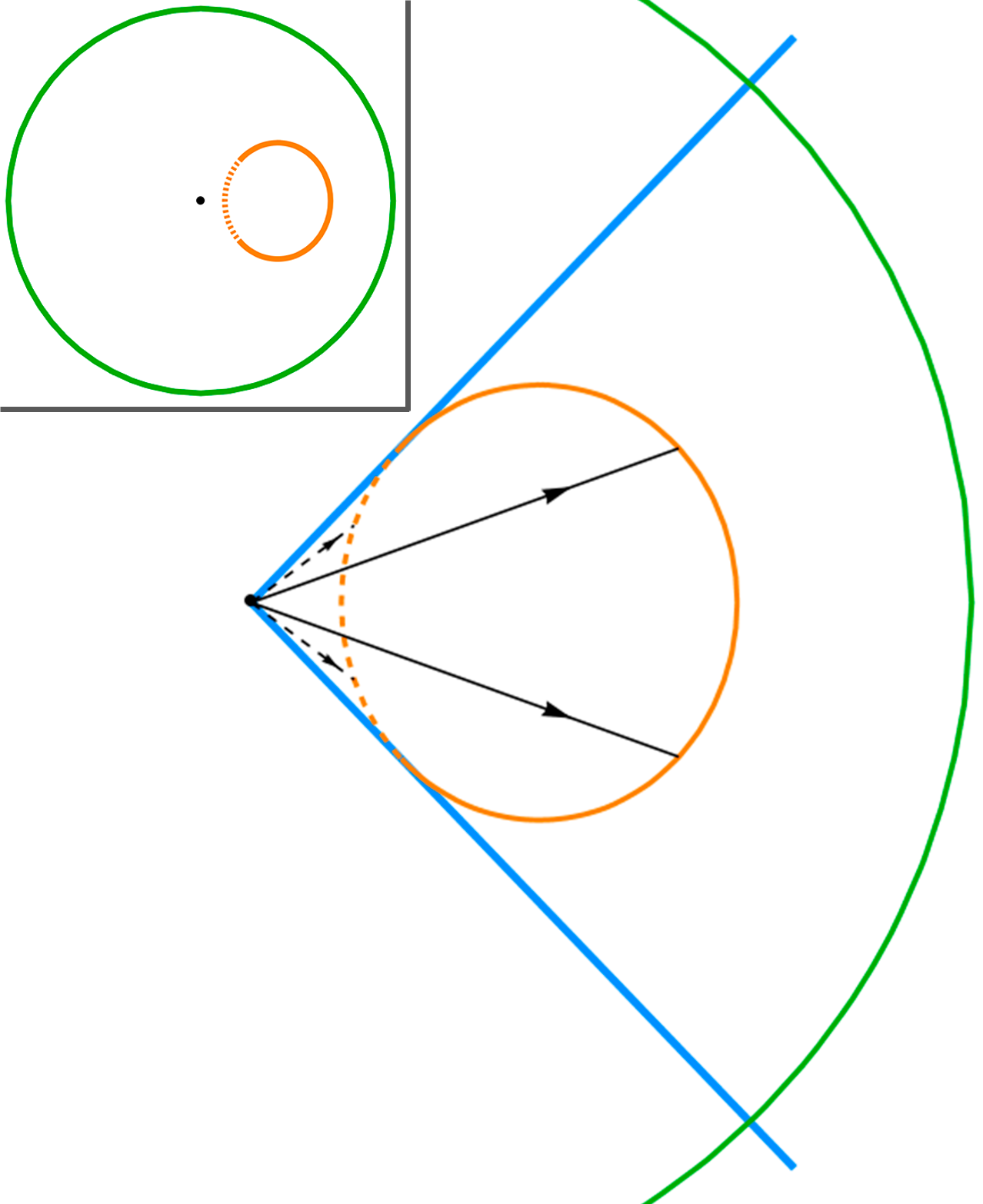}
\caption{\label{fig:phaseVelSoundHorizon}}\includegraphics[width=0.9\textwidth]{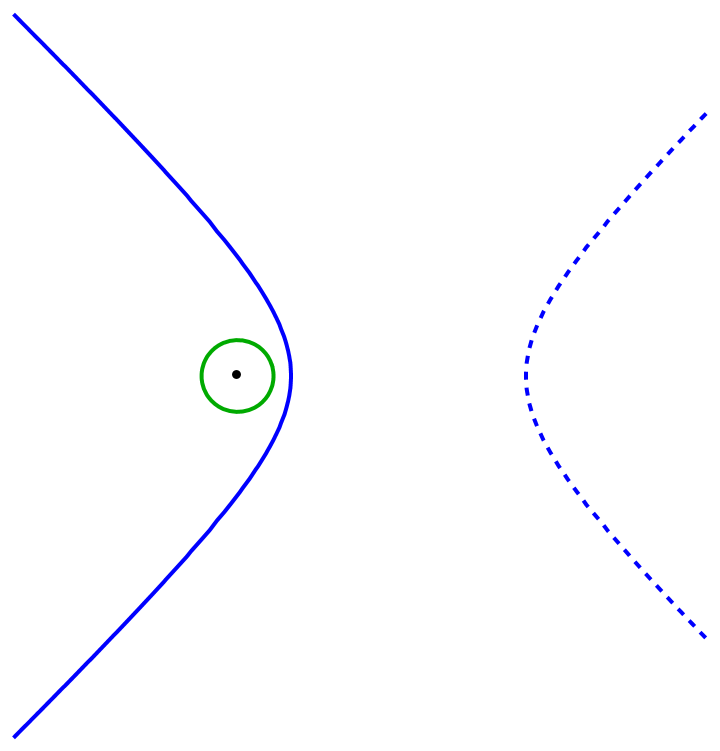}
\caption{\label{fig:dispRelSoundHorizon}}
\end{subcaptionblock}
\caption{Cone geometry in a good Cauchy frame of an observer moving supersonically with respect to an isotropic medium. Colour coding of surfaces as in figure~\ref{fig:isotropic_GoodCauchy}. (a) A subluminal ray cone (orange) is $g$-timelike and therefore there exist boosted frames in which the time direction lies outside it. The ray cone then does not cover the whole surface $\Sigma_u$ and propagation does not occur in all directions --- there is a spatial sound horizon (Mach cone). The $g$-spacelike P-cone in this frame cuts the surface $\Sigma_u$ --- mode energies are not definite. For a non-ghost, the acoustic metric maps the complete upper ray-cone nappe to the complete upper P-cone nappe (highlighted with light blue), including the part below $\Sigma_u$, so the outgoing rays have energies of both signs. The surface of the Mach cone/sound horizon is constructed by the modes with $\omega=0$. Cherenkov radiation is the emission of the negative energy modes from a source at rest in this supersonic frame. For ghosts, the acoustic metric maps the upper N-nappe to the lower P-nappe (highlighted in magenta), so ghosts can have positive energy for a supersonic observer. (b) Phase-velocity direction and magnitude for outgoing non-ghost scalar waves (orange) vs light (green). Inset shows complete wavefronts, while the graphic zooms in around the Mach cone (light blue).  The $\mho_+$ rays are constructed by both the positive energy modes (solid, $\mho_{[+]}$ branch) with phase velocities with a component parallel to the mode's momentum $k_\mu$ and the negative energy modes (dashed, $\mho_{[-]}$ branch), with an antiparallel component, see eq.~\eqref{eq:vp-dir}. (c) The wave-vector surface formed by the \emph{upper} P-nappe $\omega_+$ (i.e.~for non-ghosts) is hyperboloidal. It is constructed by two branches --- the solid corresponding to positive mode energies $\omega_{[+]}$ and mapping onto the solid part of the wavefront in figure~\ref{fig:phaseVelSoundHorizon}, and the dashed constructed by modes with negative energies $\omega_{[-]}$ and mapping onto the dashed part of the wavefront. The separatrix is conical and made out of modes with spatial momentum for which the energy is zero.\label{fig:SoundHorizon}}
\end{figure}

Inside the mach cone, the outgoing modes with $\dot{r}^\mu k_\mu <0$ have negative energies $\omega_{[-]}<0$ even though they are \emph{not} ghosts. It is interesting to note in this regard that the first direct observation of negative-frequency waves converted from positive-frequency waves in a moving medium is reported in~\cite{Rousseaux:2007is}. For both a ghost and a non-ghost, the same roots $\mho_\pm$ construct the future ray nappe, but ghost future/upper P-cones are constructed by the opposite $[\pm]$ roots to the non-ghost. One implication is that some ghost modes have positive energy when the observer is supersonic. See figure~\ref{fig:SoundHorizon} for an illustration.

The subsonicity of an observer, $\Stwo{\mu\nu}\succ0$ is a \emph{different} condition to the good Cauchy frame condition $\Ztwo{\mu\nu}\succ0$~\eqref{Z-inducedmetric}  for this observer's frame and both, neither, but also just one of them could be satisfied depending on the setup. The subsonic condition, is again quadratic in $S_{\mu\nu}$, so is satisfied for both the $(3,1)$ and $(1,3)$ signatures and it depends on the choice of observer $u^\mu$.

Just as for the case of $\Ztwo{\mu\nu}$ and the relation of its positivity to the sign of $Z^{uu}$, the sign of $S_{uu}=S_{\mu\nu}u^\mu u^\nu$ is determined by whether $\Stwo{\mu\nu}$ is positive definite or not and the overall acoustic metric signature. The transonic point is given by $S_{uu}=0$ which can be related through eq.~\eqref{eq:S_components} to the inverse metric by
\begin{equation}\label{eq:transonic}
S_{uu}=\frac{1}{Z^{uu}}\left(1-z_u^2\right), \quad \text{with} \quad z_u^2\equiv(\Ztwo{{-1}})_{\mu\nu}Z^{u\mu}Z^{u\nu}\,.
\end{equation}
Thus, whenever $z_u^2>1$, the observer $u^\mu$ is supersonic and would see negative energy modes of the fluctuations.

To extend to the global setup with the foliation~\eqref{eq:ADM_metric}, we will now identity the observer as the comoving observer of the foliation $V^\mu=(\alpha^2-\beta^2)^{-1/2}\delta^\mu_0$. We then have $S_{VV}= (\alpha^2-\beta^2)^{-1}S_{00}$ and eq.~\eqref{eq:mho_frame} for the CMO~\eqref{eq:N-cmo} gives for the roots
\begin{equation}\label{eq:mhoV}
    \frac{\mho_{V,[\pm]}}{\sqrt{\alpha^2-\beta^2}} = - \left(\frac{S_{0i}}{S_{00}}+\frac{\beta_i}{\alpha^2-\beta^2}\right)\dot{r}_V^i \mp \frac{\sqrt{\Stwoubar{ij}\dot{r}_V^i \dot{r}_V^j}} {S_{00}}\,,
\end{equation}
with $\Stwoubar{ij}$ defined in eq.~\eqref{eq:S2ubar}. One may be concerned that choosing a sufficiently large $\beta_i$ can in principle make the Abraham energy $\mho_{V,[+]}<0$. In such a case, the frame $V_\mu$ is a bad Cauchy frame. However, the good Cauchy condition for the slicing is instead related to $\mho_U$,
\begin{equation}\label{eq:mhoUrV}
    \frac{\mho_{U,[\pm]}}{\alpha} = N^0_{[\pm]} = -\frac{S_{0i}}{S_{00}} \dot{r}_V^i \mp \frac{\sqrt{\Stwoubar{ij}\dot{r}_V^i \dot{r}_V^j }}{S_{00}}\,,
\end{equation}
for which the shift does not enter. This expression is an analogue of the dispersion relation~\eqref{eq:dispersion-P0}: here we also mix variables from two distinct frames to obtain an expression for $N^0$ independent of the spacetime metric.

The energy as observed by the CMO and proportional to the would-be conserved four-momentum component $-P_0$ eq.~\eqref{eq:dispersion-P0} can be computed from eq.~\eqref{omegapm-munu},
\begin{equation}\label{eq:omVrV}
    {\sqrt{\alpha^2-\beta^2}}\ \omega_{V,[\pm]} = - S_{\mu\nu} N^\nu V^\mu = -P_{0,[\pm]} =  \pm \sqrt{\Stwoubar{ij}\dot{r}^i_V \dot{r}^j_V},
\end{equation}
The subsonicity of the CMO is related to the reality of $\mho_V$. The frame condition $\Stwo{\mu\nu}\succ 0$ becomes the matrix condition  $\Stwoubar{ij}\succ 0$. If satisfied, there are real solutions for all directions $\dot{r}^i_V$ and the modes propagate in all spatial directions since the phase velocity relative to the CMO is given by $v_{\text{p},V}^i = \dot{r}_V^i/\mho_V$ eq.~\eqref{eq:vpi-CMO}.  The CMO is then subsonic with respect to the phonons and $V^\mu$ is $S$-timelike. We may demand that this condition be satisfied at every point on the spatial slice.

The matrix $\Stwoubar{ij}$ is proportional to the Schur complement of $S_{\mu\nu}$ with respect to the component $S_{00}$. Analogously to the derivation of the expression~\eqref{eq:S_components}, it can be checked that
\begin{equation}\label{eq:ZvsS2inv}
    Z^{ij} = -S_{00} (\Stwoubar{} ^{-1} ) ^{ij},
\end{equation}
which leads to the conclusion that the comoving observer is subsonic and sound horizons are absent in its rest frame when
\begin{equation}\label{eq:SHcond}
    \Stwoubar{ij} \succ 0 \qquad \Leftrightarrow \qquad \frac{1}{S_{00}}Z^{ij} \prec 0 \,.
\end{equation}
Finally, we can evaluate eq.~\eqref{eq:transonic} for $V^\mu$ obtaining that $V^\mu$ is subsonic when
\begin{equation}\label{eq:S00_z}
    z_V^2 \equiv Z^{0i} \left( \Ztwobar{-1}  \right)_{ij} Z^{0j} < 1\,,
\end{equation}
Exactly this expression for $z_V^2$ was obtained as a condition in the discussion of the positivity of the Hamiltonian, eq.~\eqref{eq:small_z}, confirming that we have identified the correct observer for the foliation. Footnote~\ref{detS2-footnote} implies $\det\Stwoubar{ij} >0$, and therefore $S_{00}$ has the opposite sign to $\det Z^{ij}$. $S_{00}<0$ then either implies that $Z^{ij}\succ0$, and energies $-P_0$ are positive for all modes and there is no sound horizon for the CMO at this point. Alternatively $Z^{ij}$ has two negative eigenvalues and therefore there is a sound horizon and the scalar is a ghost. We recover the statement that $z_V^2=1$ is the transonic point for the CMO with $z_V^2>1$ implying the CMO is supersonic, in which case we have a single negative eigenvalue for $Z^{ij}$ for a non-ghost or one positive --- for a ghost and therefore one direction for which gradient energy is of opposite sign to the others.

We could also ask what the equivalent set of subsonicity conditions is for the slicing's NFO. Analogously to the dispersion relation~\eqref{eq:dispersion-cmo}, we can express the acoustic Lorentz factor $\mho_U$ in terms of the spatial directions as would be seen by the NFO, $\dot{r}_U^i$ by solving the $N$-cone decomposition~\eqref{eq:raycone_g},
\begin{equation}
    \mho_{U,[\pm]} = \frac{S^0_i \dot{r}^i_U}{\alpha S^{00}} \mp \frac{\sqrt{\Stwobar{ij}\dot{r}^i_U \dot{r}^j_U}}{\alpha S^{00}}\,,
\end{equation}
with $\Stwobar{ij}$ defined in eq.~\eqref{eq:S2bar}. The raised temporal index absorbs the relative velocity $\beta^i$ of the NFO, see eqs.~\eqref{eq:S00S0i}. The discussion remains as for the CMO, up to changes to the position of indices. The NFO is subsonic whenever $\Stwobar{ij}\succ 0$ which ensures that $\omega_U>0$ for all directions for a non-ghost; $U^\mu$ is $S$-timelike. Otherwise, $\Stwobar{ij}\nsucc 0$, there are directions $\dot{r}^i_U$ with negative energy $\omega_U<0$ for a non-ghost and $U^\mu$ is $S$-spacelike. The transonic condition~\eqref{eq:S00_z} for the NFO is instead $S^{00}=0$ and for a subsonic NFO $Z_{ij}$ is definite, instead of $Z^{ij}$. By choosing the shift appropriately, it is possible for either $Z_{ij}$ or $Z^{ij}$ to be definite, while the other is not.

We have thus furnished the positivity of the acoustic Hamiltonian with a more physical meaning: the acoustic Hamiltonian for fluctuations is unbounded whenever the CMO is supersonic anywhere on the slice: when there is a sound horizon for the CMO, both the locally observed $\omega_V$ and the would-be conserved energy $-P_0$ are negative for some modes even though they are not ghosts. While we identified the positivity of $\mho_U$ as the good-Cauchy condition, its \emph{reality} --- and therefore the supersonicity of the NFO --- does not appear to be relevant to the consideration of stability, similarly to the irrelevancy of the reality of $\omega_V$ to the Hamiltonian.

We reiterate here that since $\Stwo{\mu\nu}$ and $\Ztwo{\mu\nu}$ are not simply related, the choice of a bad Cauchy frame and the existence of sound horizons are in general completely independent phenomena. Depending on the metric $Z^{\mu\nu}$, any one or both can occur for a single chosen observer and their frame.

\paragraph{The sonar metric.}
We can relate $\Stwoubar{ij}$ to
the acoustic equivalent of the radar metric of ref.~\cite[pg.~84]{LandavshitzII}.  We can construct a spacetime metric
by measuring distances using proper-time elapse for an observer stationary with respect to their own coordinates --- i.e.~a CMO ---
between the emission and the return of electromagnetic radar signals bounced off reflectors located throughout the space.
Following their setup, but measuring the acoustic spacetime by sending sonar pulses, we should consider an outgoing pulse propagating with Abraham momentum $\dot{r}^i_V$ with temporal coordinate $N^0_+d\lambda$ followed by a return incoming pulse arriving with temporal coordinate $-N_-^0d\lambda$ and compare this with the change in proper time for the comoving observer $V^\mu$,  which gives us as the sonar metric $\Sigma_{ij}$
\begin{equation}
d\ell^2=-\frac{1}{2}g_{00}(N^0_{+}-N^0_{-})^{2}d\lambda^2=\Sigma_{ij}dr_V^{i}dr_V^{j}\,,\quad\quad\quad\Sigma_{ij}\equiv \frac{g_{00}\Stwoubar{ij}}{(S_{00})^{2}}\,.\label{SonarMetric}
\end{equation}
where we had to assume that we are on a good Cauchy surface, so that we can produce arbitrary pulses in the first place and $g_{00}$ converts coordinate time to the proper time of the CMO. As $V^{\mu}$ approaches the transonic point $S_{00}=0$, the time taken for the
signal to return diverges, and so does the sonar distance. Past the
transonic point, $\mathcal{S}_{2}$ is not positive definite and therefore
distances in some directions become complex --- propagation is not
allowed there. Since $S_{\mu\nu}$ appears quadratically, the sonar
distance is not sensitive to whether the scalar is a ghost or healthy.
Given the relation~\eqref{eq:ZvsS2inv}, the sonar
metric can also be written as
\begin{equation}
\Sigma_{ij}=-\frac{g_{00}}{S_{00}}\left(Z^{-1}\right)_{ij} \,.
\end{equation}

\paragraph{Multiple degrees of freedom and Cherenkov radiation.}

So far, we have discussed a frame/foliation issue with no physical implications: we have demonstrated that when a part of the ray cone is $g$-timelike, i.e.\ scalar fluctuations are subluminal in those directions, there are choices of observers for whom sound horizons exist. However, this has physical consequences once we have other fields or even just particles coupled to the scalar, not in the least gravity.

In particular, if a massive particle moving with velocity $u^\mu$ has an interaction vertex with a subluminal scalar, a sound horizon appears in the particle's rest frame the moment $u^\mu$ is outside the ray cone, $S_{\mu\nu}u^\mu u^\nu>0$ (for a non-ghost scalar). The negative energies of the scalar modes in the rest frame of the particle are now physical, meaning that it becomes kinematically allowed to conserve on-shell both energy and spatial momentum while emitting a single scalar mode with negative energy, a new tree-level three-point process which otherwise would not be permitted. This leads to Cherenkov radiation. The surface of the Mach/Cherenkov cone is formed by the scalar modes which have zero Minkowski energy and zero phase velocity in the rest frame of the particle, $\Stwo{\mu\nu} \dot{r}_*^\mu \dot{r}_*^\nu=0$, while the actual energy loss occurs into the negative energy modes inside the cone (see the example in section~\ref{sec:phys-cones} for details). Since this process is possible in the rest frame, it is of course computable in any frame. A modern derivation of the rate of this process is given in e.g.~\cite{Moore:2001bv,Elliott:2005va}. This is an instability, as a result of which, the particle extracts energy from the background and is boosted toward the medium's rest frame, effectively slowing down from the point of view of the medium. The process is arrested once $S_{\mu\nu}u^\mu u^\nu=0$.

The Hamiltonian picture we analysed in section~\ref{sec:Hamiltonia} matches this local description. If we associate the time coordinate with the time direction of a comoving observer $V^\mu$ and assume that the acoustic metric is constant in this time, the Hamiltonian is bounded provided that $V^\mu$ is S-timelike (subsonic motion). If the particle is supersonic, the Hamiltonian becomes unbounded. The possibility of spontaneous emission in the rest frame depends on the sign of the conserved energy $-P_0$ or the observed energy $\omega_V$, which are always the same in this setup. We showed in section~\ref{sec:Hamiltonia} that the unbounded term cannot be removed through boundary conditions and therefore can be exploited in this manner.

The actual instability rate depends on the details of the interaction vertex and indeed the cutoff beyond which the scalar's background configuration becomes transparent to the particle, but is finite since the phase-space volume is finite. Moreover, this is really an instability which only appears in the presence of a source: without a source, a change of frame removes the negative energy modes, so nothing can happen spontaneously.

Such a Cherenkov-like process is also kinematically allowed when instead of the supersonic particle, we have a massless mode, e.g.\ a graviton, which interacts with the scalar. In our language, we can pick a graviton with ray $l^\mu$ (i.e.\ momentum $l_\mu$) and ask if the ray is outside of the acoustic ray cone, $S_{\mu\nu}l^\mu l^\nu > 0$. If so, the graviton is kinematically allowed to lose energy by producing scalar Cherenkov radiation. In particular, if the acoustic ray cone is fully inside the lightcone, a graviton of any energy can shed it into the scalar at some finite rate determined, as for the massive particle, by the vertex, spin dependence and cutoff and realistically very small given the typical gravitational couplings. The scale independence of the massless cones means that this process does not stop until all the supersonic gravitons at energies below the cutoff decay into the slower-moving scalar. Indeed, any superluminal massless mode would be allowed to decay into the slower graviton/photon in an equivalent finite fashion. The detailed calculation of rates of these processes is model dependent (requires knowledge of the interactions) and is outside of the scope of this paper, but they are finite. In reality the validity of this description is also limited at low momenta by the curvature scale of the metrics, beyond which acoustic momentum is no longer conserved in any case, unless there is sufficient symmetry as encoded by the acoustic Killing vectors.

Let us end this section by considering two more unusual setups: (i) an  acoustic metric, in which the ray cone is partially $g$-timelike and partially $g$-spacelike (class II according to the classification of section~\ref{sec:Geometry}), and (ii) a situation where the \emph{ray} cones are completely disjoint (class Ib \emph{ibid}). In both of these cases, common exteriors of the ray cones still exist, so the Cauchy problem is well-posed in at least some frames --- we will assume we have picked such a good  frame. However, for both of these metrics it is \emph{impossible} to boost to a frame in which the medium is at rest.

In (i) (see figure~\ref{fig:C2-cones} for an illustration), the rays of the scalar which are $g$-spacelike are all outside of the lightcone, and therefore they are kinematically allowed to emit gravitons. On the other hand, the rays of the gravitons which are $S$-spacelike are outside of the acoustic ray cone and therefore they are kinematically allowed to decay into the scalar. We thus end up with a sequestration of the modes by the direction of propagation. Any acoustic P-nappe overlaps with only a single light P-nappe and therefore a frame can be chosen in which all mode energies are positive from the point of view of an external observer.  As we demonstrate in section~\ref{sec:classIIexamples}, such a background  with a class II metric can be constructed in the kinetic gravity braiding model. Since the ray cones overlap, a common $S$-timelike vector exists and, given our discussion of section~\ref{sec:otherQ}, gives a bound conserved charge. Even an external source interacting with both the scalar and gravity eventually can reach a frame in which it cannot extract energy from the medium. Despite the lack of a medium rest frame, the Cherenkov instability has an endpoint.

\label{sec:transonic}In (ii), the situation is more extreme. Since the ray cones do not contain any parts of each other, any mode of either species is kinematically allowed to emit modes of the other (see figure~\ref{fig:ConeNoIntersection}). In a good Cauchy frame, the phase space volume is still finite and therefore the rate of instability is also finite.

\enlargethispage{\baselineskip}

\looseness=-1
Nonetheless, such backgrounds suffer from multiple issues. In particular, the P-cones of the two metrics in setup (ii) both overlap in both their nappes. This means that there is no frame at all where the energies of both the degrees of freedom are all positive. A massive particle with any $g$-timelike velocity $u^\mu$ is able to emit scalar Cherenkov radiation, accelerating toward the unreachable rest frame of the medium formed by the scalar background. This process cannot stop without some limit appearing from a cutoff. However, again, this happens at a finite rate determined by the interaction vertex and only occurs in the presence of an external source.

Secondly, there are now two inequivalent choices of futures: we can declare that the upper nappe of the acoustic ray cone is the future, together the upper lightcone, or that it is the lower nappe of the acoustic ray cone. These two choices lead to disjoint sets of good Cauchy frames, but such surfaces can still be found. The proper resolution of this is to consider whether at any point in the evolution the background was such that the cones overlapped. If so, this determines the future acoustic nappe and the proper Cauchy surfaces to be used.

Whichever choice is made, it is possible to construct a process with total zero acoustic four-momentum involving only future-facing modes from both species the ray cones, since the P-cones intersect in both nappes. In this sense, this property is similar to the situation with a ghost in the rest-frame of an isotropic medium: for the correct choice of outgoing momenta of both the ghosty scalar and e.g.~gravity, total acoustic momentum can sum to zero. One could expect that a class Ib background itself would be destabilised by such spontaneous emission processes with a finite rate dependent on the interaction vertex and might not ever even form.

 This point of view is confirmed by the argument of  section~\ref{sec:otherQ}. Since there is no vector common to the interiors of both the ray cones, no bounded charge $Q_\xi$ exists. This implies that if spontaneous emission from the fluctuation-free vaccum is possible, the emission  process would not be arrested unless and until backreaction changes the background.

\begin{figure}
\centering
\begin{subfigure}[t]{0.45\textwidth}
\centering
\includegraphics[height=0.9\textwidth]{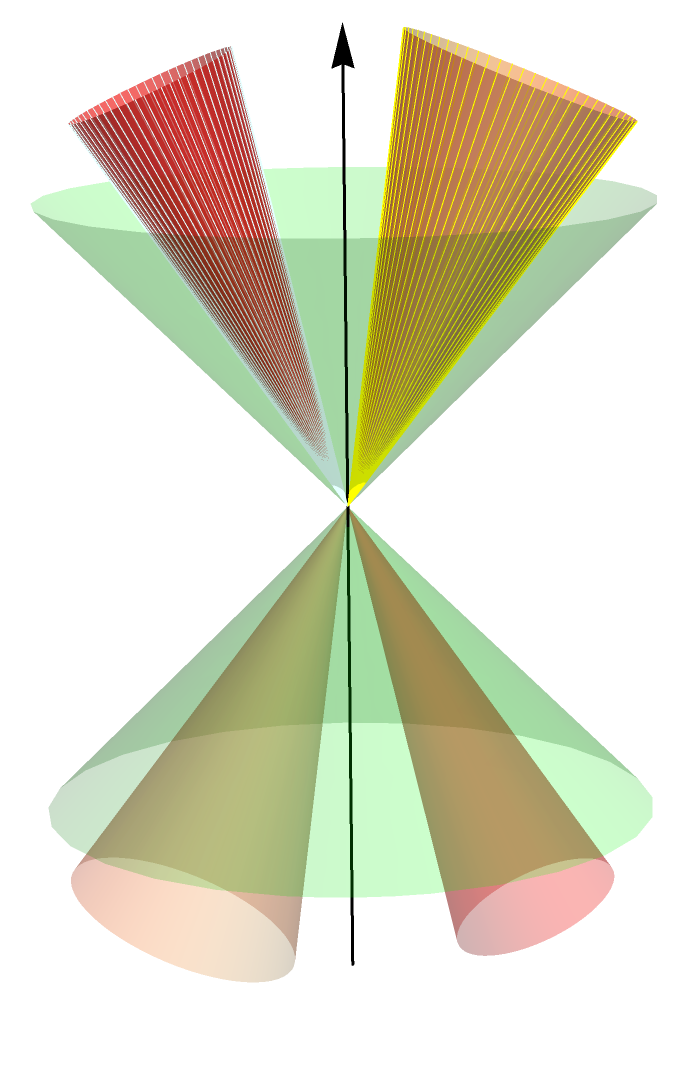}
\caption{\label{fig:2supersonic-Ncone}}
\end{subfigure}\qquad\quad\quad
\begin{subfigure}[t]{0.45\textwidth}
\centering
\includegraphics[height=0.85\textwidth, width=0.95\textwidth]
{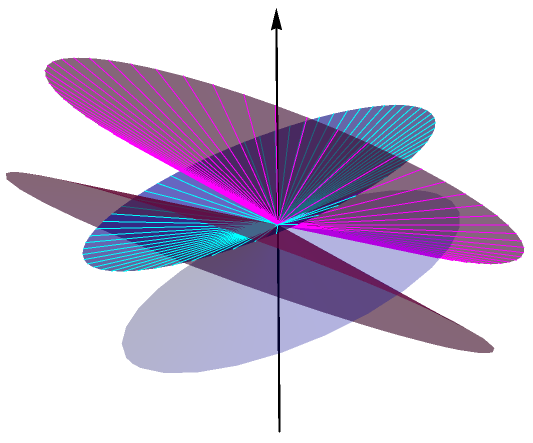}
\caption{\label{fig:2supersonicPcone}}
\end{subfigure}
\caption{A background configuration formed by two subluminal fluids moving supersonically with respect to each other. Red ray cone in panel~(a) corresponds to the purple P-cone in panel~(b), while the orange ray-cone corresponds to the blue P-cone. (a) Both the future ray cones are inside the light cone, so any $g$-spacelike hypersurface is a good Cauchy surface in the standard manner. Despite the non-intersection of the two ray cones, the larger surrounding lightcone gives an unambiguous choice of the future nappes. No vector common to the interiors of both N-cones exists and therefore there is no conserved charge which is bounded. Spontaneous emission from vacuum is. (b) The corresponding future P-nappe of the either fluid's P-cone intersects both the future and past P-nappe of the other fluid. This is a coordinate-invariant statement and therefore, in any possible frame, there are negative energy modes of at least one of the fluids. In the presence of an interaction between the fluids, decay processes into modes of both fluids with total zero acoustic momentum would be kinetically allowed and in principle would act to destabilise this supersonic background if the process were fast enough. This relative P-cone geometry is equivalent to that in the case of the scalar's ray cone being completely outside of the light cone, class Ib as in figure~\ref{fig:ConeNoIntersection}.}
\end{figure}

Nonetheless, this last problem is \emph{not} limited to class Ib metrics. It is enough to consider two weakly interacting fluids with \emph{subluminal} sound speeds moving past each other supersonically but still subluminally, i.e.~with their ray cones inside the light cone (see figure~\ref{fig:2supersonic-Ncone}). There is no common vector in the interiors of both the ray cones and the corresponding P-cone configuration (figure~\ref{fig:2supersonicPcone}) is analogous to the problematic  figure~\ref{fig:ConeNoIntersection}, with each nappe intersecting both the nappes of the other P-cone, despite the fact that neither is the light cone. Thus any such instability arising from spontaneous emission in class Ib would also appear for these supersonic fluid configurations which are not an impossible laboratory setup. The problem really lies in the relative supersonicity of the degrees of freedom as opposed to the fact that one of the cones is the light cone itself.

Let us also comment on the recent paper~\cite{Babichev:2024uro} where it is argued that Cherenkov instabilities in such setups are essentially ghost instabilities. While it is true that spontaneous production of modes is permitted and is not arrested just as in the case of ghosts, the supersonic but subluminal setup serves as a counterexample. With normal signature of the acoustic metric, the subsonic fluctuations are not ghosts with respect to other healthy degrees of freedom with the spacetime metric as their effective metric, i.e.~gravity and electromagnetism. The instability only appears through interactions of the two scalar modes. On the other hand, picking the wrong signature for the scalars' acoustic metrics, does not change the nature of the supersonic instability involving both the scalars. However, all the interactions with the usual degrees of freedom now lead to additional instabilities. These are true ghosts.

\section{Geometries of acoustic cones and dispersion relations}\label{sec:Geometry}

In section~\ref{sec:signature}, we already established that any non-singular acoustic metric with signature (3,1) or (1,3) --- representing non-ghosts and ghosts, respectively --- is hyperbolic and therefore its characteristic surface is a cone. This boils down to the requirement~\eqref{eq:detZ-updown}, $\det Z^\mu_\nu >0$.

Since the spacetime metric has indeterminate signature, it is not
always possible to diagonalise the matrix $Z_{\nu}^{\mu}$ over reals
(it is not necessarily symmetric). This happens when it is not possible to boost to the rest frame of the medium.  The eigensystems of the possible acoustic metrics allow us to classify them. We will demonstrate that the classification is determined by the relative geometry of the light cone and the acoustic cone.

The eigenvalues $\lambda_{I}$ and eigenvectors $v_{I}^{\mu}$ of the acoustic metric $Z^{\mu\nu}$,
\begin{equation}
Z_{\nu}^{\mu}v_{I}^{\nu}=\lambda_{I}v_{I}^{\mu}\,,
\end{equation}
where the capital Latin indices enumerate the eigenvectors. To obtain the eigenvalues, we solve the standard characteristic equation. In this section we present the full classification of possible acoustic metrics according to the eigensystem, discussing its physical meaning and presenting the possible types of dispersion relations and phase velocities for a dispersionless system. We will demonstrate that the metric $S_{\mu\nu}$ belongs to the same class as $Z^{\mu\nu}$ and that both have cones as characteristic surfaces (they are bi-hyperbolic) whenever condition~\eqref{eq:detZ-updown} is satisfied.

We exploit the work categorising the possible form of the energy-momentum tensor in refs.~\cite[pg.~293]{LandavshitzII} and~\cite{Hall1974}, and apply it to the different physics of the acoustic metric. A similar classification was carried for bimetric theories in~\cite{Hassan:2017ugh}. In 3+1 dimensions $Z^{\mu\nu}$ belongs to one of four classes.

\bgroup
\renewcommand\theenumi{\Roman{enumi}.}
\renewcommand\labelenumi{\theenumi}
\begin{enumerate}
\item $Z^{\mu\nu}$ is diagonalisable with a real spectrum; none of the
eigenvectors are $g$-null;
\item $Z^{\mu\nu}$ is diagonalisable with a complex spectrum; none
of the eigenvectors are $g$-null
\item There is a twice repeated eigenvalue associated to a $g$-null eigenvector.
\item There is a thrice repeated eigenvalue associated to a $g$-null eigenvector.
\end{enumerate}
\egroup{}
Only for class I do the eigenvectors form a tetrad. Nonetheless, it is always possible to choose a canonical form for the other classes of metrics using the appropriate choice of standard tetrad for the basis. For clarity, we will label the tetrad directions $(\omega,k_i)$ for $Z^{IJ}$ and $t,x^i$ for $S_{IJ}$. In a general frame $Z^{\mu\nu}$ has ten independent entries. We can perform three boosts and three rotations, fixing six of the entries. Thus in general, we should expect to obtain four free parameters for each metric class. This is true for all metrics, except those in class IV, where an additional degeneracy reduces the free parameters to three.

\begin{figure}
\begin{subcaptionblock}[T]{0.3\textwidth}
\centering
\includegraphics[height=1\textwidth]{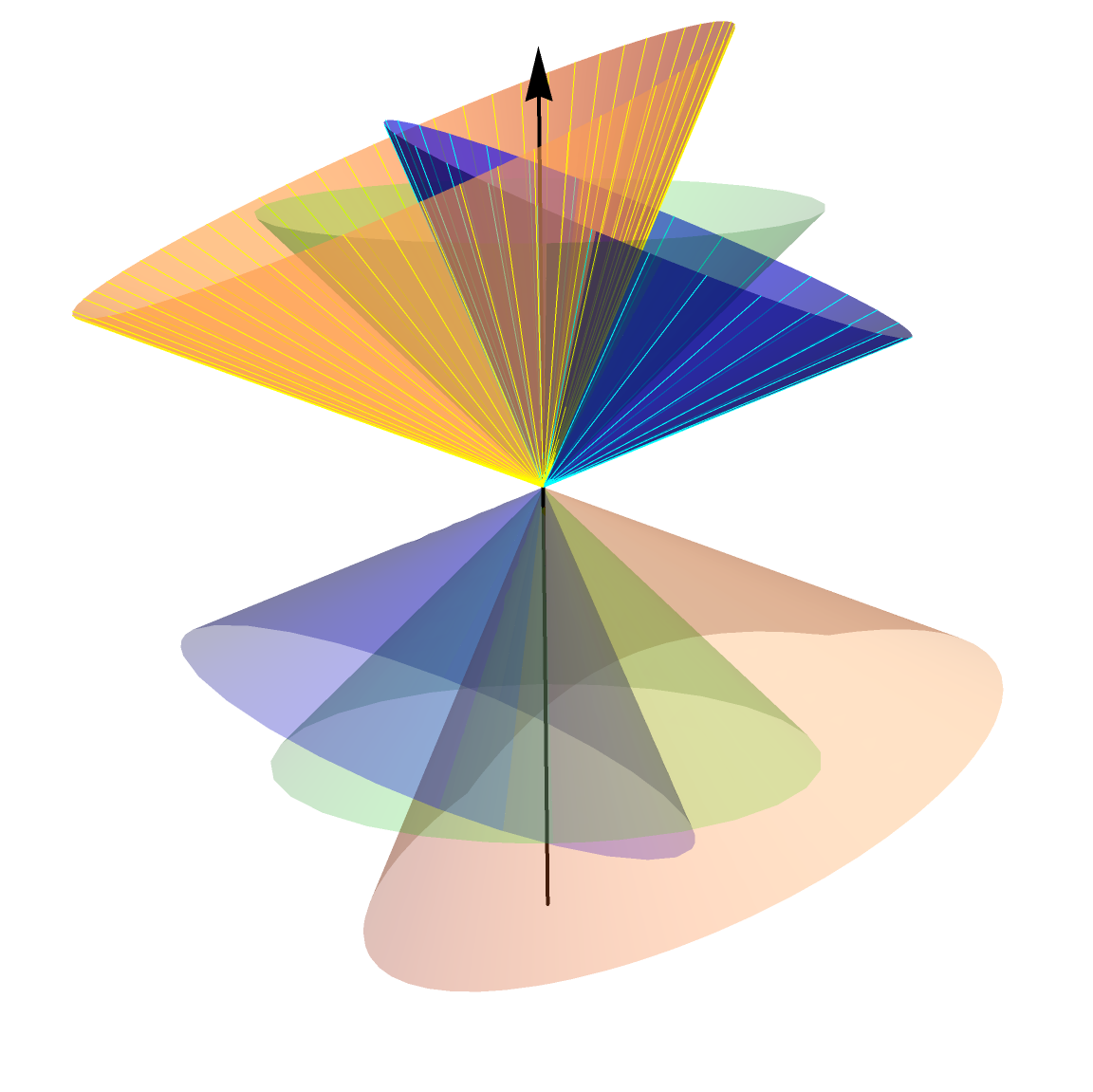}
\caption{\label{fig:C2-cones}}
\includegraphics[width=1\textwidth]{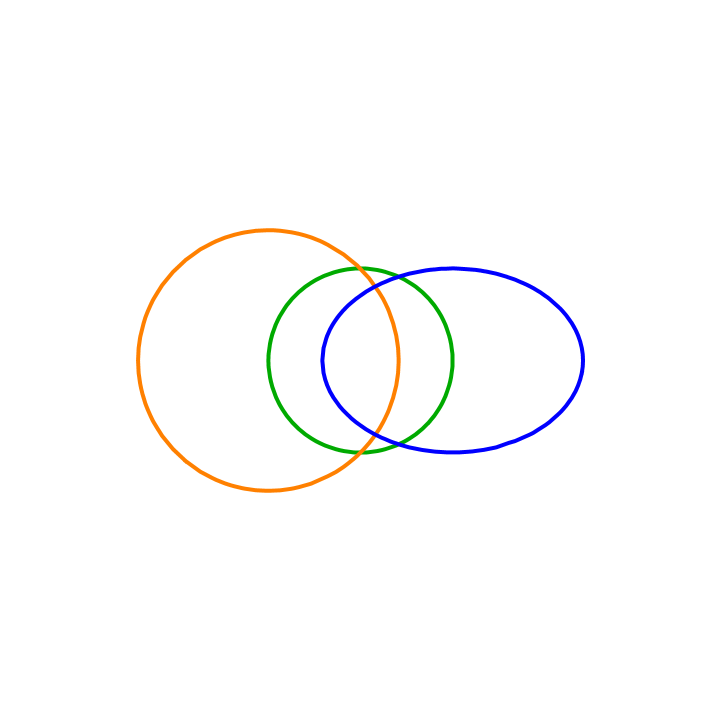}
\caption{\label{fig:C2-contours}}
\end{subcaptionblock}\quad
\begin{subcaptionblock}[T]{0.3\textwidth}
\centering
\includegraphics[height=1.\textwidth]{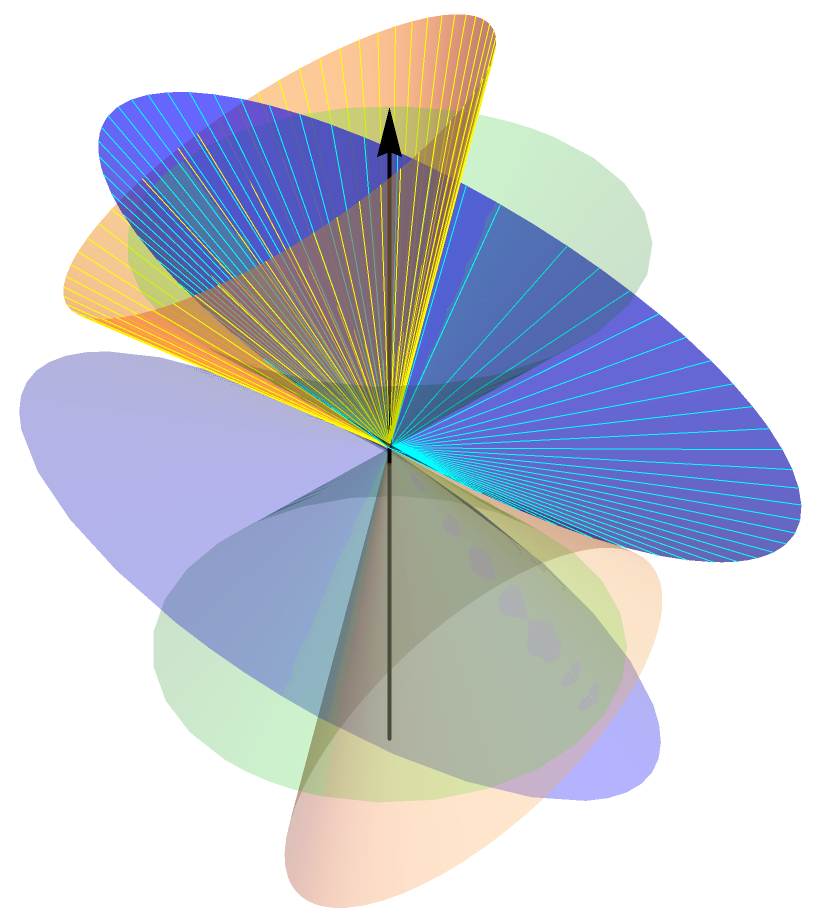}
\caption{\label{fig:C3-cones}}
\includegraphics[width=1\textwidth]{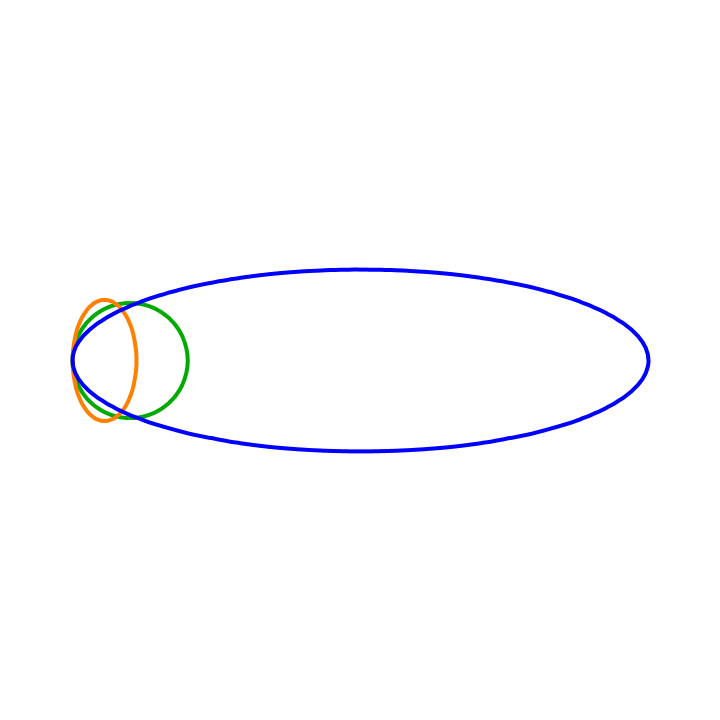}
\caption{\label{fig:C3-contours}}
\end{subcaptionblock}\quad
\begin{subcaptionblock}[T]{0.3\textwidth}
\centering
\includegraphics[height=1.\textwidth]{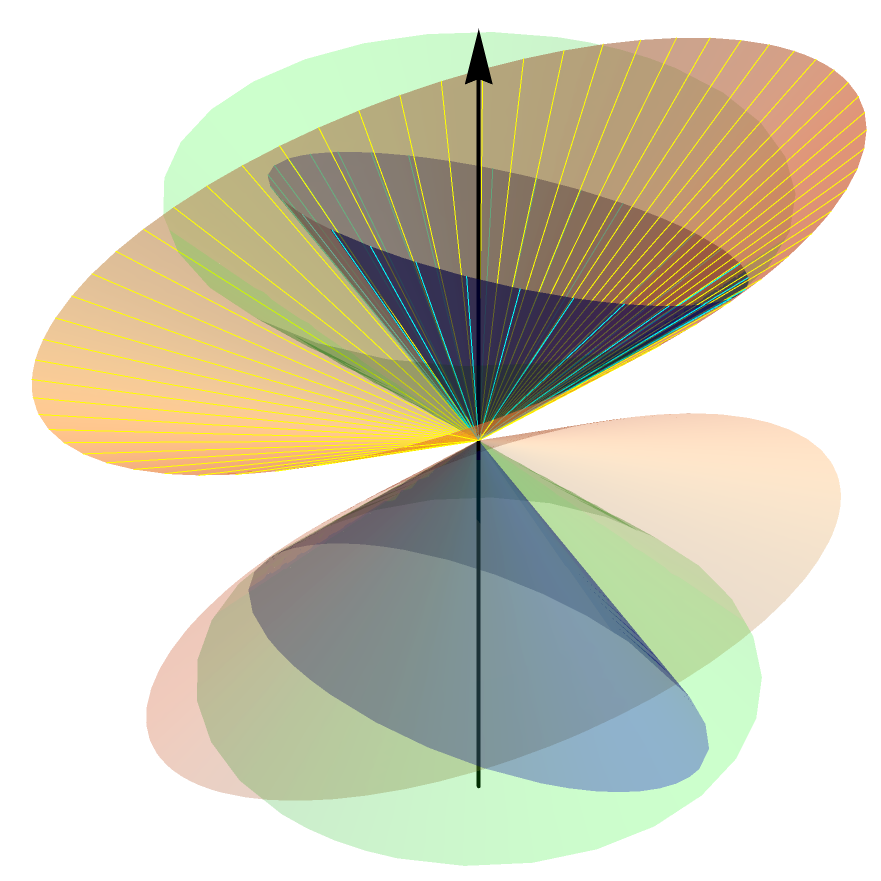}
\caption{\label{fig:C4-cones}}
\includegraphics[width=1\textwidth]{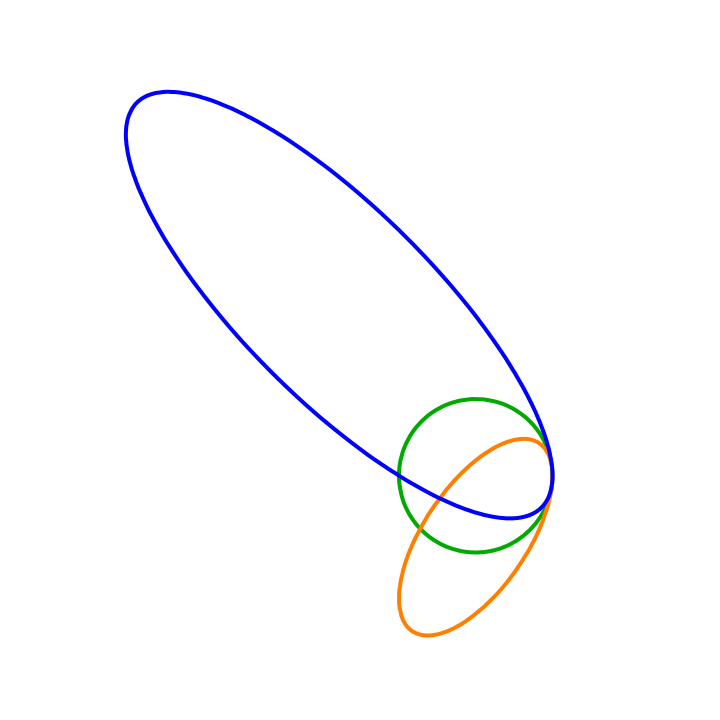}
\caption{\label{fig:C4-contours}}
\end{subcaptionblock}
\caption{Relative arrangement of acoustic cones with respect to the light cone for acoustic metrics in class II, III and IV. For class Ia see figure~\ref{fig:isotropic_GoodCauchy}, for class Ib --- figure~\ref{fig:ConesIb}. Cone colours as in figure~\ref{fig:isotropic_GoodCauchy_cones} with selected rays in future N-nappes shaded. Shading of selected momenta in P-cone based on choice of non-ghost signature. Lower panels are cuts through the cones at a constant height, demonstrating more precisely the intersection directions and relative arrangement. (a)-(b): class II metrics have cones that are both partially $g$-timelike and $g$-spacelike and cannot be brought into the rest frame through a Lorentz boost. (c)-(d): class III metrics have cones which \emph{touch} but do not intersect the light cone in exactly one direction (and can intersect in pairs of others). This is a limiting case between class I and II. (e)-(f): class IV metrics have cones which \emph{intersect} at the null eigenvector with the two cones tangent to each other at one of the intersections.\label{fig:cones2-4}}
\end{figure}

The first two classes are of most physical interest, with class III and IV limiting cases. For completeness we will consider each of the cases in turn. In the relevant $2+1$-dimensional subspace in the coordinates where the metric takes the canonical form, the relative orientation of the acoustic cone to the light cone can be described as:
\bgroup
\renewcommand\theenumi{\Roman{enumi}.}
\renewcommand\labelenumi{\theenumi}
\begin{enumerate}
\item  The acoustic P-cone is centred on the direction $\omega$ (class Ia, see figure~\ref{fig:isotropic_GoodCauchy}) or one of the other principal directions $k_i$ (class Ib, see figure~\ref{fig:ConesIb}).
\item The acoustic cone is tilted in the $\omega-k_x$ plane so that in one direction it is $g$-timelike and in the other --- $g$-spacelike. It is thus impossible to boost to the medium's rest frame where the cone would be symmetrical (see figure~\ref{fig:C2-cones}).
\item The acoustic cone nappe \emph{touches} the upper light cone nappe along the eigen-covector and is completed to either include a part of the upper light-cone nappe (class IIIa) or not (class IIIb). Limiting case between class I and class II (see figure~\ref{fig:C3-cones}).
\item The acoustic cone \emph{intersects} the light cone exactly twice. One intersection is along the $g$-null eigenvector $v_0^\mu$. The other intersection is along the $g$-null direction $v_1^\mu$ with $v_1^\mu v_{1\mu}=v_1^\mu v_{0\mu} =0$ with the surfaces of the cones tangent to each other there. (see figure~\ref{fig:C4-cones}).
\end{enumerate}
\egroup{}

In the following we will demonstrate by explicit construction that the hyperbolicity condition~\eqref{eq:detZ-updown} is equivalent to the existence of the cones, whatever the class of the metric. For some of the classes the cones will not be obvious in the canonical coordinates because of the existence of sound horizons and therefore negative energies.

\paragraph{Class I: $Z^{\mu\nu}$ diagonalisable over reals.\label{sec:classI}}

This is the most intuitive case. Here there are four real eigenvalues
$\lambda_{I}$, and $Z^{\mu}_{\nu}$ has four normalisable orthogonal
eigenvectors of which one must be $g$-timelike, e.g.~$v_{0}^{\mu}$ which is the observer's velocity for which the medium is at rest. It is only for this class that the frame can be chosen so that the medium is at rest, $q^\mu =0$.

In these coordinates, the acoustic metric is diagonal, $Z^{IJ}=\text{diag}(-\lambda_0,\lambda_1,\lambda_2,\lambda_3$) and its null surface, described by eq.~\eqref{eq:BigG-char}, is  just
\begin{equation}
0=Z^{IJ}P_{I}P_{J}=-\lambda_{0}\omega^{2}+\sum_{i}\lambda_{i}k_{i}^2 \,.\label{eq:IsotropicCone-1}
\end{equation}
This surface is a cone  for 0, 2 or all 4 of the $\lambda_{I}$ negative. Setups with an odd number of negative eigenvalues are not hyperbolic and therefore cannot be solved as an IVP. The $S_{IJ}$ acoustic metric is diagonal with eigenvalues $\lambda_{I}^{-1}$, so the P-cone and the ray cone for this class of metrics either both exist or both do not and both lie in the same (sub)class. The tensors $\Ztwo{IJ} = \diag(0,\lambda_0\lambda_1,\lambda_0\lambda_2,\lambda_0\lambda_3)$ while $\Stwo{IJ}=\lambda_0^{-1}\diag(0,\lambda_1^{-1},\lambda_2^{-1},\lambda_3^{-1})$ in these coordinates.

We subdivide the class into class Ia where the central direction of the cone is $g$-timelike (illustrated in figure~\ref{fig:isotropic_GoodCauchy}) and class Ib, where the central direction of the cone is $g$-spacelike, e.g.~$v_1^\mu$ (see figure~\ref{fig:ConeAcausal}).

\paragraph{Class Ia.}
\begin{itemize}
\item All $\lambda_{I}>0$; non-ghost (signature $(3,1)$): cone symmetric around $v_0^\mu$ with up to three distinct sound speeds corresponding to $g$-spacelike eigendirections $v_i^\mu$, $c_{\text{s},i}^{2}={\lambda_{i}}/{\lambda_{0}}$.
\item All $\lambda_{I}<0$; ghost (signature $(1,3)$): the cone is identical to the above case, but for overall sign difference giving ghost signature.
\end{itemize}
For class Ia, the tensors $\Ztwo{IJ}$ and $\Stwo{IJ}$ are both positive definite, and therefore this frame is a good Cauchy frame with no sound horizons.

\paragraph{Class Ib.}\label{def:classIbdef}
\begin{itemize}
\item $\lambda_{0,1}<0$, $\lambda_{2,3}>0$; non-ghost (signature $(3,1)$):
on its own this setup is just a mislabelling of the time and space directions, but in the presence of any other degrees of freedom, the consistency of this setup is fragile. Provided that $\lambda_0/\lambda_1<1$, this is \emph{not} acausal and a Cauchy surface can be found, i.e.~we are in the configuration of figure~\ref{fig:ConeNoIntersection}. Otherwise, no Cauchy surface exists and we have the configuration of figure~\ref{fig:ConeAcausal}.
\item $\lambda_{0}>0$, $\lambda_{1}>0$, other $\lambda_{2,3}<0$; ghost (signature $(1,3)$):
the acoustic metric differs by an overall sign from the previous and represents a ghost.
\end{itemize}
In class Ib, neither $\Ztwo{IJ}$ nor $\Stwo{IJ}$ are positive definite, so this frame is not a good Cauchy frame and sound horizons are present. When $|\lambda_0|<|\lambda_1|$, the P-cone does not overlap with the lightcone and we are in the acausal setup, figure~\ref{fig:ConeAcausal}.

In the whole of class I, the determinant~\eqref{eq:detZ-updown}
is
\begin{equation}
\det Z^I_J=\lambda_{0}\lambda_{1}\lambda_{2}\lambda_{3}>0
\end{equation}
It is easy to see that this condition is identical to the one provided
by the above cone constructions~\eqref{eq:IsotropicCone-1}.

\paragraph{Class II: $Z^{\mu\nu}$ diagonalisable with complex eigenvalues.\label{sec:classII}}

See figure~\ref{fig:C2-cones}. There are two real eigenvalues ($\lambda_{2}$ and $\lambda_{3})$
corresponding to $g$-spacelike eigenvectors and a complex conjugate
pair $\lambda'\pm i\lambda''$ with $\lambda',\lambda''\neq0$. The
acoustic metric and its inverse can then be written in a canonical
real form as
\begin{equation}
Z^{IJ}=\left(\begin{array}{cccc}
-\lambda' & \lambda'' & 0 & 0\\
\lambda'' & \lambda' & 0 & 0\\
0 & 0 & \lambda_{2} & 0\\
0 & 0 & 0 & \lambda_{3}
\end{array}\right)\,,\quad S_{IJ}=(\lambda'^{2}+\lambda''^{2})^{-1}\left(\begin{array}{cccc}
-\lambda' & \lambda'' & 0 & 0\\
\lambda'' & \lambda' & 0 & 0\\
0 & 0 & \frac{(\lambda'^{2}+\lambda''^{2})}{\lambda_{2}} & 0\\
0 & 0 & 0 & \frac{(\lambda'^{2}+\lambda''^{2})}{\lambda_{3}}
\end{array}\right)
\end{equation}
with $I=0$ corresponding to a $g$-timelike direction (see~\cite[pg.\ 293]{LandavshitzII}).
We then have
\begin{equation}
\det Z^I_J=(\lambda'^{2}+\lambda''^{2})\lambda_{2}\lambda_{3}>0.\label{detZ-nondiag}
\end{equation}
Hyperbolicity in this class requires that $\lambda_{2}$ and $\lambda_{3}$ have the same sign but does not constrain the non-diagonal block. By inspection, $S_{IJ}$ is also class II.

Then, in the canonical frame, the characteristic surface for
$Z^{IJ}$ can be written as\footnote{$P_0=-\omega$ in our convention.} \begin{equation}
Z^{IJ}P_{I}P_{J}=-\lambda'\left(\omega^{2}+\frac{2\lambda''}{\lambda'}\omega k_x-k_x^{2}-\frac{\lambda_{2}}{\lambda'}k_y^{2}-\frac{\lambda_{3}}{\lambda'}k_z^{2}\right)=0\,.
\end{equation}
The cone's opening angle in the $k_y=k_z=0$ plane is $\pi/2$ and therefore
it always includes both timelike and spacelike parts of the lightcone
in its interior, but it is never acausal and therefore a good Cauchy frame exists. We have $\Ztwo{IJ}=\diag(0,\lambda'^2 +\lambda''^2,\lambda'\lambda_2,\lambda'\lambda_3)$ while $\Stwo{IJ}= (\lambda'^2 +\lambda''^2)^{-1}\diag(0,1,\lambda'/\lambda_2,\lambda'/\lambda_3)$ The four possibilities can be categorised as
\begin{itemize}
    \item $\lambda'>0, \lambda_{2,3} >0$, non-ghost signature $(3,1)$. The canonical frame is a good Cauchy frame and there are no sound horizons.
    \item $\lambda'<0, \lambda_{2,3}<0$, ghost signature (1,3): as above, but the scalar is a ghost.
    \item $\lambda'<0, \lambda_{2,3} >0$, non-ghost signature $(3,1)$: the canonical frame is a bad Cauchy frame and has a sound horizon. $k_x$ acts as the $Z$-timelike direction for the acoustic cone. \item $\lambda'>0, \lambda_{2,3}<0$, ghost signature (1,3): as above, but the scalar is a ghost.
\end{itemize}

\paragraph{Class III: double null eigenvector.\label{sec:classIII}}

See figure~\ref{fig:C3-cones}. For this class, coordinates can be chosen in which the acoustic metric
and its inverse are both reduced to the canonical form
\begin{equation}
Z^{IJ}=\left(\begin{array}{cccc}
-\lambda-\mu & \mu & 0 & 0\\
\mu & \lambda-\mu & 0 & 0\\
0 & 0 & \lambda_{2} & 0\\
0 & 0 & 0 & \lambda_{3}
\end{array}\right)\,,\quad S_{IJ}=\lambda^{-2}\left(\begin{array}{cccc}
-\lambda+\mu & -\mu & 0 & 0\\
-\mu & \lambda+\mu & 0 & 0\\
0 & 0 & \lambda^{2}/\lambda_{2} & 0\\
0 & 0 & 0 & \lambda^{2}/\lambda_{3}
\end{array}\right)\,.\label{eq:ClassIII}
\end{equation}
$Z^{IJ}$ has two $g$-spacelike eigenvectors with eigenvalues $\lambda_{2}$
and $\lambda_{3}$ and a repeated eigenvalue $\lambda$ corresponding
to the $g$-null eigen-covector $(1,1,0,0)$ along which the acoustic cone and the lightcone touch but do not intersect. This configuration of the cones can be seen as a boundary between class Ia and class II, where the class Ia cone is tilted exactly so as to touch the lightcone, just before crossing it to become class II or when class II is tilted just before if becomes class Ib.  This canonical choice of coordinates is such that $\mu$ has the minimum possible magnitude, so the medium cannot be put in a rest frame through any boost.

$S_{IJ}$ is also class III with the mapping $\mu\rightarrow-\mu/\lambda^{2}$
and $\lambda\rightarrow\lambda^{-1}$, so either both the ray and
P-cones exist or both do not. The determinant of $Z_{J}^{I}$
is positive when
\begin{equation}
\det Z^I_J=\lambda^{2}\lambda_{2}\lambda_{3}>0\,,\label{posdetIII}
\end{equation}
i.e.\ whenever $\lambda_{2,3}$ have the same sign.  We also have $\Ztwo{IJ} = \diag(0,\lambda^2, (\lambda+\mu)\lambda_2, (\lambda+\mu)\lambda_3)$, so whenever $(\lambda+\mu)\lambda_{2,3}<0$ this canonical frame is not a good Cauchy frame. $\Stwo{IJ}= \lambda^{-2}\diag(0,1,(\lambda-\mu)/\lambda_2,(\lambda-\mu)/\lambda_3)$, so a sound horizon is present in this frame whenever $(\lambda-\mu)/\lambda_{2,3}<0$ --- constant $\omega$ slices of the P-cone are not closed. In class III, there is no frame which is both a good Cauchy frame and has no sound horizons.

We can perform a boost in the $x$ direction with parameter at least
\begin{align}
v & >\frac{\lambda-\mu}{\lambda+\mu}\,,&\frac{\mu}{\lambda}&>0\label{boostIII}\\
v & <\frac{\lambda+\mu}{\lambda-\mu}\,,&\frac{\mu}{\lambda}&<0 \notag
\end{align}
which brings the characteristic surface to the form
\begin{align}
Z^{\bar{I}\bar{J}}P_{\bar{I}}P_{\bar{J}}=-\lambda&\left(\left((1-v^{2})\left(\bar{\omega}^{2}-k^2_{\bar{x}}\right)+(1-v)^{2}\frac{\mu}{\lambda}\left(\bar{\omega}^{2}+k^2_{\bar{x}}\right)\right)\right. \\
&~~\left.+2(1-v)^2 \frac{\mu}{\lambda} \bar{\omega} k_{\bar{x}}  - \frac{\lambda_2}{\lambda}k^2_y-\frac{\lambda_{3}}{\lambda}k_z^{2}    \right)=0\,, \notag
\end{align}

Provided that $\lambda_{2}$ and $\lambda_{3}$ have a common sign,
this surface is a cone, just as in condition~\eqref{posdetIII}. Then, in the $\bar{y}=\bar{z}=0$ plane, the cone is given by
\begin{equation}
\bar{\omega}=-k_{\bar{x}},\frac{\lambda(1+v)-\mu(1-v)}{\lambda(1+v)+\mu(1-v)}k_{\bar{x}}\,, \label{eq:classIIIdirs}
\end{equation}
so it always lies on the light cone in one direction.\footnote{Since $P_0=-\omega$, the first solution in eq.~\eqref{eq:classIIIdirs} is the null eigencovector of this class.}We can now split this class into two subclasses, similarly to Class I. Subclasses are preserved under inversion of the metric.

\paragraph{Class IIIa.} the upper nappe of the acoustic cone includes a part of the upper nappe of the light-cone. This acoustic metric separates class Ia and class II:
\begin{itemize}
    \item $\lambda>0$, $\lambda_{2,3}>0$, non-ghost signature $(3,1)$.
    \item $\lambda<0$, $\lambda_{2,3}<0$, ghost signature $(1,3)$.
\end{itemize}

\paragraph{Class IIIb.} the upper nappe of the acoustic cone does not include the upper nappe of the lightcone. We have an extra condition from requiring that the N-cones are not acausal, giving $\mu<0$. In such a case, the upper P-cone nappe includes a part of the past light cone. The two possible cases are
\begin{itemize}
    \item $\lambda<0$, $\lambda_{2,3}>0$, non-ghost signature $(3,1)$.
    \item $\lambda>0$, $\lambda_{2,3}<0$, ghost signature $(1,3)$.
\end{itemize}

\paragraph{Class IV: triple null eigenvector.\label{sec:classIV}}

See figure~\ref{fig:C4-cones}. For this class, coordinates can be chosen in which the acoustic metric is reduced to the canonical form
\begin{equation}
Z^{IJ}=\lambda\left(\begin{array}{cccc}
-1 & 0 & \sigma & 0\\
0 & 1 & \sigma & 0\\
\sigma & \sigma & 1 & 0\\
0 & 0 & 0 & \rho
\end{array}\right)\,,\quad S_{IJ}=\lambda^{-1}\left(\begin{array}{cccc}
-1+\sigma^{2} & \sigma^{2} & -\sigma & 0\\
\sigma^{2} & 1+\sigma^{2} & -\sigma & 0\\
-\sigma & -\sigma & 1 & 0\\
0 & 0 & 0 & \rho^{-1}
\end{array}\right)
\end{equation}
This $Z^{IJ}$ has one $g$-spacelike eigenvector with eigenvalue
$\rho\lambda$ and the thrice-repeated eigenvalue $\lambda$ associated with the null eigenvector $v^\mu_0=(1,1,0,0)$. The acoustic cone intersects the light cone in exactly two directions. $v^\mu_0$ and $v^\mu_1=(1,-1,0,0)$, but with the acoustic and light cones tangent to each other at $v^\mu_1$.

Determinant positivity is
\begin{equation}
\det Z^I_J=\lambda^{4}\rho>0\,,\label{eq:detposIV}
\end{equation}
which is satisfied whenever $\rho>0$ and does not depend on $\sigma$. We also have
\begin{equation}
    \Ztwo{IJ}= \lambda^2 \left(\begin{array}{cccc}
    0 & 0 & 0& 0\\
    0&1 &\sigma & 0\\
    0&\sigma & 1+\sigma^2 & 0 \\
    0&0 & 0 & \rho
    \end{array}\right) \qquad
    \Stwo{IJ} = \lambda^{-2} \left(\begin{array}{cccc}
    0& 0 & 0& 0 \\
    0& 1 &-\sigma & 0\\
    0& -\sigma & 1 & 0 \\
    0& 0 & 0 & \frac{1-\sigma^2}{\rho}
    \end{array}\right)
\end{equation}
$\Ztwo{IJ}$ is positive definite for any $\sigma$, so this canonical frame is always a good Cauchy frame. On the other hand, $\Stwo{IJ}$ is only positive definite for $\sigma^2<1$. Otherwise there is a sound horizon and constant $\omega$ slices of the P-cone do not close. We can boost the canonical frame in the $k_x$ direction with speed at least
\begin{equation}
v>\frac{\sigma^{2}-1}{\sigma^{2}+1}\,.\label{eq:vIV}
\end{equation}
In these boosted coordinates the P-cone is described by
\begin{equation}
Z^{\bar{I}\bar{J}}P_{\bar{I}} P_{\bar{J}} =-\bar{\omega}^{2}+k_{\bar{x}}^{2}+k_{\bar{y}}^{2}-2\gamma_v(1-v)\sigma k_{\bar{y}}\left(\bar{\omega}-k_{\bar{x}}\right)+\rho k_{\bar{z}}^{2}=0\,,\label{eq:triplenullcone}
\end{equation}
with $\gamma_v$ the Lorentz factor. In these new coordinates with $v$ satisfying~\eqref{eq:vIV}, constant $\bar{\omega}$ hypersurfaces are ellipsoids and constant
$k_{\bar{x}}, k_{\bar{y}}, k_{\bar{z}}$ surfaces are hyperboloids, provided that $\rho>0$. Thus we explicitly have a cone for all $\sigma$ and obtain a condition equivalent to eq.~\eqref{eq:detposIV}. We can also show that
\begin{itemize}
    \item $\lambda > 0$: signature is (3,1) and the scalar is a non-ghost
    \item $\lambda < 0$: signature is (1,3) and the scalar is a ghost.
\end{itemize}

At first glance it is not clear whether $S_{IJ}$ belongs to class
IV. It can be brought to the standard form by first performing a rotation
in the $x$-$y$ plane by the angle $\sin\theta=\sigma/\sqrt{4+\sigma^{2}}$
and then boosting in the new $y$-direction with parameter $v=-\sigma/\sqrt{4+\sigma^{2}}$.
In these new coordinates, we have
\begin{equation}
S_{IJ}=\lambda^{-1}\left(\begin{array}{cccc}
-1 & 0 & -\frac{\sigma}{\sqrt{4+\sigma^{2}}} & 0\\
0 & 1 & -\frac{\sigma}{\sqrt{4+\sigma^{2}}} & 0\\
-\frac{\sigma}{\sqrt{4+\sigma^{2}}} & -\frac{\sigma}{\sqrt{4+\sigma^{2}}} & 1 & 0\\
0 & 0 & 0 & \rho^{-1}
\end{array}\right)\,,
\end{equation}
an explicit Class IV metric with replacements $\lambda\rightarrow\lambda^{-1}$,
$\sigma\rightarrow-\sigma/\sqrt{4+\sigma^{2}}$, $\rho\rightarrow\rho^{-1}$.
The constant $t$ sections are ellipsoids and constant $x$ and $y$ sections and $z$  --- hyperboloids. Thus we have shown that condition~\eqref{eq:detposIV}
is sufficient to determine if a cone exists also for this class.

We have thus demonstrated that for all possible non-singular acoustic metrics
$Z^{\mu\nu}$, an acoustic cone exists whenever $\det Z_{\nu}^{\mu}>0$.
This is a necessary and sufficient condition for $Z$ and the P-cone,
but also we are guaranteed under this condition that an acoustic ray-cone
will exist for the metric $S_{\mu\nu}$. Whether it is possible to go into the rest-frame of the medium created by the background depends on which class the metric falls into. The discussion we have presented in section~\ref{sec:twometrics} is general and applies to all the classes.

\section{Acoustic metrics: illustrative examples\label{sec:examples}}

In this section, we will give some examples of acoustic metrics and study their properties. In particular we will study the Gordon's metric~\cite{Gordon:1923qva} --- the acoustic metric for an isotropic medium, as well as two classes of scalar-tensor models, k-\emph{essence}~\cite{ArmendarizPicon:1999rj,ArmendarizPicon:2000ah} and kinetic gravity
braiding~\cite{Deffayet:2010qz,Pujolas:2011he}.

\subsection{Gordon's metric and the Mach cone \label{sec:phys-cones}}

Let us make things concrete using a well-studied example --- an isotropic
medium with phonons propagating at sound speed $\cs$. This is a metric frequently used in the analogue gravity community to model curved spacetime using superfluids (see the review~\cite{Barcelo:2005fc}). The acoustic metric is given by Gordon's metric~\cite{Gordon:1923qva},
\begin{equation}
Z^{\mu\nu} = \cs^{-2} ( \cs^2 g^{\mu\nu} - (1-\cs^2)u^\mu u^\nu)\,, \label{eq:GordonMetric}
\end{equation}
where the medium's flow velocity is given by $u^\mu$. Both $u^\mu$ and $\cs$ are in principle all functions of spacetime location. For the purpose of this section, we assume that $u^\mu$ is subluminal ($g$-timelike), while $\cs$ is arbitrary but real. This means it is possible to diagonalise $Z^{\mu\nu}$ over reals and therefore it is class I.  In the medium's rest frame, the acoustic metric is $S_{\mu\nu}= \diag \left(-\cs^2,1,1,1\right)$ while its inverse --- $Z^{\mu\nu}=\cs^{-2}\diag \left(-1, \cs^2, \cs^2, \cs^2 \right)$,
see figure~\ref{fig:isotropic_GoodCauchy} for an illustration. The dispersion relation given by the P-cone~\eqref{eq:BigG-char} is just $\omega^2 = \cs^2 \delta^{ij} k_i k_j$
while the ray cone is given by $\cs ^2 \mho^2 = \delta_{ij} \dot{r}^i \dot{r}j$. As should be expected, the rest frame is a good Cauchy frame and there is no sound horizon.

Performing a boost with speed $v$ (Lorentz factor $\gamma_v$), the metrics in the new coordinates take the form
\begin{equation}
Z^{\mu\nu}=\frac{\gamma_{v}^{2}}{\cs^2}\begin{pmatrix}\begin{array}{rrr}
-(1-v^{2}\cs^{2}) & (1-\cs^{2})v\\
(1-\cs^{2})v & \cs^{2}-v^{2}\\
 &  & \cs^{2}\gamma_{v}^{-2}
\end{array}\end{pmatrix},\, S_{\mu\nu}=\gamma_{v}^{2}\begin{pmatrix}\begin{array}{rrr}
-(\cs^{2}-v^{2}) & -(1-\cs^{2})v\\
-(1-\cs^{2})v & (1-v^{2}\cs^{2})\\
 &  & \gamma_{v}^{-2}
\end{array}\end{pmatrix}.\label{eq:isometrics}
\end{equation}
where to save space we have collapsed two dimensions into a single coordinate.

We can compute $\Ztwo{ij}$ and $\Stwo{ij}$ in this boosted frame according to eq.~\eqref{eq:GoodCauchyP} and~\eqref{eq:Stwomunu}, obtaining:
\begin{equation}
\Ztwo{ij}=\cs^{-2}\begin{pmatrix}\begin{array}{cc}
1 & 0\\
0 & \gamma_v^2(1-\cs^2 v^2)
\end{array}\end{pmatrix}\,,\quad \Stwo{ij}=\cs^{2}\begin{pmatrix}\begin{array}{cc}
1 & 0\\
0 & \gamma_v^2\left(1-\frac{v^2}{\cs^2}\right)
\end{array}\end{pmatrix}\,.\label{eq:Z2S2iso}
\end{equation}
where again we have suppressed the third dimension, identical to the second.

In the subluminal case, $\cs<1$, no boost with $v<1$ can change the sign of $Z^{00}$ or any of the eigenvalues of $\Ztwo{ij}$. All frames are good Cauchy frames. On the other hand, a supersonic boost $v>\cs$  moves the ray cone out of the time direction of the observer ($S_{00}$ changes sign). Equivalently,  eigenvalues of $\Stwo{ij}$ change sign and therefore  the wave cannot propagate in directions for which $\Srr<0$; a sound horizon has appeared. $Z^{ij}$  changes the sign of one eigenvalue, giving negative energies in the boosted frame for some modes. See figure~\ref{fig:SoundHorizon} for an illustration.

When a particle interacting with scalar moves in the boosted frame, it will now produce a shockwave --- the Cherenkov/Mach cone. Its outer surface is given by $\Srr=0$ in the particle's frame, or equivalently, by $\omega=0$ modes in this frame. In particular, we have in the particle's rest frame
\begin{align*}
N^{\mu}_v & =\left(\frac{\gamma_v v(1-\cs^2)}{\sqrt{v^2-\cs^2}},\gamma_v\sqrt{v^2-\cs^2},\cs\right)k.\\
P_{v\mu} & =\frac{\gamma_v^{-1} }{\sqrt{v^2 -\cs^2}}\left( 0, -\cs, \gamma_v \sqrt{v^2 -\cs^2} \right)k \,.
\end{align*}
The spatial vector $N^{i}$ points along the shockwave in positive 1-direction (let us call this `right'), behind the particle, while the
spatial vector $P_{i}$ is orthogonal to it and points forward (to the `left') in the direction of motion of the particle. Interpreting
this through the geodesic equations~\eqref{eq:geodeisc}, means that
the modes $P_{i}$ are created at the particle and then in its rest-frame propagate
to the right along the shockwave cone with phase speed
\begin{equation}
v_{\text{p}}=\frac{\sqrt{N^{i}N^{i}}}{N^{0}}=\gamma_{\cs}\sqrt{v^{2}-\cs^{2}}
\end{equation}
The cone opening half-angle in the particle rest frame is given by
\begin{equation}
\cos\alpha_{v}=\frac{\gamma_{\cs}\sqrt{v^2-\cs^2}}{v}\,,
\end{equation}
with the rays pointing right, while momenta point left. Momentum conservation fixes $k$ for this angle
to be zero, so no energy loss occurs at the outer surface of the
cone (although see section~\ref{sec:Hamiltonia} for a discussion of zero-energy modes living on the Mach cone). However, the modes with rays moving to the left of the fluid velocity $u^\mu$ inside the Mach cone have negative
energies from the point of view of the particle and therefore their
production is kinematically allowed, leading to energy loss and the
full Cherenkov formula.

Transforming back to the medium's rest frame, we obtain the expressions
for the vectors in the medium's frame,\footnote{Note the negative sign in $P_0$. In our convention, that is positive energy in the covector.}
\begin{align}
N^{\mu} & =\left(\frac{v^2}{\sqrt{v^2-\cs^2}},- \frac{\cs^2}{\sqrt{v^2-\cs^2}},\cs\right)k\,,\\
P_{\mu} & = \left(-\frac{\cs^2v}{\sqrt{v^2-\cs^2}},-\frac{\cs^2}{\sqrt{v^2-\cs^2}},\cs\right)k\,. \notag
\end{align}
The spatial part of these vectors is \emph{aligned} in the medium's
rest frame (although $N^{\mu}P_{\mu}=0$ as required). We recover
the standard formula for the Cherenkov cone half-angle,
\begin{align}
\cos\alpha_{\text{lab}}=\frac{\cs}{v}=\frac{1}{nv}\,,
\end{align}
with $n$ the index of refraction and the phase speed from
$N^{\mu}$ as $v_\text{p}=\cs$. Since the rays and momenta are both pointing to the left in the medium's rest-frame, the shockwave is moving together with the particle.

Let us briefly discuss the situation when the sound speed is superluminal,
$c_{\text{s}}>1$. The P-cone is now $g$-timelike, while the ray-cone
is $g$-spacelike, so their possible behaviour under boosts is now reversed.
It can be seen that a boost with $v>\cs^{-1}$ changes the sign
of $S_{11}$ in eq.~\eqref{eq:isometrics} and equivalently the eigenvalues of $\Ztwo{ij}$ in eq.~\eqref{eq:Z2S2iso}. The ray cone now intersects $\Sigma_{v}$ and \emph{in this frame} has directions with instantaneous propagation (see figure~\ref{fig:BadCauchy}) and others sending information into the coordinate past --- an apparent causal paradox \emph{for this observer} and therefore a bad Cauchy frame. Equivalently,
the P-cone no longer includes the time direction of this frame, $Z^{00}$
changes sign and the cone no longer covers all the directions on $\Sigma_{v}$ ---
there exist spatial momenta $k_{i}$ which have complex energies $\omega_{\pm}$.

While we have not derived any new properties here, we have explicitly demonstrated how our covariant approach allows us to derive the geometry of the Mach cone, phase velocities and their transformations using standard Lorentz boosts.

\subsection{k-essence\label{sec:k-Essence}}

\emph{k-essence} is a class of scalar-tensor models where with a non-canonical
kinetic term involved only first derivatives of the scalar field
$\phi$,
\begin{equation}
\mathcal{L}=K(X,\phi)\,,
\end{equation}
where $X\equiv-\phi_{,\mu}\phi^{,\mu}/2$ is the canonical kinetic
term. The properly normalised acoustic inverse metric takes the form
\begin{equation}
Z^{\mu\nu}=\frac{1}{\sqrt{D\mathcal{L}_{,X}}} \left(g^{\mu\nu}-\frac{\mathcal{L}_{,XX}}{\mathcal{L}_{,X}}\phi^{,\mu}\phi^{,\nu} \right)\,,\label{eq:Z_metric_kessence}
\end{equation}
with $D\equiv \mathcal{L}_{,X} +2X\mathcal{L}_{,XX}$. This can be inverted to give the metric for rays
\begin{equation}
S_{\mu\nu}= \sqrt{D \mathcal{L}_{,X}} \left(g_{\mu\nu}+\frac{\mathcal{L}_{,XX}}{D}\phi_{,\mu}\phi_{,\nu}\right)\,.\label{eq:N_Metrik_kessence}
\end{equation}
For both the tensors above, $\partial_{\mu}\phi$ is always an eigenvector.
Moreover, it is an eigenvector for the energy-momentum tensor (EMT)
for the \emph{k-essence} scalar field,
\begin{equation}
T_{\mu\nu}=\mathcal{L}_{,X}\phi_{,\mu}\phi_{,\nu}+g_{\mu\nu}\mathcal{L}\,.
\end{equation}
The acoustic metric represents a hyperbolic system (i.e.~the cones exist) only when $\det Z^\mu_\nu=\mathcal{L}_{,X}^{-3} D^{-1}>0$, which we will assume.

There are three separate cases:

\paragraph{Timelike $\partial_{\mu}\phi$: class Ia.\label{sec:kEssTL}}

Time-like $\partial_{\mu}\phi$ corresponds to irrotational hydrodynamics
and can be normalised to become a velocity vector, $u_{\mu}=-\partial_{\mu}\phi/\sqrt{2X}$.
The Lagrangian can be identified with the pressure $P=\mathcal{L}$, the energy density is $\mathcal{E}=2XP_{,X}-P$. We can then rewrite the metric~\eqref{eq:Z_metric_kessence} as
\begin{equation}
Z^{\mu\nu}=\frac{1}{\mathcal{E}_{,X}c_\text{s}^3} \left(-u^{\mu}u^{\nu}+c_{\text{s}}^{2}h^{\mu\nu}\right)\,,\quad S_{\mu\nu} = \mathcal{E}_{,X} c_\text{s} \left( -c_\text{s}^2 u_\mu u_\nu + h_{\mu\nu} \right),\label{eq:k-ess-TL}
\end{equation}
with $D=\mathcal{E}_{,X}$ and the sound speed given by
\begin{equation}
\cs^2= \frac{P_{,X}}{\mathcal{E}_{,X}}=\left(\frac{\partial P}{\partial\mathcal{E}}\right)_{\phi}\,.
\end{equation}
When cones exist, the signature implies we have a non-ghost for $\mathcal{E}_{,X}>0$ and a ghost whenever $\mathcal{E}_{,X}<0$. This metric can always be diagonalised with real eigenvalues, with the timelike eigenvector $u^\mu$ with eigenvalue $\mathcal{E}_{,X}^{-1}c_\text{s}^{-3}$ and three spacelike eigenvectors with shared eigenvalue $\mathcal{E}_{,X}^{-1}c_\text{s}^{-1}$, i.e.\ all k-essence metrics with timelike $\partial_\mu\phi$ are class Ia and describe an isotropic medium --- they are equivalent to the Gordon metric~\eqref{eq:GordonMetric} up to normalisation. The frame $u_\mu$ is the rest-frame of the medium and is always a good Cauchy frame with no sound horizon.

\paragraph{Spacelike $\partial_{\mu}\phi$: class Ia.\label{sec:kEssSL}}

This case is particularly interesting for static solutions, see
e.g.~ref.~\cite{ArmendarizPicon:2005nz}. For spacelike $\partial_{\mu}\phi$
we can introduce a unit vector $l_{\mu}\equiv \partial_{\mu}\phi/\sqrt{-2X}$,
so that
\begin{equation}
Z^{\mu\nu}=\frac{1}{\sqrt{D \mathcal{L}_{,X}}}\left(\mathcal{L}_{,X}g^{\mu\nu}+2X\mathcal{L}_{,XX}l^{\mu}l^{\nu} \right)\,.
\end{equation}
The signature then implies that $\mathcal{L}_{,X}>0$ is a non-ghost, while $\mathcal{L}_{,X}<0$ is a ghost. The metric is diagonalisable over reals with non-null eigenvectors and always class Ia.

In the frame of eigenvectors, the sound speed is not isotropic, but rather has a preferred direction $l^\mu$ in which it is not luminal, but rather
\begin{equation}
c_{\text{s},l}^{2}=\frac{\mathcal{L}_{,X}+2X\mathcal{L}_{,XX}}{\mathcal{L}_{,X}}\,.
\end{equation}
This is the inverse of the sound speed in timelike case of section~\ref{sec:kEssTL}, a result which was first obtained in ref.~\cite{ArmendarizPicon:2005nz}. The sound speed is luminal in the other eigendirections.

\paragraph{Null $\partial_{\mu}\phi$: class III.\label{sec:kEssNull}}

In particular, this case is relevant for plane-wave backgrounds $\phi\left(t-x\right)$
which are exact solutions for all shift-symmetric k-essence theories~\cite{Babichev:2007dw}. The gradient $\phi_{,\mu}$ is a null eigenvector
with eigenvalue $\mathcal{L}_{,X}$. Consider a timelike unit vector
$V^{\mu}$, then
\begin{equation}
r_{\mu}=\frac{\phi_{,\mu}+V_{\mu}\left(V^{\lambda}\phi_{,\lambda}\right)}{V^{\lambda}\phi_{,\lambda}}\,,
\end{equation}
is a spacelike unit vector, $r^{\mu}r_{\mu}=1$, orthogonal to $V^{\mu}$.
Two other spacelike vectors $e_{1}^{\mu}$ and $e_{2}^{\mu}$ orthogonal
to $V^{\mu}$ and $r^{\mu}$ are also orthogonal to $\phi_{,\mu}$.
These spacelike vectors are also eigenvectors with the eigenvalues
$\mathcal{L}_{,X}$. We can use $\left(V_{\mu},r_{\mu},e_{1\mu},e_{2\mu}\right)$
as a basis and rewrite the acoustic metric~\eqref{eq:N_Metrik_kessence} as
\begin{equation}
Z^{\mu\nu}=\mathcal{L}_{,X}^{-1}\left(-V^{\mu}V^{\nu} + r^{\mu}r^{\nu} + \sum_{i}e_{i}^\mu e_{i}^\nu\right) + \frac{\left(V^{\lambda}\phi_{,\lambda}\right)^{2}\mathcal{L}_{,XX}}{\mathcal{L}_{,X}^2}\left(r^{\mu}r^{\nu}-V^{\mu}V^{\nu}-V^{\mu}r^{\nu}-V^{\nu}r^{\mu}\right)\,.
\end{equation}
Given the null eigenvector, this metric is of the form of class III, eq.~\eqref{eq:ClassIII}, with
\begin{align}
\lambda= & \lambda_{2}=\lambda_{3}=\mathcal{L}_{,X}^{-1}\,,\qquad \mu= (V^{\lambda}\phi_{,\lambda})^{2}\frac{\mathcal{L}_{,XX}}{\mathcal{L}_{,X}^2}\,.
\end{align}
Note that in the null $\phi_{,\mu}$ case, $X=0$ and $\mathcal{L}_{,X}$
and $\mathcal{L}_{,XX}$ do not depend on the choice of $V^{\mu}$.
Thus only the value of $\mu$ changes when different frames are chosen.

The cone exists whenever $\mathcal{L}_{,X}\neq 0$, but $\mathcal{L}_{,X}>0$ is required for non-ghosts. These properties are independent of the sign of $\mathcal{L}_{,XX}$. However, the frame defined by $V^{\mu}$ is \emph{not} good  Cauchy frame whenever
$\mathcal{L}_{,X}+(V^\lambda \phi_{,\lambda})^2 \mathcal{L}_{,XX} < 0$ while a sound horizon is present for $\mathcal{L}_{,X}-(V^\lambda \phi_{,\lambda})^2 \mathcal{L}_{,XX} < 0$. Nonetheless, a good choice of frame always exists.

\subsection{Kinetic gravity braiding\label{sec:KGB}}
Kinetic gravity braiding~\cite{Deffayet:2010qz,Kobayashi:2010cm} is a subclass of Horndeski scalar-tensor theories~\cite{Horndeski:1974wa} in which the scalar does not derivatively couple to curvature in the action and therefore the acoustic metric of the scalar and gravity can be straightforwardly demixed~\cite{Pujolas:2011he}. The Lagrangian is given by
\begin{equation}
    \mathcal{L} = K(X) -G(X)\Box\phi \label{eq:L-KGB}
\end{equation}
where $X\equiv-\phi_{,\mu}\phi^{,\mu}/2$ is the canonical kinetic term and we have specialised to the shift symmetric case in which $K$ and $G$ only depend on $X$ and not the field $\phi$. The kinetic operator still mixes with gravity in this theory, but can be demixed and then the acoustic metric is
\begin{equation}\label{ZmunuKGB}
    \Zt^{\mu\nu} = \sqrt{-S}Z^{\mu\nu} =  \Omega \, g^{\mu\nu} + \Xi\nabla^\mu\phi \nabla^\nu\phi + 2 \nabla^{(\mu} \left( G_{3X} \nabla^{\nu)}\phi \right),
\end{equation}
where the proper normalisation can by obtained by computing the determinant of this matrix and where
\begin{eqnarray}
\Omega &=& K_{X}  - 2 G_{X} \square \phi  + G_{XX}\nabla^\rho \phi \nabla^\sigma\phi \nabla_\rho\nabla_\sigma\phi - \frac{2 }{\Mpl^2 } X^2 G_{X}^2\,, \\
\Xi &=& - K_{XX} + G_{XX}\square \phi- \frac{4 }{\Mpl^2} X G_{X}^2  \,. \notag
\end{eqnarray}
where the terms involving the reduced Planck mass $\Mpl$ are generated in the demixing process.  The essential difference with respect to the \emph{k-essence} metric~\eqref{eq:Z_metric_kessence} is the appearance of second derivatives of the background which implies that a second preferred direction appears in the acoustic metric in addition to $\partial_\mu \phi$. This implies that $Z^{\mu\nu}$ depends on the connection of the spacetime metric, but second derivatives have been removed by the demixing process.

The existence of two independent vectors in the acoustic metric means that even in the case of a $g$-timelike $u_\mu \equiv -\partial_\mu\phi/\sqrt{2X}$, and therefore a hydrodynamical interpretation for the scalar, the frame $u_\mu$ is not comoving. The constant-$\phi$ slicing usually provides natural coordinates in which to describe the scalar-field theory in a general manner using effective operators. In the below we will demonstrate that in kinetic gravity braiding it is possible to construct backgrounds which give hyperbolic $Z^{\mu\nu}$ and the fluctuations are non-ghosts and yet one of the usual assumptions about good media is violated:
\begin{itemize}
    \item The constant-$\phi$ slicing is a bad Cauchy frame and therefore this set of coordinates is not appropriate to determine how the system evolves.
    \item It is not possible to boost to a rest frame and therefore the metric is not diagonalisable over reals, i.e.\ it lies in class II.
\end{itemize}
Thus kinetic gravity braiding provides the simplest example of a concrete and consistent theory in which background solutions exist which cannot be described using the usual effective theory approach, or the medium described by the background can not be put in the rest frame and therefore the machinery of this paper must be employed to study it.

For the purpose of minimal examples, let us send $\Mpl\rightarrow\infty$ and assume that the spacetime metric is Minkowski. We now construct a spherically symmetric background with a \emph{timelike} $\partial_\mu\phi$, picking as an ansatz
\begin{equation}
    \phi(t,r) = \mu t + \varphi(r)\,. \label{eq:blob-ansatz}
\end{equation}
Under these assumptions, we have $2X=\mu^2 - \varphi'^2>0$. The only non-vanishing components of the acoustic metric are:
\begin{align}
    \Zt^t_t &= K_X+ \mu^2K_{XX} -2(G_X +XG_{XX})\varphi''\,,\label{eq:blob-Z}\\
    \Zt^r_r &= K_X- \varphi'^2 K_{XX}\,, \notag \\
    \Zt^t_r &= -\Zt^r_t =  \mu\varphi'K_{XX}\,, \notag\\
    \Zt^\theta_\theta &= \Zt^\phi_\phi = \Zt^t_t -\mu^2 K_{XX} +G_{XX}\mu^2\varphi'' +2G_X \frac{\varphi'}{r}\notag
\end{align}
This ansatz is not necessarily a stationary solution to the problem --- for our purposes, we need to it to be a valid background configuration only momentarily. As our conditions, we instead require that the model functions $K$ and $G$ and gradients of $\varphi$ are chosen in such a manner that the acoustic metric is hyperbolic everywhere, eq.~\eqref{eq:detZ-updown}, and that the fluctuations are non-ghosts everywhere (correct signature). For \emph{this form} of the acoustic metric, this reduces to
\begin{align}
    \Zt^t_t \Zt^r_r + (\Zt^t_r)^2 &>0 &&\text{cone existence}, \label{eq:KGB-good}\\
    \Zt^\theta_\theta &>0     &&\text{non-ghost}.\notag
\end{align}
For consistency with spherical symmetry, $\varphi'$ should vanish at the centre, unless the centre is hidden by a horizon. Since we have switched gravity off, we would not see this, but see the solutions in ref.~\cite{Babichev:2012re} for a similar construction. We will assume here that our background is valid beyond some minimal radius and that $\varphi' \varphi'' <0$, so that the scalar's spatial gradient decays with radius and our configuration is localised.

\paragraph{Failure of unitary gauge.}\label{sec:fail_unitary}

Here we will construct a background in which the frame of $u_\mu = -\partial_\mu\phi/\sqrt{2X}$ is not a good Cauchy frame.  By the discussion of section~\ref{sec:CauchySurface}, this happens whenever $u_\mu$ is $Z$-spacelike. Since we have already ensured that the scalar not be a ghost, conditions~\eqref{eq:KGB-good}, $u_\mu$ is a bad Cauchy frame whenever
\begin{equation}
    \Zt^{\mu\nu}u_\mu u_\nu > 0\quad\qquad \text{bad Cauchy}. \label{eq:KGB-u-badCauchy}
\end{equation}

We specialise to the specific model $K(X) = X$. Without loss of generality, we take $\varphi'>0$.  A possible background which is hyperbolic everywhere and nowhere a ghost is then given by
\begin{align}
    0 &< G_X < -X G_{XX} \label{eq:blob_d_vs_dd}\\
    \sigma &< 2(G_X+XG_{XX})\varphi'' <1
\end{align}
with $\sigma=0$. For this choice, $G_X\phi'/r>0$ and this contribution in $\Zt^\theta_\theta$ does not ever change the signature. The requirement that the frame $u_\mu$ be a bad Cauchy frame only changes the above conditions by the replacement $\sigma\rightarrow 2X/\mu^2$, so tightening the range of possible $G_X$.

It is possible to satisfy all these conditions simultaneously, even though condition~\eqref{eq:blob_d_vs_dd} does place quite an unnatural condition on the function $G_X$ --- locally it must be at least $X^{-\alpha}$ with $\alpha>1$ in the region of interest for this kind of configuration. We also have that for the metric~\eqref{eq:blob-Z} the radial sound speed is $\cs^2=(1-2(G_X+XG_{XX})\varphi'')^{-1}$. We thus see that the bad Cauchy frame occurs when either the sound speed is very large, so even a small spatial gradient $\varphi'$ makes $u_\mu$ $Z$-spacelike or in the limit of $X\rightarrow0$, a nearly null $u_\mu$, where change of frames between the static coordinates and comoving is large and the sound speed does not have to exceed that of light significantly.

This establishes the fact that it is possible in kinetic gravity braiding to construct classically consistent backgrounds on which it is not possible to write the dynamics for fluctuations in the standard effective approach of using the unitary gauge.

\paragraph{Class II acoustic spacetimes.}\label{sec:classIIexamples}

Class II metrics are not diagonalisable over reals (see section~\ref{sec:classII}). This means that for the class of background described by eq.~\eqref{eq:blob-ansatz} we need to introduce a non-zero $K_{XX}$ to provide an off-diagonal term. We then need to satisfy everywhere the conditions~\eqref{eq:KGB-good} and, in addition, if the discriminant of the eigenvalue equation for the $(tr)$ block of the acoustic metric is negative, the metric is not diagonalisable, i.e.~we need
\begin{equation}
    (\Zt^t_t - \Zt^r_r)^2 < 4(\Zt^t_r)^2\qquad\text{class II} \label{eq:KGB-cII}
\end{equation}
We specialise to a model with $G_{XX}=0$ keeping $K$ general with $K_X>0$. Picking $\varphi'>0$ and $G_X>0$ allows us to disregard the $\varphi'$ term in $Z^\theta_\theta$ and  conditions~\eqref{eq:KGB-good} are satisfied everywhere when, for example
\begin{equation}
    -\frac{K_{X}}{2X}<K_{XX}<\frac{K_X}{\varphi'^2}\ \ \text{and}\ \ 2G_X \varphi'' < 0
\end{equation}
for any $g$-timelike $\partial_\mu\phi$. On such a background, the acoustic metric is class II whenever
\begin{equation}
    -K_X < 2XK_{XX} < 0 \ \text{and}\ (\mu+\varphi')^2 K_{XX} < 2G_X\varphi'' < (\mu-\varphi')^2K_{XX}\,.
\end{equation}
Thus a small (but non-zero) $K_{XX}$ creates the possibility that the acoustic metric is class II when $\varphi''$ also of appropriate magnitude. We reiterate that for the purpose of this example, we have selected conditions which are sufficient but not necessary. Other conditions can be found even in the setup~\eqref{eq:blob-ansatz}. In general kinetic gravity braiding theories background with superluminality are very generic especially in the presence of anisotropy and therefore one can expect that such class II configuration are very generic.

We have thus shown that kinetic gravity braiding is a theory which is capable of providing backgrounds which are classically consistent but can violate properties typically assumed: that it is always possible to go to the background's rest frame or that a unitary gauge provides a good set of coordinates in which to study the evolution of the system. This was not possible in the \emph{k-essence} class of theories, on any allowed background.

\section{Discussion and summary}\label{sec:conclude}

We have presented a generally covariant description for the dynamics of small scalar fluctuations --- phonons --- propagating on general anisotropic and time-dependent backgrounds in curved spacetimes, relevant for a large class of theories. When the momentum and frequency of the phonons are much higher than the scale of variations of the background, a clear separation can be made between the two. Whenever the scalar-field part of the principal symbol for the system of equations of motion for perturbations of all fields is factorisable, the background can be seen as giving an effective --- acoustic --- metric, $S_{\mu\nu}$, for a geometric optics (acoustics) of phonons. Furthermore, if there is no mixing with other degrees of freedom, the theory of the free phonons behaves as if they were excitations of a non-interacting canonical scalar-field theory in the acoustic spacetime.
We point out that the natural acoustic connection~\eqref{eq:Chrstoffel} is $S_{\mu\nu}$-compatible, but it is not compatible with the spacetime metric $g_{\mu\nu}$. A disformation tensor~\eqref{eq:disformation} appears encoding the nonmetricity of the acoustic spacetime. The nonmetricity is not completely arbitrary --- it is produced from the background scalar configuration. We have so far proven that the Weyl-vector part of nonmetricity~\eqref{eq:WeylNonMetricity} is always a derivative of scalar, so that every acoustically conserved vector current has an associated one conserved in the usual spacetime~\eqref{eq:AcouDiv}. Moreover, the corresponding conserved charge has the same value in the acoustic and in the usual spacetimes~\eqref{eq:covariant_charge_usual}. The crucial point is that the nonmetricity appears here absolutely naturally as a part of geometric description for well-known physical phenomena. Understanding which forms of the nonmetricity can appear for acoustic geometry of a scalar field with various self-interactions and interactions with other fields is an interesting open question.

\looseness=-1
From the point of the dynamics of small fluctuations, the spacetime metric $g_{\mu\nu}$ appears only implicitly through its contribution to the acoustic metric. It is the acoustic metric that describes the properties of the fluctuations and the space in which they move. Analogously to the usual case, we have shown that the existence of acoustic cones and therefore causal evolution is related to the Lorentzian signature of the acoustic metric. We have verified that cones exists for all possible types of non-singular Lorentzian acoustic metrics, including the non-diagonalisable ones, see section~\ref{sec:Geometry}. With any other signature, the equation of motion is not hyperbolic which results in true gradient instabilities which would be seen by all observers. We have also proposed that the natural definition of a ghost fluctuation is through the signature of the Lorentzian acoustic metric $S_{\mu\nu}$ being the opposite to that of $g_{\mu\nu}$. This is also coordinate invariant and therefore \emph{all} observers would agree on the ghost nature of the fluctuations, cf.~\cite{Babichev:2024uro}.

The presence of two metrics gives a richer geometrical structure: each metric can be used to associate different covectors to a vector and two different notions of orthogonality now exist. As a result, there are in fact \emph{two distinct acoustic cones} from the point of view of the spacetime metric. One is constructed by the null vectors, $N^\mu$, of the covariant acoustic metric $S_{\mu\nu}$ given by~\eqref{eq:ray-cone} --- this \emph{ray cone} or \emph{N-cone} describes the motion of the phase or wavefronts in space,  determining the phase velocity~\eqref{AcPhaseVel}. We have shown that this ray four-vector $N^\mu$ is proportional to the momentum proposed by Abraham. The second --- the \emph{momentum cone} or \emph{P-cone} is constructed by the null covectors $P_\mu$ of the inverse (contravariant) acoustic metric $Z^{\mu\nu}$ given by~\eqref{eq:ZPP} --- describes the four-momenta of the modes and is the covariant description of the dispersion relation or a covariant notion of a refractive index~\eqref{eq:n-def}. We have identified this four-momentum $P_\mu$ with the Minkowski momentum. The ray vectors $N^\mu$ and the momenta $P_\mu$ are orthogonal~\eqref{eq:NullVecsT} in the usual spacetime $g_{\mu\nu}$ sense,\footnote{It is useful to mention that naturally the rays are vectors $N^\mu$, while the momenta $P_\mu$ are covectors and are actually exact differential forms. Thus, to be more precise, this orthogonality does not involve any metric, when the indices are in their natural positions.} which gives the on-shell relation. As a result, the phase velocity~\eqref{AcPhaseVel} is only parallel to the spatial momentum~\eqref{eq:PN-gdecomp} in the simplest case of an isotropic background in its rest frame. Otherwise, the two directions are distinct, the usual formula~\eqref{eq:usual_vp} clearly fails and should be replaced by~\eqref{eq:v_p}.

Our approach has allowed us to derive the proper description with correct transformation properties for phase velocities and refractive indices for general anisotropic media. In particular, we have pointed out that the phonon's phase velocity relative to an observer is a spatial vector while the refractive index is actually a spatial covector. Moreover both of them can be covariantly defined also for superluminal phase-velocities --- spacelike rays. Through~\eqref{eq:MhoOmega_Z} we have identified an observer-dependent spatial contravariant metric $\Ztwo{\mu\nu}$ in the space of refractive indices and similarly a dual\footnote{These metrics are not inverse to each other.} covariant one $\Stwo{\mu\nu}$ in the space of phase velocities,~\eqref{eq:MhoOmega_S}. The metrics determine the difference between the Minkowski, $\omega$, and the Abraham, $\mho$, energies of a mode, both of which enter the discussion of stability. The sign of the Abraham energy~\eqref{eq:PN-gdecomp} determines whether an outgoing mode propagates with increasing or decreasing proper time of the observer. The Minkowski energy~\eqref{eq:PN-gdecomp} is the observed energy conserved in processes and the sign for an outgoing mode is related to ghosts, Cherenkov radiation and stability. Both these energies enter quantities such as the Hamiltonian and are non-trivially related through the normalisation of the acoustic metric and Doppler corrections in moving media~\eqref{eq:more_bromance}. These findings can be useful not only for phonons, but also for description of propagation of photons in media and especially in metamaterials.

\looseness=-1
The acoustic spacetime picture is not just local --- we have shown that the ray vectors and the four-momentum covectors are both parallel transported along acoustic null geodesics~\eqref{eq:geodeisc}. Contrary to the usual light propagation, these are two different transport equations. For subluminal propagation one can identify the usual spacetime four-acceleration of the phonon~\eqref{eq:acceleration} and the four-force acting on it~\eqref{eq:Newton_Law}. These equations provide a physical meaning for the disformation tensor and nonmetricity. To the best of our knowledge this has not been pointed out in the literature before. This construction can be applied to the propagation of photons in metamaterials, where the photons can accelerate~\cite{Mendonca_BOOK}. We have demonstrated that geodesic equations are indeed invariant under Weyl transformations of the acoustic metric. Additionally, we have shown that it is the components of the Minkowski four-momentum that are conserved along acoustic geodesics in the presence of acoustic Killing vectors~\eqref{eq:P_0_conserv}. However, generically there is no similar conservation for ray/Abraham momentum, demonstrating that it is the Minkowski momentum that is related to constructing symmetries.

The phonon flux (i.e.~amplitude of fluctuations) is also conserved~\eqref{eq:eiko-charge} in this acoustic spacetime, at least when kinetic mixing is absent or negligible. The geodesic deviation equation and therefore the notion of lensing is sensitive only to the acoustic curvature~\eqref{eq:geo-dev}. Analogously to cosmography with light, observations of scalar fluctuations, if they were possible, would reconstruct the acoustic spacetime instead of the usual one.

Just as in the case of the usual null vectors of the spacetime metric where the distinction between rays and momenta does not bring new information, the geometry of each of the two cones contains exactly the same information;  it is just differently presented. However, mixing or interchanging the N-cone and P-cone, as it has sometimes happened in the literature, can be very confusing and misleading. We have demonstrated explicitly how to recover any of the information from either cone.

For fluctuations propagating subluminally, it is possible to pick a frame which is supersonic with respect to the speed of fluctuations~\eqref{eq:transonic}. This then results in the existence of a sound horizon --- the supersonic Mach cone --- beyond which the scalar fluctuations cannot move and the surface of which is delineated by phonons with zero energy, at rest in the observer's frame.
These modes correspond to  frozen or static waves~\eqref{eq:Can_be_Wave}, i.e.~they are static solutions for which the Cauchy surface itself plays a role of the spacetime one dimension lower than the actual spacetime.
We have shown that negative Minkowski energy (negative frequency) modes for non-ghosts appear in directions inside the momentum equivalent of the Mach cone (where~\eqref{eq:SHcond} is violated). For these modes the spatial Minkowski and the spatial Abraham momenta have antialigned components, see figure~\ref{fig:conesSoundHorizon}. A source at rest in a frame with a sound horizon is kinematically allowed to extract energy from the medium, creating Cherenkov radiation.

In the case of superluminal phonons, it is possible to pick a frame in which the Abraham energy for some outgoing modes is negative, i.e.~information propagates along the rays towards decreasing values of the chosen ``time'' coordinate (condition~\eqref{Z-inducedmetric} violated), see figure~\ref{fig:conesBadCauchy}. In this case, not all initial conditions are possible to set up on this spatial hypersurface of constant ``time''. Such a situation is not necessarily acausal, as has been already discussed before e.g.~in~\cite{Babichev:2007dw}, but it is an inappropriate choice of frame for the Cauchy problem, which is not well-posed in this setup --- ``bad Cauchy''. We have shown that this situation is equivalent to some spatial momentum directions being excluded from the P-cone. An attempt to include these spatial momentum directions in the dispersion relation~\eqref{eq:dispersion-munu} would result in complex energies and an apparent instability. However, this is an illusive instability entirely caused by a bad choice of coordinates.

We have proven that when the chosen frame is not a good Cauchy frame, the sign of the kinetic term for the fluctuations reverses --- non-ghosts naively look like ghosts and vice versa, see~\eqref{Hamil-Z2}. The metric for refractive indices $\Ztwo{\mu\nu}$, quadratic in the acoustic metric and therefore invariant under the change of the overall metric sign, see~\eqref{Z-inducedmetric}, needs to be positive definite for the frame to be a good Cauchy frame,~\eqref{eq:dispersion-munu}. The positivity can be checked by examining the sign of three tensor invariants,~\eqref{eq:Z2invs}, and therefore is not expensive computationally.

We have also extended this local, frame discussion to a global one, introducing a foliation through the standard ADM decomposition  for the spacetime~\eqref{eq:ADM_metric}. We have found that we can naturally extend the local good-Cauchy-frame condition~\eqref{Z-inducedmetric} to the whole spatial slice, by applying this condition to the normal frame observers of the foliation. Satisfying~\eqref{eq:GoodCauchyP} at every point of the slice means the slice is a \emph{Cauchy surface} for the scalar field fluctuations and the IVP is well posed. Similarly, the question of local subsonicity can be extended globally by applying condition~\eqref{eq:SHcond} to the spatial slice comoving observers instead. We stress that this distinction is a new observation and a subtle effect of non-luminal propagation. In the usual case without a local Lorentz invariance breaking, all velocities are timelike so that ``supersonic'' i.e.\ superluminal observers are absent.

We have also related these observations to the usual proof of weak hyperbolicity, turning it around --- usually a tacit assumption of subluminality is made and therefore any spacelike slice is equivalent. Weak hyperbolicity then implies that the acoustic metric is Lorentzian. For us, the system is hyperbolic in the first place, so a lack of weak hyperbolicity is a statement of a bad choice of coordinates in which the spatial slice is not a Cauchy surface for small fluctuations. In particular, we have shown that the eigenvalues for the principal symbol whose imaginary nature is a requirement for well-posedness~\eqref{eq:wellposed} are exactly the same as the dispersion relation for our acoustic metric~\eqref{eq:dispersion-munu}.

We have also investigated the Hamiltonian functional for perturbations~\eqref{Hamil-Z2}, finding agreement with the geometrical picture we have presented above. Choosing a foliation on which the Cauchy problem is well posed according to our condition~\eqref{eq:GoodCauchyP} and for which the slice's comoving observer is subsonic~\eqref{eq:SHcond}, is enough to give a Hamiltonian bounded from below for non-ghosts. For a supersonic comoving observer, it is unbounded from below --- negative energy modes are available just as the P-cone picture suggests. This is not a catastrophic instability, but it does mean emission of Cherenkov radiation from a source becomes kinematically allowed. If the slice is not a Cauchy surface, the Hamiltonian also is unbounded, changing the sign of the kinetic term and the gradient energy in one spatial direction. While this is clearly visible in our 3+1 dimensional analysis, in a simplified 1+1 dimensional subspace e.g.~in spherical symmetry, it is very easy to dismiss a healthy mode as a ghost, or vice versa --- declare that a ghost's Hamiltonian is bounded from below and healthy. Since on such a bad Cauchy spatial slice the points are timelike separated with respect to the acoustic metric, Hamiltonian mechanics does not allow for arbitrary initial conditions and therefore this second apparent instability is not exploitable in a predictive manner.

We have also constructed an acoustic energy-momentum tensor (EMT) quadratic in fluctuations which is covariantly conserved in the acoustic geometry on equations of motion~\eqref{eq:emtZ}. This EMT is not the perturbed version of the usual spacetime EMT formed from the fluctuations. In the high-frequency limit, the acoustic EMT describes the transport of momentum four-covectors along their acoustic rays~\eqref{T=NP}. This EMT is not symmetric, but it is the only choice yielding conserved charges invariant under the acoustic Weyl transformations~\eqref{eq:Weyl_trans}. It corresponds to the Minkowski EMT.  The acoustic energy density observed locally is proportional to a product of the Abraham energy $\mho$ for outgoing rays (positive in a good Cauchy frame), and the energy carried by the modes $\omega$ associated with these rays~\eqref{eq:epsilon_u}. This second energy is positive if the observer measuring the energy is subsonic. Even in the rest frame of the medium this product reduces to the usual $\omega^2$ only up to the acoustic metric normalisation factor, which generically is a spacetime-dependent quantity. In other frames there appears a Doppler factor separating the energies~\eqref{eq:more_bromance}.

The acoustic Hamiltonian density is proportional to the temporal component of this EMT~\eqref{eq:H_T}. However this component is \emph{not} an energy density observed by any one observer in the spacetime, but rather a mixed quantity involving both the normal frame and comoving observers of the slice. This is the origin of the two separate conditions required for Hamiltonian boundedness. These two observers become indistinguishable if the shift vector is not present in the foliation.

As we have already mentioned, we have shown that if the acoustic metric has a Killing vector then the momentum component projected onto it is conserved along acoustic geodesics~\eqref{eq:P_0_conserv}. Analogously to the usual manner, acoustic Killing vectors also give acoustically conserved currents~\eqref{eq:consJ_acoustic} when contracted with the acoustically conserved EMT. In particular, if the Killing vector is $g$-timelike, the acoustic Hamiltonian functional is a conserved charge in the acoustic spacetime. The structure of the nonmetricity tensor is such that this acoustic Hamiltonian is a conserved charge also in the usual spacetime, with even the same value~\eqref{eq:invariant_Charge}. If the velocity of the comoving observer of the slice is subsonic---$S$-timelike at every point, the Hamiltonian is also bounded. Failing this, if the Killing vector is also $S$-timelike, there nonetheless exists a bounded  charge $Q_\xi$ (positive for non-ghosts) and conserved in the usual spacetime~\eqref{eq:Qxi}. The implication is that the acoustic Hamiltonian or $Q_\xi$ can be used to bound motion even in interactions with species moving in the usual spacetime. An interesting open question is to what extent this is general, e.g.~whether acoustic Killing vectors are also necessarily spacetime Killing vectors.

It is interesting to note that the Hamiltonian describes the evolution of all possible configurations of the (massless) fluctuations, not just the high-frequency part, yet the conditions for its boundedness are exactly as those derived from the eikonal limit. This points to the fact that other non-eikonal configurations nonetheless live in the same acoustic geometry. It would be interesting to verify whether adding a mass gap to the dispersion relation would retain this eikonal limit as the limiting characteristic surface and therefore maintain the stability conditions as we have derived them here for the shift-symmetric theories.

Our overall conclusion is that if there exists at least one slicing of the spacetime such that  slice is a Cauchy surface for every degree of freedom in terms of its acoustic metric in the sense of condition~\eqref{eq:GoodCauchyP} and the slice's comoving observer is subsonic at every point, then there exists a positive-definite charge of all non-ghost degrees of freedom in total. This charge is conserved in the presence of a Killing vector timelike for all the acoustic metrics and is bounding the motion of the whole system.

In this paper we have only covered the free theory, not attempting to build interactions into this acoustic picture. Self-interaction terms could be expanded as a theory for fluctuations on top of the acoustic spacetime, and processes would conserve the Minkowski acoustic momentum and energy provided the acoustic metric were sufficiently constant. The interesting question is what would be conserved in processes involving fluctuations of fields living in different metrics. We leave this for future work.
To conclude, we stress that a lot of geometry, generally valid definitions of physical quantities and interesting physical analogies were obtained in this work from very basic assumptions. Our work demonstrates that the use of full-fledged formalism of general relativity can be fruitfully employed in more broad and applied branches of physics. We leave this for future work.

\begin{acknowledgments}

It is a pleasure to thank Eugeny Babichev, Christos Charmousis, Gilles Esposito-Farèse, Petr Hadrava, Vladimír Karas, Igor Khavkine, Guillermo Lara, Slava Mukhanov, Shinji Mukohyama, Antonio Padilla, Subodh P.~Patil, Volker Perlick, Francesco Serra, Leonardo Trombetta and Vicharit Yingcharoenrat for very useful discussions and criticisms.
I.~S.~was supported by GAČR project 24-10780S, G.~T. acknowledges support of GAČR project 20-28525S, while the work of A.~V.~was supported by the European
Structural and Investment Funds and the Czech Ministry of Education, Youth and Sports (Project FORTE CZ.02.01.01/00/22 008/0004632).

\end{acknowledgments}

\bibliographystyle{utphys}
\addcontentsline{toc}{section}{\refname}\bibliography{KGB}

\providecommand{\href}[2]{#2}\begingroup\raggedright\begin{thebibliography}{100}

\bibitem{Armendariz-Picon:1999hyi}
C.~Armendariz-Picon, T.~Damour, and V.~F. Mukhanov, ``{k - inflation},''
  \href{https://dx.doi.org/10.1016/S0370-2693(99)00603-6}{{\em Phys. Lett. B}
  {\bfseries 458} (1999) 209--218},
  \href{https://arxiv.org/abs/hep-th/9904075}{{\ttfamily
  arXiv:hep-th/9904075}}.

\bibitem{ArmendarizPicon:2000ah}
C.~Armendariz-Picon, V.~F. Mukhanov, and P.~J. Steinhardt, ``{Essentials of
  k-essence},'' \href{https://dx.doi.org/10.1103/PhysRevD.63.103510}{{\em Phys.
  Rev.} {\bfseries D63} (2001) 103510},
\href{https://arxiv.org/abs/astro-ph/0006373}{{\ttfamily
  arXiv:astro-ph/0006373}}.
%%CITATION = ASTRO-PH/0006373;%%.

\bibitem{ArmendarizPicon:2000dh}
C.~Armendariz-Picon, V.~F. Mukhanov, and P.~J. Steinhardt, ``{A dynamical
  solution to the problem of a small cosmological constant and late-time cosmic
  acceleration},'' \href{https://dx.doi.org/10.1103/PhysRevLett.85.4438}{{\em
  Phys. Rev. Lett.} {\bfseries 85} (2000) 4438--4441},
\href{https://arxiv.org/abs/astro-ph/0004134}{{\ttfamily
  arXiv:astro-ph/0004134}}.
%%CITATION = ASTRO-PH/0004134;%%.

\bibitem{Garriga:1999vw}
J.~Garriga and V.~F. Mukhanov, ``{Perturbations in k-inflation},''
  \href{https://dx.doi.org/10.1016/S0370-2693(99)00602-4}{{\em Phys. Lett.}
  {\bfseries B458} (1999) 219--225},
\href{https://arxiv.org/abs/hep-th/9904176}{{\ttfamily arXiv:hep-th/9904176}}.
%%CITATION = HEP-TH/9904176;%%.

\bibitem{Deffayet:2010qz}
C.~Deffayet, O.~Pujolas, I.~Sawicki, and A.~Vikman, ``{Imperfect Dark Energy
  from Kinetic Gravity Braiding},''
  \href{https://dx.doi.org/10.1088/1475-7516/2010/10/026}{{\em JCAP} {\bfseries
  10} (2010) 026}, \href{https://arxiv.org/abs/1008.0048}{{\ttfamily
  arXiv:1008.0048 [hep-th]}}.

\bibitem{Kobayashi:2010cm}
T.~Kobayashi, M.~Yamaguchi, and J.~Yokoyama, ``{G-inflation: Inflation driven
  by the Galileon field},''
  \href{https://dx.doi.org/10.1103/PhysRevLett.105.231302}{{\em Phys. Rev.
  Lett.} {\bfseries 105} (2010) 231302},
  \href{https://arxiv.org/abs/1008.0603}{{\ttfamily arXiv:1008.0603 [hep-th]}}.

\bibitem{Nicolis:2008in}
A.~Nicolis, R.~Rattazzi, and E.~Trincherini, ``{The galileon as a local
  modification of gravity},''
  \href{https://dx.doi.org/10.1103/PhysRevD.79.064036}{{\em Phys. Rev.}
  {\bfseries D79} (2009) 064036},
\href{https://arxiv.org/abs/0811.2197}{{\ttfamily arXiv:0811.2197 [hep-th]}}.
%%CITATION = 0811.2197;%%.

\bibitem{Deffayet:2009wt}
C.~Deffayet, G.~Esposito-Farese, and A.~Vikman, ``{Covariant Galileon},''
  \href{https://dx.doi.org/10.1103/PhysRevD.79.084003}{{\em Phys. Rev.}
  {\bfseries D79} (2009) 084003},
\href{https://arxiv.org/abs/0901.1314}{{\ttfamily arXiv:0901.1314 [hep-th]}}.
%%CITATION = 0901.1314;%%.

\bibitem{Deffayet:2009mn}
C.~Deffayet, S.~Deser, and G.~Esposito-Farese, ``{Generalized Galileons: All
  scalar models whose curved background extensions maintain second-order field
  equations and stress-tensors},''
  \href{https://dx.doi.org/10.1103/PhysRevD.80.064015}{{\em Phys. Rev.}
  {\bfseries D80} (2009) 064015},
\href{https://arxiv.org/abs/0906.1967}{{\ttfamily arXiv:0906.1967 [gr-qc]}}.
%%CITATION = 0906.1967;%%.

\bibitem{Kobayashi:2011nu}
T.~Kobayashi, M.~Yamaguchi, and J.~Yokoyama, ``{Generalized G-inflation:
  Inflation with the most general second-order field equations},''
  \href{https://dx.doi.org/10.1143/PTP.126.511}{{\em Prog. Theor. Phys.}
  {\bfseries 126} (2011) 511--529},
  \href{https://arxiv.org/abs/1105.5723}{{\ttfamily arXiv:1105.5723 [hep-th]}}.

\bibitem{Horndeski:1974wa}
G.~W. Horndeski, ``{Second-order scalar-tensor field equations in a
  four-dimensional space},'' \href{https://dx.doi.org/10.1007/BF01807638}{{\em
  Int. J. Theor. Phys.} {\bfseries 10} (1974) 363--384}.

\bibitem{Zumalacarregui:2013pma}
M.~Zumalac\'arregui and J.~Garc\'\i{}a-Bellido, ``{Transforming gravity: from
  derivative couplings to matter to second-order scalar-tensor theories beyond
  the Horndeski Lagrangian},''
  \href{https://dx.doi.org/10.1103/PhysRevD.89.064046}{{\em Phys. Rev. D}
  {\bfseries 89} (2014) 064046},
  \href{https://arxiv.org/abs/1308.4685}{{\ttfamily arXiv:1308.4685 [gr-qc]}}.

\bibitem{Gleyzes:2014dya}
J.~Gleyzes, D.~Langlois, F.~Piazza, and F.~Vernizzi, ``{Healthy theories beyond
  Horndeski},'' \href{https://dx.doi.org/10.1103/PhysRevLett.114.211101}{{\em
  Phys. Rev. Lett.} {\bfseries 114} no.~21, (2015) 211101},
  \href{https://arxiv.org/abs/1404.6495}{{\ttfamily arXiv:1404.6495 [hep-th]}}.

\bibitem{Langlois:2015cwa}
D.~Langlois and K.~Noui, ``{Degenerate higher derivative theories beyond
  Horndeski: evading the Ostrogradski instability},''
  \href{https://dx.doi.org/10.1088/1475-7516/2016/02/034}{{\em JCAP} {\bfseries
  02} (2016) 034}, \href{https://arxiv.org/abs/1510.06930}{{\ttfamily
  arXiv:1510.06930 [gr-qc]}}.

\bibitem{Motohashi:2014opa}
H.~Motohashi and T.~Suyama, ``{Third order equations of motion and the
  Ostrogradsky instability},''
  \href{https://dx.doi.org/10.1103/PhysRevD.91.085009}{{\em Phys. Rev. D}
  {\bfseries 91} no.~8, (2015) 085009},
  \href{https://arxiv.org/abs/1411.3721}{{\ttfamily arXiv:1411.3721
  [physics.class-ph]}}.

\bibitem{Langlois:2018dxi}
D.~Langlois, ``{Dark energy and modified gravity in degenerate higher-order
  scalar\textendash{}tensor (DHOST) theories: A review},''
  \href{https://dx.doi.org/10.1142/S0218271819420069}{{\em Int. J. Mod. Phys.
  D} {\bfseries 28} no.~05, (2019) 1942006},
  \href{https://arxiv.org/abs/1811.06271}{{\ttfamily arXiv:1811.06271
  [gr-qc]}}.

\bibitem{Kobayashi:2019hrl}
T.~Kobayashi, ``{Horndeski theory and beyond: a review},''
  \href{https://dx.doi.org/10.1088/1361-6633/ab2429}{{\em Rept. Prog. Phys.}
  {\bfseries 82} no.~8, (2019) 086901},
  \href{https://arxiv.org/abs/1901.07183}{{\ttfamily arXiv:1901.07183
  [gr-qc]}}.

\bibitem{Clifton:2011jh}
T.~Clifton, P.~G. Ferreira, A.~Padilla, and C.~Skordis, ``{Modified Gravity and
  Cosmology},'' \href{https://dx.doi.org/10.1016/j.physrep.2012.01.001}{{\em
  Phys. Rept.} {\bfseries 513} (2012) 1--189},
  \href{https://arxiv.org/abs/1106.2476}{{\ttfamily arXiv:1106.2476
  [astro-ph.CO]}}.

\bibitem{Joyce:2014kja}
A.~Joyce, B.~Jain, J.~Khoury, and M.~Trodden, ``{Beyond the Cosmological
  Standard Model},''
  \href{https://dx.doi.org/10.1016/j.physrep.2014.12.002}{{\em Phys. Rept.}
  {\bfseries 568} (2015) 1--98},
  \href{https://arxiv.org/abs/1407.0059}{{\ttfamily arXiv:1407.0059
  [astro-ph.CO]}}.

\bibitem{Bull:2015stt}
P.~Bull {\em et~al.}, ``{Beyond $\Lambda$CDM: Problems, solutions, and the road
  ahead},'' \href{https://dx.doi.org/10.1016/j.dark.2016.02.001}{{\em Phys.
  Dark Univ.} {\bfseries 12} (2016) 56--99},
  \href{https://arxiv.org/abs/1512.05356}{{\ttfamily arXiv:1512.05356
  [astro-ph.CO]}}.

\bibitem{Heisenberg:2018vsk}
L.~Heisenberg, ``{A systematic approach to generalisations of General
  Relativity and their cosmological implications},''
  \href{https://dx.doi.org/10.1016/j.physrep.2018.11.006}{{\em Phys. Rept.}
  {\bfseries 796} (2019) 1--113},
  \href{https://arxiv.org/abs/1807.01725}{{\ttfamily arXiv:1807.01725
  [gr-qc]}}.

\bibitem{Silverstein:2003hf}
E.~Silverstein and D.~Tong, ``{Scalar speed limits and cosmology: Acceleration
  from D-cceleration},''
  \href{https://dx.doi.org/10.1103/PhysRevD.70.103505}{{\em Phys. Rev. D}
  {\bfseries 70} (2004) 103505},
  \href{https://arxiv.org/abs/hep-th/0310221}{{\ttfamily
  arXiv:hep-th/0310221}}.

\bibitem{Creminelli:2009mu}
P.~Creminelli, G.~D'Amico, J.~Norena, L.~Senatore, and F.~Vernizzi,
  ``{Spherical collapse in quintessence models with zero speed of sound},''
  \href{https://dx.doi.org/10.1088/1475-7516/2010/03/027}{{\em JCAP} {\bfseries
  1003} (2010) 027},
\href{https://arxiv.org/abs/0911.2701}{{\ttfamily arXiv:0911.2701
  [astro-ph.CO]}}.
%%CITATION = 0911.2701;%%.

\bibitem{Sawicki:2013wja}
I.~Sawicki, V.~Marra, and W.~Valkenburg, ``{Seeding supermassive black holes
  with a non-vortical dark-matter subcomponent},''
  \href{https://dx.doi.org/10.1103/PhysRevD.88.083520}{{\em Phys. Rev. D}
  {\bfseries 88} (2013) 083520},
  \href{https://arxiv.org/abs/1307.6150}{{\ttfamily arXiv:1307.6150
  [astro-ph.CO]}}.

\bibitem{Woodard:2015zca}
R.~P. Woodard, ``{Ostrogradsky's theorem on Hamiltonian instability},''
  \href{https://dx.doi.org/10.4249/scholarpedia.32243}{{\em Scholarpedia}
  {\bfseries 10} no.~8, (2015) 32243},
  \href{https://arxiv.org/abs/1506.02210}{{\ttfamily arXiv:1506.02210
  [hep-th]}}.

\bibitem{ArkaniHamed:2003uy}
N.~Arkani-Hamed, H.-C. Cheng, M.~A. Luty, and S.~Mukohyama, ``{Ghost
  condensation and a consistent infrared modification of gravity},'' {\em JHEP}
  {\bfseries 05} (2004) 074,
\href{https://arxiv.org/abs/hep-th/0312099}{{\ttfamily arXiv:hep-th/0312099}}.
%%CITATION = HEP-TH/0312099;%%.

\bibitem{Babichev:2018twg}
E.~Babichev, S.~Ramazanov, and A.~Vikman, ``{Recovering $P(X)$ from a canonical
  complex field},''
  \href{https://dx.doi.org/10.1088/1475-7516/2018/11/023}{{\em JCAP} {\bfseries
  11} (2018) 023}, \href{https://arxiv.org/abs/1807.10281}{{\ttfamily
  arXiv:1807.10281 [gr-qc]}}.

\bibitem{Caldwell:1999ew}
R.~R. Caldwell, ``{A Phantom menace?},''
  \href{https://dx.doi.org/10.1016/S0370-2693(02)02589-3}{{\em Phys. Lett. B}
  {\bfseries 545} (2002) 23--29},
  \href{https://arxiv.org/abs/astro-ph/9908168}{{\ttfamily
  arXiv:astro-ph/9908168}}.

\bibitem{Cline:2023cwm}
J.~M. Cline, M.~Puel, T.~Toma, and Q.~S. Wang, ``{Phantom fluid cosmology:
  Impact of a phantom hidden sector on cosmological observables},''
  \href{https://dx.doi.org/10.1103/PhysRevD.108.095033}{{\em Phys. Rev. D}
  {\bfseries 108} no.~9, (2023) 095033},
  \href{https://arxiv.org/abs/2308.12989}{{\ttfamily arXiv:2308.12989
  [hep-ph]}}.

\bibitem{Cline:2024zhs}
J.~M. Cline, ``{Phantom Fluid Cosmology -or- Ghosts for Gordon},'' in {\em {}}.
\newblock 1, 2024.
\newblock \href{https://arxiv.org/abs/2401.02958}{{\ttfamily arXiv:2401.02958
  [hep-ph]}}.

\bibitem{Deffayet:2021nnt}
C.~Deffayet, S.~Mukohyama, and A.~Vikman, ``{Ghosts without Runaway
  Instabilities},''
  \href{https://dx.doi.org/10.1103/PhysRevLett.128.041301}{{\em Phys. Rev.
  Lett.} {\bfseries 128} no.~4, (2022) 041301},
  \href{https://arxiv.org/abs/2108.06294}{{\ttfamily arXiv:2108.06294
  [gr-qc]}}.

\bibitem{Deffayet:2023wdg}
C.~Deffayet, A.~Held, S.~Mukohyama, and A.~Vikman, ``{Global and local
  stability for ghosts coupled to positive energy degrees of freedom},''
  \href{https://dx.doi.org/10.1088/1475-7516/2023/11/031}{{\em JCAP} {\bfseries
  11} (2023) 031}, \href{https://arxiv.org/abs/2305.09631}{{\ttfamily
  arXiv:2305.09631 [gr-qc]}}.

\bibitem{ErrastiDiez:2024hfq}
V.~Errasti~D\'\i{}ez, J.~Gaset~Rif\`a, and G.~Staudt, ``{Foundations of Ghost
  Stability},'' \href{https://dx.doi.org/10.1002/prop.202400268}{{\em Fortsch.
  Phys.} {\bfseries 73} no.~4, (2025) 2400268},
  \href{https://arxiv.org/abs/2408.16832}{{\ttfamily arXiv:2408.16832
  [hep-th]}}.

\bibitem{Babichev:2009ee}
E.~Babichev, C.~Deffayet, and R.~Ziour, ``{k-Mouflage gravity},''
  \href{https://dx.doi.org/10.1142/S0218271809016107}{{\em Int. J. Mod. Phys.}
  {\bfseries D18} (2009) 2147--2154},
\href{https://arxiv.org/abs/0905.2943}{{\ttfamily arXiv:0905.2943 [hep-th]}}.
%%CITATION = 0905.2943;%%.

\bibitem{Vainshtein:1972sx}
A.~I. Vainshtein, ``{To the problem of nonvanishing gravitation mass},''
\href{https://dx.doi.org/10.1016/0370-2693(72)90147-5}{{\em Phys. Lett.}
  {\bfseries B39} (1972) 393--394}.
%%CITATION = PHLTA,B39,393;%%.

\bibitem{Kaloper:2011qc}
N.~Kaloper, A.~Padilla, and N.~Tanahashi, ``{Galileon Hairs of Dyson Spheres,
  Vainshtein's Coiffure and Hirsute Bubbles},''
  \href{https://dx.doi.org/10.1007/JHEP10(2011)148}{{\em JHEP} {\bfseries 10}
  (2011) 148}, \href{https://arxiv.org/abs/1106.4827}{{\ttfamily
  arXiv:1106.4827 [hep-th]}}.

\bibitem{Babichev:2012re}
E.~Babichev and G.~Esposito-Far\`ese, ``{Time-Dependent Spherically Symmetric
  Covariant Galileons},''
  \href{https://dx.doi.org/10.1103/PhysRevD.87.044032}{{\em Phys. Rev. D}
  {\bfseries 87} (2013) 044032},
  \href{https://arxiv.org/abs/1212.1394}{{\ttfamily arXiv:1212.1394 [gr-qc]}}.

\bibitem{Kobayashi:2014eva}
T.~Kobayashi and N.~Tanahashi, ``{Exact black hole solutions in shift symmetric
  scalar\textendash{}tensor theories},''
  \href{https://dx.doi.org/10.1093/ptep/ptu096}{{\em PTEP} {\bfseries 2014}
  (2014) 073E02}, \href{https://arxiv.org/abs/1403.4364}{{\ttfamily
  arXiv:1403.4364 [gr-qc]}}.

\bibitem{Babichev:2017lmw}
E.~Babichev, C.~Charmousis, G.~Esposito-Far\`ese, and A.~Leh\'ebel,
  ``{Stability of Black Holes and the Speed of Gravitational Waves within
  Self-Tuning Cosmological Models},''
  \href{https://dx.doi.org/10.1103/PhysRevLett.120.241101}{{\em Phys. Rev.
  Lett.} {\bfseries 120} no.~24, (2018) 241101},
  \href{https://arxiv.org/abs/1712.04398}{{\ttfamily arXiv:1712.04398
  [gr-qc]}}.

\bibitem{Franciolini:2018uyq}
G.~Franciolini, L.~Hui, R.~Penco, L.~Santoni, and E.~Trincherini, ``{Effective
  Field Theory of Black Hole Quasinormal Modes in Scalar-Tensor Theories},''
  \href{https://dx.doi.org/10.1007/JHEP02(2019)127}{{\em JHEP} {\bfseries 02}
  (2019) 127}, \href{https://arxiv.org/abs/1810.07706}{{\ttfamily
  arXiv:1810.07706 [hep-th]}}.

\bibitem{Hui:2021cpm}
L.~Hui, A.~Podo, L.~Santoni, and E.~Trincherini, ``{Effective Field Theory for
  the perturbations of a slowly rotating black hole},''
  \href{https://dx.doi.org/10.1007/JHEP12(2021)183}{{\em JHEP} {\bfseries 12}
  (2021) 183}, \href{https://arxiv.org/abs/2111.02072}{{\ttfamily
  arXiv:2111.02072 [hep-th]}}.

\bibitem{Khoury:2022zor}
J.~Khoury, T.~Noumi, M.~Trodden, and S.~S.~C. Wong, ``{Stability of hairy black
  holes in shift-symmetric scalar-tensor theories via the effective field
  theory approach},''
  \href{https://dx.doi.org/10.1088/1475-7516/2023/04/035}{{\em JCAP} {\bfseries
  04} (2023) 035}, \href{https://arxiv.org/abs/2208.02823}{{\ttfamily
  arXiv:2208.02823 [hep-th]}}.

\bibitem{Mukohyama:2022enj}
S.~Mukohyama and V.~Yingcharoenrat, ``{Effective field theory of black hole
  perturbations with timelike scalar profile: formulation},''
  \href{https://dx.doi.org/10.1088/1475-7516/2022/09/010}{{\em JCAP} {\bfseries
  09} (2022) 010}, \href{https://arxiv.org/abs/2204.00228}{{\ttfamily
  arXiv:2204.00228 [hep-th]}}.

\bibitem{deRham:2012fw}
C.~de~Rham, A.~J. Tolley, and D.~H. Wesley, ``{Vainshtein Mechanism in Binary
  Pulsars},'' \href{https://dx.doi.org/10.1103/PhysRevD.87.044025}{{\em Phys.
  Rev. D} {\bfseries 87} no.~4, (2013) 044025},
  \href{https://arxiv.org/abs/1208.0580}{{\ttfamily arXiv:1208.0580 [gr-qc]}}.

\bibitem{Dar:2018dra}
F.~Dar, C.~De~Rham, J.~T. Deskins, J.~T. Giblin, and A.~J. Tolley, ``{Scalar
  Gravitational Radiation from Binaries: Vainshtein Mechanism in Time-dependent
  Systems},'' \href{https://dx.doi.org/10.1088/1361-6382/aaf5e8}{{\em Class.
  Quant. Grav.} {\bfseries 36} no.~2, (2019) 025008},
  \href{https://arxiv.org/abs/1808.02165}{{\ttfamily arXiv:1808.02165
  [hep-th]}}.

\bibitem{Creminelli:2019kjy}
P.~Creminelli, G.~Tambalo, F.~Vernizzi, and V.~Yingcharoenrat, ``{Dark-Energy
  Instabilities induced by Gravitational Waves},''
  \href{https://dx.doi.org/10.1088/1475-7516/2020/05/002}{{\em JCAP} {\bfseries
  05} (2020) 002}, \href{https://arxiv.org/abs/1910.14035}{{\ttfamily
  arXiv:1910.14035 [gr-qc]}}.

\bibitem{Barcelo:2005fc}
C.~Barcelo, S.~Liberati, and M.~Visser, ``{Analogue gravity},''
  \href{https://dx.doi.org/10.12942/lrr-2005-12}{{\em Living Rev. Rel.}
  {\bfseries 8} (2005) 12},
  \href{https://arxiv.org/abs/gr-qc/0505065}{{\ttfamily arXiv:gr-qc/0505065}}.

\bibitem{Coviello:2024vht}
C.~Coviello, M.~L. Chiofalo, D.~Grasso, S.~Liberati, M.~Mannarelli, and
  S.~Trabucco, ``{Gravitational Waves and Black Hole perturbations in Acoustic
  Analogues},'' \href{https://arxiv.org/abs/2410.00264}{{\ttfamily
  arXiv:2410.00264 [gr-qc]}}.

\bibitem{Adams:2006sv}
A.~Adams, N.~Arkani-Hamed, S.~Dubovsky, A.~Nicolis, and R.~Rattazzi,
  ``{Causality, analyticity and an IR obstruction to UV completion},'' {\em
  JHEP} {\bfseries 10} (2006) 014,
\href{https://arxiv.org/abs/hep-th/0602178}{{\ttfamily arXiv:hep-th/0602178}}.
%%CITATION = HEP-TH/0602178;%%.

\bibitem{Nicolis:2009qm}
A.~Nicolis, R.~Rattazzi, and E.~Trincherini, ``{Energy's and amplitudes'
  positivity},'' \href{https://dx.doi.org/10.1007/JHEP05(2010)095}{{\em JHEP}
  {\bfseries 05} (2010) 095},
\href{https://arxiv.org/abs/0912.4258}{{\ttfamily arXiv:0912.4258 [hep-th]}}.
%%CITATION = 0912.4258;%%.

\bibitem{Bellazzini:2017fep}
B.~Bellazzini, F.~Riva, J.~Serra, and F.~Sgarlata, ``{Beyond Positivity Bounds
  and the Fate of Massive Gravity},''
  \href{https://dx.doi.org/10.1103/PhysRevLett.120.161101}{{\em Phys. Rev.
  Lett.} {\bfseries 120} no.~16, (2018) 161101},
  \href{https://arxiv.org/abs/1710.02539}{{\ttfamily arXiv:1710.02539
  [hep-th]}}.

\bibitem{Tolley:2020gtv}
A.~J. Tolley, Z.-Y. Wang, and S.-Y. Zhou, ``{New positivity bounds from full
  crossing symmetry},'' \href{https://dx.doi.org/10.1007/JHEP05(2021)255}{{\em
  JHEP} {\bfseries 05} (2021) 255},
  \href{https://arxiv.org/abs/2011.02400}{{\ttfamily arXiv:2011.02400
  [hep-th]}}.

\bibitem{Camanho:2014apa}
X.~O. Camanho, J.~D. Edelstein, J.~Maldacena, and A.~Zhiboedov, ``{Causality
  Constraints on Corrections to the Graviton Three-Point Coupling},''
  \href{https://dx.doi.org/10.1007/JHEP02(2016)020}{{\em JHEP} {\bfseries 02}
  (2016) 020}, \href{https://arxiv.org/abs/1407.5597}{{\ttfamily
  arXiv:1407.5597 [hep-th]}}.

\bibitem{CarrilloGonzalez:2022fwg}
M.~Carrillo~Gonzalez, C.~de~Rham, V.~Pozsgay, and A.~J. Tolley, ``{Causal
  effective field theories},''
  \href{https://dx.doi.org/10.1103/PhysRevD.106.105018}{{\em Phys. Rev. D}
  {\bfseries 106} no.~10, (2022) 105018},
  \href{https://arxiv.org/abs/2207.03491}{{\ttfamily arXiv:2207.03491
  [hep-th]}}.

\bibitem{Serra:2023nrn}
F.~Serra and L.~G. Trombetta, ``{Five-point superluminality bounds},''
  \href{https://dx.doi.org/10.1007/JHEP06(2024)117}{{\em JHEP} {\bfseries 06}
  (2024) 117}, \href{https://arxiv.org/abs/2312.06759}{{\ttfamily
  arXiv:2312.06759 [hep-th]}}.

\bibitem{Serra:2024tmz}
F.~Serra and L.~G. Trombetta, ``{IR Bounds on Theories with
  Spontaneously-Broken Lorentz Symmetry},''
  \href{https://arxiv.org/abs/2412.19745}{{\ttfamily arXiv:2412.19745
  [hep-th]}}.

\bibitem{Evslin:2011rj}
J.~Evslin, ``{Stability of Closed Timelike Curves in a Galileon Model},''
  \href{https://dx.doi.org/10.1007/JHEP03(2012)009}{{\em JHEP} {\bfseries 03}
  (2012) 009}, \href{https://arxiv.org/abs/1112.1349}{{\ttfamily
  arXiv:1112.1349 [hep-th]}}.

\bibitem{Babichev:2007dw}
E.~Babichev, V.~Mukhanov, and A.~Vikman, ``{k-Essence, superluminal
  propagation, causality and emergent geometry},''
  \href{https://dx.doi.org/10.1088/1126-6708/2008/02/101}{{\em JHEP} {\bfseries
  02} (2008) 101},
\href{https://arxiv.org/abs/0708.0561}{{\ttfamily arXiv:0708.0561 [hep-th]}}.
%%CITATION = 0708.0561;%%.

\bibitem{Burrage:2011cr}
C.~Burrage, C.~de~Rham, L.~Heisenberg, and A.~J. Tolley, ``{Chronology
  Protection in Galileon Models and Massive Gravity},''
  \href{https://dx.doi.org/10.1088/1475-7516/2012/07/004}{{\em JCAP} {\bfseries
  07} (2012) 004}, \href{https://arxiv.org/abs/1111.5549}{{\ttfamily
  arXiv:1111.5549 [hep-th]}}.

\bibitem{Kaplan:2024qtf}
D.~E. Kaplan, S.~Rajendran, and F.~Serra, ``{Wrong signs are alright},''
  \href{https://dx.doi.org/10.1007/JHEP03(2025)031}{{\em JHEP} {\bfseries 03}
  (2025) 031}, \href{https://arxiv.org/abs/2406.06681}{{\ttfamily
  arXiv:2406.06681 [hep-th]}}.

\bibitem{Liberati:2001sd}
S.~Liberati, S.~Sonego, and M.~Visser, ``{Faster than c signals, special
  relativity, and causality},''
  \href{https://dx.doi.org/10.1006/aphy.2002.6233}{{\em Annals Phys.}
  {\bfseries 298} (2002) 167--185},
  \href{https://arxiv.org/abs/gr-qc/0107091}{{\ttfamily arXiv:gr-qc/0107091}}.

\bibitem{Bruneton:2006gf}
J.-P. Bruneton, ``{On causality and superluminal behavior in classical field
  theories. Applications to k-essence theories and MOND-like theories of
  gravity},'' \href{https://dx.doi.org/10.1103/PhysRevD.75.085013}{{\em Phys.
  Rev.} {\bfseries D75} (2007) 085013},
\href{https://arxiv.org/abs/gr-qc/0607055}{{\ttfamily arXiv:gr-qc/0607055}}.
%%CITATION = GR-QC/0607055;%%.

\bibitem{Bruneton:2007si}
J.-P. Bruneton and G.~Esposito-Farese, ``{Field-theoretical formulations of
  MOND-like gravity},''
  \href{https://dx.doi.org/10.1103/PhysRevD.76.124012}{{\em Phys. Rev.}
  {\bfseries D76} (2007) 124012},
\href{https://arxiv.org/abs/0705.4043}{{\ttfamily arXiv:0705.4043 [gr-qc]}}.
%%CITATION = 0705.4043;%%.

\bibitem{Kang:2007vs}
J.~U. Kang, V.~Vanchurin, and S.~Winitzki, ``{Attractor scenarios and
  superluminal signals in k-essence cosmology},''
  \href{https://dx.doi.org/10.1103/PhysRevD.76.083511}{{\em Phys. Rev.}
  {\bfseries D76} (2007) 083511},
\href{https://arxiv.org/abs/0706.3994}{{\ttfamily arXiv:0706.3994 [gr-qc]}}.
%%CITATION = 0706.3994;%%.

\bibitem{Geroch:2010da}
R.~Geroch, ``{Faster Than Light?},''
\href{https://arxiv.org/abs/1005.1614}{{\ttfamily arXiv:1005.1614 [gr-qc]}}.
%%CITATION = 1005.1614;%%.

\bibitem{Babichev:2018uiw}
E.~Babichev, C.~Charmousis, G.~Esposito-Far\`ese, and A.~Leh\'ebel,
  ``{Hamiltonian unboundedness vs stability with an application to Horndeski
  theory},'' \href{https://dx.doi.org/10.1103/PhysRevD.98.104050}{{\em Phys.
  Rev. D} {\bfseries 98} no.~10, (2018) 104050},
  \href{https://arxiv.org/abs/1803.11444}{{\ttfamily arXiv:1803.11444
  [gr-qc]}}.

\bibitem{Esposito-Farese:2019vlh}
G.~Esposito-Farese, ``{Hamiltonian vs stability in alternative theories of
  gravity},'' in {\em {54th Rencontres de Moriond on Gravitation}}.
\newblock 5, 2019.
\newblock \href{https://arxiv.org/abs/1905.04586}{{\ttfamily arXiv:1905.04586
  [gr-qc]}}.

\bibitem{Raetzel:2010je}
D.~Raetzel, S.~Rivera, and F.~P. Schuller, ``{Geometry of physical dispersion
  relations},'' \href{https://dx.doi.org/10.1103/PhysRevD.83.044047}{{\em Phys.
  Rev. D} {\bfseries 83} (2011) 044047},
  \href{https://arxiv.org/abs/1010.1369}{{\ttfamily arXiv:1010.1369 [hep-th]}}.

\bibitem{Khavkine:2012jf}
I.~Khavkine, ``{Characteristics, Conal Geometry and Causality in Locally
  Covariant Field Theory},'' \href{https://arxiv.org/abs/1211.1914}{{\ttfamily
  arXiv:1211.1914 [gr-qc]}}.

\bibitem{LandavshitzII}
L.~D. Landau and E.~M. Lifshitz, {\em {Course of Theoretical Physics, Vol. 2,
  Classical Theory of Fields}}.
\newblock Pergamon Press, 1980.

\bibitem{Courant}
R.~Courant and D.~Hilbert, {\em {Methods of Mathematical Physics, Vol. 2}}.
\newblock Wiley-Interscience, 1989.

\bibitem{Moncrief:1980}
V.~{Moncrief}, ``{Stability of stationary, spherical accretion onto a
  Schwarzschild black hole},'' \href{https://dx.doi.org/10.1086/157707}{{\em
  ApJ} {\bfseries 235} (1980) 1038}.

\bibitem{Unruh:1980cg}
W.~G. Unruh, ``{Experimental black hole evaporation},''
  \href{https://dx.doi.org/10.1103/PhysRevLett.46.1351}{{\em Phys. Rev. Lett.}
  {\bfseries 46} (1981) 1351--1353}.

\bibitem{Perlick_BOOK}
V.~{Perlick}, {\em {Ray Optics, Fermat's Principle, and Applications to General
  Relativity, Lecture Notes in Physics}}, vol.~61.
\newblock Springer-Verlag Berlin Heidelberg, 2000.

\bibitem{ArmendarizPicon:1999rj}
C.~Armendariz-Picon, T.~Damour, and V.~F. Mukhanov, ``{k-Inflation},''
  \href{https://dx.doi.org/10.1016/S0370-2693(99)00603-6}{{\em Phys. Lett.}
  {\bfseries B458} (1999) 209--218},
\href{https://arxiv.org/abs/hep-th/9904075}{{\ttfamily arXiv:hep-th/9904075}}.
%%CITATION = HEP-TH/9904075;%%.

\bibitem{Itzykson:1980rh}
C.~Itzykson and J.~B. Zuber, {\em {Quantum Field Theory}}.
\newblock International Series In Pure and Applied Physics. McGraw-Hill, New
  York, 1980.

\bibitem{Mironov:2024idn}
S.~Mironov, A.~Shtennikova, and M.~Valencia-Villegas, ``{Reviving Horndeski
  after GW170817 by Kaluza-Klein compactifications},''
  \href{https://dx.doi.org/10.1016/j.physletb.2024.139058}{{\em Phys. Lett. B}
  {\bfseries 858} (2024) 139058},
  \href{https://arxiv.org/abs/2405.02281}{{\ttfamily arXiv:2405.02281
  [hep-th]}}.

\bibitem{Babichev:2024kfo}
E.~Babichev, C.~Charmousis, B.~Muntz, A.~Padilla, and I.~D. Saltas,
  ``{Horndeski speed tests with scalar-photon couplings},''
  \href{https://arxiv.org/abs/2407.20339}{{\ttfamily arXiv:2407.20339
  [gr-qc]}}.

\bibitem{Synge:1960ueh}
J.~L. Synge, {\em {Relativity: The General theory}}.
\newblock North-Holland, 1960.

\bibitem{deRham:2014wfa}
C.~de~Rham and R.~H. Ribeiro, ``{Riding on irrelevant operators},''
  \href{https://dx.doi.org/10.1088/1475-7516/2014/11/016}{{\em JCAP} {\bfseries
  11} (2014) 016}, \href{https://arxiv.org/abs/1405.5213}{{\ttfamily
  arXiv:1405.5213 [hep-th]}}.

\bibitem{Gangopadhyay:2012dz}
D.~Gangopadhyay and G.~Manna, ``{The Hawking temperature in the context of dark
  energy},'' \href{https://dx.doi.org/10.1209/0295-5075/100/49001}{{\em EPL}
  {\bfseries 100} no.~4, (2012) 49001},
  \href{https://arxiv.org/abs/1211.1268}{{\ttfamily arXiv:1211.1268
  [astro-ph.CO]}}.

\bibitem{McCrea:1992wa}
J.~D. McCrea, ``{Irreducible decompositions of non-metricity, torsion,
  curvature and Bianchi identities in metric-affine spacetimes},''
  \href{https://dx.doi.org/10.1088/0264-9381/9/2/018}{{\em Class. Quant. Grav.}
  {\bfseries 9} (1992) 553--568}.

\bibitem{Hehl:1994ue}
F.~W. Hehl, J.~D. McCrea, E.~W. Mielke, and Y.~Ne'eman, ``{Metric affine gauge
  theory of gravity: Field equations, Noether identities, world spinors, and
  breaking of dilation invariance},''
  \href{https://dx.doi.org/10.1016/0370-1573(94)00111-F}{{\em Phys. Rept.}
  {\bfseries 258} (1995) 1--171},
  \href{https://arxiv.org/abs/gr-qc/9402012}{{\ttfamily arXiv:gr-qc/9402012}}.

\bibitem{Leonhardt}
U.~{Leonhardt}, ``{Optics: Momentum in an uncertain light},''
  \href{https://dx.doi.org/10.1038/444823a}{{\em Nature} {\bfseries 444}
  no.~7121, (Dec., 2006) 823--824}.

\bibitem{Pfeifer:2007zz}
R.~N.~C. Pfeifer, T.~A. Nieminen, N.~R. Heckenberg, and H.~Rubinsztein-Dunlop,
  ``{Colloquium: Momentum of an electromagnetic wave in dielectric media},''
  \href{https://dx.doi.org/10.1103/RevModPhys.79.1197}{{\em Rev. Mod. Phys.}
  {\bfseries 79} (2007) 1197--1216},
  \href{https://arxiv.org/abs/0710.0461}{{\ttfamily arXiv:0710.0461
  [physics.class-ph]}}. [Erratum: Rev.Mod.Phys. 81, 443 (2009)].

\bibitem{Enigma}
S.~M. Barnett and R.~Loudon, ``The enigma of optical momentum in a medium,''
  \href{https://dx.doi.org/10.1098/rsta.2009.0207}{{\em Philosophical
  Transactions of the Royal Society A: Mathematical, Physical and Engineering
  Sciences} {\bfseries 368} no.~1914, (2010) 927--939}.

\bibitem{Minkowski}
H.~Minkowski, ``Die grundgleichungen für die elektromagnetischen vorgänge in
  bewegten körpern,'' {\em Nachrichten von der Gesellschaft der Wissenschaften
  zu Göttingen, Mathematisch-Physikalische Klasse.} (1908) 53–111.
  \url{https://de.wikisource.org/wiki/Die_Grundgleichungen_f%C3%BCr_die_elektromagnetischen_Vorg%C3%A4nge_in_bewegten_K%C3%B6rpern}.

\bibitem{Abraham}
M.~Abraham, ``Zur elektrodynamik bewegter körper,''
  \href{https://dx.doi.org/https://doi.org/10.1007/BF03018208}{{\em Rendiconti
  del Circolo Matematico di Palermo} {\bfseries 28} (1909) 1–28}.

\bibitem{Nicolis:2004qq}
A.~Nicolis and R.~Rattazzi, ``{Classical and quantum consistency of the DGP
  model},'' {\em JHEP} {\bfseries 06} (2004) 059,
\href{https://arxiv.org/abs/hep-th/0404159}{{\ttfamily arXiv:hep-th/0404159}}.
%%CITATION = HEP-TH/0404159;%%.

\bibitem{Primakoff:1951iae}
H.~Primakoff, ``{Photoproduction of neutral mesons in nuclear electric fields
  and the mean life of the neutral meson},''
  \href{https://dx.doi.org/10.1103/PhysRev.81.899}{{\em Phys. Rev.} {\bfseries
  81} (1951) 899}.

\bibitem{Gertsenshtein:1962kfm}
M.~E. Gertsenshtein and V.~I. Pustovoit, ``{On the Detection of Low Frequency
  Gravitational Waves},'' {\em Sov. Phys. JETP} {\bfseries 16} (1962) 433.

\bibitem{Bilic:1999sq}
N.~Bilic, ``{Relativistic acoustic geometry},''
  \href{https://dx.doi.org/10.1088/0264-9381/16/12/312}{{\em Class. Quant.
  Grav.} {\bfseries 16} (1999) 3953--3964},
  \href{https://arxiv.org/abs/gr-qc/9908002}{{\ttfamily arXiv:gr-qc/9908002}}.

\bibitem{Mironov:2023bdq}
S.~Mironov and V.~Volkova, ``{DPSV trick for spherically symmetric
  backgrounds},''
  \href{https://dx.doi.org/10.1016/j.nuclphysb.2024.116550}{{\em Nucl. Phys. B}
  {\bfseries 1004} (2024) 116550},
  \href{https://arxiv.org/abs/2306.17791}{{\ttfamily arXiv:2306.17791
  [hep-th]}}.

\bibitem{Dobre:2017pnt}
D.~A. Dobre, A.~V. Frolov, J.~T. G\'alvez~Ghersi, S.~Ramazanov, and A.~Vikman,
  ``{Unbraiding the Bounce: Superluminality around the Corner},''
  \href{https://dx.doi.org/10.1088/1475-7516/2018/03/020}{{\em JCAP} {\bfseries
  03} (2018) 020}, \href{https://arxiv.org/abs/1712.10272}{{\ttfamily
  arXiv:1712.10272 [gr-qc]}}.

\bibitem{thorne_modern_2017}
K.~S. Thorne and R.~D. Blandford, {\em Modern classical physics: optics,
  fluids, plasmas, elasticity, relativity, and statistical physics}.
\newblock Princeton University Press, Princeton, 2017.

\bibitem{Ezquiaga:2020dao}
J.~M. Ezquiaga and M.~Zumalac\'arregui, ``{Gravitational wave lensing beyond
  general relativity: birefringence, echoes and shadows},''
  \href{https://dx.doi.org/10.1103/PhysRevD.102.124048}{{\em Phys. Rev. D}
  {\bfseries 102} no.~12, (2020) 124048},
  \href{https://arxiv.org/abs/2009.12187}{{\ttfamily arXiv:2009.12187
  [gr-qc]}}.

\bibitem{Menadeo:2024uoq}
N.~Menadeo and M.~Zumalac\'arregui, ``{Gravitational wave propagation beyond
  General Relativity: geometric optic expansion and lens-induced dispersion},''
  \href{https://arxiv.org/abs/2411.07164}{{\ttfamily arXiv:2411.07164
  [gr-qc]}}.

\bibitem{Anile_BOOK}
A.~M. {Anile}, {\em {Relativistic Fluids and Magneto-fluids with Applications
  in Astrophysics and Plasma Physics}}.
\newblock Cambridge University Press, 2005.

\bibitem{Ehlers_I}
R.~A. {Breuer} and J.~{Ehlers}, ``{Propagation of High-Frequency
  Electromagnetic Waves Through a Magnetized Plasma in Curved Space-Time. I},''
  \href{https://dx.doi.org/10.1098/rspa.1980.0040}{{\em Proceedings of the
  Royal Society of London Series A} {\bfseries 370} no.~1742, (Mar., 1980)
  389--406}.

\bibitem{Ehlers_II}
R.~A. {Breuer} and J.~{Ehlers}, ``{Propagation of high-frequency
  electromagnetic waves through a magnetized plasma in curved space-time. II -
  Application of the asymptotic approximation},''
  \href{https://dx.doi.org/10.1098/rspa.1981.0011}{{\em Proceedings of the
  Royal Society of London Series A} {\bfseries 374} no.~1756, (Jan., 1981)
  65--86}.

\bibitem{Bicak_Hadrava}
J.~{Bicak} and P.~{Hadrava}, ``{General-relativistic radiative transfer theory
  in refractive and dispersive media.},'' {\em Astronomy and Astrophysics}
  {\bfseries 44} no.~2, (Nov., 1975) 389--399.

\bibitem{deBoer:2017ing}
J.~de~Boer, J.~Hartong, N.~A. Obers, W.~Sybesma, and S.~Vandoren, ``{Perfect
  Fluids},'' \href{https://dx.doi.org/10.21468/SciPostPhys.5.1.003}{{\em
  SciPost Phys.} {\bfseries 5} no.~1, (2018) 003},
  \href{https://arxiv.org/abs/1710.04708}{{\ttfamily arXiv:1710.04708
  [hep-th]}}.

\bibitem{Landafshitz_V8}
L.~D. Landau and E.~M. Lifshitz, {\em {Course of Theoretical Physics, Vol. 8,
  Electrodynamics of Continuous Media}}.
\newblock Pergamon Press, {Second English Edition, Revised and enlarged}~ed.,
  1984.

\bibitem{deRham:2010kj}
C.~de~Rham, G.~Gabadadze, and A.~J. Tolley, ``{Resummation of Massive
  Gravity},'' \href{https://dx.doi.org/10.1103/PhysRevLett.106.231101}{{\em
  Phys. Rev. Lett.} {\bfseries 106} (2011) 231101},
  \href{https://arxiv.org/abs/1011.1232}{{\ttfamily arXiv:1011.1232 [hep-th]}}.

\bibitem{Hassan:2011zd}
S.~F. Hassan and R.~A. Rosen, ``{Bimetric Gravity from Ghost-free Massive
  Gravity},'' \href{https://dx.doi.org/10.1007/JHEP02(2012)126}{{\em JHEP}
  {\bfseries 02} (2012) 126}, \href{https://arxiv.org/abs/1109.3515}{{\ttfamily
  arXiv:1109.3515 [hep-th]}}.

\bibitem{Gangopadhyay:2011iu}
D.~Gangopadhyay, G.~Manna, and S.~S. Choudhury, ``{Masking singularities with
  $k-$essence fields in an emergent gravity metric},''
  \href{https://arxiv.org/abs/1103.3380}{{\ttfamily arXiv:1103.3380 [gr-qc]}}.

\bibitem{Callaway}
J.~Callaway, {\em {Quantum Theory of the Solid State}}.
\newblock Academic Press, San Diego, {Second Edition}~ed., 1991.

\bibitem{Arnowitt:1962hi}
R.~L. Arnowitt, S.~Deser, and C.~W. Misner, ``{The Dynamics of general
  relativity},'' \href{https://dx.doi.org/10.1007/s10714-008-0661-1}{{\em Gen.
  Rel. Grav.} {\bfseries 40} (2008) 1997--2027},
  \href{https://arxiv.org/abs/gr-qc/0405109}{{\ttfamily arXiv:gr-qc/0405109}}.

\bibitem{Poisson}
E.~Poisson, {\em A Relativist's Toolkit: The Mathematics of Black-Hole
  Mechanics}.
\newblock Cambridge University Press, 2009.

\bibitem{Dubovsky:2005xd}
S.~Dubovsky, T.~Gregoire, A.~Nicolis, and R.~Rattazzi, ``{Null energy condition
  and superluminal propagation},''
  \href{https://dx.doi.org/10.1088/1126-6708/2006/03/025}{{\em JHEP} {\bfseries
  0603} (2006) 025}, \href{https://arxiv.org/abs/hep-th/0512260}{{\ttfamily
  arXiv:hep-th/0512260 [hep-th]}}.

\bibitem{Sawicki:2012pz}
I.~Sawicki and A.~Vikman, ``{Hidden Negative Energies in Strongly Accelerated
  Universes},'' \href{https://dx.doi.org/10.1103/PhysRevD.87.067301}{{\em Phys.
  Rev. D} {\bfseries 87} no.~6, (2013) 067301},
  \href{https://arxiv.org/abs/1209.2961}{{\ttfamily arXiv:1209.2961
  [astro-ph.CO]}}.

\bibitem{weinberg_quantum_2013}
S.~Weinberg, \href{https://dx.doi.org/10.1017/CBO9781139644167}{{\em The
  quantum theory of fields. 1: {Foundations}}}.
\newblock Cambridge University Press, Cambridge, 2013.

\bibitem{Cheung:2007st}
C.~Cheung, P.~Creminelli, A.~L. Fitzpatrick, J.~Kaplan, and L.~Senatore, ``{The
  Effective Field Theory of Inflation},''
  \href{https://dx.doi.org/10.1088/1126-6708/2008/03/014}{{\em JHEP} {\bfseries
  03} (2008) 014}, \href{https://arxiv.org/abs/0709.0293}{{\ttfamily
  arXiv:0709.0293 [hep-th]}}.

\bibitem{Gubitosi:2012hu}
G.~Gubitosi, F.~Piazza, and F.~Vernizzi, ``{The Effective Field Theory of Dark
  Energy},'' \href{https://dx.doi.org/10.1088/1475-7516/2013/02/032}{{\em JCAP}
  {\bfseries 02} (2013) 032}, \href{https://arxiv.org/abs/1210.0201}{{\ttfamily
  arXiv:1210.0201 [hep-th]}}.

\bibitem{Sarbach:2012pr}
O.~Sarbach and M.~Tiglio, ``{Continuum and Discrete Initial-Boundary-Value
  Problems and Einstein's Field Equations},''
  \href{https://dx.doi.org/10.12942/lrr-2012-9}{{\em Living Rev. Rel.}
  {\bfseries 15} (2012) 9}, \href{https://arxiv.org/abs/1203.6443}{{\ttfamily
  arXiv:1203.6443 [gr-qc]}}.

\bibitem{wald_general_2009}
R.~M. Wald, {\em General relativity}.
\newblock Univ. of Chicago Press, Chicago, repr.~ed., 2009.

\bibitem{Kovacs:2020ywu}
A.~D. Kov\'acs and H.~S. Reall, ``{Well-posed formulation of Lovelock and
  Horndeski theories},''
  \href{https://dx.doi.org/10.1103/PhysRevD.101.124003}{{\em Phys. Rev. D}
  {\bfseries 101} no.~12, (2020) 124003},
  \href{https://arxiv.org/abs/2003.08398}{{\ttfamily arXiv:2003.08398
  [gr-qc]}}.

\bibitem{Takahashi:2019oxz}
K.~Takahashi, H.~Motohashi, and M.~Minamitsuji, ``{Linear stability analysis of
  hairy black holes in quadratic degenerate higher-order scalar-tensor
  theories: Odd-parity perturbations},''
  \href{https://dx.doi.org/10.1103/PhysRevD.100.024041}{{\em Phys. Rev. D}
  {\bfseries 100} no.~2, (2019) 024041},
  \href{https://arxiv.org/abs/1904.03554}{{\ttfamily arXiv:1904.03554
  [gr-qc]}}.

\bibitem{Kourkoulou:2022doz}
I.~Kourkoulou, A.~Nicolis, and K.~Parmentier, ``{Low-temperature thermal
  corrections to a superfluid's equation of state},''
  \href{https://arxiv.org/abs/2212.12555}{{\ttfamily arXiv:2212.12555
  [hep-th]}}.

\bibitem{Mukohyama:2013ew}
S.~Mukohyama and J.-P. Uzan, ``{From configuration to dynamics: Emergence of
  Lorentz signature in classical field theory},''
  \href{https://dx.doi.org/10.1103/PhysRevD.87.065020}{{\em Phys. Rev. D}
  {\bfseries 87} no.~6, (2013) 065020},
  \href{https://arxiv.org/abs/1301.1361}{{\ttfamily arXiv:1301.1361 [hep-th]}}.

\bibitem{Kehayias:2014uta}
J.~Kehayias, S.~Mukohyama, and J.-P. Uzan, ``{Emergent Lorentz Signature,
  Fermions, and the Standard Model},''
  \href{https://dx.doi.org/10.1103/PhysRevD.89.105017}{{\em Phys. Rev. D}
  {\bfseries 89} no.~10, (2014) 105017},
  \href{https://arxiv.org/abs/1403.0580}{{\ttfamily arXiv:1403.0580 [hep-th]}}.

\bibitem{Matrix_Analysis}
C.~R. Horn, Roger A.;~Johnson, {\em Matrix Analysis}.
\newblock Cambridge University Press, 1985.

\bibitem{Smilga:2013vba}
A.~V. Smilga, ``{Supersymmetric field theory with benign ghosts},''
  \href{https://dx.doi.org/10.1088/1751-8113/47/5/052001}{{\em J. Phys. A}
  {\bfseries 47} no.~5, (2014) 052001},
  \href{https://arxiv.org/abs/1306.6066}{{\ttfamily arXiv:1306.6066 [hep-th]}}.

\bibitem{Damour:2021fva}
T.~Damour and A.~Smilga, ``{Dynamical systems with benign ghosts},''
  \href{https://dx.doi.org/10.1103/PhysRevD.105.045018}{{\em Phys. Rev. D}
  {\bfseries 105} no.~4, (2022) 045018},
  \href{https://arxiv.org/abs/2110.11175}{{\ttfamily arXiv:2110.11175
  [hep-th]}}.

\bibitem{Gross:2020tph}
C.~Gross, A.~Strumia, D.~Teresi, and M.~Zirilli, ``{Is negative kinetic energy
  metastable?},'' \href{https://dx.doi.org/10.1103/PhysRevD.103.115025}{{\em
  Phys. Rev. D} {\bfseries 103} no.~11, (2021) 115025},
  \href{https://arxiv.org/abs/2007.05541}{{\ttfamily arXiv:2007.05541
  [hep-th]}}.

\bibitem{Deffayet:2025lnj}
C.~Deffayet, A.~Held, S.~Mukohyama, and A.~Vikman, ``{Ghostly interactions in
  (1+1) dimensional classical field theory},''
  \href{https://arxiv.org/abs/2504.11437}{{\ttfamily arXiv:2504.11437
  [hep-th]}}.

\bibitem{Rousseaux:2007is}
G.~Rousseaux, C.~Mathis, P.~Maissa, T.~G. Philbin, and U.~Leonhardt,
  ``{Observation of negative phase velocity waves in a water tank: A classical
  analogue to the Hawking effect?},''
  \href{https://dx.doi.org/10.1088/1367-2630/10/5/053015}{{\em New J. Phys.}
  {\bfseries 10} (2008) 053015},
  \href{https://arxiv.org/abs/0711.4767}{{\ttfamily arXiv:0711.4767 [gr-qc]}}.

\bibitem{Moore:2001bv}
G.~D. Moore and A.~E. Nelson, ``{Lower bound on the propagation speed of
  gravity from gravitational Cherenkov radiation},''
  \href{https://dx.doi.org/10.1088/1126-6708/2001/09/023}{{\em JHEP} {\bfseries
  09} (2001) 023}, \href{https://arxiv.org/abs/hep-ph/0106220}{{\ttfamily
  arXiv:hep-ph/0106220}}.

\bibitem{Elliott:2005va}
J.~W. Elliott, G.~D. Moore, and H.~Stoica, ``{Constraining the new Aether:
  Gravitational Cerenkov radiation},''
  \href{https://dx.doi.org/10.1088/1126-6708/2005/08/066}{{\em JHEP} {\bfseries
  08} (2005) 066}, \href{https://arxiv.org/abs/hep-ph/0505211}{{\ttfamily
  arXiv:hep-ph/0505211}}.

\bibitem{Babichev:2024uro}
E.~Babichev, ``{Cherenkov radiation as ghost instability},''
  \href{https://dx.doi.org/10.1103/bgwl-dwbd}{{\em Phys. Rev. D} {\bfseries
  112} no.~8, (2025) 084007},
  \href{https://arxiv.org/abs/2412.20093}{{\ttfamily arXiv:2412.20093
  [gr-qc]}}.

\bibitem{Hall1974}
G.~S. Hall, ``The classification of the ricci tensor in general relativity
  theory,'' \href{https://dx.doi.org/10.1088/0305-4470/9/4/010}{{\em J. Phys.
  A} {\bfseries 9} no.~4, (1976) 541}.

\bibitem{Hassan:2017ugh}
S.~F. Hassan and M.~Kocic, ``{On the local structure of spacetime in ghost-free
  bimetric theory and massive gravity},''
  \href{https://dx.doi.org/10.1007/JHEP05(2018)099}{{\em JHEP} {\bfseries 05}
  (2018) 099}, \href{https://arxiv.org/abs/1706.07806}{{\ttfamily
  arXiv:1706.07806 [hep-th]}}.

\bibitem{Gordon:1923qva}
W.~Gordon, ``{Zur Lichtfortpflanzung nach der Relativit\"atstheorie},''
  \href{https://dx.doi.org/10.1002/andp.19233772202}{{\em Annalen Phys.}
  {\bfseries 377} no.~22, (1923) 421--456}.

\bibitem{Pujolas:2011he}
O.~Pujolas, I.~Sawicki, and A.~Vikman, ``{The Imperfect Fluid behind Kinetic
  Gravity Braiding},'' \href{https://dx.doi.org/10.1007/JHEP11(2011)156}{{\em
  JHEP} {\bfseries 11} (2011) 156},
  \href{https://arxiv.org/abs/1103.5360}{{\ttfamily arXiv:1103.5360 [hep-th]}}.

\bibitem{ArmendarizPicon:2005nz}
C.~Armendariz-Picon and E.~A. Lim, ``{Haloes of k-essence},''
  \href{https://dx.doi.org/10.1088/1475-7516/2005/08/007}{{\em JCAP} {\bfseries
  0508} (2005) 007}, \href{https://arxiv.org/abs/astro-ph/0505207}{{\ttfamily
  arXiv:astro-ph/0505207 [astro-ph]}}.

\bibitem{Mendonca_BOOK}
J.~T. {Mendonça}, {\em {Theory of Photon Acceleration, Series in Plasma
  Physics}}.
\newblock Institute of Physics Publishing Bristol and Philadelphia, 2001.

\end{thebibliography}\endgroup

\clearpage
\appendix

\section{Summary tables} \label{app:tables}

\begin{table}[h]
\centering
\scriptsize

\renewcommand{\arraystretch}{1.5}

\setlength{\tabcolsep}{2pt}

\begin{tabular}{ lll }
\hline
 \textbf{Spacetime quantity} & \textbf{Name} & \textbf{Defined in}  \\
\hline
$\phi = \bar\phi + \pi$  &  field split into background and perturbation & \eqref{eq:eikonal} \\
$\mathcal{S}=\text{const}$ surfaces  &  fluctuation phase/wavefronts & \eqref{eq:eikonal} \\
$\mathcal{A}$ & fluctuation amplitude & \eqref{eq:eikonal}\\
$P_\mu\equiv\partial_\mu\mathcal{S}$ & momentum covector for a mode& \eqref{eq:Pdef} \\
$Z^{\mu\nu}$ & contravariant (inverse) acoustic metric & \eqref{eq:BigG-char} \\
$S_{\mu\nu} = (Z^{-1})_{\mu\nu}$ & covariant acoustic metric & \eqref{eq:Sdef}\\
$Z^{\mu\nu}P_\mu P_\nu=0$ & characteristic equation & \eqref{eq:ZPP}  \\
$ N^\mu =  Z^{\mu\nu} P_\nu$ & ray vector & \eqref{eq:NullVecsT} \\
$S_{\mu\nu}N^\mu N^\nu=0$ & acoustic null ray surface & \eqref{eq:ZPP}  \\
$ \omega = -u^\mu P_\mu $ & Minkowski energy of a phonon mode & \eqref{eq:PN-gdecomp} \\
$ \mho= - u_\mu N^\mu $ & Abraham energy of a phonon mode& \eqref{eq:PN-gdecomp} \\
$ k_\mu = h^\nu_\mu P_\nu $ & Minkowski spatial momentum & \eqref{eq:PN-gdecomp} \\
$ \dot{r}^\mu = h^\mu_\nu N^\nu$ & spatial ray/Abraham momentum & \eqref{eq:PN-gdecomp} \\
$\T^\mu_{\nu}=Z^{\mu\lambda}\partial_\lambda \pi \,\partial_\nu \pi - \frac{1}{2} \delta^{\mu}_{\nu} \,\, Z^{\alpha\beta}\partial_\alpha \pi \partial_\beta \pi$ & conserved, acoustic EMT for fluctuations & \eqref{eq:T_correct} \\
$\left<\T^\mu_\nu\right> = |\mathcal{A}|^2 N^\mu P_\nu$& average acoustic EMT in eikonal limit & \eqref{T=NP}\\
$\xi^\mu$ & acoustic Killing vector & \eqref{eq:Killing_V} \\
$\GZ_{\mu\nu}^{\alpha}=\frac{1}{2}Z^{\alpha\beta}\left(\partial_{\mu}S_{\beta\nu}+\partial_{\nu}S_{\mu\beta}-\partial_{\beta}S_{\mu\nu}\right)$ & connection coefficients of acoustic covariant derivative $\nZ_\mu$ & \eqref{eq:Chrstoffel} \\
$L_{\phantom{\alpha}\mu\nu}^{\alpha}=\GZ_{\mu\nu}^{\alpha}-\Gamma^{\alpha}_{\mu\nu}$ & acoustic disformation & \eqref{eq:disformation} \\

\hline

\end{tabular}
\caption{Summary of main notation and symbols introduced in this paper. The table contains definitions of recurring objects.}
\end{table}

\clearpage

\begin{table}
\centering
\scriptsize

\renewcommand{\arraystretch}{1.5}

\setlength{\tabcolsep}{1.3pt}

\begin{tabular}{ lll }
\hline
 \textbf{Observer-frame  quantity} & \textbf{Name} & \textbf{Defined in}  \\
\hline
$P_\mu = \omega u_\mu + k_\mu$ & $P_\mu$-decomposition in frame $u^\mu$ & \eqref{eq:PN-gdecomp}\\
$N^\mu= \mho u^\mu + \dot{r}^\mu$ & $N^\mu$-decomposition for observer $u^\mu$& \eqref{eq:PN-gdecomp}\\
$ v_\text{p}^\mu  = -\, h^\mu_\nu N^\nu / N^\alpha u_\alpha $ & phase velocitiy for observer $u^\mu$ & \eqref{AcPhaseVel}\\
$n_\mu  = -\, h^\nu_\mu P_\nu / u^\alpha P_\alpha $ & refractive index & \eqref{eq:n-def}\\
$v_\text{p}^\mu n_\mu =1$ & duality of $v_\text{p}^\mu$ and $n_\mu$ & \eqref{eq:n-def} \\
$Z^{uu}=Z^{\mu\nu}u_\mu u_\nu=-D$ & coefficient of kinetic term  & \eqref{eq:D-def} \\
$\mathcal{Z}_2^{\mu\nu}= D \DZ^{\mu\nu} = Z^{\mu u} Z^{\nu u} - Z^{uu} Z^{\mu\nu}$ & metric in space of refractive indices in the frame $u_\mu$ & \eqref{Z-inducedmetric}\\
$z^2=\left(\mathcal{Z}_2^{-1}\right)_{\mu\nu}Z^{u\mu}Z^{u\nu}$  & sonicity parameter & \eqref{eq:transonic}\\
$\mho_+(k_\mu)/\mho_-(k_\mu) $  & Abraham energy for future/past moving rays & \eqref{eq:mhopm-munu} \\
$ \omega_+(k_\mu)/\omega_-(k_\mu)$  & Minkowski energy for future/past moving modes  & \eqref{eq:dispersion-munu}\\
$S_{uu} = \frac{1}{Z^{uu}}\left( 1 - z^2 \right) = 0$  & transonic point for observer $u^\mu$  & \eqref{eq:S00_z}\\
$\mathcal{S}_{2\mu\nu} = -S_{uu}\DS_{\mu\nu} = S_{\mu u} S_{\nu u} - S_{uu} S_{\mu\nu}$  & metric in the space of phase velocities for observer $u^\mu$ & \eqref{eq:Stwomunu}\\
$ \mho_{[+]}(\dot{r}^\mu)/\mho_{[-]}(\dot{r}^\mu)$  & Abraham energy of $+$ve/$-$ve Minkowski energy modes  & \eqref{eq:mho_frame}\\
$\omega_{[+]}(\dot{r}^\mu)/\omega_{[-]}(\dot{r}^\mu)$  & positive/negative Minkowski energy for rays & \eqref{omegapm-munu}\\
$\mathfrak{n}=\sqrt{\Ztwo{\mu\nu}n_\mu n_\nu}$ & refractive index norm relating Minkowski and Abraham energies
 & \eqref{eq:square_nfrac} \\
$\mathfrak{v}_p= \sqrt{\Stwo{\mu\nu}v^\mu_{\text{p}} v^\nu_{\text{p}}}$ & phase velocity norm relating Minkowski and Abraham energies & \eqref{eq:MhoOmega_S} \\
\hline

\end{tabular}
\caption{Summary of main notation and symbols introduced in this paper. The table contains a decomposition of quantities with respect to a generic observer $u^\mu$.}
\end{table}

\clearpage

\begin{table}
\centering
\footnotesize
\begin{tabular}{llll}
\hline
  & \textbf{N-cone} &  \textbf{P-cone} & \textbf{pg.}  \\
\hline
  \textbf{hyperbolicity}  & $\det S^\mu_\nu >0 $  &    $\det Z^\mu_\nu >0$ & \pageref{eq:detZ-updown} \\
cone existence &  &     &  \\
\hline
  \textbf{gradient instability} & $\det S^\mu_\nu \leq 0 $ &    $\det Z^\mu_\nu \leq0 $ & \pageref{eq:detZ-updown} \\
   non-hyperbolicity &  &     &  \\
\hline
  acausality & \makecell[l]{either N-nappe overlaps with \\ future and past lightcone}  & \makecell[l]{P-cone does not overlap \\ with lightcone}  &  \pageref{fig:ConeAcausal} \\
\hline
  necessarily transonic & \makecell[l]{N-cone does not overlap \\ with lightcone}   & \makecell[l]{either P-nappe overlaps with \\ future and past lightcone}  & \pageref{fig:ConeAcausal} \\

\hline
 \textbf{ghost} = signature mismatch & $S_{\mu\nu}$ has signature (1,3) & $Z^{\mu\nu}$ has signature (1,3)  &  \pageref{def:ghosts} \\
    future for non-ghost & upper N-nappe &  upper P-nappe  &  \\

future for ghost & upper N-nappe &  lower P-nappe  &  \\
\hline
 subluminal phase speed   & $N^\mu$ $g$-timelike & $P_\mu$ $g$-spacelike  & \pageref{fig:isotropic_GoodCauchy}  \\
  superluminal phase speed   & $N^\mu$ $g$-spacelike & $P_\mu$ $g$-timelike &   \\
\hline

\end{tabular}
\caption{Summary of conditions valid for all frames. These instabilities are real and seen by all observers.}
\end{table}
\clearpage

\begin{table}
\scriptsize

\renewcommand{\arraystretch}{1.5}

\setlength{\tabcolsep}{2pt}

\begin{tabular}{llll}
\hline
  & \textbf{N-cone} &  \textbf{P-cone} & \textbf{pg.}  \\
\hline

\textbf{Cauchy surface} & future N-nappe above $\Sigma_t$  & $U_\mu$ $Z$-timelike & \pageref{fig:BadCauchy} \\
 &  outgoing $\mho_U>0$ all modes & $\Ztwobar{ij}\succ 0$, all $\omega_{U,\pm}$ real & \pageref{eq:GoodCauchyP} \\
 &  closed wavefronts  &  all modes $k^U_i$ covered &  \\
& $\mho_{U,+}$ ---   future N-nappe  &  $\omega_{U,+}$ --- future P-nappe & \pageref{eq:omegadifference}  \\
non-ghost &  $S_{ij}\succ 0$  & $Z^{00}<0$  & \pageref{Conditions_Well_IVP}   \\
ghost &  $S_{ij}\prec 0$  & $Z^{00}>0$  & \\
Hamiltonian mechanics well posed &  &  &  \\
\hline
\textbf{Not Cauchy surface} & upper N-nappe cuts $\Sigma_t$  & $U_\mu$ $Z$-spacelike & \pageref{fig:BadCauchy} \\
  &   exists mode with $\mho_U\leq0$  & $\Ztwobar{ij}\not\succ 0$, some $\omega_{U,\pm}$ complex & \pageref{eq:GoodCauchyP} \\
only for superluminality & non-closed wavefronts  & \makecell[l]{no modes $k^U_i$ with $\Ztwobar{ij} k^U_ik^U_j <0$}  &  \\
& both $\mho_{U,\pm}$ form future N-nappe   &  both $\omega_{U,\pm}$ form future P-nappe &  \pageref{eq:omegadifference}  \\
non-ghost &  $S_{ij}\not\succ 0$  & $Z^{00}>0$ & \pageref{Conditions_Well_IVP} \\
ghost &  $S_{ij}\not\prec 0$  & $Z^{00}<0$ &  \\
Hamiltonian mechanics ill posed &   \multicolumn{2}{l}{\emph{apparent} instability in one spatial and the time direction}  &  \\
\hline
  \textbf{No sound horizon}  & $V^\mu$ $S$-timelike & P-cone does not cut $\Sigma_t$   & \pageref{fig:SoundHorizon}\\
     &  $\Stwoubar{ij}\succ 0$, all $\mho_{V,[\pm]}$ real  & $\omega_V>0$ for non-ghosts ($<0$ for ghosts) &  \pageref{eq:SHcond}\\
   subsonic $V^\mu$ ($z_V^2<1$) &  propagation in all directions $\dot{r}_V^i$  & wave-vector surface elipsoidal  &  \\
     & $\mho_{V,[+]}$ ---   future non-ghost N-nappe  & $\omega_{V,[+]}$ --- future non-ghost P-nappe  &  \\
    non-ghost & $S_{00}<0$ & $Z^{ij}\succ 0$ & \pageref{eq:S00_z} \\
    ghost & $S_{00}>0$  & $Z^{ij}\prec 0$  &  \\
Hamiltonian bounded & & &  \pageref{text:boundedHam} \\
\hline
\textbf{Sound horizon}  & $V^\mu$ $S$-spacelike  & P-cone cuts $\Sigma_t$    & \pageref{fig:SoundHorizon}\\
     &  $\Stwoubar{ij}\not\succ 0$, some $\mho_{V,[\pm]}$ complex  &  $\omega_V$ not definite&  \pageref{eq:SHcond}\\
   supersonic $V^\mu$ ($z_V^2>1$) &  \makecell[l]{no propagation in directions \\ $\dot{r}_V^i$ with $\Stwoubar{ij} \dot{r}_V^i \dot{r}_V^j <0$}  & wave-vector surface hyperboloidal  &  \\
     & both $\mho_{V,[\pm]}$ form future N-nappe  & both $\omega_{V,[\pm]}$ form future P-nappe &  \\
    non-ghost & $S_{00}>0$  & $Z^{ij}\not\succ 0$  & \pageref{eq:S00_z} \\
    ghost & $S_{00}<0$  & $Z^{ij}\not\prec 0$  &  \\
    Hamiltonian unbounded & \multicolumn{2}{l}{\emph{apparent} instability in one spatial direction}  & \pageref{text:boundedHam}  \\

\hline

\end{tabular}
\caption{Summary of conditions for foliation with spatial slice $\Sigma_t$. The conditions on well-posedness are  constructed with the slice's normal frame observers (vector $U^\mu$) and need to be satisfied at every point. The conditions concerning the presence of sound horizons relate to the coordinates' comoving observers (associated to vector $V^\mu$) and need to be satisfied at every point on $\Sigma_t$ for Hamiltonian boundedness.}
\end{table}

\end{document}